\documentclass{report}
\usepackage{latexsym}
\usepackage{amssymb}
\usepackage{pdfpages}
\usepackage{graphics}
\begin{document}
\newcommand{\ja}{Jakubassa-Amundsen}

\newcommand{\bfx}{\mbox{\boldmath $x$}}
\newcommand{\bfq}{\mbox{\boldmath $q$}}
\newcommand{\bfnabla}{\mbox{\boldmath $\nabla$}}
\newcommand{\bfsigma}{\mbox{\boldmath $\sigma$}}
\newcommand{\bfSigma}{\mbox{\boldmath $\Sigma$}}
\newcommand{\bfsigmas}{\mbox{{\scriptsize \boldmath $\sigma$}}}
\newcommand{\bfzetas}{\mbox{{\scriptsize \boldmath $\zeta$}}}
\newcommand{\bfGamma}{\mbox{\boldmath $\Gamma$}}
\newcommand{\bfalpha}{\mbox{\boldmath $\alpha$}}
\newcommand{\bfeps}{\mbox{\boldmath $\epsilon$}}
\newcommand{\bfA}{\mbox{\boldmath $A$}}
\newcommand{\bfP}{\mbox{\boldmath $P$}}
\newcommand{\bfF}{\mbox{\boldmath $F$}}
\newcommand{\bfL}{\mbox{\boldmath $L$}}
\newcommand{\bfR}{\mbox{\boldmath $R$}}
\newcommand{\bfe}{\mbox{\boldmath $e$}}
\newcommand{\bfd}{\mbox{\boldmath $d$}}
\newcommand{\bfes}{\mbox{{\scriptsize \boldmath $e$}}}
\newcommand{\bfn}{\mbox{\boldmath $n$}}
\newcommand{\bfW}{{\mbox{\boldmath $W$}_{\!\!rad}}}
\newcommand{\bfM}{\mbox{\boldmath $M$}}
\newcommand{\bfK}{\mbox{\boldmath $K$}}
\newcommand{\bfI}{\mbox{\boldmath $I$}}
\newcommand{\bfJ}{\mbox{\boldmath $J$}}
\newcommand{\bfQ}{\mbox{\boldmath $Q$}}
\newcommand{\bfY}{\mbox{\boldmath $Y$}}
\newcommand{\bfp}{\mbox{\boldmath $p$}}
\newcommand{\bfk}{\mbox{\boldmath $k$}}
\newcommand{\bfks}{\mbox{{\scriptsize \boldmath $k$}}}
\newcommand{\bfps}{\mbox{{\scriptsize \boldmath $p$}}}
\newcommand{\bfqs}{\mbox{{\scriptsize \boldmath $q$}}}
\newcommand{\bfxs}{\mbox{{\scriptsize \boldmath $x$}}}
\newcommand{\bfLs}{\mbox{{\scriptsize \boldmath $L$}}}
\newcommand{\bfRs}{\mbox{{\scriptsize \boldmath $R$}}}
\newcommand{\bfs}{\mbox{\boldmath $s$}}
\newcommand{\bfv}{\mbox{\boldmath $v$}}
\newcommand{\bfw}{\mbox{\boldmath $w$}}
\newcommand{\bfb}{\mbox{\boldmath $b$}}
\newcommand{\bfxi}{\mbox{\boldmath $\xi$}}
\newcommand{\bfzeta}{\mbox{\boldmath $\zeta$}}
\newcommand{\bfr}{\mbox{\boldmath $r$}}
\newcommand{\bfrs}{\mbox{{\scriptsize \boldmath $r$}}}
\newcommand{\bfbs}{\mbox{{\scriptsize \boldmath $b$}}}

\newcommand{\re}{\mbox{ Re }}
\newcommand{\im}{\mbox{ Im }}

\newcommand{\kslash}{{k\!\!\!/}}
\newcommand{\pslash}{{p\!\!\!/}}
\newcommand{\Pslash}{{P\!\!\!\!/}}
\newcommand{\Eslash}{{E\!\!\!\!/}}

\renewcommand{\theequation}{\arabic{chapter}.\arabic{section}.\arabic{equation}}
\renewcommand{\thesection}{\arabic{chapter}.\arabic{section}}
\renewcommand{\thefigure}{\arabic{chapter}.\arabic{section}.\arabic{figure}}

{\huge\bf Advances in Bremsstrahlung: A Review}

\vspace{1cm}

\centerline{\large D.~H.~Jakubassa-Amundsen}

\vspace{0.5cm} 

\centerline{Mathematics Institute, University of Munich, Theresienstrasse 39, 80333 Munich, Germany}

\vspace{0.5cm}

{\bf Abstract}

Recent developments in bremsstrahlung from electrons colliding with atoms and nuclei at energies
between 0.1 MeV and 500 MeV are reviewed.
Considered are cross sections differential in the photon degrees of freedom, including coincidence geometries of photon and scattered electron.
Also spin asymmetries and polarization transfer for polarized electron beams
are investigated. An interpretation of the measurements in terms of the current bremsstrahlung theories is furnished.

\vspace{0.5cm} 

{\Large\bf Contents} 
\begin{enumerate}
\item[1.] Introduction \hfill p. 2
\item[2.] Theory for electron bremsstrahlung \hfill p. 2
\begin{enumerate}
\item[2.1]
Plane-wave Born approximation \hfill 2
\item[2.2] 
Sommerfeld-Maue approximation \hfill 9
\item[2.3]
Higher-order analytical theories \hfill 12
\item[2.4]
Relativistic partial-wave theory \hfill 17
\item[2.5]
The Dirac-Sommerfeld-Maue (DSM) model \hfill 25
\item[2.6]
Screening effects \hfill 32
\item[2.7]
Nuclear and QED effects \hfill 36
\end{enumerate}
\item[3.]
Polarization \hfill p. 37
\begin{enumerate}
\item[3.1]
Definition of the electron-photon polarization correlations \hfill 38
\item[3.2]
Triply differential cross section in coplanar geometry \hfill 41
\item[3.3]
Outlook into noncoplanar geometry \hfill 44
\item[3.4]
Sum rules for the polarization correlations \hfill 45
\item[3.5]
Correspondence to the spin asymmetries in elastic scattering \hfill 47
\end{enumerate}
\item[4.]
Positron bremsstrahlung \hfill p. 49
\begin{enumerate}
\item[4.1]
Positron theory \hfill 50
\item[4.2]
Results for positron versus electron impact \hfill 52
\end{enumerate}
\item[5.] 
Experiment in comparison with theory \hfill p. 54
\begin{enumerate}
\item[5.1]
Cross sections \hfill 54
\item[5.2]
Spin asymmetries \hfill 61
\end{enumerate}
\item[6.] 
Summary \hfill p. 70
\item[]
References \hfill 77
\end{enumerate}

{\Large\bf 1. Introduction}
\setcounter{chapter}{1}

\vspace{0.2cm}

The interaction of charged particles by means of electromagnetic potentials and their coupling to weak photon fields are basically well-understood processes.
Nevertheless, the electronic and atomic collision physics has kept its fascination all over the years.
In particular, the interplay between theory and experiment is crucial in this field.

The subject of this review, the radiation of photons by polarized or unpolarized electrons while being decelerated in the electromagnetic field of the collision partner, has attracted much interest since the middle of last century.
First experiments on doubly differential cross sections were performed in 1955 by Motz \cite{Mo55} and in 1956 by Starfelt and Koch \cite{SK56}.
An overview of the early experiments and theories is provided by Koch and Motz \cite{KM59}.

Theoretical work on bremsstrahlung from relativistic collisions started with the plane-wave Born approximation (PWBA)
introduced in 1934 by Bethe and Heitler \cite{BH34,H54}.
Progress for heavier projectiles was made with the introduction of the semirelativistic Sommerfeld-Maue wavefunctions by Bethe and Maximon in 1954 \cite{BM54}, and a generalization of this bremsstrahlung theory was provided by Elwert and Haug \cite{EH69}.
Higher-order (in $Z_T \alpha)$ approaches were considered in the following period \cite{RDP72}. These analytical theories are reviewed by Mangiarotti and Martins \cite{MM17}.
The modern state-of-the-art bremsstrahlung theory, the relativistic Dirac partial-wave theory which is based on exact solutions of the Dirac equation, was  in the early 1970' put forth by Tseng and Pratt \cite{TP71}. An optimization of this theory is provided by Yerokhin and Surzhykov \cite{YS10}.

Great experimental progress was made by Nakel and his group with investigating the elementary process of bremsstrahlung by means of a coincident detection of the bremsstrahlung photon and the scattered electron \cite{Na66}. The use of polarized electrons allowed to investigate the polarization transfer from the projectile to the photon.
The respective experiments up to 2003, including the theoretical approaches, are collected in the comprehensive book by Haug and Nakel \cite{HN04}.

The present review focuses on the experimental and theoretical high-energy developments during  the last twenty years, being designed as an update of  the survey by Haug and Nakel.
An overview of the current bremsstrahlung theories for the doubly and triply differential cross sections (section 2) and for the corresponding spin asymmetries in the case that electrons and photons are polarized (section 3) is provided.
Also positron projectiles are considered in order to exploit the difference to electron bremsstrahlung (section 4).
An extensive comparison with the available experimental data is provided in section 5, followed by a short summary (section 6).
All plots of differential cross sections refer to unpolarized particles.
Atomic units $(\hbar=m=e=1)$ are used throughout (unless indicated otherwise).

\vspace{1.0cm} 

{\Large\bf 2. Theory}

\setcounter{chapter}{2}
\setcounter{equation}{0}

\section{Plane-wave Born approximation (PWBA)}

A great advantage of the PWBA is its simplicity and its feasibility to give predictions at arbitrarily high collision energies. Moreover, it allows to account easily for perturbation effects, at least to first order.
Such perturbation effects involve the screening of the nuclear field by the atomic electrons or the modification of the pure Coulomb interaction by means of the finite charge distribution of the nucleus.
Furthermore, effects on the photon emission by the recoiling nucleus, such as energy loss and photon deflection (the so-called kinematical recoil) can be accounted for, as well as additional radiation emitted by the recoiling nucleus (the dynamical recoil). Also the influence of nuclear current densities in spinning nuclei at ultrahigh velocities can be assessed.

When formulating a bremsstrahlung theory it must be noted that photon emission by a free particle (which is described by a plane wave) is not possible, due to the requirement of simultaneous energy and momentum conservation,
\begin{equation}\label{2.1.1}
\bfp_i\;-\;\bfp_f\;-\;\bfk\;=\;\bfq\;\neq \;0,\qquad E_i\;-\;E_f\;-\;\omega\;=\;0,
\end{equation}
where $\bfp_i\;\bfp_f$ and $\bfk$ are the momenta of incoming electron, scattered electron and photon, respectively, and $E_i,\;E_f$ and $\omega=ck$ are the respective total energies.
Instead, it must always be allowed to transfer a certain momentum $\bfq$ to a second collision partner, which in the present case is the nucleus.
This means that bremssstrahlung can be interpreted as a second-order process which involves two electron couplings, one
to the photon field and one to the field of the target nucleus.
If the mass $M_T$ of the target nucleus is set to infinity, $
\bfq$ is simply absorbed, while a nucleus with a finite mass acquires a slight motion.
If the nucleus carries spin, this motion can in turn
lead to photon emission by the nucleus. The intensity of such photons is in general reduced (by the ratio $qZ_T/M_Tc$ \cite{Dr52}, where $Z_T$ is the nuclear charge number) as compared to the  photon intensity originating from the beam electrons of mass $m$.

Since electron bremsstrahlung in PWBA is  a second-order process, there occur two contributions to the transition amplitude: one, where the photon is emitted prior to the electron-nucleus scattering, the other when the photon is emitted after the interaction with the nucleus. The corresponding Feynman diagrams are shown in Fig.2.1.1.

\begin{figure}
\vspace{-1.5cm}
\includegraphics[width=11cm,angle=90]{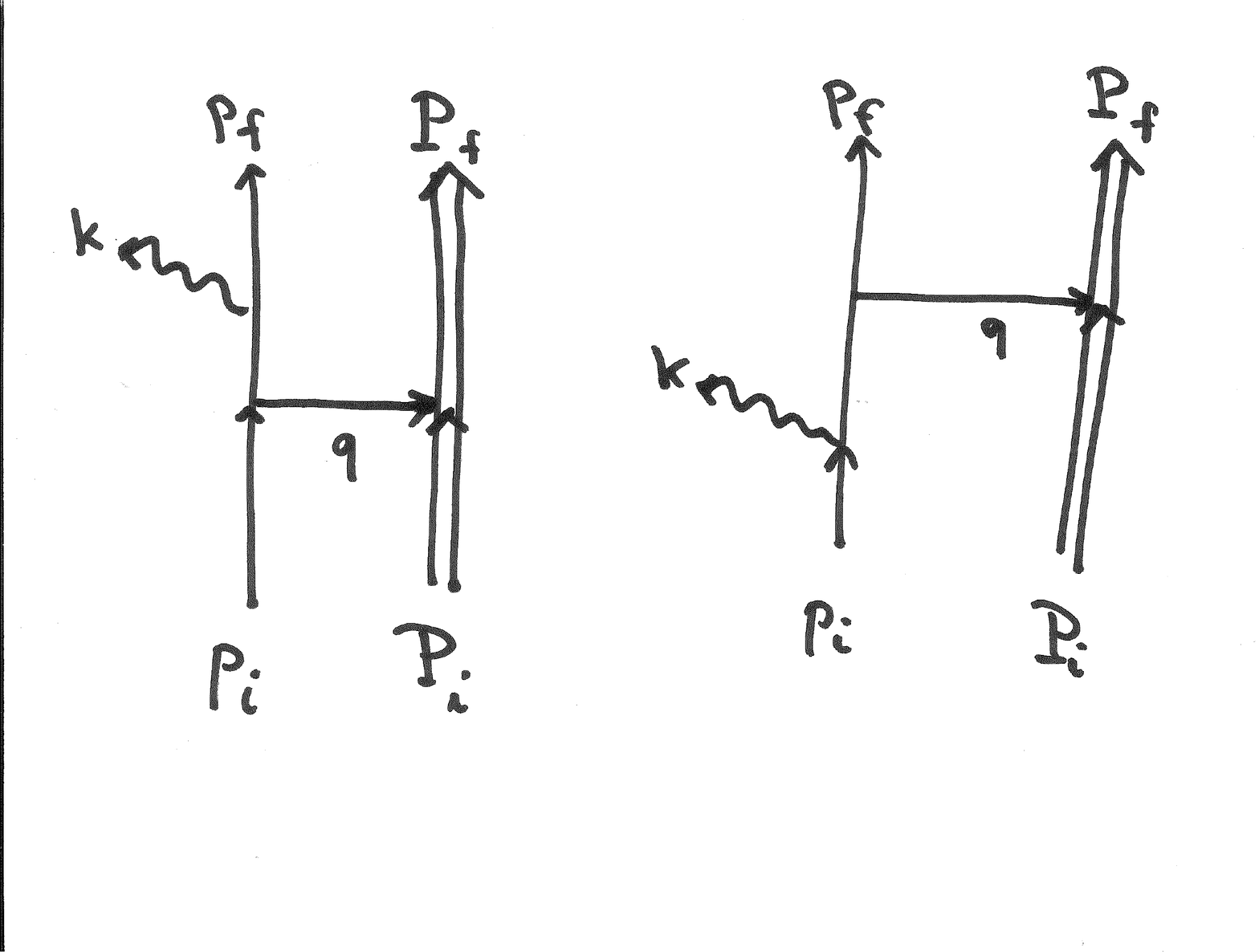}
\caption{Lowest-order Feynman diagrams for electron bremsstrahlung. The nucleus is shown by a double line.}
\end{figure}

This section  starts with the derivation of the Bethe-Heitler formula according to these Feynman diagrams. Subsequently, perturbative effects such as modifications of the nuclear potential or recoil effects will be discussed. Finally the influence of the nuclear current densities will be considered.

\subsection{The Bethe-Heitler formula}

To begin with, let us introduce 4-vectors for the electron, $\bar{p}_i=(E_i/c,\bfp_i),\;\bar{p}_f=(E_f/c,\bfp_f)$, as well as for the nucleus, 
$\bar{P}_i=(E_{nuc,i}/c,\bfP_i),\;\bar{P}_f=(E_{nuc,f}/c,\bfP_f)$, where the total initial and final energies of the nucleus are, respectively, given by
$E_{nuc,i}=M_Tc^2$  and $E_{nuc,f}=\sqrt{P_f^2c^2+M_T^2c^4}$. The photon 4-vector is given by $\bar{k}=(\omega/c, \bfk)$.
Since the nucleus is initially at rest ($\bfP_i=0$), one has $\bar{q}=\bar{P}_f-\bar{P}_i=(q_0,\bfq)$ with $\bfq=\bfP_f$ and $q_0=(E_{nuc,f}-E_{nuc,i})/c$.

For exact electronic scattering states $\psi_i$ and $\psi_f$ the relativistic radiation matrix element is to first order in the photon field given by
\begin{equation}\label{2.1.2}
W_{\rm rad}(\bfzeta_f,\bfzeta_i)\;=\;\int d\bfr\;\psi_f^+(\bfr,\bfzeta_f)\;(\bfalpha \bfe_\lambda^\ast)\;e^{-i\bfks\bfrs}\;\psi_i(\bfr,\bfzeta_i),
\end{equation}
where
\begin{equation}\label{2.1.3}
\bfalpha=(\alpha_1,\alpha_2,\alpha_3),\;\; \alpha_i\;=\;\left( \begin{array}{ll}
0&\sigma_i\\
\sigma_i &0
\end{array} \right), \quad \mbox{and}\quad \sigma_1=
\left( \begin{array}{ll}
0&1\\
1&0
\end{array} \right),\;\;\sigma_2=\left( \begin{array}{ll}
0&-i\\
i&0
\end{array} \right),\;\;\sigma_3=\left( \begin{array}{ll}
1&0\\
0&-1
\end{array} \right)
\end{equation}
are the three Pauli matrices. The polarization vector of the photon is denoted by $\bfe_\lambda$, while $\bfzeta_i$ and $\bfzeta_f$ are the respective spin polarization vectors  of the electron.

Rather than using the coordinate-space representation (\ref{2.1.2}) we adopt the representation by abstract state vectors, which will then allow us to switch to the simpler representation in momentum space.

Let us expand the  states $\psi_f$ and $\psi_i$ up to first order in the nuclear field $V_T$,
\begin{equation}\label{2.1.4}
\psi_i\;=\;\phi_{i0}+G_0V_T\phi_{i0}\;+\;\cdots,\qquad \psi_f\;=\;\phi_{f0}+G_0V_T\phi_{f_0}\;+\;\cdots,
\end{equation}
where $G_0$ is the free electron propagator, and $\phi_{i0}$ and $\phi_{f0}$ are plane waves.
The plane-wave Born approximation to the radiation matrix element is linear in $V_T$ and is given by  \cite{BH34,HN04}
$$W_{\rm rad}^{B1}(\bfzeta_f,\bfzeta_i) \;=\;\langle \phi_{f0}|\,\bfalpha \bfe_\lambda^\ast\;e^{-i\bfks\bfrs}\,G_0V_T|\,\phi_{i0}\rangle\;+\;\langle \phi_{f_0}|\,V_TG_0\,\bfalpha \bfe_\lambda^\ast\,e^{-i\bfks\bfrs}|\phi_{i,0} \rangle
$$
\begin{equation}\label{2.1.5}
=\;W_{\rm rad}^{B1}(1)\;+\;W_{\rm rad}^{B1}(2).
\end{equation}

From its derivation it follows that the PWBA is only valid if the electronic scattering states are not too different from plane waves.
This implies that the electron-nucleus potential $V_T$ should be weak enough such that the second-order contributions in $V_T$ are negligible.
As a measure of the validity of the PWBA serves the Sommerfeld parameter $\eta=Z_TE/(pc^2)$  which should obey $2\pi \eta \ll 1$. 
For relativistic energies, one has $E \approx pc$, such that $\eta \approx Z_T/c$ which requires target atoms
with low nuclear charge number. At low velocities, $p \ll c$, one has $E \approx c^2$, and $\eta \approx Z_T/p$ becomes very large.
As a consequence, the PWBA breaks down near the high-energy end of the bremsstrahlung spectrum,
leading to a vanishing cross section at $p_f=0$ in contrast to more elaborate theories.

On the other hand, the PWBA may nevertheless be applicable, irrespective of $Z_T$ and $E$, if the radiation is emitted at very large electron-nucleus distances.
This is the case for soft photons and small-angle scattering, which can be traced back to small momentum transfers $q$ attributed to such processes (see section 2.4.3). 

In  momentum-space representation, the perturbation of the initial or final plane wave  is easily calculated for a Coulomb potential, $V_T(r)\,=\,-Z_T/r.$
Identifying $\phi_{f0}=e^{i\bfps_f\bfrs}\;u_{p_f}/(2\pi)^{3/2}$ with $u_{p_f}$ a free 4-spinor, and introducing the Fourier transform $(G_0V_T\phi_{i0})(\bfq)$ of $G_0V_T\phi_{i0}$, one obtains 
for the first term in (\ref{2.1.5})  \cite{HN04}
$$W_{\rm rad}^{B1}(1)\;=\;u_{p_f}^+\;(\bfalpha \bfe_\lambda^\ast)\;(G_0V_T\phi_{i0})(\bfp_f+\bfk)$$
\begin{equation}\label{2.1.6}
=\;\frac{Z_T}{2\pi^2c^2}\;u_{p_f}^+\;(\bfalpha \bfe_\lambda^\ast )\;\frac{1}{q^2}\;\frac{\bfalpha c(\bfp_f+\bfk)+\beta c^2+E_i}{(\bfp_f+\bfk)^2-p_i^2-i\epsilon}\;u_{p_i},
\end{equation}
where $\beta=\left(\begin{array}{ll} 1&0\\
0&-1
\end{array}
\right)\;$ is a $4 \times 4$ Dirac matrix and $\epsilon=+0$.
Correspondingly,
\begin{equation}\label{2.1.7}
W_{\rm rad}^{B1}(2)\;=\;\frac{Z_T}{2\pi^2c^2}\;u_{p_f}^+\;\frac{1}{q^2}\;\frac{\bfalpha c(\bfp_i-\bfk)+\beta c^2+E_f}{(\bfp_i-\bfk)^2-p_f^2-i\epsilon}\;(\bfalpha \bfe_\lambda^\ast)\;u_{p_i}.
\end{equation}
Identifying the electron spin polarization vector $\bfzeta_i$ with a helicity eigenstate $(\pm)$, aligned with $\bfp_i,$ the two initial free-electron spinors  are given by \cite{BD}
\begin{equation}\label{2.1.8}
u_{p_i}^{(+)}\;=\;\sqrt{\frac{E_i+c^2}{2E_i}}\left(\begin{array}{l}
1\\
0\\
cp_i/(E_i+c^2)\\
0
\end{array}\right),\qquad u_{p_i}^{(-)}\;=\;\sqrt{\frac{E_i+c^2}{2E_i}}\;\left(\begin{array}{l} 0\\
1\\
0\\
-cp_i/(E_i+c^2)\end{array} \right).
\end{equation}
For the final free-electron spinor $u_{p_f}^{(+)}$, the third and forth components read $cp_{fz}/(E_f+c^2),\;cp_{f+}/(E_f+c^2)$
and for $u_{p_f}^{(-)}$ they are $cp_{f-}/(E_f+c^2),\;-cp_{fz}/(E_f+c^2)$,
where $p_{f\pm}=p_{fx}\pm i p_{fy}$, while in the prefactor, $E_i$ has to be replaced with $E_f$. 

The triply differential cross section for photon emission into the solid angle $d\Omega_k$ and electron scattering into the solid angle $d\Omega_f$ is given by
\begin{equation}\label{2.1.9}
\frac{d^3\sigma^{B1}}{d\omega d\Omega_k d\Omega_f}(\bfzeta_f,\bfzeta_i,\bfe_\lambda^\ast)\;=\;\frac{4\pi^2 \omega\,p_fE_iE_f}{c^5p_i}\;\left| W_{\rm rad}^{B_1}(1)+W_{\rm rad}^{B1}(2)\right|^2.
\end{equation}
The Bethe-Heitler formula for unpolarized particles is obtained by averaging (\ref{2.1.9}) over the two projections of the initial electron spin and by summing
over the photon polarization directions and the final electron spin projections.
We define the auxiliary quantities,
$$d_i\;=\;E_i/c\,-\,p_i\cos \theta_k,\qquad d_f\;=\;E_f/c\,-\,p_f\cos \theta_f,$$
\begin{equation}\label{2.1.10}
\delta_i\;=\;p_i\sin \theta_k,\qquad \delta_f\;=\;p_f\sin \theta_f,
\end{equation}
where $\theta_k$ is the angle between $\bfp_i$ and $\bfk$, $\theta_f$ is the angle between $\bfp_f$ and $\bfk$, and $\varphi$
is the azimuthal angle of $\bfp_f$ with respect to $\bfp_i$.
When taking the $z$-axis along $\bfp_i$, as done later on, the scattering angle $\vartheta_f$ (between $\bfp_f$ and $\bfp_i$) will be introduced instead of $\theta_f$.
The Bethe-Heitler formula reads \cite{BH34,Lan4}
$$\left( \frac{d^3\sigma^{B1}}{d\omega d\Omega_k d\Omega_f}\right)_0\;=\;\frac{1}{4\pi^2}\;\frac{Z_T^2\,p_f}{c^3\omega \,p_i}\;\frac{1}{q^4}\left\{
\frac{\delta_f^2}{d_f^2}\;[4(E_i/c)^2-q^2]\;+\;
\frac{\delta_i^2}{d_i^2}\;[4(E_f/c)^2-q^2]\right.$$
\begin{equation}\label{2.1.11}
\left.-\;2\,\frac{\delta_i\delta_f\cos \varphi}{d_id_f}\;[2(E_i/c)^2\,+\,2(E_f/c)^2-q^2]\;+\;2k^2\;\frac{\delta_i^2+\delta_f^2}{d_id_f} \right\},
\end{equation}
where the subscript 0 indicates that all particles are unpolarized.

\subsection{Consideration of modified potentials}

Potential modifications resulting from static screening by the atomic electrons (for collision energies below about 3 MeV or for extreme forward photon angles)
are easily incorporated into the PWBA cross section.
Since the momentum-space representation of the potential enters linearly into the radiation matrix element,
one simply has to make the substitution,
$$V_T(q)\;=\;-\;\sqrt{\frac{2}{\pi}}\;\frac{Z_T}{q^2}\;\mapsto\;\frac{1}{(2\pi)^{3/2}}\int d\bfr \;V_T(\bfr)\;e^{i\bfqs\bfrs}$$
\begin{equation}\label{2.1.12}
=\;\sqrt{\frac{2}{\pi}}\;\int_0^\infty r^2dr\;V_T(r)\;\frac{\sin(qr)}{qr},
\end{equation}
where $q=|\bfp_i-\bfk-\bfp_f|$. The second line of (\ref{2.1.12}) holds for  spherical potentials only.

When $V_T$ is generated from a charge disribution $\varrho$, such that
\begin{equation}\label{2.1.13}
V_T(\bfr)\;=\;\int d\bfr'\;\frac{\varrho(\bfr')}{|\bfr-\bfr'|},
\end{equation}
one can express its Fourier transform by means of
\begin{equation}\label{2.1.14}
V_T(\bfq)\;=\;\frac{1}{(2\pi)^{3/2}}\;\int d\bfr' \,\varrho(\bfr')\;\int d\bfr\;e^{i\bfqs \bfrs}\;\frac{1}{|\bfr-\bfr'|}\;=\;\sqrt{\frac{2}{\pi}}\;\frac{1}{q^2}\;\int d\bfr'\;\varrho(\bfr')\;e^{i\bfqs \bfrs'}.
\end{equation}
For a target atom, $\varrho$ is the sum of its nuclear and electronic charge distributions.
For a target nucleus, the integral over its charge distribution $\varrho_N$ defines
 the charge form factor $F_c(\bfq)$, 
\begin{equation}\label{2.1.15}
F_c(\bfq)\;=\;\frac{1}{Z_T}\int d\bfr\;\varrho_N(\bfr)\;e^{i\bfqs\bfrs},
\end{equation}
with the property $F_c(0)=1$.
Hence, in PWBA, the radiation matrix element (\ref{2.1.5}) for a pointlike nucleus is simply multiplied by the form factor $F_c(\bfq)$ (which depends only on the modulus $q$ for spherically symmetric
charge distributions). 

\subsection{Recoil effects}

Let us first investigate the kinematical recoil which results from consideration of the finite nuclear mass $M_T$ (which in atomic units is 1836$\,A$ with $A$ the mass number).
Then the energy gain $q_0c$ of the nucleus is no longer set to zero, and the momentum transfer $\bfq$ (occurring, for example, in the denominators of (\ref{2.1.6}) and (\ref{2.1.7})) 
has to be replaced by the 4-momentum $\bar{q}$, such that $q^2 \mapsto -\bar{q}^2 = q^2-q_0^2$  \cite{BL58}.
A further consequence of the finite residual motion of the nucleus is a slight decrease of the energy $\omega$ of the radiated photon.
This results from the conservation of the 4-momentum,
\begin{equation}\label{2.1.16}
\bar{P}_f\;=\;\bar{p}_i\,+\,\bar{P}_i\;-\;\bar{p}_f\;-\;\bar{k}.
\end{equation}
Upon squaring $\bar{P}_f$ and using that $\bar{P}_f^2=E_{\rm nuc,f}^2/c^2-P_f^2=M_T^2c^2=\bar{P}_i^2$
and $\bar{k}^2=\omega^2/c^2-k^2=0$, one obtains
\begin{equation}\label{2.1.17}
k\;=\;\frac{c^2+M_T(E_i-E_f)-E_iE_f/c^2+\bfp_i\bfp_f}{(E_i-E_f)/c+M_Tc-\bfp_i\hat{\bfk}+\bfp_f\hat{\bfk}},
\end{equation}
where $\hat{\bfk}=\bfk/k$, and which depends on the angles $\vartheta_f, \;\theta_f$ and $\theta_k$ defined below (\ref{2.1.10}).
In particular, for $p_f=0$, the maximum radiated energy is now
\begin{equation}\label{2.1.18}
\omega_{\rm max}\;=\;\frac{(M_T-1)c(E_i-c^2)}{E_i/c+(M_T-1)c-p_i\cos \theta_k},
\end{equation}
which, for low $M_T$ and large $\theta_k$, can be considerably smaller than $E_i-c^2$.
For $M_T \to \infty$, (\ref{2.1.17}) reduces to $k=(E_i-E_f)/c$, in agreement with (\ref{2.1.1}).

Recoil induces also a slight modification of the prefactor of the triply differential cross section.
For a given $\omega$, the dependence on the final energy $E_f$ is trivial, due to the energy-conserving $\delta$-function.
For finite nuclear mass, the integration over $p_f$ involves an integral of the type
\begin{equation}\label{2.1.19}
I\;\equiv\; \int_0^\infty p_f^2dp_f\;\delta(E_i+E_{\rm nuc,i}-E_f-E_{\rm nuc,f}-\omega)\;=\;\frac{p_f^2}{\frac{d}{dp_f}(E_f+E_{\rm nuc,f})}.
\end{equation}
Since $E_{\rm nuc,f}$ depends on $\bfq$ and hence on $p_f$, the rhs of (\ref{2.1.19}) acquires a recoil factor $f_{\rm re}$, such that
\begin{equation}\label{2.1.20}
I\;=\;\frac{p_fE_f}{c^2}\;\frac{1}{f_{\rm re}},\qquad f_{\rm re}\;=\;1\,-\,\frac{\hat{\bfp}_f \bfq \,E_f}{p_fE_{\rm nuc,f}}.
\end{equation}
Usually $f_{\rm re}$ is  close to unity.

Now we turn to the dynamical recoil, present for nuclei carrying spin, which plays some role for high collision energies and large photon angles.
In PWBA it again consists of two contributions, one where the nucleus emits a a photon prior to the interaction with the beam electron, 
the other where photoemission takes place after the electron-nucleus scattering.
The Feynman diagrams for these processes are displayed in Fig.2.1.2.

\begin{figure}
\vspace{-1.5cm}
\includegraphics[width=11cm,angle=90]{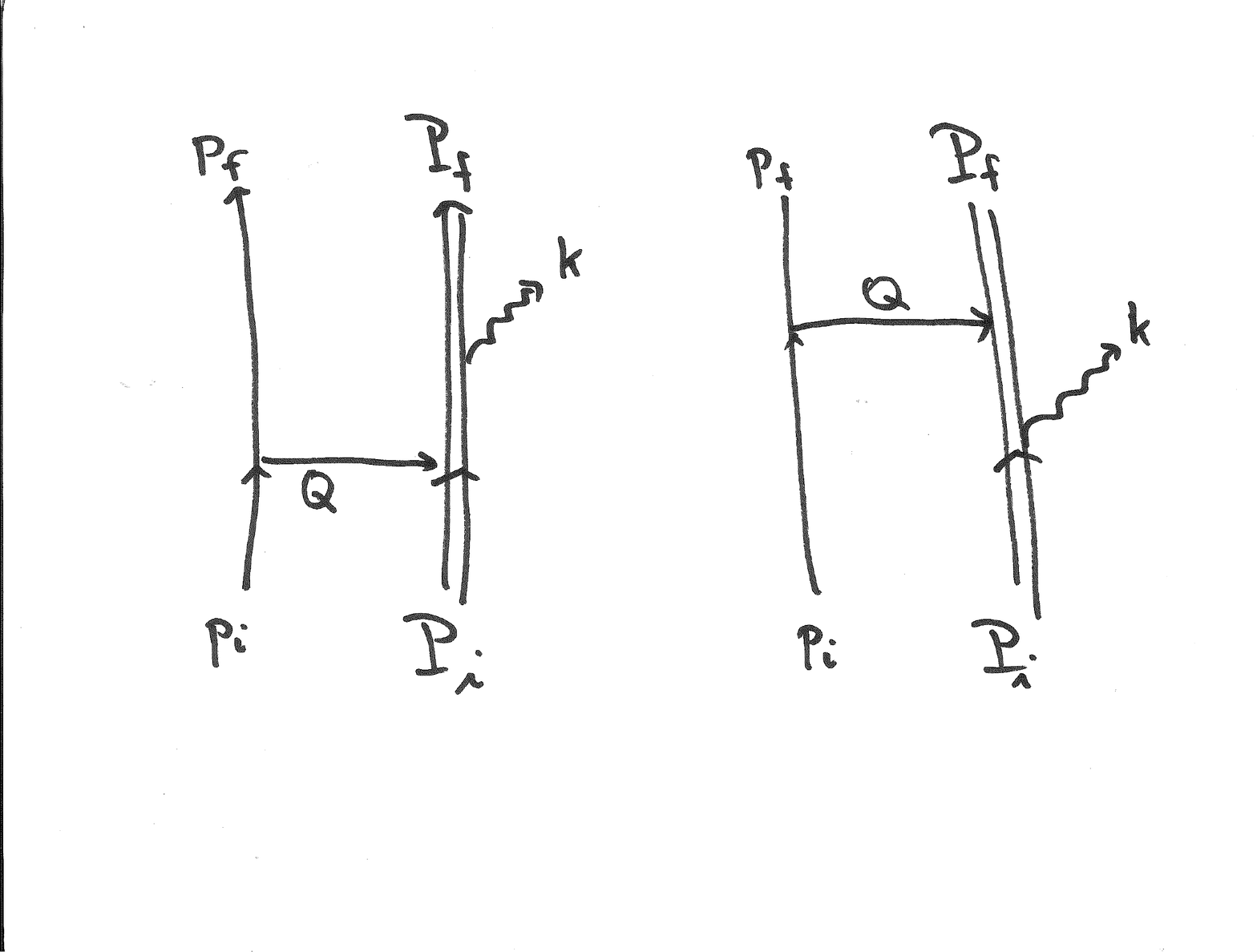}
\caption{Lowest-order Feynman diagrams for nuclear bremsstrahlung (also called virtual Compton scattering).}
\end{figure}

For this process the momentum transfer is $\bar{Q}=\bar{p}_i-\bar{p}_f.$
The photon emission by the nucleus involves the coupling constant $+Z_Te$ (as compared to $-e$ for photon emission by an electron).
 In correspondence to (\ref{2.1.6}) and (\ref{2.1.7}), the matrix element for nuclear bremsstrahlung from potential scattering is therefore given by  \cite{BL58}
$$W_{\rm nuc}^{B1}(s_f,s_i)\;=\;\frac{Z_T}{2\pi^2c^2}\;\frac{(-Z_T)}{\bar{Q}^2}\;\left\{U_{P_f}^{(s_f)+}(\bfalpha \bfe_\lambda^\ast)\;
\frac{\bfalpha c(\bfp_i-\bfp_f)+\beta M_Tc^2+(E_i-E_f+M_Tc^2)}{(\bfp_i -\bfp_f)^2-[(E_i-E_f+M_Tc^2)^2/c^2-M_T^2c^2]-i\epsilon} \;U_{P_i}^{(s_i)}\right.$$
\begin{equation}\label{2.1.21}
\left. +\;U_{P_f}^{(s_f)+}\;\frac{-\bfalpha c\bfk+(\beta +1) M_Tc^2-\omega}{k^2-[(M_Tc^2-\omega)^2/c^2-M_T^2c^2] -i\epsilon} (\bfalpha \bfe_\lambda^\ast)\;U_{P_i}^{(s_i)}\right\},
\end{equation}
where $U_{P_i}^{(s_i)}$ and $U_{P_f}^{(s_f)}$ are, respectively, the initial and final nuclear states with spin projections $s_i$ and $s_f$.

This matrix element has to be considered in addition to the radiation matrix element from electron bremsstrahlung in order to provide the corresponding cross section (see below).
 An experimental investigation of the $(e,e'\gamma)$ reaction on C  at collision energies in the GeV region
and angles at $6^\circ$ \cite{Si69}
gives no conclusive evidence for the presence of nuclear bremsstrahlung. For that collision  geometry, its effect is predicted to be at most 3 \% of the Bethe-Heitler cross section.

\subsection{Magnetic effects for nuclei with spin}

In this subsection we provide the complete PWBA formula for bremsstrahlung,
including nuclear structure effects due to finite nuclear size and magnetic moment, as well as recoil effects.

For nuclei carrying spin, the electron-nucleus interaction consists not only of the charge interaction (mediated by the potential $V_T$),
but also of the current interaction between the collision partners, which depends on the magnetic structure of the nucleus  \cite{GP64,HR66}
We will restrict ourselves to the consideration of the simplest case,  the spin-$\frac12$ nuclei.
This implies the replacement of the scalar form factor $F_c({q})$ with the 4-vector-valued form factor $\gamma_0 \Gamma_\nu(\bar{q}),\; \nu=$ 0-3,
which is defined by  \cite{BD}
\begin{equation}\label{2.1.22}
\gamma_0 \Gamma_0(\bar{q})\;=\;F_1(\bar{q}) 
 \;+\;\frac{\kappa}{2M_Pc\,Z_T}\;F_2(\bar{q})
\;\gamma_0 \,(\bfq \bfalpha)
\end{equation}
and
\begin{equation}\label{2.1.23}
\gamma_0\Gamma_\nu(\bar{q})\;=\;-\bfalpha \;F_1(\bfq)\;+\;\frac{\kappa}{2M_PcZ_T}\;F_2(\bar{q})\;\gamma_0\;\left[
-\bfalpha q_0\;+\;i\;(\bfq \times \bfSigma)\right],\qquad \nu=1,2,3,
\end{equation}
where $\bfSigma= {\bfsigma\;0 \choose 0\;\bfsigma}$
with $\bfsigma$ the vector of Pauli matrices (defined in (\ref{2.1.3})),
and $\gamma_0=\beta = {1\;\;\;0 \choose 0\;-1}$.
The Dirac form factor $F_1(\bar{q})$ is approximately equal to $F_c({q})$.
If the nucleus carries spin and has an anomalous magnetic moment $\kappa$, an extra term has to be added, which is governed by a second form factor,
the Pauli form factor $F_2(\bar{q}$).
In the simplest case of a hydrogen target ($Z_T=1$), one has $F_2(\bar{q}) \approx F_1(\bar{q})$  \cite{Ho58}.
$\kappa$ is measured in units of the Bohr magneton, $e\hbar/(2 M_Pc)$, where
$M_P$ is the proton mass.

The explicit treatment of the nucleus as a collision partner requires the introduction of its wavefunctions into the radiation amplitude $W_{\rm rad}^{B1}$.
In the special case of spin-$\frac12$ nuclei, they can, like for electrons, be represented in terms of free 4-spinors.
Hence $U_{P_i}^{(s_i)}$ and $U_{P_f}^{(s_f)}$ are of the form given below (\ref{2.1.8}), where $E_f$ is replaced by $E_{{\rm nuc,i,f}}$ and $c^2$ by $M_Tc^2$.

Also the dynamical recoil term is modified by the nuclear structure effects.
Photon emission is now induced by $\gamma_0 \Eslash$ instead of $\bfalpha \bfe_\lambda^\ast$, where
\begin{equation}\label{2.1.24}
\Eslash \;=\;-\bfGamma(\bar{K})\,\bfe_\lambda^\ast,\qquad \bar{K}\;=\;(-\omega/c,\bfk),
\end{equation}
where $\bfGamma(\bar{K})$ is defined in (\ref{2.1.23}) for $\nu=1,2,3$, using that $F_1(\bar{K}) = F_2(\bar{K})=1$ (since $\bar{K}^2= 0$ and $F_1(0)=F_2(0)=1$).

Collecting results, the total bremsstrahlung transition amplitude reads  \cite{BL58}, see also \cite{Jaku13}),
\begin{equation}\label{2.1.25}
W_{\rm tot}^{B1}(\bfzeta_f,\bfzeta_i,s_f,s_i)\;=\;W_{\rm rad}^{B1}(\bfzeta_f,\bfzeta_i,s_f,s_i)\;+\;W_{\rm nuc}^{B1}(\bfzeta_f,\bfzeta_i,s_f,s_i)
\end{equation}
with
$$W_{\rm rad}^{B1}(\bfzeta_f,\bfzeta_i,s_f,s_i)\;=\;\frac{Z_T}{2\pi^2c^2}\;\frac{1}{\bar{q}^2}  
\sum_{\nu=0}^3 \left( u_{p_f}^+\left[ (\bfalpha \bfe_\lambda^\ast)\;\frac{1}{\pslash_f+\kslash-mc}\;\gamma^\nu \right.\right.$$
\begin{equation}\label{2.1.26}
\left. \left. +\;\gamma_0 \gamma^\nu\;\frac{1}{\pslash_i-\kslash -mc}\;\gamma_0(\bfalpha \bfe_\lambda^\ast) \right] u_{p_i}\right)\left( U_{P_f}^{(s_f)+}\; \gamma_0 \Gamma_\nu(\bar{q})\;U_{P_i}^{(s_i)}\right),
\end{equation}
where $\gamma^0=\beta$,  $\gamma^\nu=\beta\alpha_\nu,\;\,\nu=1,2,3$, and
$$W_{\rm nuc}^{B1}(\bfzeta_f,\bfzeta_i, s_f,s_i)\;=\;-\frac{Z_T^2}{2\pi^2c^2}\;\frac{1}{\bar{Q}^2}
\sum_{\nu=0}^3 \left( u_{p_f}^+\;\gamma_0 \gamma^\nu\;u_{p_i}\right)$$
\begin{equation}\label{2.1.27}
\times\; \left( U_{P_f}^{(s_f)+}\left[ \gamma_0 \Eslash \;\frac{1}{\Pslash_f+\kslash -M_Tc}\;\Gamma_\nu(\bar{Q})\;+\;\gamma_0 \Gamma_\nu(\bar{Q})\;\frac{1}{\Pslash_i-\kslash-M_Tc}\;\Eslash \right] U_{P_i}^{(s_i)}\right).
\end{equation}
Thereby the following abbreviation is used,
\begin{equation}\label{2.1.28}
\frac{1}{\pslash \pm \kslash -mc}\;=\;-\frac{\bfalpha c (\bfp \pm \bfk)+\beta mc^2 +(E_p\pm \omega)}{(\bfp\pm \bfk)^2\,-\,\left( (E_p\pm \omega)^2/c^2\,-\,m^2c^2\right)}\;\gamma_0,
\end{equation}
with $E_p=\sqrt{p^2c^2+m^2c^4}$.

The complete PWBA cross section for unpolarized nuclei results in
\begin{equation}\label{2.1.29}
\frac{d^3\sigma^{B1}}{d\omega d\Omega_k d\Omega_f}(\bfzeta_f,\bfzeta_i,\bfe_\lambda^\ast)\;=\;\frac{4\pi^2\omega\,p_fE_iE_f}{c^5\,p_i}\;\frac{1}{f_{\rm re}}\;\frac12 \sum_{s_i,s_f}\left| W_{\rm tot}^{B1}(\bfzeta_f,\bfzeta_i,s_f,s_i)\right|^2.
\end{equation}

For spinless nuclei, $W_{\rm nuc}^{B1}$ is absent and only $\nu=0$ occurs in (\ref{2.1.26}),
provided the nucleus remains in  its ground state during the collision. 
In that case  the Bethe-Heitler amplitudes (\ref{2.1.6}) and (\ref{2.1.7}) are readily retrieved, apart from a multiplication with  the form factor $F_1(q)$.
In fact, since $\kappa=0$ for spinless nuclei, we have
$\gamma_0\Gamma_0(\bar{q})=F_1(\bar{q})$, such that the nuclear matrix element in (\ref{2.1.26})
reduces to a simple product of nuclear wavefunctions, multiplied by $F_1(\bar{q}).$
When forming the square of $W_{\rm rad}^{B1}$ and summing over the nuclear spin projections according to (\ref{2.1.29}) the following property of the free 4-spinors, valid in the recoil-free case where $\bfP_f=\bfP_i={\bf 0}$, can be used,
\begin{equation}\label{2.1.30}
\frac12 \sum_{s_i,s_f} \left( U_{P_f}^{(s_f)+}\,F_1(\bar{q})\,U_{P_i}^{(s_i)}\right)\left( U_{P_i}^{(s_i)+}\,F_1^\ast(\bar{q})\,U_{P_f}^{(s_f)}\right)\;=\;|F_1(\bar{q})|^2.
\end{equation}
The cross section is therefore equal to the one given by (\ref{2.1.9}) multiplied by $|F_1(\bar{q})|^2$ (see section 2.1.2).
However, at extremely high energies, intermediate {\it excited} nuclear states may be populated in the electron-nucleus encounter. If such states carry spin, then nuclear bremsstrahlung will nevertheless exist. This is studied  by Hubbard and Rose \cite{HR66} 
for an $^{16}$O target at a collision energy of 51 MeV and angles up to $40^\circ$.

In the general case of nuclei with arbitrary spin ($J_i>\frac12$), the nuclear matrix element can still be described
in terms of two form factors, a longitudinal one (corrresponding to $F_1(\bar{q})$)
and a transversal one (generating $F_2(\bar{q})$).
These form factors can for example be calculated from microscopic nuclear models  \cite{DS84}.

\setcounter{equation}{0}

\section{Sommerfeld-Maue approximation}

The analytical Sommerfeld-Maue (SM) approximation to the solution of the Dirac equation for a Coulomb potential, $V(r)=-Z_T/r$, was discovered by Furry \cite{Fu34} and elaborated by Sommerfeld and Maue \cite{SM35}.
This was done by transforming the Dirac equation to the total energy $E$,
\begin{equation}\label{2.2.1}
(-ic\bfalpha\bfnabla\;+\;\beta c^2\;+\;V(r)\;-\;E)\;\psi(\bfr)\;=\;0,
\end{equation}
where $\bfalpha$ and $\beta$ denote Dirac matrices,
via multiplication by $(-ic\bfalpha\bfnabla +\beta c^2-V(r)+E)$
into a second-order differential equation,
\begin{equation}\label{2.2.2}
(c^2 \bfnabla^2\;+\;p^2c^2\;-\;2EV(r))\;\psi(\bfr)\;=\;(-ic\bfalpha\;(\bfnabla V(r))\;-\;V^2(r))\;\psi(\bfr),
\end{equation}
where $p^2c^2=E^2-c^4$. Its solution 
$\psi(\bfr)$  is expanded in increasing powers of $Z_T/c$,
\begin{equation}\label{2.2.3}
\psi(\bfr)\;=\;\psi_a(\bfr)\;+\;\psi_b(\bfr)\;+\;\psi_c(\bfr).
\end{equation}
The lowest-order term, $\psi_a$, is defined by the lhs of (\ref{2.2.2}) (with the rhs set to zero), since this function is known analytically from the nonrelativistic
theory \cite{Lan4}. The function $\psi_b$ includes the first-order term of the rhs by means of setting $\psi=\psi_a$  in the rhs of (\ref{2.2.2}),
\begin{equation}\label{2.2.4}
\left(c^2\bfnabla^2 + p^2c^2 \,-\,2EV(r)\right)\;\psi_b(r)\;=\;-ic\bfalpha \;(\bfnabla V(r))\;\psi_a(r),
\end{equation}
 while the term quadratic in $V$ is neglected.
It can also be represented in closed form, in contrast to the remainder $\psi_c$.

For a Coulomb field, $V(r)=-Z_T/r$, the sum of $\psi_a$ and $\psi_b$ defines the Sommerfeld-Maue wavefunction,
$\psi^{SM}=\psi_a+\psi_b$. For the impinging electron, this function reads  \cite{BM54,HN04},  see also \cite{Lan4})
$$\psi_i^{SM}(\bfr)\;=\;N_i^{SM}\;e^{i\bfps_i \bfrs}\;\left( 1\;-\;\frac{ic}{2E_i}\;\bfalpha \bfnabla\right)\;_1F_1(i\eta_i,1,i(p_ir -\bfp_i\bfr))\;u_{p_i},$$
\begin{equation}\label{2.2.5}
N^{SM}\;=\;e^{\pi \eta/2}\;\Gamma(1-i\eta)/(2\pi)^{3/2},
\end{equation}
where $\psi_a$ corresponds to the first term in (\ref{2.2.5}) and $\psi_b$ to the gradient term.
$_1F_1(a,b,z)$ is a confluent hypergeometric function \cite{AS64}, $\eta=Z_TE/(pc^2)$ is the Sommerfeld parameter, $\Gamma(z)$ is the Gamma function and $u_p$ is a free 4-spinor (see, e.g. (\ref{2.1.8})).
The function $\psi^{SM}$ is accurate to first order in $Z_T/c$ and hence approximates the exact Dirac solution the better, the smaller $Z_T$. In the limit of $Z_T \to 0$, $\psi_i^{SM}$  turns into the free solution $e^{i\bfps_i\bfrs} u_{p_i}/(2\pi)^{3/2}$. 
For further use, we also give the SM function for the (adjoint of the) scattered electron,
\begin{equation}\label{2.2.6}
\psi_f^{SM+}(\bfr)\;=\;N_f^{SM}\;e^{-i\bfps_f\bfrs}\;\left( 1\;+\;\frac{ic}{2E_f}\;\bfalpha\bfnabla\right)\;_1F_1(i\eta_f,1,i(p_fr + \bfp_f\bfr))\;u_{p_f}^+.
\end{equation}

Apart from the $Z_T \to 0$ limit, there are other situations where $\psi^{SM}$ approaches the exact solution of the Dirac equation for the Coulomb field.
They are accessible by inserting $\psi^{SM}$ into the Dirac equation and subsequently estimating the remainder $R^{SM}$  \cite{HN04},
$$|R^{SM}|\;=\;\left|N^{SM}\; \frac{Z_T^2}{2cr}\;_1F_1(1+i\eta,2,i(pr-\bfp \bfr))\;(\bfalpha (\hat{\bfr}-\hat{\bfp}))\;u_p\right|
$$
\begin{equation}\label{2.2.7}
\leq\; C^{SM}\;\sqrt{\frac{\eta}{1-e^{-2\pi \eta}}}\;\;\frac{Z_T^2}{cr}\;\left| \sin \frac{\theta}{2}\right|,
\end{equation}
where $C^{SM}$ is some constant. This follows from the fact that the confluent hypergeometric function is bounded since its third entry is purely imaginary.
The $Z_T^2$-dependence of $R^{SM}$ verifies that $\psi^{SM}$ is exact to first order in $Z_T/c$.
A consequence of (\ref{2.2.7}) is that $\psi^{SM}$ becomes also  exact when the angle $\theta$ between $\hat{\bfp}$ and $\hat{\bfr}$ tends to zero, or when $r \to \infty$, provided $p\neq 0$.
Thus the conditions for the applicability of $\psi^{SM}$ are
\begin{equation}\label{2.2.8}
\theta\;\ll \;1\qquad \mbox{or } \qquad cr\;\gg\;1.
\end{equation}

A slightly different approach for studying the accuracy of the SM function is provided in \cite{Lan4}. One can obtain $\psi^{SM}$ by solving the Dirac equation in the ultrarelativistic case,
i.e. by requiring from the outset
\begin{equation}\label{2.2.9}
E\;\gg\;c^2 \quad \mbox{ and } E\;\gg\;|V(r)|.
\end{equation}
With $V(r)=-Z_T/r$ and $E\approx pc$ this leads to the condition $ pr \gg Z_T/c$
for the validity of $\psi^{SM}$.
Comparing (\ref{2.2.9}) with (\ref{2.2.8}) one notes that the requirement $E\gg c^2$ is substituted with $\theta \ll 1.$
For the bremsstrahlung process, these two restrictions are, however, found to be equivalent, as far as the maximum of the radiation is concerned  \cite{BM54}.
It also follows that, provided $r$ is large enough, $\psi^{SM}$ is accurate irrespective of $Z_T$.
More precisely, one has convergence of the SM function to the exact solution of the Dirac equation for $E \to \infty$ or for $r \to \infty$, but this convergence is not uniform. This is shown in Appendix A.

With the Sommerfeld-Maue wavefunctions at hand, we calculate the radiation matrix element,
\begin{equation}\label{2.2.10}
W_{\rm rad}^{SM}(\bfzeta_f,\bfzeta_i)\;=\;\int d\bfr\;\psi^{SM +}_f(\bfr,\bfzeta_f)\;(\bfalpha \bfe_\lambda^\ast)\;e^{-i\bfks \bfrs}\;\psi_i^{SM}(\bfr,\bfzeta_i),
\end{equation}
where we refer explicitly to the dependence of $u_{p_i}$ and $u_{p_f}$ on the spin polarization.
Integrals involving two confluent hypergeometric functions can be evaluated analytically with the help of Nordsieck's formula \cite{No54},
$$W(\epsilon,\bfs_0)\;\equiv\;\int d\bfr\;_1F_1(i\eta_i,1,i(s_0r-\bfs_0\bfr))\;e^{i\bfqs \bfrs}\;\frac{e^{-\epsilon r}}{r}\;_1F_1(i\eta_f,1,i(p_fr+\bfp_f \bfr))$$
\begin{equation}\label{2.2.11}
=\;\frac{2\pi}{\tilde{\alpha}}\;e^{-\pi \eta_i}\left( \frac{\tilde{\alpha}}{\tilde{\gamma}}\right)^{i\eta_i}\left( \frac{\tilde{\gamma}+\tilde{\delta}}{\tilde{\gamma}}\right)^{-i\eta_f}\;_2F_1(1-i\eta_i,i\eta_f,1,\frac{\tilde{\alpha}\tilde{\delta}-\tilde{\beta}\tilde{\gamma}}{\tilde{\alpha}(\tilde{\gamma}+\tilde{\delta})}),
\end{equation}
where $_2F_1$ is a hypergeometric function \cite{AS64}, $\epsilon =+0$. The parameters are defined by
$$
\tilde{\alpha}\;=\;\frac12 (q^2+\epsilon^2),\qquad \tilde{\beta}\;=\;\bfp_f \bfq -i\epsilon p_f,$$
\begin{equation}\label{2.2.12}
\tilde{\gamma}\;=\;\bfs_0\bfq+i\epsilon s_0-\tilde{\alpha},\qquad \tilde{\delta}\;=\; s_0p_f+\bfs_0\bfp_f-\tilde{\beta},
\end{equation}
where $\bfq=\bfp_i-\bfp_f-\bfk$ is the momentum transferred to the nucleus.
When inserting (\ref{2.2.5}) and (\ref{2.2.6}) into
the radiation matrix element, the first three terms
(containing at most one gradient) can be evaluated analytically by employing derivatives of $W(\epsilon,\bfs_0)$ and afterwards identifying $\bfs_0$ with $\bfp_i$.
The remaining term, proportional to the product of gradients,  cannot be represented in closed form. It is, however, of the order of $(Z_T/c)^2$
and is therefore neglected in consistency with other
second-order terms neglected in $\psi^{SM}$.
In addition, for small angles, this term is of the order of $c^2/E_i$ as compared to the three leading terms \cite{BM54}.

The Sommerfeld-Maue theory for bremsstrahlung is thus
based on the following approximation for the radiation matrix element,
$$W_{\rm rad}^{SM}(\bfzeta_f,\bfzeta_i)\;=\;N_i^{SM}N_f^{SM}u_{p_f}^+\left\{ \left[ 1\,+\,\frac{c}{2E_f}(\bfalpha \bfq)\right] (\bfalpha \bfe_\lambda^\ast)\;\left(-\lim_{\epsilon \to 0}\frac{\partial}{\partial \epsilon}\;W(\epsilon,\bfs_0)\right) \right.$$
\begin{equation}\label{2.2.13}
\left.+\;i\;\frac{cp_i}{2}\left[ \frac{2}{E_f}\left( \lim_{\epsilon \to 0}\bfnabla_{s_0} W(\epsilon,\bfs_0)\;\bfe_\lambda^\ast\right)\;-\;\left( \frac{1}{E_f}\,-\,\frac{1}{E_i}\right)(\bfalpha \bfe_\lambda^\ast)\;\left( \bfalpha \lim_{\epsilon \to 0}\bfnabla_{s_0} W(\epsilon,\bfs_0)\right)\right]\right\}_{\bfs_0=\bfps_i}u_{p_i}.
\end{equation}
Explicit formulas are given in  \cite{BM54,EH69,HN04}.

The argument $x=\frac{\tilde{\alpha} \tilde{\delta} -\tilde{\beta} \tilde{\gamma}}{\tilde{\alpha} ( \tilde{\gamma}+\tilde{\delta})}$ of the hypergeometric function 
contained in $W(\epsilon,\bfs_0)$ can for $\bfs_0=\bfp_i$  be expressed as  \cite{BM54}
$$ x\;=\;1\,-\,y,$$
\begin{equation}\label{2.2.14}
y\;=\;\frac{4\omega^2}{q^2c^2}\;\frac{(E_i/c\,-p_i\cos \theta_k)\;(E_f/c\,- p_f\cos \theta_f)}{(p_i+p_f)^2\,-\,\omega^2/c^2}\;=\;
\frac{4\omega^2}{q^2c^2}\;\frac{d_i\;d_f}{(p_i+p_f)^2-\omega^2/c^2},
\end{equation}
where  $\omega=kc$ is the photon frequency and $d_i,d_f$ are defined in (\ref{2.1.10}).
It is seen that the angle $\theta$ between $\hat{\bfp}$ and $\hat{\bfr}$ in the confluent hypergeometric functions (\ref{2.2.5}) and (\ref{2.2.6}) 
has transformed into the angles $\theta_k$ and $\theta_f$ which the electron momenta $\bfp_i$ and $\bfp_f$ form with the photon momentum $\bfk$.
Thus the condition $\theta \ll 1$ for the validity of the SM functions transforms into the conditions $\theta_k \ll 1$ and $\theta_f \ll 1$ for the applicability of the SM bremsstrahlung theory.
It should, however, be kept in mind that at ultrarelativistic energies $(E_i \to \infty,\;E_f \to \infty)$ where $\psi^{SM}_i$ and $\psi^{SM}_f$ become exact for all angles, the SM bremsstrahlung theory does not necessarily
 because of the further approximation inherent in (\ref{2.2.13}).
In particular, for  large  photon emission angles (as well as large $Z_T$), the double-gradient term has to be taken into account \cite{Jaku12}.  

With $W^{SM}_{\rm rad}$ from (\ref{2.2.13}), the triply differential bremsstrahlung cross section is obtained from
 \cite{HN04},
\begin{equation}\label{2.2.15}
\frac{d^3\sigma^{SM}}{d\omega d\Omega_k d\Omega_f}(\bfzeta_f,\bfzeta_i,\bfe_\lambda^\ast)\;=\;\frac{4\pi^2\omega\,p_fE_iE_f}{c^5p_i}\;|W_{\rm rad}^{SM}(\bfzeta_f,\bfzeta_i)|^2.
\end{equation}
For unobserved electron spin, one has to average over the initial and sum over the final spin projections.
In that case, the spinors $u_{p_i}$ and $u_{p_f}$ can be identified with the free spinors corresponding to the helicity eigenstates (see (\ref{2.1.8})).

For unobserved scattered electrons, the doubly differential cross section of the SM theory,
\begin{equation}\label{2.2.16}
\frac{d^2\sigma^{SM}}{d\omega d\Omega_k}(\bfzeta_i,\bfe_\lambda^\ast)\;=\;\sum_{\zeta_f} \int d\Omega_f\;\frac{d^3\sigma^{SM}}{d\omega d\Omega_k d\Omega_f}(\bfzeta_f,\bfzeta_i,\bfe_\lambda^\ast),
\end{equation}
is obtained with the help of a numerical integration over the solid angle $d\Omega_f$, 
including the sum over the final spin projections.

\setcounter{equation}{0}
\setcounter{figure}{0}

\section{Higher-order analytical theories}

In this section we discuss two prescriptions which allow for simple analytical formulae for the brems\-strah\-lung differential cross section, that include contributions of higher order in $Z_T\alpha$ beyond the Sommerfeld-Maue approximation.
However, the validity of these prescriptions is restricted to high energies of the scattering electron, $E\gg mc^2.$
The first model is a quantum mechanical one, while the second one is based on the semiclassical theory.

\subsection{Quantum mechanical theory}
 
We will  derive an approximation for the remainder $\psi_c$ from (\ref{2.2.3}) which is disregarded in the Sommerfeld-Maue theory, following the work of Roche et al \cite{RDP72}.
We insert the expansion $\psi=\psi_a+\psi_b+\psi_c$ into the transformed Dirac equation (\ref{2.2.2}),
 recalling that $\psi_b$ is defined by (\ref{2.2.4}). For the Coulomb potential one has
\begin{equation}\label{2.3.1}
\bfnabla V(r)\;=\;Z_T\;\frac{\bfr}{r^3},
\end{equation}
such that $\psi_c$ has to satisfy
\begin{equation}\label{2.3.2}
\left( c^2\bfnabla^2+p^2c^2\,+\,2E\;\frac{Z_T}{r}\right)\;\psi_c(r)\;=\;-icZ_T\frac{\bfalpha\bfr}{r^3}\;(\psi_b(r)+\psi_c(r))\;
-\;\frac{Z_T^2}{r^2}\;(\psi_a(r)+\psi_b(r)+\psi_c(r)).
\end{equation}
We recall that the expansion of $\psi$ uses the assumption that $\frac{Z_T}{c}$ or $\frac{1}{cr}$ is small, see (\ref{2.2.7}) and (\ref{2.2.8}).
Hence we approximate $\psi_c$ by its leading contribution if retaining on the rhs of (\ref{2.3.2}) only the term
proportional to $\psi_a$, while disregarding the contributions of $\psi_b$ and $\psi_c$. 
Consequently, also the term proportional to $Z_T/r$ on the lhs has to be ignored.
This leaves the defining equation for the approximate $\psi_c$,
\begin{equation}\label{2.3.3}
(c^2\bfnabla^2+p^2c^2)\;\psi_c(r)\;=\;-\frac{Z_T^2}{r^2}\;\psi_a(r),
\end{equation}
from which it follows consistently that $\psi_c$ comprises all second-order terms in $\frac{Z_T}{c}$ and in $\frac{1}{cr}$.

Instead of looking for an explicit solution to this equation, it is sufficient for our purpose to find an expression for the radiation matrix element (\ref{2.1.2}),
$$W_{\rm rad} \;=\;\int d\bfr\;(\psi_{fa}^++\psi_{fb}^+ + \psi_{fc}^+)\;(\bfalpha \bfe_\lambda^\ast)\;e^{-i\bfks\bfrs}\;(\psi_{ia}+\psi_{ib}+\psi_{ic})$$
\begin{equation}\label{2.3.4}
\approx \;W_{\rm rad}^{SM}\;+\;W_{\rm rad}(fa,ic)\;+\;W_{\rm rad}(fc,ia),
\end{equation}
where $W_{\rm rad}(fa,ic)$ is the matrix element between $\psi_{fa}$ and $\psi_{ic}$, and $W_{\rm rad}^{SM}=W_{\rm rad}(fa,ia)+W_{\rm rad}(fa,ib)+W_{\rm rad}(fb,ia)$ as discussed in section 2.2.
One can show  \cite{BM54} that all ignored contributions, including $W_{\rm rad}(fb,ib)$, decrease faster with energy than $c^2/E_i$, such that the resulting approximation will only be valid in the high-energy limit
(or at small $\frac{Z_T}{c}$ where $\psi_c$ is irrelevant and $W_{\rm rad}^{SM}$ becomes exact).

For evaluating $W_{\rm rad}(fa,ic)$ we start from the auxiliary integral
\begin{equation}\label{2.3.5}
I\;=\;\int d\bfr\;\psi_{fa}^+\;(\bfalpha \bfe_\lambda^\ast)\;e^{-i\bfks\bfrs}\;(\bfnabla^2+p_i^2)\;\psi_{ic}
\end{equation}
and transform it by means of two partial integrations.
Thereby we make a further approximation to $\psi_{fa}$ when forming the derivative,
\begin{equation}\label{2.3.6}
_1F_1(i\eta_f,1,i(p_fr+\bfp_f\bfr))\;\approx\;_1F_1(0,1,i(p_fr+\bfp_f\bfr))\;=\;1,
\end{equation}
trivially valid for small $\eta_f \sim \frac{Z_T}{c}$.
However, replacing $_1F_1$ by  a constant is equivalent to neglecting  $\bfnabla _1F_1$ and $\bfnabla^2 _1F_1$ which are of higher order in $\frac{1}{cr}$.
The result for $I$ after the two partial integrations is
$$ I\approx\int d\bfr\;N_f^{SM}\,_1F_1(i\eta_f,1,i(p_fr+\bfp_f\bfr)\;u_{p_f}^+\left[\bfnabla^2(e^{-i(\bfps_f+\bfks)\bfrs})\right](\bfalpha \bfe_\lambda^\ast)\;\psi_{ic}
\;+\;p_i^2\int d\bfr\;\psi_{fa}^+\;(\bfalpha\bfe_\lambda^\ast)\;e^{-i\bfks\bfrs}\;\psi_{ic}$$
\begin{equation}\label{2.3.7}
=\left[ p_i^2-(\bfp_f+\bfk)^2\right]\;W_{\rm rad}(fa,ic),
\end{equation}
where the boundary terms are omitted.
Thus $W_{\rm rad}(fa,ic)$ is directly related to $I$.

Now we evaluate $I$ in a different way by making use of (\ref{2.3.3}),
\begin{equation}\label{2.3.8}
I\;=\;\int d\bfr\;\psi_{fa}^+ (\bfalpha \bfe_\lambda^\ast)\;e^{-i\bfks\bfrs}\;\left( -\frac{Z_T^2}{c^2r^2}\;\psi_{ia}\right).
\end{equation}
We note that $I\sim \frac{Z_T^2}{c^2}$,
which mirrors  the $Z_T$-dependence of $\psi_{ic}$. Hence any further $Z_T$-dependence in (\ref{2.3.8}) can be neglected. Therefore,
as in (\ref{2.3.6}), $\eta_i$ and $\eta_f$ are replaced by zero, resulting in
\begin{equation}\label{2.3.9}
_1F_1(i\eta_f,1,i(p_f+\bfp_f\bfr))\;_1F_1(i\eta_i,1,i(p_ir-\bfp_i\bfr))\;\approx\;1.
\end{equation}
With
\begin{equation}\label{2.3.10}
\int d\bfr\;e^{i\bfqs\bfrs}\;\frac{1}{r^2}\;=\;\frac{4\pi}{q}\int_0^\infty dr\;\frac{\sin (qr)}{r}\;=\;\frac{2\pi^2}{q},
\end{equation}
where $\bfq=\bfp_i-\bfp_f-\bfk$ as before, $W_{\rm rad}(fa,ic)$ is approximately given by
\begin{equation}\label{2.3.11}
W_{\rm rad}(fa,ic)\;\approx\; \frac{I}{p_i^2-(\bfp_f+\bfk)^2}\;\approx\; -\;\frac{1}{p_i^2-(\bfp_f+\bfk)^2}\;\frac{2\pi^2 Z_T^2}{c^2q}
\;N_i^{SM}N_f^{SM}\;\left(u_{p_f}^+(\bfalpha \bfe_\lambda^\ast)\;u_{p_i}\right).
\end{equation}

In a similar way, $W_{\rm rad}(fc,ia)$ can be evaluated,
using $\tilde{I}=\int d\bfr\,\psi_{fc}^+(\bfnabla^2+p_f^2)e^{-i\bfks \bfrs}\psi_{ia}$, with the result 
\begin{equation}\label{2.3.12}
W_{\rm rad}(fc,ia)\;\approx\;-\;\frac{1}{p_f^2-(\bfp_i-\bfk)^2}\;\frac{2\pi^2Z_T^2}{c^2q}\;N_i^{SM}N_f^{SM}\;\left( u_{p_f}^+ (\bfalpha \bfe_\lambda^\ast) \,u_{p_i}\right).
\end{equation}

The cross section corresponding to this next-to-leading-order (NLO) SM theory is given by
$$\frac{d^3\sigma^{NLO-SM}}{d\omega d\Omega_k d\Omega_f}(\bfzeta_f,\bfzeta_i,\bfe_\lambda^\ast)\;=\;\frac{4\pi^2\omega\,p_fE_iE_f}{c^5\;p_i}$$
\begin{equation}\label{2.3.13}
\times\;\left[ \;|W_{\rm rad}^{SM}\,|^2\;+\;2\mbox{ Re}\,\left\{ W_{\rm rad}^{SM\ast} \;(W_{\rm rad}(fa,ic)+W_{\rm rad}(fc,ia))\right\}\,\right].
\end{equation}
In consistency with disregarding terms in the radiation matrix element which are of higher order in $\frac{Z_T}{c}$, the contribution proportional to $|W_{\rm rad}(fa,ic)+W_{\rm rad}(fc,ia)|^2$ has to be discarded.

We can write (\ref{2.3.13}) formally in the following way,
\begin{equation}\label{2.3.14}
\frac{d^3\sigma^{NLO-SM}}{d\omega d\Omega_k d\Omega_f}\;=\;\frac{d^3\sigma^{SM}}{d\omega d\Omega_k d\Omega_f}\;+\;\frac{d^3\sigma^{\rm corr}}{d\omega d\Omega_k d\Omega_f}.
\end{equation}

A comparison of photon spectra from electrons colliding with Cu (and unobserved scattered electrons) obtained within different theoretical approaches
is provided in Fig.2.3.1. As a guideline serves the result from  accurate bremsstrahlung calculations within the Dirac partial-wave formalism (to be discussed in section 2.4).
It is seen that the NLO-SM theory gives a quite reasonable representation of the photon spectrum,
while the Sommerfeld-Maue theory underpredicts the intensity at photon angles in the backward hemisphere, even for a light target such as copper $(Z_T=29)$.
In contrast, were the quadratic correction term arising from the absolute square of (\ref{2.3.4}) kept in the cross section, the result (termed NNLO-SM in Fig.2.3.1) 
would be largely in error for such angles.
Hence it is crucial to omit this term, which is erroneously included in both theory and results of Roche et al \cite{RDP72}, as well as in all subsequent publications on this subject before the year 2019.

\begin{figure}
\vspace{-1.5cm}
\includegraphics[width=8cm]{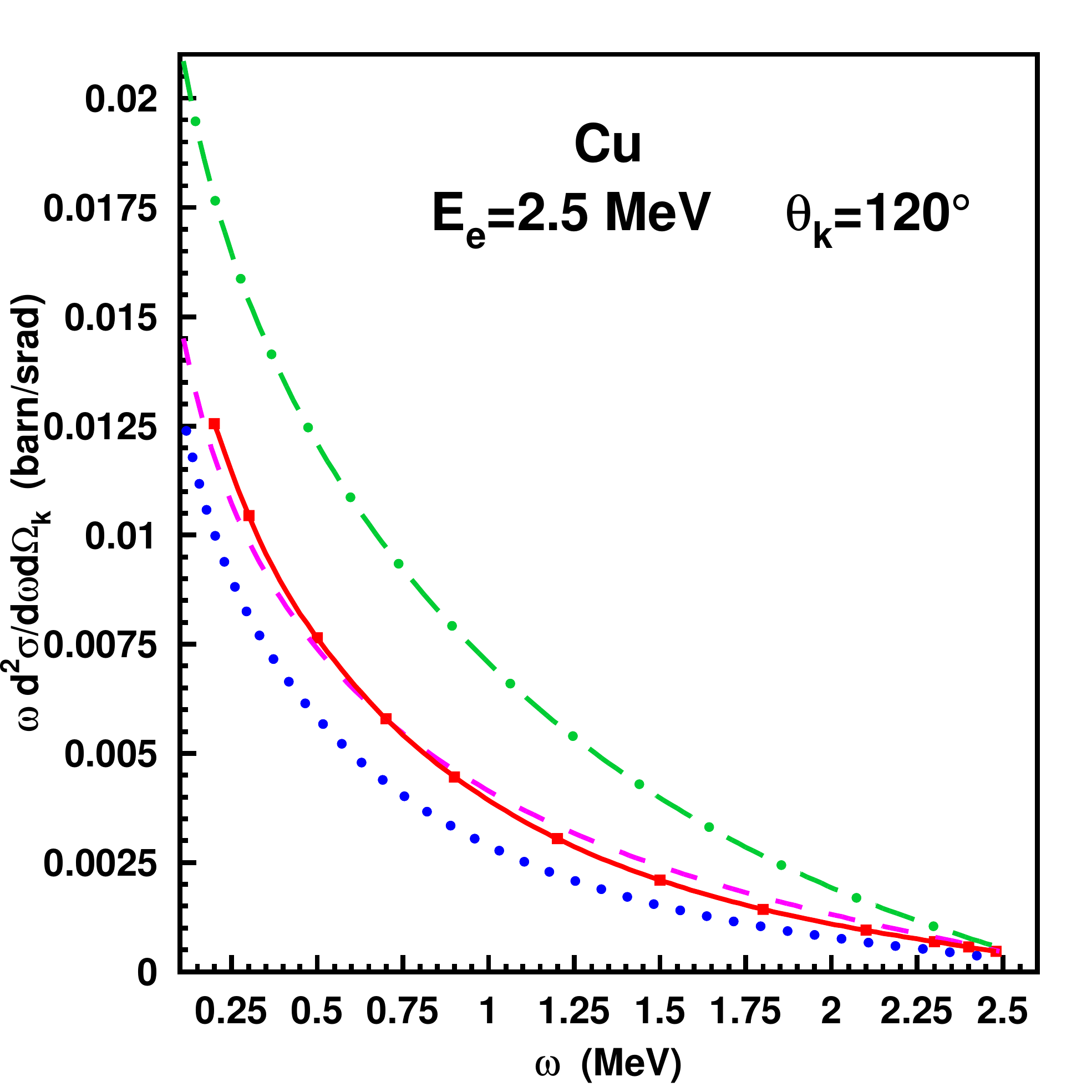}
\caption{
Doubly differential bremsstrahlung cross section times frequency for 2.5 MeV electrons colliding with Cu at a photon angle $\theta_k=120^\circ$, as function of
photon frequency $\omega$.
Shown are results from the SM theory (\ref{2.2.16}) $(\cdots\cdots)$,
from the NLO-SM theory ((\ref{2.3.13}), integrated over electron angles) $(-----$ \cite{JM19}), from the Dirac partial-wave theory (\ref{2.4.26}) (--------------), as well as from the (inconsistent) NNLO-SM approximation $(-\cdot - \cdot -$ \cite{M20}).
All particles are unpolarized. Screening effects play no role at this angle.
}
\end{figure}

\subsection{Quasiclassical theory}

Semiclassical methods are applicable when the de Broglie wavelength $\lambda = 2\pi \hbar/p$ is small compared to the distance over which the scattering potential varies appreciably.
This means that the electron is sufficiently localized to adjust, like a classical particle, to the spatial changes
of the potential. Given the potential $V(\bfr)$, the respective condition is
\begin{equation}\label{2.3.15}
\frac{1}{p} \,\cdot\, \frac{|\bfnabla V|}{V}\;\ll\;1.
\end{equation}
The semiclassical prescription of the electronic wavefunction leads to the Eikonal approximation \cite{Jo83}.
The terminology 'quasiclassical' has been introduced to include approximations which rely on the semiclassical assumption,
but go beyond the Eikonal approximation \cite{KLM16}.
The advantage of the quasiclassical prescription is the representation of the wavefunction in an integral form, rather than in terms of a differential equation. This allows, in
principle, to proceed to arbitrarily high orders in the
parameters $mc^2/E$, respectively, $\bfnabla V/(pV)$.
However, the evaluation of such integrals requires some skill for finding refined approximation techniques which are in concord
with the semiclassical condition.
We describe here the basic techniques, but otherwise refer to the original literature.

In order to construct the quasiclassical wavefunction, the quasiclassical Greens function $G(\bfr_2,\bfr_1)$ for the Dirac equation will be derived.
To this aim, we first calculate the Greens function $G_0(\bfr_2,\bfr_1)$ for the Klein-Gordon equation, which is formally given by
$$G_0(\bfr_2,\bfr_1)\;=\;\langle \bfr_2|\,\frac{1}{H_{KG}+i\epsilon}\,|\bfr_1\rangle,$$
\begin{equation}\label{2.3.16}
H_{KG}\;=\;p^2c^2+c^2\bfnabla^2-2pc\,\phi,\qquad \phi\;=\;\frac{1}{pc}\;(EV-V^2/2).
\end{equation}

The operator $H_D$ of the transformed Dirac equation (\ref{2.2.2}), $H_D=H_{KG}+ic\bfalpha \bfnabla V$,
differs from $H_{KG}$ only by a term proportional to $\bfnabla V$, which is small according to (\ref{2.3.15}).
Therefore the Greens function $G(\bfr_2,\bfr_1)$ can  be obtained from $G_0(\bfr_2,\bfr_1)$ by means of the expansion 
$$G(\bfr_2,\bfr_1)\;=\; \langle \bfr_2 | \,\frac{1}{H_{KG}+i\epsilon}\;-\;\frac{1}{H_{KG}+i\epsilon}\;(ic\bfalpha \bfnabla V)\;\frac{1}{H_{KG}+i\epsilon}$$
\begin{equation}\label{2.3.17}
+\;\frac{1}{H_{KG}+i\epsilon}\;(ic\bfalpha \bfnabla V)\;\frac{1}{H_{KG}+i\epsilon}\;(ic\bfalpha \bfnabla V)\;\frac{1}{H_{KG}+i\epsilon} \;+\;\cdots |\bfr_1\rangle.
\end{equation}
The wavefunction $\psi(\bfr)$ to the momentum $\bfp$ is then calculated from \cite{KM15,Lee00}
\begin{equation}\label{2.3.18}
\lim_{r_1 \to \infty}\,G(\bfr,\bfr_1)\;u_p\;=\;-\;\frac{e^{ipr_1}}{4\pi r_1}\;\psi(\bfr),
\end{equation}
with $u_p$ the free 4-spinor (defined  below (\ref{2.1.8})).

In the following we derive the quasiclassical approximation to $G_0(\bfr_2,\bfr_1)$ up to second order, following the work of Lee and coworkers \cite{Lee00}.
Starting point is the defining equation for the Greens function,
\begin{equation}\label{2.3.19}
\left[ p^2c^2\,+\,\bfnabla_{r_2}^2 c^2\,-\,2pc\,\phi(\bfr_2)\right] \, G_0(\bfr_2,\bfr_1)\;=\;c^2\;\delta(\bfr_2-\bfr_1).
\end{equation}
For vanishing potential $V$, i.e. $\phi=0$, the solution to (\ref{2.3.19}) is \cite{Jo83}
\begin{equation}\label{2.3.20}
G_0^{(0)}(\bfr)\;=\;-\;\frac{e^{ipr}}{4\pi r},\qquad \bfr\;=\;\bfr_2-\bfr_1,
\end{equation}
which is just the prefactor of $\psi$ in (\ref{2.3.18}).
We therefore make the ansatz
\begin{equation}\label{2.3.21}
G_0(\bfr_1+\bfr,\bfr_1)\;=\;G_0^{(0)}(\bfr)\;F(\bfr,\bfr_1)
\end{equation}
and insert it into (\ref{2.3.19}). Making use of
$(p^2c^2+\bfnabla^2c^2)\,G_0^{(0)}(\bfr)=c^2\delta(\bfr)$
and $\bfnabla^2_r(G_0^{(0)}\,F)=\\
\frac{1}{r^2}\,\frac{\partial}{\partial r}(r^2 \,\frac{\partial(G_0^{(0)}F)}{\partial r})\,-\,\frac{1}{r^2}\,G_0^{(0)}\,\bfL^2\,F$, 
where $\bfL$ is the angular momentum operator \cite{Ed60}, this leads to an equation for $F$,
\begin{equation}\label{2.3.22}
\left( ic\frac{\partial}{\partial r}\;-\;\phi(\bfr_1+\bfr)\;-\;\frac{c}{2pr^2}\;\bfL^2\right)F(\bfr,\bfr_1)\;=\;-\;\frac{c}{2p}\;\frac{\partial^2}{\partial r^2}\;F(\bfr,\bfr_1),
\end{equation}
under the condition that $F(0,\bfr_1)=1\;$ (such that $F\delta(\bfr)=\delta(\bfr)$).

In a first step to solve this equation, we assume that $F$ is a slowly varying function (in concord with the assumption that $V$ is slowly varying)
and neglect the rhs as well as the angular-momentum dependent term.
Calling $F_0$ the solution of the remaining equation,
\begin{equation}\label{2.3.23}
ic\;\frac{\partial F_0}{\partial r}\;=\;\phi(\bfr_1+\bfr)\;F_0,
\end{equation}
it is given by
\begin{equation}\label{2.3.24}
F_0(\bfr,\bfr_1)\;=\;e^{-\frac{i}{c}\int_0^r dr \,\phi(\bfrs_1+\bfrs)}\;=\; e^{-\frac{ir}{c} \int_0^1 dx\,\phi(\bfrs_1+x\bfrs)},
\end{equation}
and obviously fulfills $F_0(0,\bfr_1)=1.$
This yields, together with (\ref{2.3.21}), the Eikonal approximation to $G_0$.

In the next step, we retain all terms on the lhs of (\ref{2.3.22}) and introduce a perturbative function $g$ by setting
\begin{equation}\label{2.3.25}
F(\bfr,\bfr_1)\;=\;e^{iA}\;F_0(\bfr,\bfr_1)\;(1+g)
\end{equation}
to be inserted into the lhs of (\ref{2.3.22}), while $g=0$ is set in its rhs.
The exponent function $A=(\frac{1}{r}-\frac{1}{a_1})\,\frac{\bfLs^2}{2p}\,$ takes care of the angular momentum operator in (\ref{2.3.22}).
$a_1<r$ is an auxiliary parameter related to the maximum gradient of the potential.

With this choice, the resulting equation for $g$ contains no linear terms. In fact, with $\frac{\partial F_0}{\partial r}$ from (\ref{2.3.23}) and $\frac{\partial A}{\partial r} =-\frac{\bfLs^2}{2pr^2}$, it reads
\begin{equation}\label{2.3.26}
ic\;e^{iA}\;F_0\;\frac{\partial g}{\partial r}\;+\;\left( ic\;\frac{\partial}{\partial r}\,-\,\phi\,-\;\frac{c}{2pr^2}\;\bfL^2\right)\;e^{iA}\;F_0
\;=\;-\;\frac{c}{2p}\;\frac{\partial^2}{\partial r^2}\;\left( e^{iA}\,F_0\right)
\end{equation}
and hence can be solved for $g$ by a mere integration.

The expression $e^{iA}F_0$ in (\ref{2.3.25}) can be evaluated by means of the approximate formula \cite{Lee00},
\begin{equation}\label{2.3.27}
e^{-i\beta \bfLs^2}\,f(\bfr)\;\approx\;-\;\frac{i}{\pi}\int d^2\bfq_\perp\;e^{iq_\perp^2}\;f(\bfr+2\sqrt{\beta} r \,\bfq_\perp)
\end{equation}
for $\beta>0$, where $\bfq_\perp$ is a 2-dimensional vector perpendicular to $\bfr$.
In the above approximation, the result for the Klein-Gordon Greens function is, with $\phi$ expressed in terms of $V$ \cite{KLM16},
$$G_0(\bfr_1+\bfr,\bfr_1)\;=\;\frac{i}{\pi}\;\frac{e^{ipr}}{4\pi r}\int d^2\bfq_\perp\;e^{i q_\perp^2}\;e^{-icr \int_0^1 dx\,V(\bfRs_x)}$$
\begin{equation}\label{2.3.28}
\times \;\left[ 1\;+\;\frac{ir^3}{2pc^2}\int_0^1 dx \int_0^x dy\;(x-y)\;\bfnabla_\perp V(\bfR_x)\;\bfnabla_\perp V(\bfR_y)\right],
\end{equation}
$$\qquad\qquad \bfR_x\;=\;\bfr_1\,+\,x\bfr \,+\,\bfq_\perp \sqrt{\frac{2r_1r_2}{pr}},$$
where the high-energy relation $E=pc$ has been applied, and $\bfnabla_\perp$ is the gradient component perpendicular to $\bfr$.

Following Krachkov and Milstein \cite{KM15}, $G_0$ can be used to obtain the higher-order terms in the expansion (\ref{2.3.17}) of $G$, and hence the corresponding expansion of the wavefunction (\ref{2.3.18}).
To first order in $\bfnabla V$, one finds
\begin{equation}\label{2.3.29}
G(\bfr_2,\bfr_1)\;=\;G_0(\bfr_2,\bfr_1)\;-\;\frac{ic}{2E}\;\bfalpha\;(\bfnabla_{r_1}\,+\,\bfnabla_{r_2})\;G_0(\bfr_2,\bfr_1).
\end{equation}
With the help of this formula (and applying some suitable approximations), it can be shown that (to first order in $\bfnabla V$) the quasiclassical wavefunction for the Coulomb field $V=-\frac{Z_T}{r}$ agrees with the Sommerfeld-Maue function (\ref{2.2.5}), however with the Sommerfeld parameter replaced by its high-energy limit $\eta = \frac{Z_T}{c}$.
The next-order term of the wavefunction (i.e. of second order in $\bfnabla V$) also exists explicitly in case of the Coulomb field \cite{KM15}.

As an application of the quasiclassical theory  we mention the calculation of the photon spectrum, $\frac{d\sigma}{d\omega}$.
For the Coulomb field, this theory allows for a simple analytical formula \cite{Lee04}.
Its straightforward evaluation has to be contrasted to the result of the quantum mechanical theory, which only provides an analytical expression for the {\it triply} differential bremsstrahlung cross section,
 and which necessitates a numerical integration over the electron and photon angles.
It has been shown \cite{Ma19} that at a collision energy of 500 MeV, the quasiclassical approach leads to the same results for $\frac{d\sigma}{d\omega}$ as the quantum mechanical one in the lower half of the photon spectrum.
However, there are severe discrepancies near the short-wavelength limit where the quasiclassical theory fails.

A further advantage of the quasiclassical theory in its range of applicability is the straightforward implementation of screening effects caused by the atomic electrons.
This relies on the fact that the atomic potential enters explicitly into the formulae for the Greens function and hence into the radiation matrix element. 

\setcounter{equation}{0}

\section{Relativistic partial-wave theory}

If the target electrons are accounted for by means of a spherical screened atomic potential $V(r)$, the bremsstrahlung process can, within the one-photon approximation, be described
rigorously by the relativistic partial-wave theory.
In this theory, the electronic scattering states are decomposed into partial waves.
Each of them is a product of a radial function, which is an exact solution of the radial Dirac equation containing $V(r)$, and an analytically known angular function.
For a pure Coulomb field, the partial-wave expansion of the scattering states was first given by Darwin \cite{Da28}.

\subsection{Dirac wavefunctions}

Let us start with describing the continuum wavefunctions in terms of  exact solutions of the Dirac equation. 
Early applications of such functions in  bremsstrahlung calculations can be found in Rozics and Johnson \cite{RJ64} and Brysk et al \cite{BZP69} as well as in a series of papers by Pratt and coworkers.
Their first rigorous calculations \cite{TP71} covered collision energies up to 1 MeV,
and were later extended up to 4.54 MeV  \cite{T97}.
Recent refined integration methods  \cite{YS10} allowed to extend the  results  in some cases even to 30 MeV \cite{Jaku16}.
When restriction is made to the short-wavelength limit, where the energy of the scattered electron is close to zero and hence only the lowest partial waves are required for the description of $\psi_f$,
further investigations for collision energies of several MeV \cite{PT75,H10,JS11}
do exist.

The reduction of the Dirac equation in three-dimensional space to a pair of radial Dirac equations, combined with the derivation of the partial-wave structure of its unbound solution, can be found in several textbooks \cite{Lan4,Ros61,HN04} and will not be repeated here. 

Let us  assume that the spin of the scattered electron is recorded. The corresponding  spin polarization vector $\bfzeta_f$ is defined by the polarization spinor $w=\sum\limits_{m_s=\pm \frac12} b_{m_s} \chi_{m_s}$,
 where $\chi_{1/2} = {1 \choose 0}$ and $\chi_{-1/2}= {0 \choose 1}$ are the two basis states of the spinor space, and the coefficients $b_{m_s}$ relate to the coordinates of $\bfzeta_f$ as discussed in detail in section 3.
Then the  outgoing electron is described by the scattering state
\begin{equation}\label{2.4.1}
\psi_f^+(\bfr,\bfzeta_f)\;=\;\sum_{m_s=\pm \frac12} b_{m_s}^\ast \sum_{\kappa_f m_f} (l_fm_l\,\frac12 m_s|\,j_fm_f)\;(-i)^{l_f}\;e^{i\delta_{\kappa_f}}\;Y_{l_fm_l}(\hat{\bfp}_f)\;\psi_{\kappa_f m_f}^{+}(\bfr),
\end{equation}
where $m_l=m_f-m_s$ and where each partial wave is characterized by the angular momentum quantum numbers $j_f,\;l_f$ and their respective projections $m_f,\;m_l$. The spin-orbit coupling coefficients $(l_fm_l\,\frac12 m_s|\,j_fm_f)$
are Clebsch-Gordan coefficients as described in \cite{Ed60}. The angular functions $Y_{l_fm_l}(\hat{\bfp}_f),$
depending on the direction $\hat{\bfp}_f$ of the final momentum $\bfp_f$, are the spherical harmonic functions \cite{Ed60}.
The quantum number $\kappa_f$ relates to $j_f$ and $l_f$ by means of $\kappa_f=(j_f+\frac12)(-1)^{j_f+l_f+\frac12}$ and runs over all integers except zero.
The reverse relations are $j_f=|\kappa_f|-\frac12$ and $l_f=|\kappa_f+\frac12|-\frac12$.
The partial-wave 4-spinor $\psi_{\kappa_f m_f}$ is defined by
\begin{equation}\label{2.4.2}
\psi_{\kappa\,m}(\bfr)\;=\;{g_\kappa(r)\;Y_{jlm}(\hat{\bfr}) \choose i\,f_\kappa (r)\;Y_{jl'm}(\hat{\bfr}) },
\end{equation}
where $Y_{jlm}(\hat{\bfr})$ is a spherical harmonic spinor \cite{Ed60}, and where $l'=|\kappa -\frac12|-\frac12$.
The functions $g_\kappa$ and $f_\kappa$ are, respectively,
the large and small components of the radial Dirac 4-spinor, and are solutions to the two coupled radial Dirac equations,
$$\frac{dg_\kappa}{dr} \;+\;\frac{1+\kappa}{r}\;g_\kappa\;-\;\frac{1}{c}\;(E+c^2-V(r))\;f_\kappa\;=\;0$$
\begin{equation}\label{2.4.3}
\frac{df_\kappa}{dr}\;+\;\frac{1-\kappa}{r}\;f_\kappa\;+\;\frac{1}{c}\;(E-c^2-V(r))\;g_\kappa\;=\;0,
\end{equation}
where $E=\sqrt{p^2c^2+c^4}$ is the total energy of the electron.
The phase shift $\delta_{\kappa_f}$ is the phase difference between $g_{\kappa_f}(r)$ for $r \to \infty$ and a plane-wave solution to angular momentum $l_f$.

The incoming electron is described
in terms of
\begin{equation}\label{2.4.4}
\psi_i(\bfr,\bfzeta_i)\;=\;\sum_{m_s=\pm \frac12} a_{m_s} \sum_{\kappa_i m_i} (l_im_l\frac12 m_s|\,j_im_i)\;i^{l_i}\;e^{i \delta_{\kappa_i}}\;Y_{l_im_l}(\hat{\bfp}_i)\;\psi_{\kappa_i m_i}(\bfr),
\end{equation}
where $a_{1/2}$ and $a_{-1/2}$ are the coefficients of the initial  polarization spinor, and $m_l=m_i-m_s$.
When the quantization axis $\bfe_z$ is taken along the
momentum $\bfp_i$ of the incident electron, implying that the spherical angles $\theta_i$ and $\varphi_i$ are zero, the spherical harmonic function reduces to
\begin{equation}\label{2.4.5}
Y_{l_im_l}(\hat{\bfp}_i)\;=\;\sqrt{\frac{(2l_i+1)\;(l_i-m_l)!} {4\pi\;(l_i+m_l)!}}\;P_{l_i}^{m_l}(\cos \theta_i)\;e^{i m_l\varphi_i}\;=\;\sqrt{\frac{2l_i+1}{4\pi}}\;\delta_{m_l,0},
\end{equation}
since the Legendre function $P_{l_i}^{m_l}(1)$ is only nonvanishing for $m_l=0$ \cite{AS64}.
Then (\ref{2.4.4}) is simplified to
\begin{equation}\label{2.4.6}
\psi_i(\bfr,\bfzeta_i)\;=\; \sum_{m_i=\pm \frac12} a_{m_i}\sum_{\kappa_i} \sqrt{\frac{2l_i+1}{4\pi}}\;(l_i 0 \,\frac12 m_i|\,j_i m_i)\;i^{l_i}\;e^{i\delta_{\kappa_i}}\;\psi_{\kappa_im_i}(\bfr).
\end{equation}

Our goal is to evaluate the radiation matrix element,
$$
 W_{\rm rad}(\bfzeta_f,\bfzeta_i)\;=\; \int d\bfr \;\psi_f^+(\bfr,\bfzeta_f)\;(\bfalpha \bfe_\lambda^\ast) \;e^{-i \bfks \bfrs}\;\psi_i(\bfr,\bfzeta_i)
$$
\begin{equation}\label{2.4.7}
\;\equiv\; \sum_{m_s=\pm \frac12} b_{m_s}^\ast\;F_{fi}(m_s,\bfzeta_i).
\end{equation}
To this aim, the photon operator is also partial-wave expanded,
\begin{equation}\label{2.4.8}
e^{-i\bfks \bfrs}\;=\;4\pi \sum_{l=0}^\infty (-i)^l\;j_l(kr)\;\sum_{\mu=-l}^l Y_{l\mu}^\ast(\hat{\bfk})\;Y_{l\mu}(\hat{\bfr}),
\end{equation}
where $j_l$ is a spherical Bessel function \cite{AS64}.

The vector spherical harmonics in (\ref{2.4.2}) are decomposed according to
\begin{equation}\label{2.4.9}
Y_{jlm}(\hat{\bfr})\;=\;\sum_{\mu=-l}^l \sum_{m_s=\pm \frac12} Y_{l\mu}(\hat{\bfr})\;\chi_{m_s}\;(l\mu\,\frac12 m_s|\,jm).
\end{equation}

Moreover, the photon polarization vector $\bfe_\lambda^\ast$ is expanded in a basis of spherical unit vectors, $\bfe_0=\bfe_z,\;\bfe_{+1}=-\,\frac{1}{\sqrt{2}}(\bfe_x+i\,\bfe_y)$ and $\bfe_{-1}=\,\frac{1}{\sqrt{2}}(\bfe_x-i\,\bfe_y)$,
\begin{equation}\label{2.4.10}
\bfe_\lambda^\ast\;=\;\sum_{\varrho=0,\pm 1} c_\varrho^{(\lambda)}\;\bfe_\varrho,
\end{equation}
such that use can be made of the relation
\begin{equation}\label{2.4.11}
\chi_{m_{s_f}}^+(\bfsigma \bfe_\varrho)\;\chi_{m_{s_i}}\;=\;\sqrt{3}\;\;(\frac12 m_{s_i} 1 \varrho\,|\,\frac12 m_{s_f}),
\end{equation}
where 
 $\bfsigma$ is the vector of Pauli matrices.

We define the angular part of the radiation matrix element,
\begin{equation}\label{2.4.12}
M_{fi}(r)\;\equiv\; \int d\Omega\;\psi_{\kappa_f m_f}^+(\bfr)\;(\bfalpha \bfe_\varrho)\;\psi_{\kappa_i m_i}(\bfr)\;Y_{l\mu}(\hat{\bfr}).
\end{equation}
With the help of
\begin{equation}\label{2.4.14}
Y_{l_i\mu_i}(\hat{\bfr})\;Y_{l \mu}(\hat{\bfr})\;=\;\sum_{LM} \sqrt{\frac{(2l_i+1)(2l+1)}{4\pi\,(2L+1)}}\;(l_i\mu_il\mu\,|\,LM)\;(l_i0\,l\,0\,|\,L\,0)\;Y_{LM}(\hat{\bfr})
\end{equation}
and the orthogonality of the spherical harmonic functions \cite{Ed60}, $M_{fi}$ can be evaluated analytically,
\begin{equation}\label{2.4.15}
M_{fi}(r)\;=\;i\;\left[ g_{\kappa_f}(r)\;f_{\kappa_i}(r)\;W_{12}^B(l,l_f,l_i')\;-\;
f_{\kappa_f}(r)\;g_{\kappa_i}(r)\;W_{12}^B(l,l_f',l_i) \right],
\end{equation}
with
$$W_{12}^B(l,l_f,l_i')\;=\;\int d\Omega\;Y_{j_fl_fm_f}^+(\hat{\bfr})\;(\bfsigma \bfe_\varrho)\;Y_{j_il_i'm_i}(\hat{\bfr})\;Y_{l\mu}(\hat{\bfr})$$
\begin{equation}\label{2.4.16}
=\;\sqrt{\frac{3}{4\pi}}\;\sqrt{2l+1}\;\sqrt{\frac{2l_i'+1}{2l_f+1}}\;(l_i'0\,l\,0\,|\,l_f 0)\;\sum_{m_{s_i}=\pm \frac12} (l_f\mu_f \frac12\,m_{s_f}|\,j_fm_f)
\end{equation}
$$\times \;(l_i'\mu_i\frac12\,m_{s_i}|\,j_im_i)\;(\frac12\,m_{s_i} 1 \varrho\,|\,\frac12 m_{s_f})\;(l_i'\mu_il\,\mu\,|\,l_f\mu_f).$$
Hence (\ref{2.4.15}) involves only the sums over $\varrho$ and $m_{s_i}$, while the remaining magnetic quantum numbers are, due to the selection rules of the Clebsch-Gordan coefficients, determined by
\begin{equation}\label{2.4.17}
m_{s_f}\;=\;m_{s_i} + \varrho,\qquad \mu_i\;=\;m_i-m_{s_i},\qquad \mu_f\;=\;m_f-m_{s_f},\qquad \mu\;=\;\mu_f-\mu_i.
\end{equation}
Then the reduced matrix element $F_{fi}(m_s,\bfzeta_i)$ from (\ref{2.4.7}) turns into
$$ F_{fi}(m_s,\bfzeta_i)\;=\;\sqrt{4 \pi}\sum_{\kappa_f m_f}(l_f m_l \frac12 m_s|\,j_fm_f)\;(-i)^{l_f}\;e^{i\delta_{\kappa_f}}\;Y_{l_f m_l}(\hat{\bfp}_f) $$
\begin{equation}\label{2.4.13}
\times \;\sum_{\kappa_i} \sum_{m_i=\pm \frac12} a_{m_i}\;\sqrt{2 l_i+1}\;(l_i 0 \,\frac12 m_i|\,j_i m_i)\;i^{l_i}\;e^{i\delta_{\kappa_i}}
\end{equation}
$$ \times \;\sum_{l=0}^\infty (-i)^l\;\sum_\varrho c_\varrho \; Y_{l\mu}^\ast(\hat{\bfk})\;\int_0^\infty r^2dr\;j_l(kr)\;M_{fi}(r),$$
with $\mu=m_f-m_i-\varrho$.
Inserting (\ref{2.4.15}) into (\ref{2.4.13}) there occur two kinds of radial integrals,
$$R_{fi}(l)\;=\;\int_0^\infty r^2dr\;g_{\kappa_f}(r)\;f_{\kappa_i}(r)\;j_l(kr),$$
\begin{equation}\label{2.4.18}
R_{if}(l)\;=\;\int_0^\infty r^2dr\; f_{\kappa_f}(r)\;g_{\kappa_i}(r)\;j_l(kr).
\end{equation}
Details for the evaluation of the radial integrals are provided in Appendix B.

From the selection rules of the Clebsch-Gordan coefficient $(l_i' 0 \,l\,0\,|\,l_f0)$ in the term proportional to $R_{fi}(l)$, one has the requirement that $l_i'+l+l_f$ is even and that $l$ runs from $|l_i'-l_f|$ to $l_i'+l_f$ in steps of 2.
Analogously, the sum in the term proportional to $R_{if}(l)$ runs from $|l_i-l_f'|$ to $l_i+l_f'$ in steps of 2.

Choosing the $x$ and $y$ axes according to $\bfe_y = \bfp_i \times \bfk/|\bfp_i \times \bfk|$ and $\bfe_x = \bfe_y \times \hat{\bfp}_i$ such that $\hat{\bfk}=(\sin \theta_k,0,\cos \theta_k)$ lies in 
the $(x,z)$ reaction plane, the spherical harmonic function $Y_{l \mu}^\ast (\hat{\bfk})$ is real and depends only on the photon polar angle $\theta_k$ (see its representation in (\ref{2.4.5})).

Because of numerical cancellations the sum over $\kappa_f$ should  be replaced by a sum over $l_f$ with  two values of $j_f$ to each $l_f$ for $l_f>0$. It is also of advantage to sum over the magnetic quantum numbers $m_l$ and $m_{s_f}$, replacing the pair $m_f$ and $\varrho$.
The final result for $F_{fi}$ can then be written in the following way,
$$ F_{fi}(m_s,\bfzeta_i)\;=\;i\sum_{l_f=0}^\infty \sum_{m_l=-l_f}^{l_f} (-i)^{l_f}\;Y_{l_f m_l}(\hat{\bfp}_f)\sum_{j_f=l_f \pm \frac12}(l_fm_l\frac12 m_s|\,j_fm_f)$$
\begin{equation}\label{2.4.19}
\times\; \sum_{m_i=\pm \frac12} a_{m_i} \sum_{\kappa_i} \sqrt{2l_i+1}\;i^{l_i}\;e^{i(\delta_{\kappa_i}+\delta_{\kappa_f})}\;(l_i 0\,\frac12\,m_i|\,j_im_i)\;S_{fi}
\end{equation}
with
$$S_{fi}\;=\; \sum_{l=|l_i'-l_f|}^{l_i'+l_f}(-i)^l\;R_{fi}(l)\sum_{m_{s_f},m_{s_i}=\pm \frac12}\sqrt{\frac{(2l+1)\,(l-\mu)!}{(l+\mu)!}}\;P_l^\mu(\cos \theta_k)\;c_\varrho^{(\lambda)} \;\tilde{W}_{12}^B(l,l_f,l_i')$$
\begin{equation}\label{2.4.20}
-\;\sum_{l=|l_i-l_f'|}^{l_i+l_f'}(-i)^l\;R_{if}(l) \sum_{m_{s_f},m_{s_i}=\pm \frac12} \sqrt{\frac{(2l+1)\,(l-\mu)!}{(l+\mu)!}}\;P_l^\mu(\cos \theta_k)\;c_\varrho^{(\lambda)}\;\tilde{W}_{12}^B(l,l_f',l_i),
\end{equation} 
where $\tilde{W}_{12}^B(l,l_f,l'_i)$ is given by (\ref{2.4.16}) with the sum over $m_{s_i}$ omitted.
For high $l_f$, the summation over $m_l$ can be restricted to the lowest values, say, $|m_l| \,\leq 30$ when $l_f >30$.

If the spin of the emitted electron remains unobserved, one has to sum over its final spin projections.
In that  case one can take $b_{m_s}$ to be either 1 or 0, such that
\begin{equation}\label{2.4.21}
\sum_{\zeta_f}|W_{\rm rad}(\bfzeta_f,\bfzeta_i)|^2\;=\; \sum_{m_s=\pm \frac12}|F_{fi}(m_s,\bfzeta_i)|^2.
\end{equation}
With this, the triply differential bremsstrahlung cross section for a polarized initial electron and  a polarized photon reads
\begin{equation}\label{2.4.22}
\frac{d^3\sigma}{d\omega d\Omega_k d\Omega_f}(\bfzeta_i,\bfe_\lambda^\ast)\;=\;\frac{4\pi^2 \omega \,p_f E_i E_f}{c^5\;p_i}\sum_{m_s=\pm \frac12}|F_{fi}(m_s,\bfzeta_i)|^2.
\end{equation}

An essential  advantage of the partial-wave  formalism becomes evident in the calculation of the doubly differential cross section for unobserved electrons.
This implies an additional integration over the electron's solid angle $d\Omega_f$. If recoil effects are neglected, this integration is straightforward.

In order to derive that result, let us abbreviate $F_{fi}$ in the following way,
\begin{equation}\label{2.4.23}
F_{fi}(m_s,\bfzeta_i)\;=\;\sum_{\kappa_f m_f}(-i)^{l_f}\;(l_f,m_f-m_s,\frac12 m_s|\,j_fm_f)\;Y_{l_f,m_f-m_s}(\hat{\bfp}_f)\;T_{fi}(\kappa_f,m_f),
\end{equation}
where $T_{fi}$ comprises all terms of (\ref{2.4.13}) not considered explicitly in (\ref{2.4.23}).
Making use of the unitarity of the Clebsch-Gordan coefficients,
\begin{equation}\label{2.4.24}
\sum_{m_s}(\frac12 m_s\, l_f,m_f-m_s|\,j_fm_f)\;(\frac12 m_s\, l_f,m_f-m_s|\,j'_f m_f)\;=\;\delta_{j_fj'_f},
\end{equation}
we have
$$\int d\Omega_f\sum_{m_s}|F_{fi}(m_s,\bfzeta_i)|^2\;=\;\sum_{m_s}\sum_{\kappa_f m_f} i^{l_f}\;(l_f,m_f-m_s,\frac12 m_s|\,j_fm_f)\;T_{fi}^\ast(\kappa_f,m_f)$$
$$\times \sum_{\kappa_f' m_f'} (-i)^{l_f'}\;(l_f',m_f'-m_s,\frac12 m_s|\,j'_f m_f')\;T_{fi}(\kappa_f'm_f')\;\delta_{l_f l_f'}\delta_{m_f m_f'}$$
\begin{equation}\label{2.4.25}
=\;\sum_{\kappa_f m_f} |T_{fi}(\kappa_f, m_f)|^2.
\end{equation}
Thus, while the triply differential cross section involves a coherent sum over the final-state partial waves, this sum turns into an incoherent one for the doubly differential cross section,
$$\frac{d^2\sigma}{d\omega d\Omega_k}(\bfzeta_i,\bfe_\lambda^\ast)\;=\;\frac{4\pi^2 \omega\, p_f E_iE_f}{c^5\;p_i}\sum_{l_f=0}^\infty \sum_{j_f=l_f\pm \frac12} \sum_{m_f=-j_f}^{j_f}$$
\begin{equation}\label{2.4.26}
\times \;\left| \sum_{m_i=\pm \frac12} a_{m_i} \sum_{\kappa_i} \sqrt{2l_i+1}\;i^{l_i}\;e^{i(\delta_{\kappa_i} +\delta{\kappa_f})}\;(l_i 0 \frac12 m_i|\,j_im_i)\;S_{fi}\right|^2,
\end{equation}
with $S_{fi}$ from (\ref{2.4.20}).

The other case where bremsstrahlung is calculated in terms of a doubly differential cross section,
concerns the situation where   it is not the electron but the photon which remains unobserved. This  case cannot be treated in an easy way,  because the dependence on the photon emission angle is not only given by $P_l^\mu$, but is also inherent in the components $c_\varrho^{(\lambda)}$ of the photon polarization vector. 
Therefore the two additional integrals (over $\theta_k$, as well as over the azimuthal angle $\varphi_f$ of the electron with respect to the photon emission plane) have to be performed numerically. 
This challenge has to our knowledge not yet been met.

\subsection{Relation to the Sommerfeld-Maue wavefunctions}

When the potential  is the Coulomb field of a point nucleus, $V(r)=-Z_T/r$  with nuclear charge number $Z_T$, the solutions to the radial Dirac equations (\ref{2.4.3}), the Coulomb-Dirac waves, are known in closed form \cite{Lan4}.
Taking the $z$-axis along the momentum $\bfp_i$, the Coulomb-Dirac wave describing the incoming electron can be written in the following way  \cite{JY13},
\begin{equation}\label{2.4.27}
\psi_i^C(\bfr,\bfzeta_i)\;=\;\sum_{m_i=\pm \frac12} a_{m_i}\sum_{\kappa_i} \sqrt{\frac{2l_i+1}{4\pi}}\;(l_i0\,\frac12 m_i|\,j_im_i)\;i^{l_i}\;e^{i\delta_{\kappa_i}^C}\;\psi_{\kappa_i m_i}^C(\bfr),
\end{equation}
with
\begin{equation}\label{2.4.28}
e^{i\delta_{\kappa_i}^C}\;=\;\sqrt{\frac{i\eta_i c^2/E_i\,-\kappa_i}{\gamma_i +i \eta_i}}\;e^{-i\, {\rm arg } \Gamma(\gamma_i+i\eta_i)\,+i(l_i+1-\gamma_i)\pi/2},
\end{equation}
where $\Gamma$ is the Gamma function, $\eta_i=Z_TE_i/(p_ic^2)$ the Sommerfeld parameter, $\gamma_i=\sqrt{\kappa_i^2 -(Z_T/c)^2}$, and
$${g_{\kappa_i} \choose f_{\kappa_i}}^C\;=\;{ -i \sqrt{\frac{E_i+c^2}{E_i}} \choose \sqrt{\frac{E_i-c^2}{E_i}} } \;N_i^C\;e^{-i(\gamma_i+\frac12)\pi/2}\;(2 p_ir)^{-3/2}$$
\begin{equation}\label{2.4.29}
\times\; \left[ (\kappa_i\,-\,i\eta_ic^2/E_i)\;M_{-i\eta_i-\frac12,\gamma_i}(2ip_ir)\;\mp\;(\gamma_i-i\eta_i)\;M_{-i\eta_i+\frac12, \gamma_i}(2ip_ir)\right],
\end{equation}
where the upper sign in the square bracket corresponds to $g_{\kappa_i}$ and the lower sign to $f_{\kappa_i}$.
The normalization constant is
\begin{equation}\label{2.4.30}
N_i^C\;=\;\frac{1}{\sqrt{\pi}}\;e^{\pi \eta_i/2}\;\frac{|\Gamma(\gamma_i+1+i\eta_i)|}{\Gamma(2\gamma_i+1)}\;\frac{\mbox{\rm sign } \kappa_i}{\sqrt{\kappa_i-i\eta_ic^2/E_i}\;\sqrt{\gamma_i -i\eta_i}},
\end{equation}
and $M_{\alpha,\gamma}(z)$ is a Whittaker function of the first kind \cite{AS64}.

As discussed in Section 2.2, the Sommerfeld-Maue (SM) function is an approximate analytical solution to the Dirac equation for the Coulomb field. The incoming SM scattering state (choosing $\bfe_z=\hat{\bfp}_i$) is according to (\ref{2.2.5}) given by
$$\psi_i^{SM}(\bfr,\bfzeta_i)\;=\;N_i^{SM}\;e^{i p_i r_z}\;\left\{ _1F_1(i\eta_i, 1,ip_i(r-r_z))\right.$$
\begin{equation}\label{2.4.31}
 +\;\frac{iZ_T}{2c}\;(\bfalpha \hat{\bfr}\,-\alpha_z)\;_1F_1(i\eta_i+1,2,ip_i(r-r_z) \;\}\;u_{p_i}(\bfzeta_i),
\end{equation}
where $_1F_1$ is a confluent hypergeometric function \cite{AS64}, $r_z=\bfr \bfe_z$ and $\alpha_z=\bfalpha \bfe_z$. 
The normalization constant is
\begin{equation}\label{2.4.32}
N_i^{SM}\;=\;e^{\pi \eta_i/2}\;\Gamma(1-i\eta_i)/(2\pi)^{3/2},
\end{equation}
and $u_{p_i}(\bfzeta_i)$ is the free Dirac spinor relating to the spin polarization $\bfzeta_i$,
\begin{equation}\label{6.4.33}
u_{p_i}(\bfzeta_i)\;=\;\sum_{m_i=\pm \frac12} a_{m_i}\;u_{p_i}^{({\rm sign }\,m_i)} \;=
\;\sum_{m_i=\pm \frac12} a_{m_i}\;\sqrt{\frac{E_i+c^2}{2E_i}}\;{1 \choose \frac{\bfsigmas \bfps_i c}{E_i+c^2}}\;\chi_{m_i},
\end{equation}
the helicity states being identical to the ones in (\ref{2.1.8}).

In order to interrelate the Coulomb-Dirac and the Sommerfeld-Maue functions we investigate their behaviour at small distances.
We make use of the partial-wave expansion of the SM function which for the first term in (\ref{2.4.31}) is given by
(setting $r_z=r\cos \theta)$  \cite{Go28,BM54},
$$N_i^{SM}\;e^{ip_ir \cos \theta}\;_1F_1(i\eta_i,1,ip_ir(1-\cos \theta))\;=\; \frac{1}{(2\pi)^{3/2}}\; \sum_{l=0}^\infty P_l(\cos \theta)\;i^l$$
\begin{equation}\label{2.4.34}
\times\;e^{\pi \eta_i/2}\;\Gamma(l+1-i\eta_i)\;\frac{1}{(2l)!}\;(2p_ir)^l\;e^{-ip_ir}\;_1F_1(l+1+i\eta_i,2l+2,2ip_ir).
\end{equation}
 With $_1F_1(a,b,z)=1$ for $z=0$, these partial waves behave like $r^l$ for $r \to 0$, which means that $\psi^{SM}_i(\bfr,\bfzeta_i) \to $ const for $r \to 0$.

On the other hand, representing the Whittaker function in terms of $_1F_1$ according to \cite{AS64}
\begin{equation}\label{2.4.35}
M_{\alpha,\gamma}(z)\;=\;e^{-z/2}\;z^{\frac12 + \gamma}\;_1F_1(\frac12 -\alpha +\gamma, 2\gamma+1,z),
\end{equation}
it follows that the Coulomb-Dirac partial waves behave like $r^{\gamma_i -1} $ for $r \to 0$, such that (with $\kappa_i=-1$ for the lowest partial wave) 
$\psi_i^C(\bfr,\bfzeta_i)$ diverges like $r^{-(1-\sqrt{1-(Z_T/c)^2})}$. In the SM approximation, this divergence is avoided by omitting the term $(Z_T/c)^2$ in $\gamma_i$.

Bethe and Maximon \cite{BM54} argued that the replacement
\begin{equation}\label{2.4.36}
\gamma_i \;=\;\sqrt{\kappa_i^2-(Z_T/c)^2}\;\mapsto\; |\kappa_i|
\end{equation}
in  all Dirac-Coulomb partial waves is sufficient to reproduce the partial waves of the Sommerfeld-Maue function. The result  
$\psi_i^{SM}(\bfr,\bfzeta_i)=\psi_i^C(\bfr,\bfzeta_i)|_{\gamma_i \mapsto |\kappa_i|}$
was verified
by Yerokhin both analytically and numerically when starting
from the representation (\ref{2.4.27})-(\ref{2.4.29}). The choice of this representation (among others available in the literature) is, however, crucial for (\ref{2.4.36}) to work
 \cite{JY13}.
In fact, according to Bethe and Maximon \cite{BM54}, the correspondence between the Dirac-Coulomb functions and the SM functions holds
for arbitrary directions of the electron momentum.
 
We recall from Section 2.2 that $\psi^{SM}$ approaches the exact Coulomb-Dirac wavefunction for large distances
and high energies. This, however, is just the situation where large angular momenta come into play (according to the classical formula $l=|\bfr \times \bfp|\,=rp|\sin \theta |$),
for which $\gamma \approx |\kappa|$.
From the classical relation it also follows that for a fixed angular momentum $l$ in a given  physical process, the corresponding angle has to be the smaller, the higher the energy. 

\subsection{The Born limit}

In Section 2.2 it was argued that not only the Sommerfeld-Maue wavefunctions, but also the analytical Sommerfeld-Maue bremsstrahlung theory becomes exact for arbitrary nuclear charge number $Z_T$,
when the energy of incoming and scattered electron is high (as compared to the rest energy) and when the photon and electron
emission angles (with respect ot the beam axis) are sufficiently small.
Moreover, it was shown  \cite{BM54} that under these conditions, the Sommerfeld-Maue theory reduces to the relativistic Born approximation as formulated by Bethe and Heitler \cite{BH34}.

In fact, with $E_i \gg c^2$ and $E_f \gg c^2$ such that one can set $\eta_f = \eta_i=Z_T/c$, the triply differential SM bremsstrahlung cross section for unpolarized particles turns into  \cite{BM54}
$$
\frac{d^3\sigma^{SM}}{d\omega d\Omega_k d\Omega_f}\;=\;R(x)\;\frac{d^3\sigma^{B1}}{d\omega d\Omega_k d\Omega_f},
$$
\begin{equation}\label{2.4.37}
R(x)\;=\;\frac{1}{V^2(1)}\;\left[ V^2(x)\;+\;\eta_i^2\,(1-x)^2\;W^2(x)\right],
\end{equation}
where 
\begin{equation}\label{2.4.38}
V(x)\;=\;_2F_1(-i\eta_i,i\eta_i,1,x),\qquad W(x)\;=\;_2F_1(1-i\eta_i,1+i\eta_i,2,x)
\end{equation}
are real hypergeometric functions, and their argument $x$ is given in (\ref{2.2.14}).

The prefactor $R$ has the property,
\begin{equation}\label{2.4.39}
\lim_{x \to 1}\;R(x)\;=\;1,
\end{equation}
in which case the SM theory coincides with the Born approximation.
For small emission angles (in addition to high energies),
$d_f$ from (\ref{2.1.10}) can be expanded in terms of $\theta_f$ and
$c^2/E_f$ to give
\begin{equation}\label{2.4.42}
d_f\;\approx\; \frac{E_f}{2c}\;\left( \frac{c^4}{E_f^2}\;+\;\theta_f^2 \right),
\end{equation}
with an analogous expression for $d_i$.
Therefore,  $x$ is approximated by \cite{BM54,Lan4}
\begin{equation}\label{2.4.40}
x\;=\;1\;-\;\frac{\omega^2}{4q^2c^2}\;\left( \frac{c^4}{E_iE_f}\right)^2\;\left( 1\;+\;\frac{\theta_k^2E_i^2}{c^4}\right)\;\left(1\;+\;\frac{\theta_f^2E_f^2}{c^4}\right),
\end{equation}
where $\bfq=\bfp_i-\bfp_f-\bfk$ is the momentum transferred to the nucleus, and $\theta_k, \;\theta_f$ are, respectively, the angles between $\bfk$ and $\bfp_i, \;\bfp_f$.
For $\theta_k \lesssim c^2/E_i$ and $\theta_f \lesssim c^2/E_f$, the argument $x$ is mostly close to unity,
except for collinear emission $(\theta_k=\theta_f=0)$ where $q$ attains its minimum value
and becomes proportional to the photon frequency $\omega$.

In particular, $x \to 1$ for $\omega \to 0$ provided
the scattering angle is nonzero. Therefore we have the following result, valid for arbitrarily strong Coulomb potentials,
\begin{equation}\label{2.4.41}
\frac{d^3\sigma}{d\omega d\Omega_k d\Omega_f}\;=\;\frac{d^3\sigma^{SM}}{d\omega d\Omega_k d\Omega_f}\;=\;\frac{d^3\sigma^{B1}}{d\omega d\Omega_k d\Omega_f}\qquad \mbox { for } E_i \to \infty,\; \omega \to 0,
\end{equation}
if the emission is noncollinear. 
For a fixed collision energy, the Born limit is approached the faster by the exact theory, the smaller the emission angles and the smaller the photon frequency.
This is shown in  figure Fig.2.4.1 where the photon spectrum for 3.5 MeV electrons colliding with a gold target, obtained from the exact theory and from the SM theory, is compared to the Born approximation. We have chosen the case where photon ($\theta_k=20^\circ$) and electron ($\vartheta_f=30^\circ$) are emitted in coplanar geometry to the same side of the beam axis (where the intensity is high, see, e.g. \cite{HN04,Jaku18}. For emission to opposite sides, see Fig.4.2.2).

Let us now proceed from the triply differential to the doubly differential bremsstrahlung cross section.
In the Born approximation, the dominant contribution to the bremsstrahlung intensity originates from very small  angles, if the collision energy is ultrarelativistic.
This follows from the Bethe-Heitler formula (\ref{2.1.11}) containing the denominators $d_i$ and $d_f$.
According to (\ref{2.4.42}), they are smallest and hence the photon intensity maximum,
when $\theta_k \lesssim c^2/E_i$ and $\theta_f \lesssim c^2/E_f$.
In the case $E_f \approx E_i \gg c^2,$ i.e. $\theta_f \approx \theta_k$, the angular distribution has the shape \cite{AB62}
\begin{equation}\label{2.4.43}
\frac{d\sigma^{B1}}{d\theta_k} \;=\;A\;\frac{\theta_k}{[\theta_k^2+(c^2/E_i)^2]^2}\;\left[\, \ln (1\,+\,\theta_k^2c^4/E_i^2)\;+\;B\right],
\end{equation}
where $A$ and $B$ are constants.
From this formula it is obvious that the forward focusing increases with $E_i$.
The zero at $\theta_k=0$, predicted by (\ref{2.4.43}), is, however, changed to a shallow minimum when the collision energy is lower (such that the PWBA becomes inappropriate).

From these considerations it becomes clear that only small electron angles
give a significant contribution to the doubly differential cross section. Hence
  the validity criteria for (\ref{2.4.41}) remain satisfied when the integration over the solid angle of the emitted electron is performed. As a consequence, the equalities in (\ref{2.4.41}) hold also for the doubly differential cross section,
\begin{equation}\label{2.4.44}
\frac{d^2\sigma}{d\omega d\Omega_k}\;=\;\frac{d^2\sigma^{SM}}{d\omega d\Omega_k}\;=\;\frac{d^2\sigma^{B1}}{d\omega d\Omega_k} \qquad \mbox{ for } E_i \to \infty, \;\omega \to 0,
\end{equation}
for forward (nonzero) photon angles. The approach of these limits with decreasing $\omega$ is also demonstrated in figure Fig.2.4.1.

\vspace{0.5cm}

\setcounter{figure}{0}
\begin{figure}
\vspace{-1.5cm}
\centering
\begin{tabular}{cc}
\hspace{-1cm}\includegraphics[width=.7\textwidth]{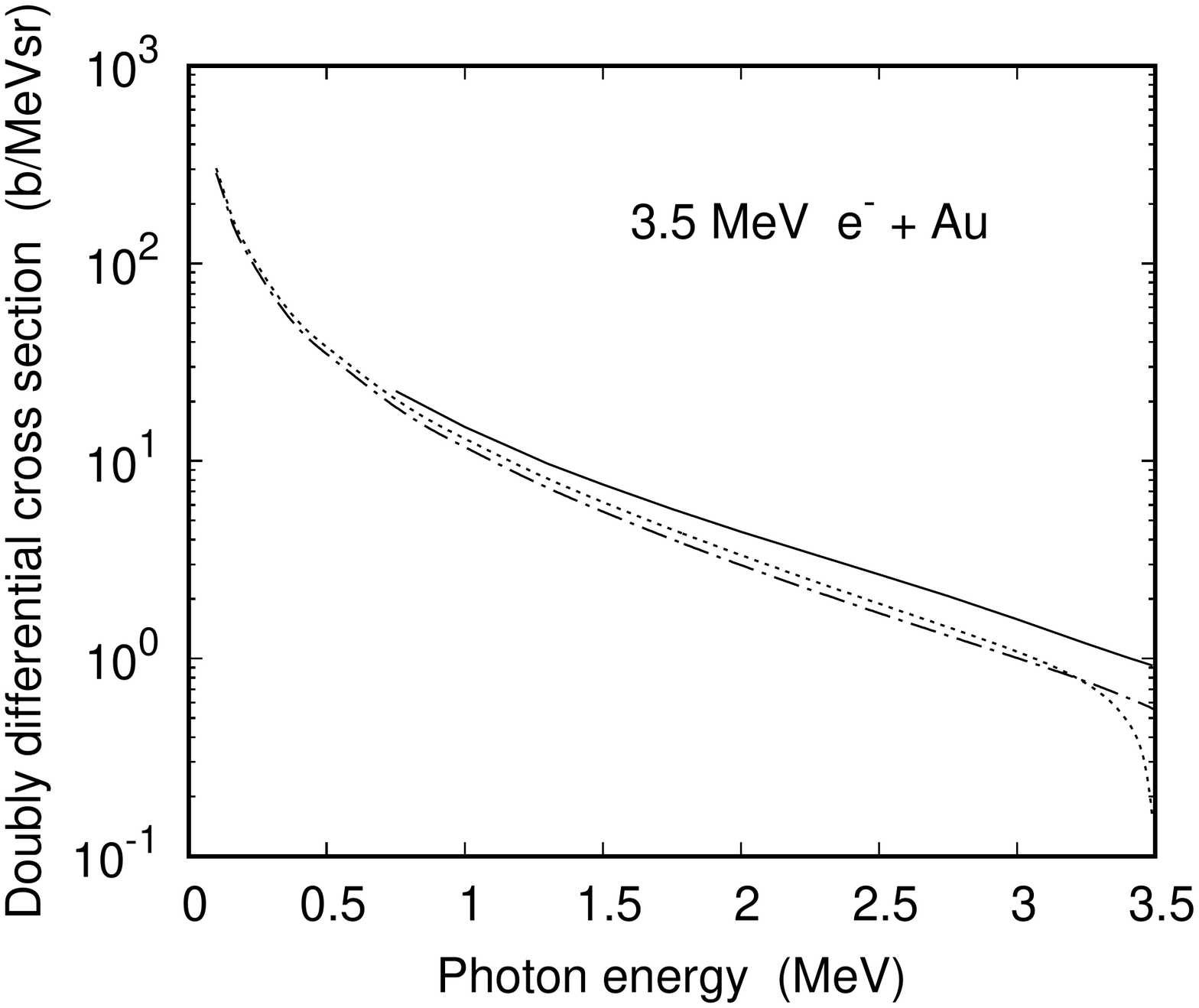}&
\hspace{-3cm} \includegraphics[width=.7\textwidth]{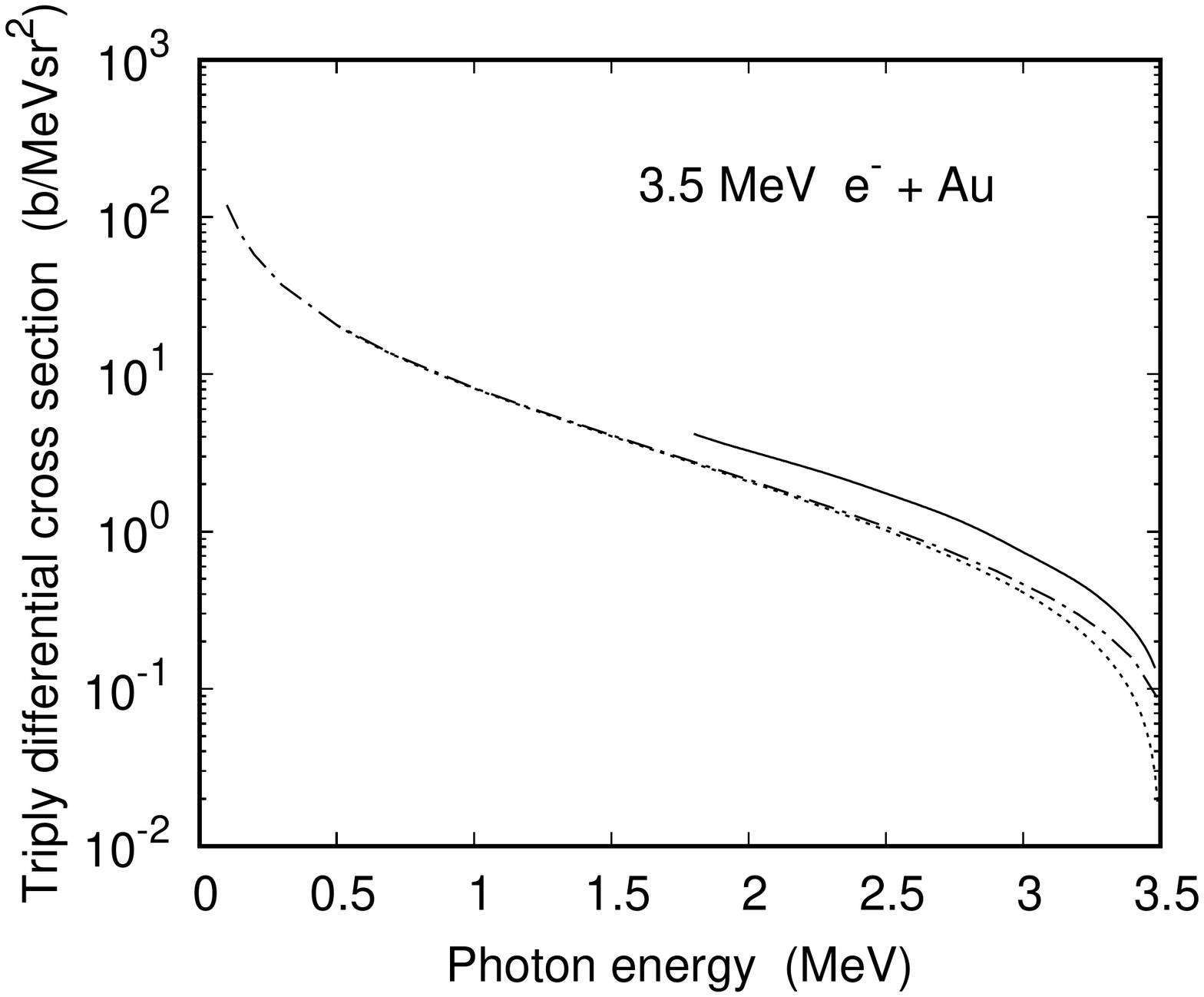}
\end{tabular}
\caption{
Doubly differential (left) and triply differential (right) cross section for bremsstrahlung emission
 from 3.5 MeV electrons colliding with gold as a function of photon frequency $\omega$. Shown are results from the partial-wave theory (--------------),
 the SM theory ($-\cdot -\cdot -)$ and the PWBA ($\cdots\cdots)$.
Left: Photon angle $\theta_k=21^\circ$.
Right: Photon angle $\theta_k=20^\circ$, electron angle $\vartheta_f=30^\circ$, azimuthal angle $\varphi_f = 0$.
}
\end{figure}

It should be pointed out that the feature of extreme forward peaking of the ultrarelativistic bremsstrahlung cross section
is far more general than the PWBA. In fact, it is a consequence of the Lorentz transformation from the emitter frame (the rest system of the beam electron, which is referred to by primed quantities) to the
laboratory frame (the rest frame of the target atom, described by unprimd quantities).
The momenta of scattered electron and photon transform according to
$$ E_f\;=\;\gamma\;(E_f'+vp_f'\cos \vartheta_f'),$$
\begin{equation}\label{2.4.45}
p_f \cos \vartheta_f\;=\;\gamma\;\left( \frac{vE_f'}{c^2}\;+\,p_f'\cos \vartheta_f'\right)
\end{equation}
and
$$\omega\;=\;\gamma \omega'\;\left( 1\;+\;\frac{v}{c}\;\cos \theta_k'\right)$$
\begin{equation}\label{2.4.46}
\cos \theta_k\;=\;\frac{\cos \theta_k'+\frac{v}{c}}{1+\frac{v}{c}\cos \theta_k'}\,,
\end{equation}
where $\gamma = (1-v^2/c^2)^{-1/2}$ and $v=\frac{c^2p_i}{E_i}$ is the collision velocity   \cite{Ha63}.
The emission of a photon by a slowly moving particle can be described within the nonrelativistic dipole approximation.
For small scattering angles, the photon angular distribution is proportional to $\sin^2 \vartheta_k'$, which is peaked at $\vartheta_k'=\pi/2 $  \cite{HN04}.
Using the relation $\theta_k'=\pi-\vartheta_k'$ which is a consequence of the reversed quantization axes in emitter and target frames,
 we obtain from (\ref{2.4.46}) for this angle,
\begin{equation}\label{2.4.47}
\cos \theta_k\;=\;\frac{v}{c}\;=\;\frac{\sqrt{E_i^2-c^4}}{E_i}\;=\;1\;-\;\frac{c^4}{2\,E_i^2}\;+\;O(c^4/E_i^2)^2.
\end{equation}
For $E_i \gg c^2$ it follows that $\cos \theta_k \approx 1$.
Using the expansion of the cosine at small arguments, $\cos \theta_k = 1 - \frac12 \theta_k^2 +O(\theta_k^4),$
we get the result
\begin{equation}\label{2.4.48}
\theta_k\;\approx\; \frac{c^2}{E_i}
\end{equation}
for the angle corresponding to the maximum photon intensity in the laboratory frame.
This agrees with the earlier result that the scattering angles of importance are inversely proportional to the collision energy.

We note that from the Lorentz-invariance of the phase-space elements $c^2d\bfp_f/E_f$ and $c^2d\bfk/\omega$  \cite{BD},
the triply differential bremsstrahlung cross section transforms according to
\begin{equation}\label{2.4.49}
\frac{d^3\sigma}{d\omega d\Omega_k d\Omega_f}\;=\;\frac{\omega}{\omega'}\;\frac{p_f}{p_f'}\;\frac{dE_f}{dE_f'}\;\frac{d^3\sigma'}{d\omega' d\Omega_k' d\Omega_f'},
\end{equation}
where, from (\ref{2.4.46}), $\frac{\omega}{\omega'}=\gamma$ for $\theta'_k=\frac{\pi}{2}$, while
the remaining prefactors on the rhs are independent of photon angle.
Thus the angular position of the bremsstrahlung peak is not affected by this transformation. 

\section{The Dirac-Sommerfeld-Maue (DSM) model}
\setcounter{equation}{0}

The DSM model is a hybrid model which covers extremely high collision energies (up to 500 MeV), made possible by the choice of a Sommerfeld-Maue wavefunction for the initial state. 
However, it is only
applicable at the short-wavelength limit (SWL) of the bremsstrahlung spectrum, due to the (time-consuming) partial-wave representation of the electronic final state even at low kinetic energies.
This hybrid model was put forth in a quantum mechanical prescription, to be presented below \cite{Jaku10}, and in the quasiclassical theory, based on the Greens function method, 
where the angle-integrated spectrum 
averaged over the electron spin polarization and summed over the photon polarization,
\begin{equation}\label{2.5.11}
\frac{d\sigma}{d\omega}\;=\;\frac12\;\sum_{\zeta_i,\lambda} \int d\Omega_k\;\frac{d^2\sigma}{d\omega d\Omega_k}(\bfzeta_i,\bfe_\lambda^\ast),
\end{equation}
was considered \cite{PM10}.

From section 2.2 we know that the SM functions become exact, even for heavy nuclei, when the total projectile energy is large, $E_i \gg c^2$  (typically $E_i \gtrsim 20$ MeV), and
at large internuclear distances, which are related to photon emission into the forward hemisphere.
For photons  near the SWL, the scattered electron
is slow and the corresponding Dirac wavefunction requires only a few partial waves. Typically, the inclusion of the $j=\frac12$ and $j=\frac{3}{2}$ states 
provides an accuracy in the percent region, except possibly in the extreme forward direction  \cite{JP63} (see also section 2.5.3). This has been verified by a comparison with SWL 
results from the partial-wave theory at a moderate collision energy,  $25-35$ MeV.

\subsection{The DSM formalism}

In the DSM theory the exact radiation matrix element (\ref{2.4.7}) is replaced by
\begin{equation}\label{2.5.1}
W_{\rm rad}^{\rm DSM} \;=\;\int d\bfr\;\psi_f^+(\bfr,\bfzeta_f)\;(\bfalpha \bfe_\lambda^\ast)\;e^{-i\bfks \bfrs}\;\psi_i^{SM}(\bfr,\bfzeta_i),
\end{equation}
with the Sommerfeld-Maue function $\psi_i^{SM}$  given by (\ref{2.2.5}).
The function $\psi_f$ is taken as the exact solution to the Dirac equation, with a partial-wave expansion according to (\ref{2.4.1}).

In order to profit from an analytical representation of the scattering states, the nuclear potential is chosen to be the Coulomb field of a point nucleus 
in concord with the choice of the Coulombic $\psi_i^{SM}$. 
This means that nuclear size effects have to be negligible, which implies that the photon emission has to
take place at large electron-nucleus distances,
corresponding to small photon angles at ultrahigh energies.

At the short-wavelength limit where the scattered electron is left with zero momentum $\bfp_f$, the radial Coulomb-Dirac functions have a simple
representation in terms of Bessel functions $J_\nu$
 \cite{BZP69},
$$g_{\kappa_f}(r)\;=\;-\;\frac{1}{r}\,(\mbox{sign }\kappa_f)\;\frac{1}{\sqrt{p_fZ_T}}
\left[ \sqrt{2Z_Tr}\,J_{2 \gamma_f-1}(\sqrt{8 Z_Tr})\;-\;(\gamma_f+\kappa_f)\;J_{2\gamma_f}(\sqrt{8Z_Tr})\right],$$
\begin{equation}\label{2.5.2}
f_{\kappa_f}(r)\;=\;\frac{1}{r}\;(\mbox{sign }\kappa_f)\;\frac{1}{c}\;\sqrt{\frac{Z_T}{p_f}}\;J_{2\gamma_f}(\sqrt{8Z_Tr}),
\end{equation}
where $\gamma_f=\sqrt{\kappa_f^2-(Z_T/c)^2}.$

Since the angular integration can no longer be performed analytically, it is convenient to introduce spherical coordinates, $\bfr=(r,\theta,\varphi)$,
and to take the $z$-axis along $\bfp_i$.
Then $\psi^{SM}$ from (\ref{2.4.31}) can be written in the following way,
$$\psi^{SM}_i(\bfr,\bfzeta_i)\;=\;\tilde{\psi}_i(\bfr)\;u_{p_i}(\bfzeta_i),$$
$$\tilde{\psi}_i(\bfr)\;=\;N_i^{SM}e^{ip_ir \cos \theta}\left\{ _1F_1(i\eta_i, 1,ip_ir(1-\cos \theta))\right.$$
\begin{equation}\label{2.5.3}
\left. +\;\frac{iZ_T}{2c}\left[ \alpha_z(\cos \theta-1)+\frac12 \alpha_- \sin \theta e^{i\varphi}+\frac12 \alpha_+ \sin \theta e^{-i\varphi}\right] {_1F_1}(1+i\eta_i,2,ip_ir(1-\cos \theta))\right\},\end{equation}
where $\alpha_\pm = \alpha_x \pm i\alpha_y$ and
\begin{equation}\label{2.5.4}
u_{p_i}(\bfzeta_i)\;=\;a_\frac12 u_{p_i}^{(+)}\;+\;a_{-\frac12} u_{p_i}^{(-)},
\end{equation}
with $u_{p_i}^{(+)}$ and $u_{p_i}^{(-)}$ the free-electron 4-spinors from (\ref{2.1.8}) describing the two helicity eigenstates.

The insertion of (\ref{2.5.3}) and (\ref{2.4.1}) with (\ref{2.5.2}) into the radial matrix element (\ref{2.5.1}), and the use of the decomposition (\ref{2.4.9}) of the final vector spherical harmonics $Y_{j_fl_fm_f}(\hat{\bfr})$, involves the following spatial integrals,
$$R_1\;=\;\int d\bfr\;g_{\kappa_f}(r)\;Y^\ast_{l_fm_l}(\hat{\bfr})\;e^{-i\bfks\bfrs}\tilde{\psi}_i(\bfr),$$
\begin{equation}\label{2.5.5}
R_2\;=\;\int d\bfr\;f_{\kappa_f}(r)\;Y^\ast_{l_f'm_l}(\hat{\bfr})\;e^{-i\bfks\bfrs}\tilde{\psi}_i(\bfr),
\end{equation}
with $m_l=m_f-m_s$.

With the representation of $Y_{l_fm_l}(\hat{\bfr})$ in terms of $P_{l_f}^{m_l}(\cos \theta) e^{im_l\varphi}$ according to (\ref{2.4.5}),
and writing $\bfk \bfr=kr(\cos \theta \cos \theta_k+\sin \theta \sin \theta_k \cos \varphi)$,
the azimuthal integrals can be performed analytically with the help of
\begin{equation}\label{2.5.6}
\int_0^{2\pi} d\varphi\; e^{-i\mu\varphi} e^{-ikr \sin \theta \sin \theta_k \cos \varphi}\;=\;2\pi (-i)^{|\mu|}J_{|\mu|}(kr \sin \theta \sin \theta_k),
\end{equation}
where $\mu=m_l -s$ and $s \in \{0,1,-1\}$.

The double integrals over the remaining degrees of freedom, $r$ and $\theta$, have to be performed numerically.

\subsection{Numerics and examples}

The particular spatial integral which has the weakest decay at $r \to \infty$ as compared to the other integrals involved in (\ref{2.5.5}), and hence has the poorest convergence property, is the following,
$$I_1\;=\;\int_0^\infty r^{3/2}dr\;J_{2\gamma_f-1}(\sqrt{8Z_Tr})\int_{-1}^1 d(\cos \theta) \;P_{l_f}^{m_l}(\cos \theta)\;J_{|\mu|}(kr\sin \theta \sin \theta_k)$$
\begin{equation}\label{2.5.7}
\times\;e^{i(p_i-k\cos \theta_k)r\cos \theta} {_1F_1}(i\eta_i,1,ip_ir(1-\cos \theta)).
\end{equation}
The other double integrals result from replacing $J_{2\gamma_f-1}(\sqrt{8Z_Tr})$ by $r^{-1/2}J_{2\gamma_f}(\sqrt{8Z_Tr})$ or from replacing the confluent hypergeometric function in (\ref{2.5.7}) by
$h(\theta)_1F_1(1+i\eta_i,2,ip_ir(1-\cos \theta))$ where $h(\theta) \in \{1,\cos \theta,\sin \theta\}.$
Paired with the weak decay in $r$, all integrands are strongly oscillating and hence cannot be evaluated in a straightforward way.
In contrast to the case of the single integrals occurring in the relativistic partial-wave theory,
the complex-plane rotation method is not applicable for the present radial integrals, because of the additional dependence on $\theta$.
However, integrals of the type $\int_0^\infty r^\lambda dr\,J_\mu(r)\;$ do converge for $\lambda +\mu > -1$,
which can be shown by introducing a convergence generating funtion $e^{-\epsilon r}$
and by considering the limit $\epsilon \to 0 $ \cite{Grad},
\begin{equation}\label{2.5.8}
\lim_{\epsilon \to 0} \int_0^\infty r^\lambda dr\;J_\mu(ar)\;e^{-\epsilon r}\;=\;\frac{2^\lambda}{a^{\lambda +1}}\;\frac{\Gamma(\frac{\mu+\lambda+1}{2})}{\Gamma(\frac{\mu-\lambda+1}{2})},\qquad a>0.
\end{equation}

Using the fact that the angular integral in (\ref{2.5.7}) is a smooth function  of $r$, to be termed $M_1(r)$, one can apply this technique by evaluating
\begin{equation}\label{2.5.9}
I_1(\epsilon)\;=\;\int_0^\infty r^{3/2}dr\;J_{2\gamma_f-1}(\sqrt{8Z_Tr})\;M_1(r)\;e^{-\epsilon r},
\end{equation}
where $\epsilon$ (which depends on $Z_T$ and on the photon momentum) is chosen small enough such that $I_1(\epsilon)$ changes in a negligible way when $\epsilon$ is further decreased.
The inclusion of four final-state partial waves ($\kappa_f=\pm1,\pm2$) necessitates the simultaneous evaluation of 23 double integrals, which are all handled as indicated in (\ref{2.5.9}).
The consideration of higher partial waves would lead to a prohibitively long  computation time.

In Fig.2.5.1 the angular dependence of the doubly differential bremsstrahlung cross section,
\begin{equation}\label{2.5.10}
\frac{d^2\sigma^{DSM}}{d\omega d\Omega_k}\;=\;\frac{4\pi^2 \omega \,p_f E_iE_f}{c^5p_i}\;
\sum_{\kappa_f m_f}\left| \int d\bfr {g_{\kappa_f}(r)\,Y_{j_fl_fm_f}(\hat{\bfr}) \choose if_{\kappa_f}(r)\,Y_{j_fl'_fm_f}(\hat{\bfr})}^+\;(\bfalpha \bfe_\lambda^\ast)\;e^{-i\bfks \bfrs}\;\psi_i^{SM}\right|^2,
\end{equation}
at $\omega=E_e\equiv E_i-c^2$ from 500 MeV electrons colliding with Cu and Au is shown in comparison with the result from the Sommerfeld-Maue theory
as derived in section 2.2.
The agreement of the DSM and the SM results for Cu at angles above $0.1^\circ$ proves the validity of the SM wavefunctions for
weak potentials as well as the correctness of the numerical integrals in the DSM theory.

\setcounter{figure}{0}
\begin{figure}
\vspace{-1.5cm}
\includegraphics[width=11cm]{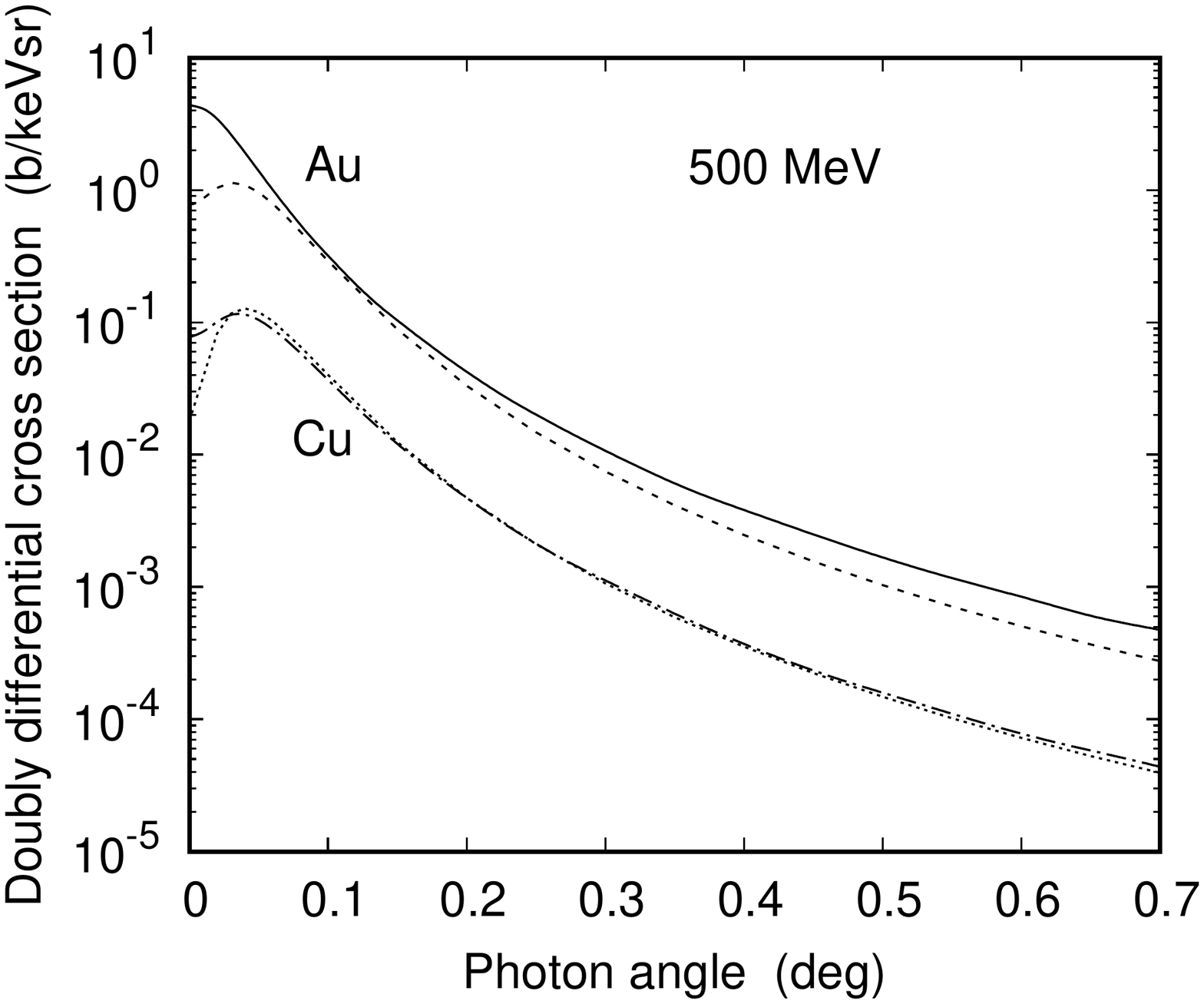}
\caption{
Doubly differential cross section for bremsstrahlung emission from 500 MeV electrons colliding with Cu $(Z_T=29$, lower curves) and Au ($Z_T=79$, upper curves) at the SWL ($\omega = 500 $ MeV)
as a function of photon angle. 
DSM: ---------, Au; $-\cdot - \cdot -$, Cu. SM: $-----$, Au; $\cdots\cdots$, Cu.
}
\end{figure}

\subsection{Bremsstrahlung at ultrarelativistic energies}

When the energy of the impinging electron is extremely large, $E_i \gg c^2$,
a simple approximation to the Sommerfeld-Maue function $\psi_i^{SM}$ is possible, thus leading to a more tractable form of the transition matrix element (\ref{2.5.1}).
In turn, this allows for an approximate analytical formula for the doubly differential cross section \cite{PM12}.
 Even more, this choice of initial state results in  an exact analytical expression for the
angle-integrated singly differential cross section
(\ref{2.5.11}),
see  \cite{JP63}. 
We start with deriving the asymptotic form of $\psi_i^{SM}$  which is used in these approaches, following Pratt \cite{P60}.

Starting point is the second-order differential equation (\ref{2.2.2}) which was derived from the Dirac equation.
We split off the plane-wave factor by making the substitution $\psi(\bfr)=e^{i\bfps_i\bfrs} F(\bfr) u_{p_i}$,
where $u_{p_i}$ is the free 4-spinor.
Upon insertion into (\ref{2.2.2}) we obtain the following equation for $F(\bfr)u_{p_i}$,
\begin{equation}\label{2.5.12}
\left[ 2ic^2\bfp_i\bfnabla\,+\,c^2\bfnabla^2\,-\,2E_iV\,+\,V^2\,+\,ic\bfalpha (\bfnabla V)\right]\;F(\bfr) u_{p_i}\;=\;0.
\end{equation}
We recall that neglecting $V^2$ results, upon iteration,  in the Sommerfeld-Maue function (\ref{2.2.5}), and further neglecting the gradient term $ic\bfalpha (\bfnabla V)$ results in the function $\psi_a$ of (\ref{2.2.3}), which corresponds
to the first term of the SM function (\ref{2.2.5}).
In concord with neglecting slowly varying (gradient) terms, we now omit $c^2\bfnabla^2$ in addition, and are thus left with a scalar equation
for $F(\bfr)$,
\begin{equation}\label{2.5.13}
(ic^2\bfp_i\bfnabla\,-\,E_iV)\;F(\bfr)\;=\;0.
\end{equation}
The solution is of exponential form, $F(\bfr)=e^{i\chi(\bfrs)}$, where
\begin{equation}\label{2.5.14}
\chi(\bfr)\;=\;-\frac{E_i}{c^2p_i}\int^z dz'\;V(r)\;+\;\mbox{const},
\end{equation}
and we have chosen the $z$-direction along $\bfp_i$, i.e. $\bfp_i =p_i\bfe_z$. For a pure Coulomb field $V(r)=-Z_T/r$, we obtain
\begin{equation}\label{2.5.15}
\psi(\bfr)\;=\;e^{i\bfps_i\bfrs}\;e^{-i\eta_i\ln(r-z)\,+i\,{\rm const}}\;u_{p_i}
\end{equation}
for an incoming wave, where we have profited from the fact that $V(r)$ is even in $z$ to make the substitution $z'=-z''$ in the integrand.

In order to determine the constant in (\ref{2.5.15}), we consider the asymptotic form of the confluent hypergeometric function $_1F_1(i\eta_i,1,i(p_ir-\bfp_i\bfr))$, the essential part of $\psi_a$, for large arguments $|p_ir-\bfp_i\bfr|$.
In fact, this function can be split into the sum of an incoming wave and a scattered wave.
Retaining only the incoming part, one gets  \cite[\S6.1]{Jo83}
\begin{equation}\label{2.5.16}
_1F_1(i\eta_i,1,i(p_ir-\bfp_i\bfr))\;=\;\frac{e^{-\pi\eta_i/2}}{\Gamma(1-i\eta_i)}\;e^{-i\eta_i \ln(p_ir-\bfps_i\bfrs)}
\end{equation}
$$+\;O\left(\frac{1}{p_ir-\bfp_i\bfr}\right).$$
Hence one obtains the asymptotic function, using the first term of (\ref{2.2.5}),
\begin{equation}\label{2.5.17}
\psi_i(\bfr)\;\approx\; \frac{1}{(2\pi)^{3/2}}\;e^{i\bfps_i\bfrs}\;e^{-i\eta_i\ln(p_ir-p_iz)}\,u_{p_i}\;=\;\frac{p_i^{-\eta_i}}{(2\pi)^{3/2}}\;e^{i\bfps_i\bfrs}\;(r-z)^{-i\eta_i}\;u_{p_i},
\end{equation}
such that the constant in (\ref{2.5.15}) can be identified with
\begin{equation}\label{2.5.18}
e^{i{\rm const}}\;=\;\frac{p_i^{-i\eta_i}}{(2\pi)^{3/2}}\;.
\end{equation}
For the validity of (\ref{2.5.17}) we must require that the gradient term neglected in (\ref{2.5.13}) is small,
more precisely, that
\begin{equation}\label{2.5.19}
\frac{c^2\bfnabla^2 F}{2E_iVF}\;\ll\;1.
\end{equation}
A straightforward calculation leads to
\begin{equation}\label{2.5.20}
 \frac{\eta_i}{p_i(r-z)}\;\ll\;1,
\end{equation}
in concord with the condition $|p_i(r-z)|\gg 1$, which is required for the validity of the asymptotic expansion (\ref{2.5.16}).

With the function $\psi_i(\bfr,\bfzeta_i)$ from (\ref{2.5.17}) with (\ref{2.5.4}),
 the singly differential cross section for bremsstrahlung can be evaluated according to Jabbur and Pratt \cite{JP63}.
Table 2.5.1 gives a comparison of the results for the singly differential cross section (\ref{2.5.11}) for a lead target, obtained with the DSM on one hand and using the asymptotic wavefunction 
in an analytical and a numerical approach as calculated  in \cite{JP64}, on the other hand. For their results it is assumed that $d\sigma/d\omega$ scales with the inverse collision energy, $E_e^{-1}$.

\vspace{0.5cm}

Table 2.5.1.
Singly differential cross section $d\sigma/d\omega$ (in b/MeV) for tip bremsstrahlung emission from electrons colliding with Pb at  collision energies from 35-500 MeV.
Shown are the DSM results ($2^{\rm nd}$ column) and the numerical results from Jabbur and Pratt \cite{JP64}
 ($4^{\rm th}$ column) by using the asymptotic wavefunction (and including the $s_{1/2},\;p_{1/2},\;p_{3/2}$ and $d_{3/2}$ partial waves). The third column gives Jabbur and Pratt's analytical results (without $d_{3/2}$ partial waves).
The last column gives the percent deviation of the DSM results from $d\sigma^{JP}/d\omega({\rm numer})$.

\vspace{0.2cm}
\begin{tabular}[t]{r||c|c|c|c|}  
 $E_e$& $\frac{d\sigma^{DSM}}{d\omega}$& $\frac{d\sigma^{JP}}{d\omega}({\rm analyt})$ & $\frac{d\sigma^{JP}}{d\omega}({\rm numer})$&$\Delta^{DSM}(\%)$\\
&&&\\ \hline
&&&\\ 
 500& $1.750\times 10^{-2}$&$1.804 \times 10^{-2}$ & $1.762 \times 10^{-2}$&0.7\\ 
200&$4.381 \times 10^{-2}$&$ 4.510 \times 10^{-2}$&$ 4.405 \times 10^{-2}$&0.6\\
100&$8.782 \times 10^{-2}$&$9.022 \times 10^{-2}$&$ 8.810 \times 10^{-2}$&0.3\\
35&0.2551&0.2577&0.2517&1.4\\
\end{tabular}

\vspace{0.5cm} 

It is seen that there is  very good agreement between the different approaches, validating the asymptotic
approach, as well as the scaling with $1/E_e$, down to 35 MeV. 

For collision energies beyond, say, 50 MeV the asymptotic wavefunction (\ref{2.5.17}) can be used to extend the DSM theory to lower photon frequencies.
This is done by replacing in the cross section formula (\ref{2.5.10}) the SM function $\psi_i^{SM}$ by (\ref{2.5.17}) , and by replacing the radial zero-energy Dirac functions $g_{\kappa_f}$ and $f_{\kappa_f}$ with solutions to the Dirac equation for a given finite kinetic energy.
However, the required number of partial waves increases strongly with $E_f$, such that the energy of the scattered electron has to be
restricted to at most a few MeV.

The method of evaluating the radiation matrix element in (\ref{2.5.10}) is the same as in the DSM theory.
Considering explicitly the spin degrees of freedom, the doubly differential cross section
is calculated from this DaSM theory,
\begin{equation}\label{2.5.21}
\frac{d^2\sigma^{\rm DaSM}}{d\omega d\Omega_k}(\bfzeta_i,\bfe^\ast_\lambda)\;=\;\frac{\omega \,p_fE_iE_f}{2\pi\,c^5\,p_i}\sum_{\kappa_f m_f}\left| A_{fi}(\bfzeta_i,\bfe^\ast_\lambda)\right|^2
\end{equation}
with
$$A_{fi}(\bfzeta_i,\bfe^\ast_\lambda)\;=\;\frac{1}{4\pi} \sum_{m_{s_f}=\pm \frac12}
\left\{ \sqrt{2l_f+1}\;\sqrt{\frac{(l_f-\mu_f)!}{(l_f+\mu_f)!}}\;(l_f\mu_f\frac12\,m_{s_f}|j_fm_f)\,(\chi^+_{m_{s_f}}(\bfsigma \bfe_\lambda^\ast) u_l)\;R_{if}(l_f)\right.$$
\begin{equation}\label{2.5.22}
\left. -i\;\sqrt{2l'_f+1}\;\sqrt{\frac{(l'_f-\mu_f)!}{(l'_f+\mu_f)!}}\;(l_f'\mu_f\frac12\,m_{s_f}|j_fm_f)\,(\chi^+_{m_{s_f}}(\bfsigma \bfe_\lambda^\ast)u_u)\;R_{fi}(l_f')\right\},
\end{equation}
where $\mu_f=m_f-m_{s_f}$ and where the definitions of section 2.4 are used.
The upper and lower components of $u_{p_i}$ are termed $u_u$ and $u_l$, respectively.
The two-dimensional spatial integrals are, like the ones occurring in the DSM theory, evaluated with the help of a convergence generating function $e^{-\epsilon r}$.
They are defined by
$$R_{if}(l_f)\;=\;2\pi \,(-i)^{|\mu_f|}
\;\lim_{\epsilon \to 0} \int_0^\infty r^2dr\;e^{-\epsilon r}\,g_{\kappa_f}(r)\,r^{-i\eta_i}\;I(r,l_f)$$
\begin{equation}\label{2.5.23}
R_{fi}(l_f')\;=\;2\pi\;(-i)^{|\mu_f|}
\;\lim_{\epsilon \to 0}\int_0^\infty r^2dr\;e^{-\epsilon r}\,f_{\kappa_f}(r)\;r^{-i\eta_i}\;I(r,l'_f),
\end{equation}
where the momentum transfer components are given by
\begin{equation}\label{2.5.24}
q_z\;=\;p_i\,-\,k\,\cos \theta_k\qquad\quad q_\perp\;=\; k\,\sin \theta_k,
\end{equation}
which agree with those from $\bfq=\bfp_i-\bfp_f-\bfk$ for $p_f=0$ in our choice of coordinate system (section 2.4.1).
$I(r,l)$ denotes the integral over the polar angle $\theta$ (with $x=\cos \theta$),
\begin{equation}\label{2.5.25}
I(r,l)\;= \;\int_{-1}^1 dx\;P_{l}^{\mu_f}(x)\;(1-x)^{-i\eta_i}\;e^{iq_zrx}\,J_{|\mu_f|}(q_\perp r\sqrt{1-x^2}).
\end{equation}
In the pathological case $\mu_f=0$ (which actually is the only case occurring for $\theta_k=0$), the branch point at $x=1$ can be handled by means of a logarithmic substitution, $x=1-e^y$.
It should also be noted that the required cutoff-parameter $\epsilon$, necessary for obtaining stability, is particularly small for high $E_f$, such that the upper limit of the radial integral (and hence the step numbers in both integrals) have to be taken very large.

\vspace{0.2cm}

\begin{figure}
\vspace{-1.5cm}
\centering
\begin{tabular}{cc}
\hspace{-1cm}\includegraphics[width=.7\textwidth]{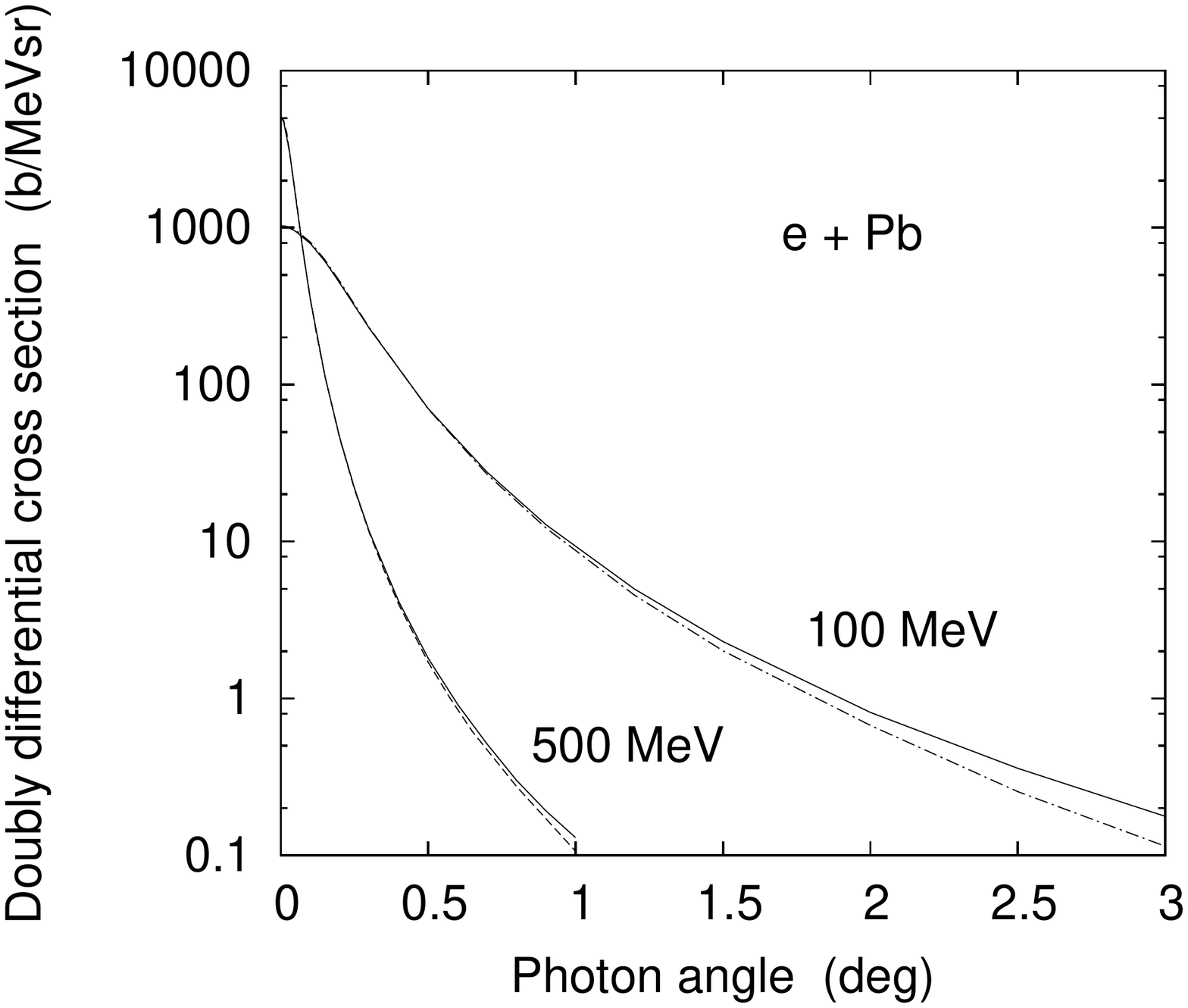}&
\hspace{-3cm} \includegraphics[width=.7\textwidth]{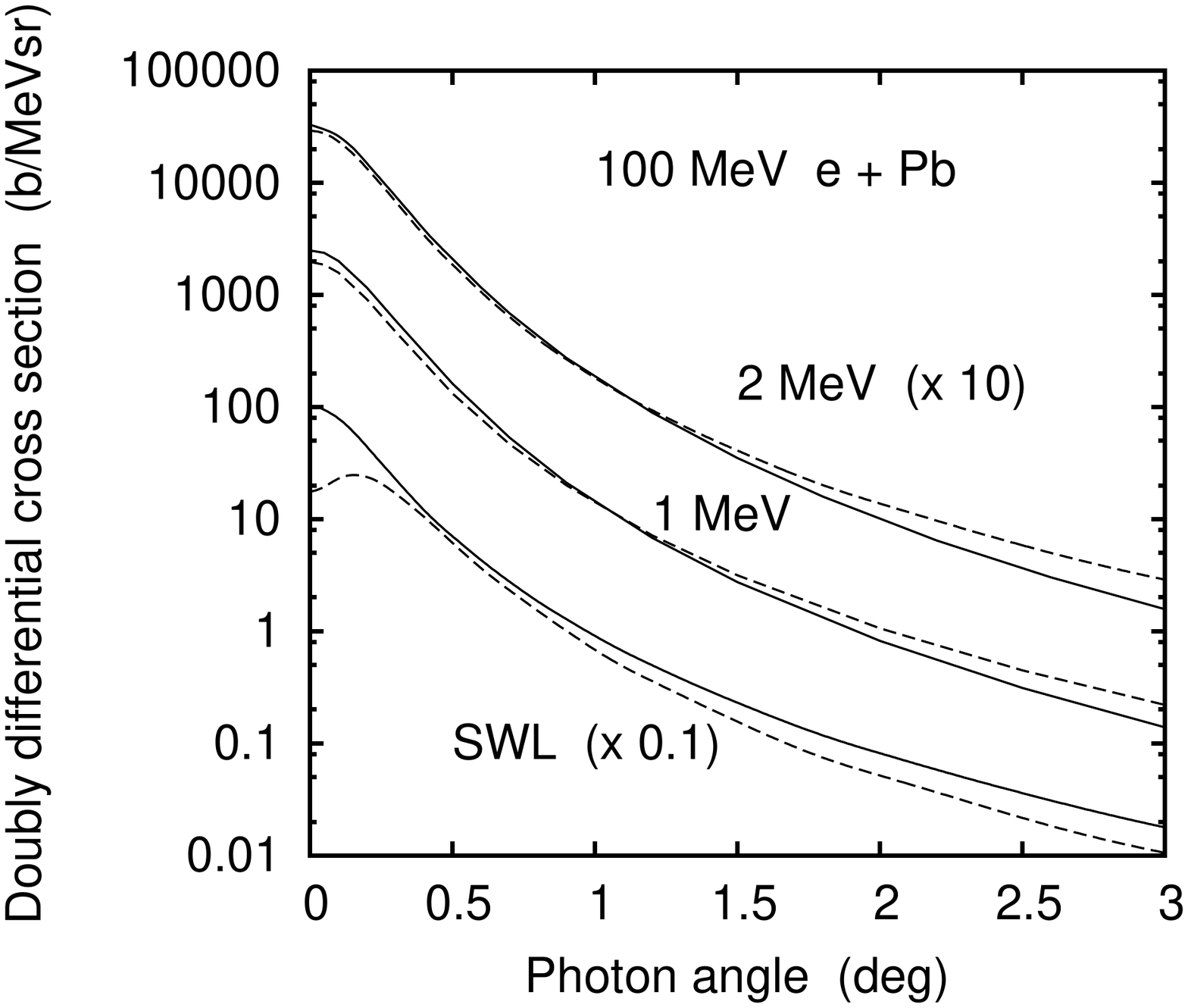}
\end{tabular}
\caption{
Left panel: Doubly differential bremsstrahlung cross section from 100 MeV and 500 MeV electrons colliding with Pb ($Z_T=82$) as a function
of photon angle $\theta_k$.
Shown are SWL results from the DSM theory (-------------: upper curve, 100 MeV; lower curve, 500 MeV) at $E_{f,kin}=0$  in comparison with the DaSM results ($-----$, 500 MeV; $-\cdot -\cdot -$, 100 MeV) at $E_{f,kin}=0.003$ MeV.
Right panel:  Bremsstrahlung from 100 MeV electrons colliding with Pb at final kinetic energies of $0-2$ MeV. ----------, DSM theory at the SWL, DaSM theory at higher $E_f$; $-----$, SM results. Uppermost curves, $E_{f,kin}=2$ MeV, multiplied by a factor of 10.
Middle curves, $E_{f,kin}=1$ MeV. Lowermost curves, SWL (multiplied by a factor of 0.1).
}
\end{figure}

\vspace{0.2cm}

In order to prove the validity of the asymptotic SM function, we compare in
the left panel of Fig.2.5.2  the cross section results  for 100 MeV
and 500 MeV electrons colliding with Pb as obtained from the DSM theory and from the DaSM theory (in which the asymptotic SM function is used) at the short-wavelength limit. There is good agreement between both theories at not too large photon angles.

We can  make use of the scaling property with $E_i$ for the singly differential cross section to derive a similar scaling for the doubly differential cross section. Let us take the cross section at the collision energy of $E_e=100$ MeV 
as a reference value. Then the doubly differential cross section for an arbitrary value of $E_e$ and a given photon angle $\theta_k$  can approximately be calculated from the following substitution \footnote{There is a misprint in Eqs.(4.1) and (4.7) of \cite{Jaku20}, where $\frac{100}{E_{i,kin}}\,\theta_k$ should be replaced by $\frac{E_{i,kin}}{100} \,\theta_k$ on the rhs.},
\begin{equation}\label{2.5.26}
\frac{d^2\sigma}{d\omega d\Omega_k}(E_e,\theta_k)\;=\;\frac{E_e}{100\,{\rm MeV}}\times \frac{d^2\sigma}{d\omega d\Omega_k}(100\, {\rm MeV},\frac{E_e}{100\,{\rm MeV}}\times \theta_k).
\end{equation}
The scaling property 
can be derived from the fact that the photon intensity is basically determined by the momentum transfer. For small $\theta_k$ and $E_i \approx p_ic,\;\;q_z\approx p_i-k \approx E_f/c$ is constant in $E_i$ and $\theta_k$, while
\begin{equation}\label{2.5.27}
q_\perp\;\approx\;k\,\theta_k\;\approx\;(p_i\,-\,E_f/c)\,\theta_k\;\approx\;p_i\,\theta_k
\end{equation}
is constant if $p_i\theta_k$ is constant, i.e. if $\theta_k \sim \,\frac{1}{p_i}\,\approx\,\frac{c}{E_i}.$
Hence the angle scales like $E_i^{-1}$ (note that $E_i\approx E_e$).
A further dependence of the cross section on $\theta_k$ (via the polarization basis vector $\bfe_{\lambda_2}$) is negligibly small.
The scaling of the photon intensity with $E_i$ derives from the normalization constant in (\ref{2.5.21}) for $\omega$ near the SWL,
\begin{equation}\label{2.5.28}
\frac{\omega \,p_fE_iE_f}{2\pi c^5\,p_i}\;\approx\; \frac{E_i^2p_fE_f}{2\pi c^4\,E_i}\;\sim\;E_i.
\end{equation}

For the DSM theory, this scaling formula holds for all angles considered and for all energies $E_i \gtrsim 50$ MeV.
The same is true for the SM results, see Fig.2.5.3.
 For the DaSM approach, it is strictly valid only for the smallest angles, but it improves with collision energy \cite{Jaku20}. This also indicates that the
slightly faster decrease  with angle (as compared to the DSM or SM results) is a deficiency of the DaSM theory.

Having justified the DaSM approach at the SWL, we can employ it to check the validity of the Sommerfeld-Maue theory at finite $E_f$.
We recall that the description of the final state by a SM function requires a sufficiently high $E_f$.
In the right panel of Fig.2.5.2 the  DaSM results are compared to those from the SM theory for final kinetic energies of 1 MeV and 2 MeV. It is found that the deviations between these two theories in fact diminish with $E_f$.
  At 2 MeV the SM theory provides already a quantitative description of the bremsstrahlung process for small angles, despite the fact that the final energy is not yet ultrarelativistic.

If only the dominant terms in the sum over $m_f$ in (\ref{2.5.21}), $m_f=\pm \frac12$, are retained, and if the electron energy is so high that the photons are basically emitted into the forward direction, such that also a small-angle approximation for the angle $\theta$ in the final wavefunction can be applied, the spatial integrals can be performed. To do all three of them analytically, the upper limit of the integral over the polar angle has to be extended to infinity. This approximation causes an unphysical divergence at $\theta_k=0$.
For the evaluation of the doubly differential cross section for unpolarized particles, there
remains thus only the sum over $l_f$ to be carried out numerically \cite{PM12}.

\begin{figure}
\vspace{-1.5cm}
\includegraphics[width=11cm]{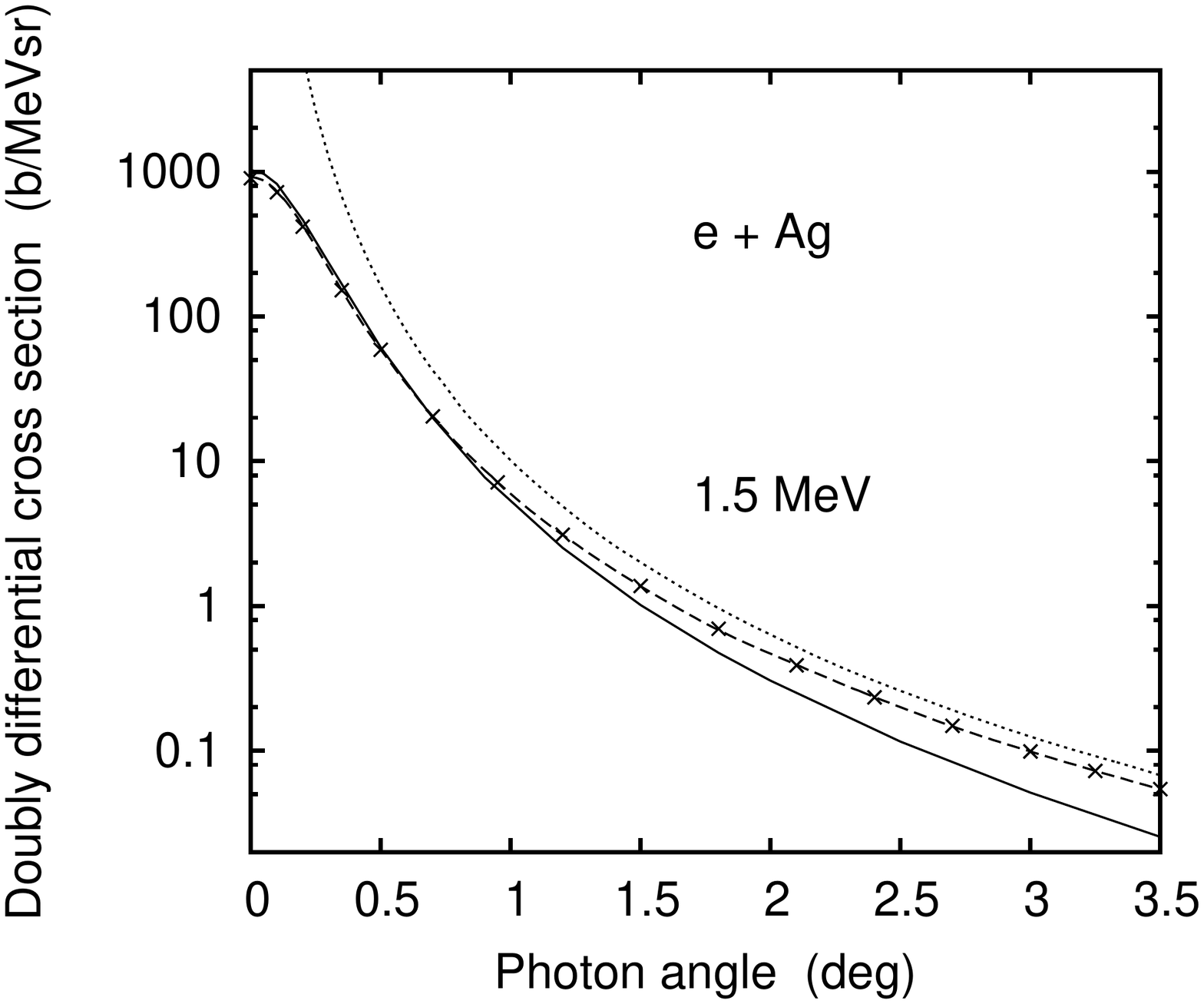}
\caption{
Doubly differential cross section for bremsstrahlung emission from 100 MeV electrons colliding with Ag $(Z_T=47)$
as a function of photon angle. The final energy of the electron is 1.5 MeV.
DaSM: ---------; analytical theory from \cite{PM12}: $\cdots  \cdots $;  SM: $-----$. Included are SM results at 500 MeV, scaled down to 100 MeV according to (\ref{2.5.26}): $\times\times\times$.
}
\end{figure}

Fig.2.5.3 compares this analytical result with the DaSM and the Sommerfeld-Maue theory for 100 MeV e + Ag collisions. It is seen that at the larger angles, it approaches
the SM results, while it fails at those angles which provide the main contribution to the singly differential cross section.

\setcounter{equation}{0}
\section{Screening effects}

The presence of atomic electrons leads to a screening of the electron-nucleus interaction potential, and hence to a reduction of the bremsstrahlung intensity.
It affects the cross section basically for collision energies below a few MeV. Only at very small scattering angles, photon angles and frequencies can screening effects occur at still higher beam energies.
In this section we concentrate on this so-called passive screening. On the other hand, the presence of target electrons may  lead to electron-electron bremsstrahlung, which enhances the photon intensity.
This process, known as active screening, will not be considered here. For the treatment of electron-electron bremsstrahlung, see the book by Haug and Nakel \cite{HN04}.

We start by discussing different types of screened potentials, and subsequently describe methods for incorporating these potentials into the bremsstrahlung theories.

\subsection{Screening potentials}

We will concentrate here on collision energies near or above 1 MeV, which means that the action of the passive target electrons can be accounted for in terms of an effective static electron-atom potential.
While at collision energies around 1 keV the exchange interaction between beam electron and the bound target electrons may still be important, as well as the loss of beam intensity due to inelastic
processes (see, e.g. the detailed discussion in \cite{H16,H18}), such effects can   safely be neglected at the higher energies.

Different kinds of static potentials are used in the literature, and usually they are  spherically symmetric.
In their bremsstrahlung calculations, Tseng and Pratt \cite{TP71} have applied the Thomas-Fermi potential,
and Borie \cite{Bo72} has used the Moli\`{e}re approximation. In \cite{TP71}, also  a self-consistent Hartree-Fock-Slater potential is employed.
 With the help of a numerically obtained  charge density $\varrho_{el}(r)$ of the electron cloud, this  potential is calculated
 from
\begin{equation}\label{2.6.1}
V_T(r)\;=\;-\frac{Z_T}{r}\;+\;\frac{1}{r}\int_0^r dx\,n_{el}(x)\;+\;\int_r^\infty \frac{dx}{x}\;n_{el}(x)\;-\;\frac{2}{3r}\left( \frac{81\,r}{32 \pi^2}\;n_{el}(r)\right)^{1/3},
\end{equation}
where $n_{el}(r)=4\pi r^2\varrho_{el}(r)$ is introduced.
Analytical representations are of the form
\begin{equation}\label{2.6.2}
V_T(r)\;=\;-\;\frac{Z_T}{r}\sum_{i=1}^n a_i e^{-b_ir},
\end{equation}
with $a_i,\;b_i$ and $n$ fit parameters which have to obey
$\sum_{i=1}^n a_i=1$ and $b_i>0$.
A tabulation can be found in  \cite{Sal87}.

Yerokhin and Surzhykov \cite{YS10} have employed the
multiconfigurational Dirac-Fock method for calculating $\varrho_{el}(r)$.
Keller and Dreizler \cite{KD97} have used a relativistic Kohn-Sham potential which in addition accounts for the shell structure of the atomic target.

\subsection{Analytical implementation of screening}

We have already seen in section 2.1.2 that in the Born approximation, screening can be accounted for in terms of a form factor $F(q)$,
\begin{equation}\label{2.6.3}
\frac{d^3\sigma^{B1,{\rm screen}}}{d\omega d\Omega_k d\Omega_f}\;=\;F^2(q)\;\frac{d^3\sigma^{B1}}{d\omega d\Omega_k d\Omega_f},
\end{equation}
where $q=|\bfp_i-\bfp_f-\bfk|$ is the momentum transfer. This formula is valid irrespective of the consideration of polarization.
For electron screening, the form factor is calculated from the total charge density, which is the sum of the nuclear and the electronic contribution. For a pointlike nucleus and  spherical symmetry, one has
$\varrho(\bfr)=-Z_T\delta(\bfr) +\varrho_{el}(r)$, and the
corresponding form factor is according to (\ref{2.1.15})
\begin{equation}\label{2.6.4}
F(q)\;=\;-1\;+\;F_{el}(q),
\end{equation}
where $F_{el}(q)$ is given by
\begin{equation}\label{2.6.5}
F_{el}(q)\;=\;\frac{4\pi}{Z_T}\int_0^\infty r^2dr\;\varrho_{el}(r)\;\frac{\sin (qr)}{qr},
\end{equation}
keeping in mind that for neutral atoms $\int d\bfr \varrho_{el}(r)=Z_T$.
Since $F_{el}(q)=1$ for $q=0$, formula (\ref{2.6.3}) implies a strong reduction of the cross section for small momentum transfer.
For large $q$, on the other hand, $F_{el}(q)$ is small, such that $F^2(q) \approx 1$ and the Bethe-Heitler result is recovered.

In cases where the PWBA is no longer applicable, one can use an approximation which allows for a straightforward consideration of screening.
Thereby one makes use of the fact that screening
affects bremsstrahlung particularly if it is emitted at  large distances from the atomic center.
However, in this region the influence of the target field is weak such that screening may be estimated within the PWBA.
On the other hand, the distortion of the scattering states by the action of $V_T$ is most effective at electron-atom distances which are small enough
to lie well  inside the atomic shells where screening plays no role. This suggests a separation of screening and distortion effects, which has first been contemplated by Olsen, Maximon and Wergeland (OMW \cite{OMW57}).
If the collision energy is high and the scattered electron not observed, they have found that the effects of screening and Coulomb correction are nearly independent,
in particular, that they are additive. 
Consequently, this OMW additivity rule can be applied to implement screening in the doubly differential cross section, calculated within any bremsstrahlung theory,
acccording to
\begin{equation}\label{2.6.6}
\frac{d^2\sigma^{\rm screen}}{d\omega d\Omega_k}\;=\;\frac{d^2\sigma}{d\omega d\Omega_k}\;+\;\left[ \frac{d^2\sigma^{B1,{\rm screen}}}{d\omega d\Omega_k}\;-\;\frac{d^2\sigma^{B1}}{d\omega d\Omega_k}\right].
\end{equation}

\subsection{Numerical implementation of screening}

Like in PWBA, an exact consideration of screening is  possible in the relativistic partial-wave theory. Here, the screening potential $V_T(r)$
 has to substitute the pure nuclear Coulomb field $-Z_T/r$ in the Dirac equation for the electronic scattering states. 
The basic difference in the numerical implementation is the fact that the asymptotic solutions for a
fully screened atom are Bessel functions, which have to substitute the Coulomb waves

Table 2.6.1 shows the bremsstrahlung results for a neutral Pb atom calculated within the Dirac partial-wave (DW) theory for a bare nucleus,
where subsequently the prescription (\ref{2.6.6}) is used for implementing screening. As a test for the OMW additivity rule,
these results are compared with calculations within the DW theory for the screened potential.

\vspace{0.2cm}

Table 2.6.1 \quad
Doubly differential bremsstrahlung cross section (in b/MeVsr) from electrons with  energy $E_e$ from 50 keV to 1 MeV colliding with $^{208}$Pb at $\theta_k=10^\circ$ and $\omega/E_e=0.2.$\\
Second row, partial-wave results for $V_T=-Z_T/r.$
Third row, partial-wave results for the static Pb potential provided by Haque et al \cite{H18b}.
Fourth row, OMW results from applying (\ref{2.6.6}) to the second row.
The last row gives the percentage deviation of $\frac{d^2\sigma^{\rm OMW}}{d\omega d\Omega}$ from the exact result in the $3^{\rm rd}$ row.

\vspace{0.2cm}

\begin{tabular}[t]{l|l|l|l|l|l|l|l|}  
$E_e$ (MeV)& 0.05 & 0.1 & 0.2 & 0.4 & 0.6 & 0.8 & 1.0\\
&&&&&&&\\ \hline
&&&&&&&\\
${\displaystyle \frac{d^2\sigma^{\rm coul}}{d\omega d\Omega_k}}$ &5314&2313&1204& 773.0&646.4 & 577.6& 528.6\\
&&&&&&&\\ \hline
&&&&&&&\\
${\displaystyle \frac{d^2\sigma^{\rm screen}}{d\omega d\Omega_k}}$& 3150 & 1564 & 884.6& 595.1& 509.3& 461.7& 425.5\\
&&&&&&&\\ \hline
&&&&&&&\\
${\displaystyle \frac{d^2\sigma^{\rm OMW}}{d\omega d\Omega_k}}$& 3650 & 1749 & 955.4 & 624.1 & 523.6 & 469.1 & 431.4\\
&&&&&&&\\ \hline
&&&&&&&\\
$\Delta^{\rm OMW}$ (\%)&15.9&11.8&8.0&4.9&2.8&1.6&1.4\\
&&&&&&&\\ \hline
\end{tabular}

\vspace{0.2cm} 

We observe that for a constant forward photon angle $(10^\circ)$ and fixed frequency ratio $(\omega/E_e=0.2)$, the screening effect decreases with collision energy, from 68.7\% at 50 keV to 24.2\% at 1 MeV. 
Above 0.5 MeV, the OMW prescription is accurate on the percent level.

The influence of screening on the photon spectrum
is displayed in Fig.2.6.1 for 1 MeV electrons colliding with Sn at a forward angle of $10^\circ$.
It is seen that screening gains importance if $\omega < E_e/2$, and for $\omega/E_e=0.1$ it leads to a considerable reduction of the cross section,
in agreement with the experimental data.
This figure shows also nicely the validity of the OMW additivity rule at this angle down to $\omega=200$ keV.

\vspace{0.5cm}
\setcounter{figure}{0}

\begin{figure}
\vspace{-1.5cm}
\includegraphics[width=8cm]{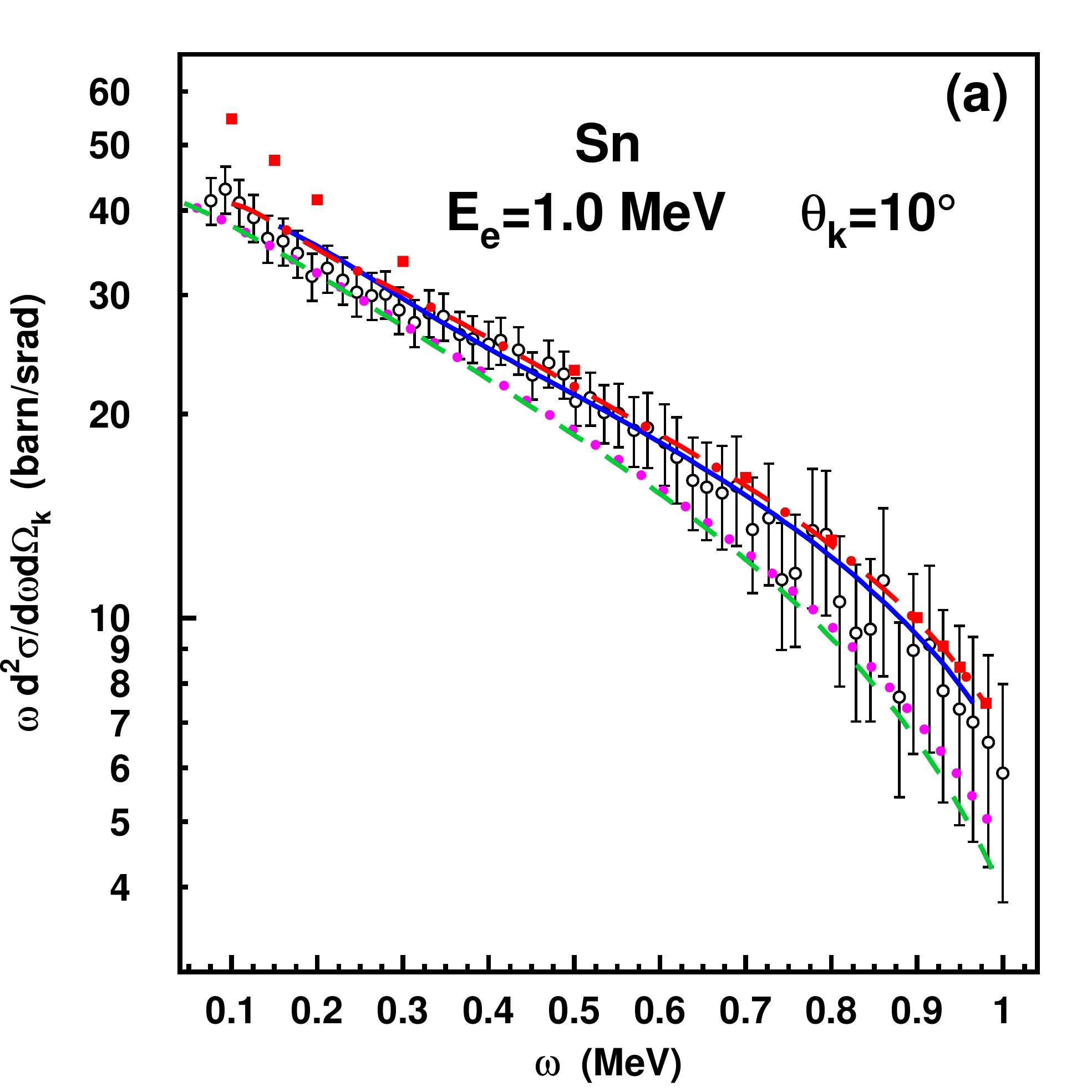}
\caption{
Bremsstrahlung intensity $\omega \times \frac{d^2\sigma}{d\omega d\Omega_k}$ for 1 MeV electrons scattering from Sn $(Z=50)$ at $\theta_k=10^\circ$ as a function of photon frequency $\omega$. Shown are the
 partial-wave results with (----------) and without ($\blacksquare$) screening, as well as those from the OMW additivity rule applied to the (unscreened) partial-wave results ($-\cdot-\cdot-$),
to the Sommerfeld-Maue (SM) results $(\cdots\cdots)$,
as well as to the  next-to-leading order SM results ($-----$ \cite{JM19}). The experimental data are from Rester and Dance \cite{RD67}.
}
\end{figure}

In Fig.2.6.2 we compare the results from the screened partial-wave theory with new experimental data for Te  at the short-wavelength limit and a photon angle of  $35^\circ$ and $131^\circ$  \cite{Ga17}. Theory explains nicely the energy dependence of the doubly differential bremsstrahlung cross
section. 
It is seen that screening is quite unimportant at the larger angle, while it leads to a considerable reduction of
the cross section  at $35^\circ$.

\vspace{0.5cm}

\begin{figure}
\vspace{-1.5cm}
\includegraphics[width=13cm]{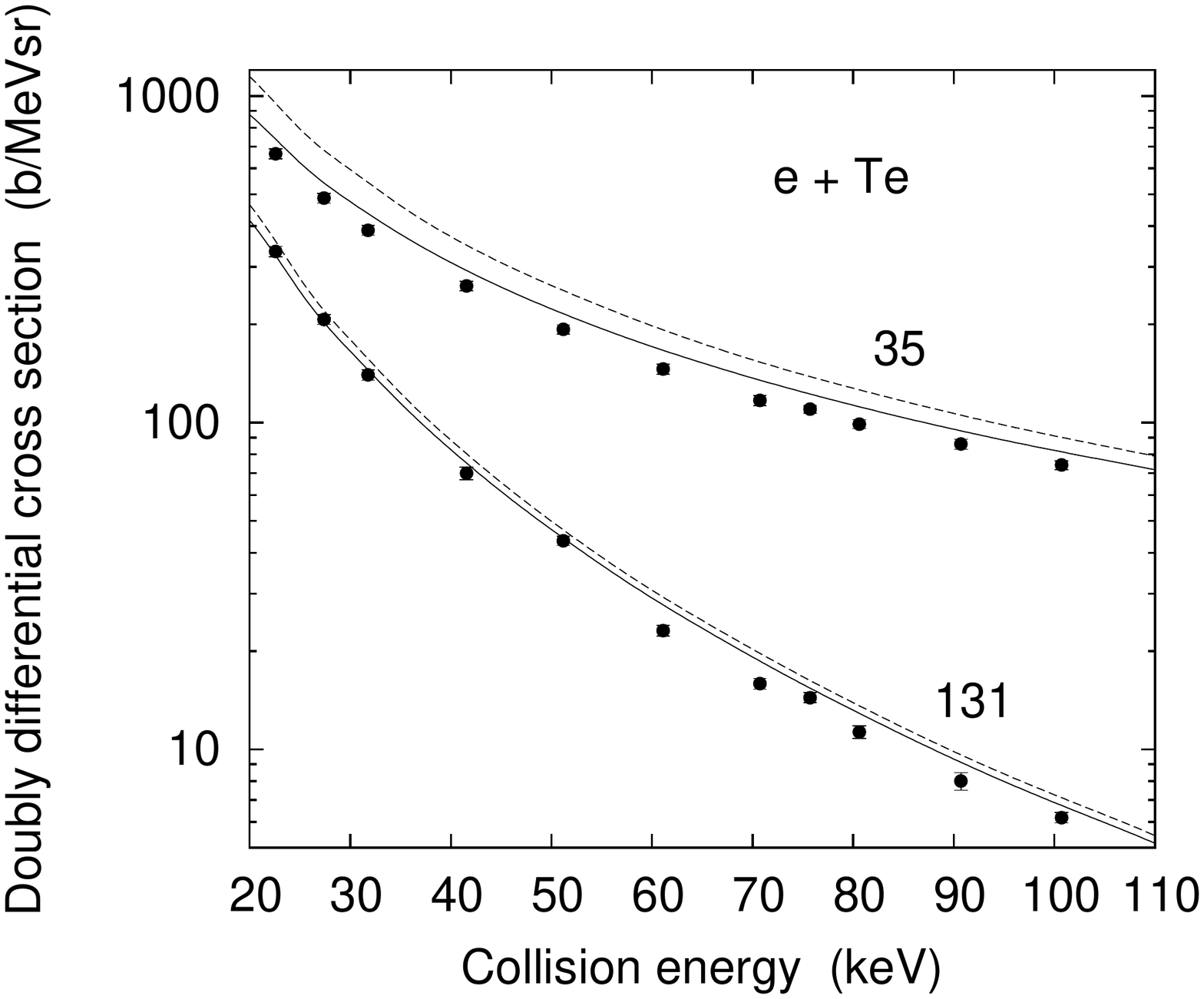}
\caption{
Doubly differential cross section for tip bremsstrahlung from electrons colliding with Te ($Z=52$) at photon
angles $35^\circ$ (upper curves) and $131^\circ$ (lower curves) as a function of collision energy $E_e$.
Shown are partial-wave results including screening (-----------) as well as the pure Coulombic DW results ($-----$).
The experimental data are from Garc\'{i}a-Alvarez et al \cite{Ga17}.
}
\end{figure}

We mention that there exist tabulated partial-wave results for the doubly differential bremsstrahlung cross section at collision energies of $1-500$ keV, covering the whole region of photon angles and
frequencies for a variety of neutral atoms  \cite{Ki83}.

For the singly differential cross section, $\frac{d\sigma}{d\omega}$, there are early tabulations  \cite{P77} for collision energies between 1 keV and 2 MeV.
A new tabulation comprises singly and doubly differential cross sections from 10 eV to 3 MeV for all stable neutral elements \cite{Po19}.
Numerical interpolations of the existing partial-wave results at low and high energy have led to a comprehensive tabularization across the periodic table for collision energies from 1 keV to 10 GeV  \cite{SB85}.

\section{Nuclear and QED effects in the partial-wave approximation}
\setcounter{equation}{0}

We have seen that it is quite simple to account for nuclear structure effects as long as the PWBA is valid. But even in the partial-wave formalism such studies are straightforward as long as one can implement these effects in an effective
potential, such that they can be treated exactly by solving the Dirac equation in this modified potential.

\subsection{Finite nuclear size}

Provided the nuclear charge density $\varrho_N(r)$ is known from nuclear structure calculations or from elastic scattering experiments, the electron-nucleus interaction potential is calculated from (\ref{2.1.13})
with $\varrho=-\varrho_N$, which in the spherically symmetric case reduces to
\begin{equation}\label{2.7.1}
V_T(r)\;=\;-4\pi \left( \frac{1}{r}\int_0^r x^2dx\;\varrho_N(x)\;+\;\int_r^\infty xdx\;\varrho_N(x)\right).
\end{equation}
For finite nuclear size effects to come into play, the collision energy has to be so high that the influence of the bound target electrons can be disregarded in $V_T$.
The deviation between the results for the potential  (\ref{2.7.1}) and the point-Coulomb field $-Z_T/r$ is most prominent when the radiation is emitted at electron-nucleus distances which are comparable to the nuclear radius,
which in turn corresponds to a large momentum transfer.

For a lead nucleus and backward electron scattering, nuclear size effects become visible if the photons are emitted into the backward hemisphere, and if the collision energy exceeds 5 MeV. This is displayed in Fig.2.7.1.
In the plot the azimuthal angle $\varphi_f=0$ is kept fixed, while $\theta_k$ runs clockwise from $0^\circ$ (i.e. aligned with the beam axis) to $360^\circ$.
The angles between $360^\circ$ and $180^\circ$ correspond to angles between $0^\circ$ and $180^\circ$ at $\varphi_f=180^\circ$,
where photon and electron are observed on opposite sides of the beam axis.

The reduction of the cross section becomes more prominent the higher the collision energy.
 Moreover, at energies well above 50 MeV, an extended charge density may lead to diffraction structures in the cross section \cite{Jaku13},
in a similar way as for elastic electron scattering \cite{Lu97}.

\setcounter{figure}{0}

\begin{figure}
\vspace{-1.5cm}
\includegraphics[width=13cm]{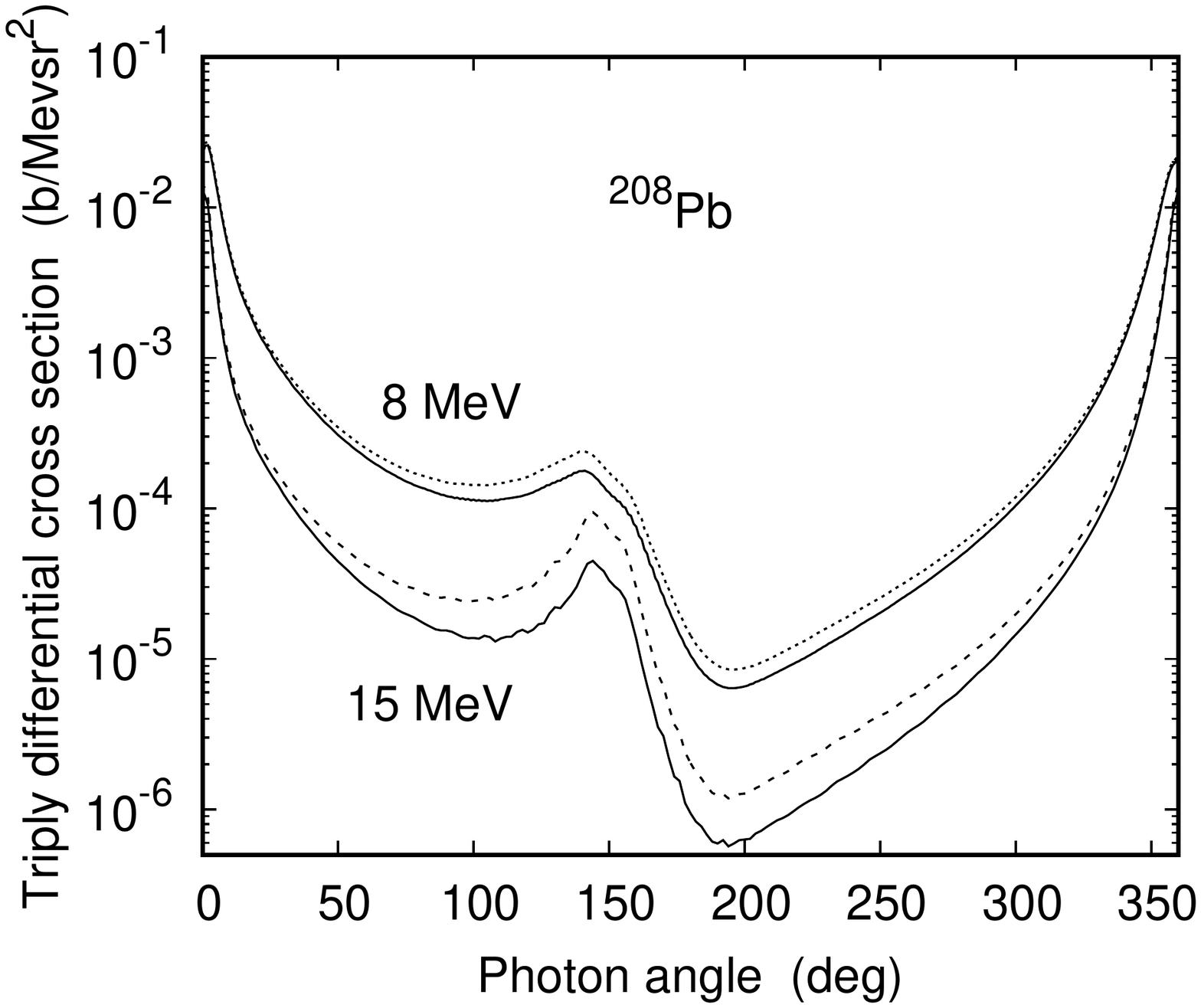}
\caption{
Triply differential bremsstrahlung cross section from 8 MeV (upper curves) and 15 MeV (lower curves) electrons
colliding with $^{208}$Pb at $\omega/E_e=0.75$, a scattering angle of $\vartheta_f=150^\circ$ and $\varphi_f=0$ as a function of photon angle $\theta_k$.
Shown are the partial-wave results for an extended nucleus
(----------) and a point-like nucleus (8 MeV, $\cdots\cdots$; 15 MeV, $-----$). The wiggles are due to numerics.
}
\end{figure}

\vspace{0.5cm}

\subsection{QED effects}

The lowest-order QED effects to bremsstrahlung result from the vacuum polarization and the self-energy \cite{BD}. The nonperturbative consideration of the self-energy is challenging and  is still an unsolved problem to date  for the continuum electrons. However, the vacuum polarization can to a good approximation be described in terms of an additional potential, the Uehling potential  \cite{Ue35,Kl77}. This potential is given in terms of the nuclear charge density,
$$U_e(r)\;=\;-\frac{2}{3c^2r}\int_0^\infty xdx\;\varrho_N(x)\;\left[ \chi_2(2c|r-x|)\,-\,\chi_2(2c|r+x|)\right],$$
\begin{equation}\label{2.7.2}
\chi_2(y)\;=\;\int_1^\infty dt\;e^{-yt}\;\frac{1}{t^2}\;\left( 1+\,\frac{1}{2t^2}\right) \left( 1-\,\frac{1}{t^2}\right)^\frac12,
\end{equation}
which has to be added to $V_T(r)$ when solving the Dirac equation.
From investigations by Keller and Dreizler \cite{KD97} it follows that for $E_e=0.3$ MeV, its effect on the bremsstrahlung intensity is around 1\%, being 
largest at the high-energy end of the photon spectrum. From elastic scattering estimates, vacuum polarization effects remain near or below 1\% for
collision energies up to 50 MeV.

There exists an experimental investigation on QED effects in bremsstrahlung \cite{SL68}. At an electron energy of 5.15 GeV, a 5\% effect was discovered at the high-energy end of
the bremsstrahlung spectrum, well above the experimental uncertainty.

\vspace{1cm}

{\Large\bf 3. Polarization}
\setcounter{equation}{0}
\setcounter{chapter}{3}
\setcounter{section}{0}

\vspace{0.5cm}

Working with polarized electron beams and detecting the polarization of the outgoing particles provides much more information on the collision dynamics
and on the target properties than the measurement of cross sections where the polarization variables remain unobserved.
An overview of the polarization phenomena can be found in the book by Balashov et al \cite{BGK}.
In principle, all participating particles can be polarized, including target nuclei with spin.
However, in the usual experimental situation, only the polarization transfer from the electron to the emitted photon is considered.
We start by restricting ourselves to the case of unpolarized nuclei and unobserved polarization of the scattered electron.
Later, we will also consider the polarization of the final electron.

\section{Definition of the electron-photon polarization correlations}

The first model-independent classification of the spin correlation between beam electron and emitted photon is given by Tseng and Pratt \cite{TP73} for the doubly differential, and later by Tseng \cite{T02} for the triply differential cross section.
In the latter case, the parameters $C_{jk0}$ describing the polarization correlations between incident electron (index $j$) and photon (index $k$) are defined by
$$\frac{d^3\sigma}{d\omega d\Omega_k d\Omega_f}(\bfzeta_i,\bfe^\ast_\lambda)\;=\;\frac12 \left( \frac{d^3\sigma}{d\omega d\Omega_k d\Omega_f}\right)_0\;\left[ 1\;+\;C_{030}\;\xi_3\;+\;(C_{110}\;\xi_1\;-\;C_{120}\;\xi_2)\;(\bfzeta_i\bfe_x)\right.$$
\begin{equation}\label{3.1.1}
\left. -\;(C_{230}\;\xi_3\;+\;C_{200})\;(\bfzeta_i \bfe_y)\;-\;(C_{310}\;\xi_1\;-\;C_{320}\;\xi_2)\;(\bfzeta_i\bfe_z)\right]
\end{equation}
in our coordinate system with $\bfe_z=\hat{\bfp}_i,\;\bfe_y$  along $\bfp_i \times \bfk$ and $\bfe_x = \bfe_y \times \hat{\bfp}_i$.
The sign changes in (\ref{3.1.1}) as compared to the literature
 \cite{BM54,TP73,HN04} result from a different choice of
coordinate system in the earlier work, where the $z$-axis was directed along $\bfk$ like for photoionization. 
From an experimental point of view it is, however, of advantage to take $\bfe_z$ along the beam axis as done in more recent work  \cite{YS10,MYS}. 
The factor $\left( \frac{d^3\sigma}{d\omega d\Omega_k d\Omega_f}\right)_0$ denotes the unpolarized cross section which is averaged over the spin projections of the beam electron, and summed over the photon polarization directions $\bfe_\lambda$ and over the spin polarization of the scattered electron.
The three terms in (\ref{3.1.1}) depending on the electron spin polarization vector $\bfzeta_i$ relate to its projection
along the three coordinate axes. Therefore, if the beam electron is polarized along one of the axes, the cross section (\ref{3.1.1}) simplifies considerably.

Conveniently, the unit vector $\bfzeta_i$ is described in spherical coordinates, involving polar ($\alpha_s)$ and azimuthal ($\varphi_s$) angles,
$\bfzeta_i=(\sin \alpha_s \cos \varphi_s, \sin \alpha_s \sin \varphi_s, \cos \alpha_s)$. In this representation, the coefficients $a_{m_s}$ of the polarization spinor $w=\sum_{m_s} a_{m_s} \chi_{m_s}$ entering into the wavefunction $\psi_i$
(see Section 2.4.1) are given by \cite{Ros61}
\begin{equation}\label{3.1.2}
a_\frac12\;=\;\cos \frac{\alpha_s}{2}\;e^{-i \varphi_s/2},\qquad a_{-\frac12}\;=\;\sin \frac{\alpha_s}{2}\;e^{i\varphi_s/2}.
\end{equation}
For unpolarized electrons, the average over the two opposite directions of $\bfzeta_i$ eliminates the three aforementioned terms, so that only $C_{030}$ survives.

The second vector $\bfxi=(\xi_1,\xi_2,\xi_3)$ appearing in  (\ref{3.1.1}) is related to the photon polarization coefficients. The photon polarization vector can be represented in the basis $\bfe_{\lambda_1}=(0,1,0)$ and $\bfe_{\lambda_2}=(-\cos \theta_k,0,\sin \theta_k)$ for linear polarization,
\begin{equation}\label{3.1.3}
\bfe_\lambda\;=\;\beta_1\,\bfe_{\lambda_1}\;+\;\beta_2\, \bfe_{\lambda_2}.
\end{equation}
These basis vectors are orthonormal and perpendicular to the photon momentum $\bfk=k(\sin \theta_k,0,\cos \theta_k).$
Then $\bfxi$ is given by
\begin{equation}\label{3.1.4}
\bfxi\;=\;(2\mbox{ Re }(\beta_1\beta_2^\ast),\,2\mbox{ Im }(\beta_1 \beta_2^\ast),\,|\beta_2|^2\,-\,|\beta_1|^2).
\end{equation}

Photon detectors can specify between linearly and circularly polarized photons. A linearly polarized photon is characterized by $\bfe_\lambda(\varphi_\lambda)=\sin \varphi_\lambda \bfe_{\lambda_1}+\cos \varphi_\lambda \bfe_{\lambda_2}$ with $0\leq \varphi_\lambda \leq \pi$, such that $\beta_1=\sin \varphi_\lambda$ and $\beta_2=\cos \varphi_\lambda$.

Circularly polarized photons ($\bfe_\lambda = \bfe_\pm$) are characterized by the complex coefficients $\beta_2=1/\sqrt{2}$ and $\beta_1=-i/\sqrt{2}$ for right-handed (+) and $\beta_1=+i/\sqrt{2}$ for left-handed (-) photons.
Accordingly, for circularly polarized photons, one has $|\beta_2|\,=\,|\beta_1|$, while  $\beta_1 \beta_2^\ast$ is purely imaginary.
Hence $\xi_1=\xi_3=0$ and
\begin{equation}\label{3.1.5}
\xi_2\;=\;2\mbox{ Im }(\mp \,\frac{i}{\sqrt{2}}\;\frac{1}{\sqrt{2}})\;=\;\mp\;1,
\end{equation}
such that, for circularly polarized photons, the cross section reduces to
$$\frac{d^3\sigma}{d\omega d\Omega_k d\Omega_f}(\bfzeta_i,\bfe^\ast_\pm)\;=\;\frac12\;\left(\frac{d^3\sigma}{d\omega d\Omega_k d\Omega_f}\right)_0\;\left[ 1\;-\;C_{120}\;\xi_2\;(\bfzeta_i \bfe_x)\right.$$
\begin{equation}\label{3.1.6}
\left. -\;C_{200}\;(\bfzeta_i \bfe_y)\;+\;C_{320}\;\xi_2\;(\bfzeta_i \bfe_z)\right],
\end{equation}
with $\xi_2=\pm 1$ for $\bfe_\pm^\ast$.
If, in addition,  the photon polarization is unobserved, the terms proportional to $\xi_2$ vanish, and a spin asymmetry can only be observed if the electron is polarized perpendicular to the $(x,z)$-reaction plane, which is spanned by $\bfp_i$ and $\bfk$.
 The parameter characterizing this spin asymmetry is $C_{200}$.

For linearly polarized photons, on the other hand, $\beta_1$ and $\beta_2$ are real such that $\xi_2=0$.
Moreover,
$$\xi_1\;=\;2\beta_1\beta_2\;=\;2\;\sin \varphi_\lambda \cos \varphi_\lambda\;=\;\sin 2\varphi_\lambda,$$
\begin{equation}\label{3.1.7}
\xi_3\;=\;\beta_2^2-\beta_1^2\;=\;\cos^2 \varphi_\lambda\;-\;\sin^2 \varphi_\lambda\;=\;\cos 2\varphi_\lambda,
\end{equation}
such that the corresponding cross section reads
$$\frac{d^3\sigma}{d\omega d\Omega_k d\Omega_f}(\bfzeta_i,\bfe_\lambda^\ast(\varphi_\lambda))\;=\;\frac12\;\left( \frac{d^3\sigma}{d\omega d\Omega_k d\Omega_f}\right)_0\;\left[ 1\;+\;C_{030}\;\cos 2\varphi_\lambda\;+\;C_{110}\;\sin 2\varphi_\lambda\;(\bfzeta_i \bfe_x)\right.$$
\begin{equation}\label{3.1.8}
\left. -\;(C_{230}\;\cos 2 \varphi_\lambda\;+\;C_{200})\;(\bfzeta_i \bfe_y)\;-\;C_{310}\;\sin 2\varphi_\lambda\;(\bfzeta_i \bfe_z)\right].
\end{equation}
Averaging over the photon polarization eliminates all terms containing angular functions. Only the term proportional to $C_{200}$ is left over, in concord with the case for circular polarization.

For the calculation of the triply differential cross section and the 
polarization correlations within the partial-wave theory (according to (\ref{2.4.19}) - (\ref{2.4.22})), the
expansion coefficients $c_\varrho^{(\lambda)}$ of the photon polarization vectors in (\ref{2.4.10}) are needed.
For right (+) and left (-) circular polarization, they are given by
\begin{equation}\label{3.1.9}
c_\varrho^{(+)}\;=\;\left\{ \begin{array}{cl}
\frac12 \;(\cos \theta_k -1), & \varrho =1\\
-\frac12\;(\cos \theta_k+1),& \varrho = -1\\
\sin \theta_k/\sqrt{2},& \varrho =0
\end{array} \right.
\end{equation}
$$c_\varrho^{(-)}\;=\;(-1)^\varrho\;c_{-\varrho}^{(+)}.$$
For the linear polarization vector $\bfe_\lambda(\varphi_\lambda)$ one has
\begin{equation}\label{3.1.10}
c_\varrho^{(\lambda)} \;=\;\left\{ \begin{array}{cl}
(\cos \varphi_\lambda \cos \theta_k \,+\,i\sin \varphi_\lambda)/\sqrt{2},& \varrho =1\\
(-\cos \varphi_\lambda \cos \theta_k \,+\,i\sin \varphi_\lambda)/\sqrt{2},& \varrho =-1\\
\cos \varphi_\lambda\;\sin \theta_k,& \varrho =0.
\end{array} \right.
\end{equation}

An experimental determination of each of the seven nonvanishing polarization correlations $C_{jk0}$ in (\ref{3.1.1}) can be  achieved by measuring the spin asymmetries in terms of relative cross section differences.
For linear polarization we consider
\begin{equation}\label{3.1.11}
P(\alpha_s,\varphi_s,\varphi_\lambda)\;=\;\frac{d^3\sigma(\bfzeta_i,\bfe^\ast_\lambda(\varphi_\lambda))\;-\;d^3\sigma(\bfzeta_i,\bfe^\ast_\lambda(\varphi_\lambda + \pi/2))}{d^3\sigma(\bfzeta_i,\bfe^\ast_\lambda(\varphi_\lambda))\;+\;d^3\sigma(\bfzeta_i,\bfe^\ast_\lambda(\varphi_\lambda + \pi/2))}.
\end{equation}
Setting $\varphi_\lambda =0$ and $\varphi_s=0$ (i.e. $\xi_3=1$ and $\bfzeta_i \bfe_y=0$), one obtains from (\ref{3.1.8})
\begin{equation}\label{3.1.12}
P(\alpha_s,0,0)\;\equiv\;P_1\;=\;\frac{(1+C_{030})\;-\;(1-C_{030})}{(1+C_{030})\;+\;(1-C_{030})}\;=\;C_{030},
\end{equation}
independent of $\alpha_s$.
With the linearly independent choice $\varphi_\lambda=\frac{\pi}{4}$ (i.e. $\xi_1=1$) and $\varphi_s=0$ one finds
\begin{equation}\label{3.1.13}
P(\alpha_s,0,\frac{\pi}{4})\;\equiv\;P_2(\alpha_s)\;=\;C_{110}\;(\bfzeta_i \bfe_x)\;-\;C_{310}\;(\bfzeta_i \bfe_z)\;=\;C_{110}\sin \alpha_s\;-\;C_{310}\cos \alpha_s,
\end{equation}
such that $P_2(0)=-C_{310}$ and $P_2(\frac{\pi}{2})=\;C_{110}$.

The polarization correlation $C_{230}$ is only indirectly accessible for unpolarized final electrons. We calculate
$$P(\uparrow)\;\equiv\;\frac{d^3\sigma(\uparrow,\varphi_\lambda=0)\;-\;d^3\sigma(\uparrow,\varphi_\lambda=\pi/2)}{d^3\sigma_0}\;=\;C_{030}\;+\;C_{230},$$
\begin{equation}\label{3.1.14}
P(\downarrow)\;\equiv\;\frac{d^3\sigma(\downarrow,\varphi_\lambda =0)\;-\;d^3\sigma(\downarrow,\varphi_\lambda=\pi/2)}{d^3\sigma_0}\;=\;C_{030}\;-\;C_{230},
\end{equation}
where spin up $(\uparrow)$ is defined by $\alpha_s=\frac{\pi}{2},\;\varphi_s=-\frac{\pi}{2}$ and spin down $(\downarrow)$ by $\alpha_s=\frac{3\pi}{2},\;\varphi_s=-\frac{\pi}{2}$, in concord with $\bfzeta_i$ polarized perpendicular to the $(x,z)$-plane. 
Moreover, $d^3\sigma_0\,=\,\frac12\,[d^3\sigma(\uparrow, \varphi_\lambda=0)\,+\,d^3\sigma(\downarrow,\varphi_\lambda =0)\,+\,d^3\sigma(\uparrow,\varphi_\lambda = \frac{\pi}{2})\,+\,d^3\sigma(\downarrow,\varphi_\lambda =\frac{\pi}{2})]$ relates to the unpolarized cross section.
From (\ref{3.1.14}) it follows that  $C_{230}$ is obtained from
\begin{equation}\label{3.1.15}
C_{230}\;=\;\frac12\;[P(\uparrow)\;-\;P(\downarrow)].
\end{equation}

For circular polarization we consider
\begin{equation}\label{3.1.16}
P(\alpha_s,\varphi_s,\pm)\;=\;\frac{d^3\sigma(\bfzeta_i,\bfe_+^\ast)\,-\,d^3\sigma(\bfzeta_i,\bfe_-^\ast)}{d^3\sigma(\bfzeta_i,\bfe_+^\ast)\,+\,d^3\sigma(\bfzeta_i,\bfe_-^\ast)},
\end{equation}
in correspondence to (\ref{3.1.11}).
Taking $\varphi_s=0$ one obtains
\begin{equation}\label{3.1.17}
P(\alpha_s,0,\pm)\;\equiv\;P_3(\alpha_s)\;=\;C_{320}\cos \alpha_s\;-\;C_{120}\sin \alpha_s,
\end{equation}
upon which it follows that $P_3(0)=C_{320}$ and $P_3(\frac{\pi}{2})=-C_{120}$.

The spin asymmetry $C_{200}$ is accessible from (\ref{3.1.6}) as well as from (\ref{3.1.8}) since it does not depend on the photon polarization.
It can be determined from
\begin{equation}\label{3.1.18}
C_{200}\;=\;\frac{\sum_\lambda d^3\sigma(\bfzeta_i,\bfe_\lambda^\ast)\,-\,\sum_\lambda d^3\sigma(-\bfzeta_i,\bfe_\lambda^\ast)}{\sum_\lambda d^3\sigma(\bfzeta_i,\bfe_\lambda^\ast)\,+\,\sum_\lambda d^3\sigma(-\bfzeta_i,\bfe_\lambda^\ast)},
\end{equation}
where $\bfzeta_i$ has to be polarized perpendicular to the $(x,z)$-plane (i.e. $\varphi_s=-\frac{\pi}{2}$ with $\alpha_s=\frac{\pi}{2}$ for $\bfzeta_i=(0,-1,0)$ and
$\alpha_s=\frac{3\pi}{2}$ for $-\bfzeta_i=(0,1,0))$. The photon polarization remains undetected (i.e. is summed over).
We note that the definition 'spin up' for $\bfzeta_i$ antiparallel to $\bfe_y$ has again historical reasons, being related to the different choice of  coordinate system in early work.

The representation of the polarization correlations in terms of relative cross section differences has the advantage that it is independent of any model which is used for calculating these cross sections.
The only condition is the linearity of the transition matrix element in the electron spin polarization vectors and in the photon polarization vector.
Consequently, this definition holds also for the polarization correlations pertaining to the doubly differential cross section where it is integrated over the solid angle $d\Omega_f$ of the scattered electron.
Omitting the last index of $C_{jk0}$ which pertains to an unpolarized, but observed electron, (\ref{3.1.1}) is replaced by
$$\frac{d^2\sigma}{d\omega d\Omega_k}(\bfzeta_i,\bfe_\lambda^\ast)\;=\;\frac12\;\left( \frac{d^2\sigma}{d\omega d\Omega_k}\right)_0\;\left[ 1\;+\;C_{03}\xi_3\right.$$
\begin{equation}\label{3.1.19}
\left. +\;(C_{11} \xi_1-C_{12}\xi_2)\;(\bfzeta_i \bfe_x)\;-\;(C_{23}\xi_3+C_{20})\;(\bfzeta_i \bfe_y)\;-\;(C_{31}\xi_1-C_{32}\xi_2)\;(\bfzeta_i \bfe_z)\right].
\end{equation}
Therefore the formulas (\ref{3.1.6}) - (\ref{3.1.18})
remain true for the $C_{jk}$ if the triply differential cross sections are replaced by the doubly differential ones.
In particular, the basic equation for circularly polarized photons reads
\begin{equation}\label{3.1.20}
\frac{d^2\sigma}{d\omega d\Omega_k}(\bfzeta_i,\bfe^\ast_\pm)\;=\;\frac12\;\left(\frac{d^2\sigma}{d\omega d\Omega_k }\right)_0\;\left[ 1\;-\;C_{12}\;\xi_2\;(\bfzeta_i \bfe_x)
\; -\;C_{20}\;(\bfzeta_i \bfe_y)\;+\;C_{32}\;\xi_2\;(\bfzeta_i \bfe_z)\right],
\end{equation}
and for linearly polarized photons,
$$\frac{d^2\sigma}{d\omega d\Omega_k }(\bfzeta_i,\bfe_\lambda^\ast(\varphi_\lambda))\;=\;\frac12\;\left( \frac{d^2\sigma}{d\omega d\Omega_k}\right)_0\;\left[ 1\;+\;C_{03}\;\cos 2\varphi_\lambda\;+\;C_{11}\;\sin 2\varphi_\lambda\;(\bfzeta_i \bfe_x)\right.$$
\begin{equation}\label{3.1.21}
\left. -\;(C_{23}\;\cos 2 \varphi_\lambda\;+\;C_{20})\;(\bfzeta_i \bfe_y)\;-\;C_{31}\;\sin 2\varphi_\lambda\;(\bfzeta_i \bfe_z)\right].
\end{equation}

\section{Triply differential cross section  in coplanar geometry}
\setcounter{equation}{0}

The situation where the correlation between the polarizations of the beam electron, the photon as well as the scattered electron is considered, albeit for angle-integrated cross sections, was already treated by Olsen and Maximon \cite{OM59}.
The formulation of these correlations within the Sommerfeld-Maue theory is given by Haug \cite{Ha69}. However, in his numerical calculations it is still summed over the spin projections of the scattered electron.

In the following we derive the general polarization correlations $C_{jkl}$, where the subscript $l$ accounts for the polarization of the scattered electron.
Hence we  omit the sum over $\zeta_f$ in (\ref{2.4.21}). Recalling that the electronic states $\psi_i$ and $\psi_f$ from
(\ref{2.4.6}) and (\ref{2.4.1}) are linear in the spin vectors,
and representing the photon polarization vector in the basis of the circular states $\bfe_\pm$,
\begin{equation}\label{3.2.1}
\bfe_\lambda^\ast \;=\; \sum_{\sigma=\pm} f_\sigma\;\bfe_\sigma^\ast,
\end{equation}
the radiation matrix element $W_{\rm rad}$ from (\ref{2.4.7}) can be written as a triple sum over the polarization coefficients, such that
the cross section is expressed in the general form,
$$ \frac{d^3\sigma}{d\omega d\Omega_k d\Omega_f}(\bfzeta_f,\bfzeta_i,\bfe_\lambda^\ast)\;=\; \frac{4\pi^2 \omega\, p_f E_i E_f}{c^5p_i}\;|W_{\rm rad}(\bfzeta_f,\bfzeta_i)|^2$$
\begin{equation}\label{3.2.2}
=\;\frac{4\pi^2 \omega \,p_f E_iE_f}{c^5p_i}\;\left| \sum_{m_i=\pm \frac12} a_{m_i} \sum_{m_s=\pm \frac12} b_{m_s}^\ast \sum_{\sigma=\pm} f_\sigma\;\tilde{M}_{fi}(\bfe_\sigma^\ast,m_i,m_s)\right|^2.
\end{equation}
In the partial-wave formalism, the transition matrix element $\tilde{M}_{fi}$ is obtained from $F_{fi}$ defined in (\ref{2.4.7}) and (\ref{2.4.19})
by omitting the sum over $m_i$ and by identifying $\bfe_\lambda^\ast$ with $\bfe_\sigma^\ast$.
However, in the derivation of the cross section in terms of the $C_{jkl}$ the explicit form of $\tilde{M}_{fi}$ is not needed.

For the sake of simplicity we restrict ourselves to the coplanar geometry where $\bfp_i, \; \bfp_f$ and $\bfk$ lie in one plane, the only case considered so far in experiments. Then one has the symmetry relation,
\begin{equation}\label{3.2.3}
\tilde{M}_{fi}(\bfe_\mp^\ast, -m_i, -m_s)\;=\;(-1)^{m_i-m_s}\;\tilde{M}_{fi}(\bfe_\pm^\ast,m_i,m_s),
\end{equation}
which relies on time reversal invariance. It is readily derived by making use of the partial-wave formalism \cite{Jaku17}.
In particular, the equality
\begin{equation}\label{3.2.4}
Y_{l_fm_l}(\bfp_f)\;\stackrel{!}{=}\;Y_{l_fm_l}^\ast(\bfp_f)\;=\;Y_{l_fm_l}(\bfp_f)\;e^{-2im_l\varphi_f}
\end{equation}
is needed, which implies that the azimuthal angle $\varphi_f$ of $\bfp_f$
has to be 0 or $\pi$, in concord with the requirement for coplanar geometry.

With (\ref{3.2.3}) the eight different matrix elements (according to the eight possible values of the triplet $(\sigma,m_i,m_s)$) reduce to four, which will be abbreviated in the following way,
\begin{equation}\label{3.2.5}
\tilde{M}_{fi}(\bfe_\pm^\ast,\frac12,\frac12)\;=\;J_\pm,\qquad \tilde{M}_{fi}(\bfe_\pm^\ast,-\frac12,\frac12)\;=\;S_\pm.
\end{equation}
Accordingly, the sum in (\ref{3.2.2}) can be written in the following way,
$$\left| \sum_{m_i=\pm \frac12} a_{m_i} \sum_{m_s=\pm \frac12} b_{m_s}^\ast\sum_{\sigma=\pm} f_\sigma\;\tilde{M}_{fi}(\bfe_\sigma^\ast, m_i,m_s)\right|^2$$
\begin{equation}\label{3.2.6}
=\left| (a_{\frac12} b_{\frac12}^\ast f_+\;+\;a_{-\frac12} b_{-\frac12}^\ast f_-)\;J_+\;+\;(a_{\frac12}b_{\frac12}^\ast f_-\;+\;a_{-\frac12} b_{-\frac12}^\ast f_+)\;J_- \right.
\end{equation}
$$\left. +\;(a_{-\frac12} b_{\frac12}^\ast f_+\;-\;a_{\frac12}b_{-\frac12}^\ast f_-)\;S_+\;+\;(a_{-\frac12}b_{\frac12}^\ast f_-\;-\;a_{\frac12}b_{-\frac12}^\ast f_+)\;S_-\right|^2.$$
Using the representation (\ref{3.1.2}) of $a_\frac12$ and $a_{-\frac12}$ in terms of the spherical angles $\alpha_s$ and $\varphi_s$ of $\bfzeta_i=(\zeta_{ix},\zeta_{iy},\zeta_{iz})$, one finds
$$|a_\frac12|^2\;=\;\frac12\;(1+\zeta_{iz}),\qquad |a_{-\frac12}|^2\;=\;\frac12\;(1-\zeta_{iz}),$$
\begin{equation}\label{3.2.7}
a_\frac12 a_{-\frac12}^\ast \;=\;\frac12\;(\zeta_{ix}-i\zeta_{iy}),\quad a_\frac12^\ast a_{-\frac12}\;=\;\frac12\;(\zeta_{ix} +i \zeta_{iy}),
\end{equation}
with equivalent relations holding for $b_{m_s}$, e.g. $b_\frac12^\ast b_{-\frac12}=\frac12(\zeta_{fx} +i \zeta_{fy}).$

Let us first consider linearly polarized photons. Upon inversion of (\ref{3.1.3}) for circularly polarized photons one finds the representation of the basis vectors
$\bfe_{\lambda_1}=\frac{i}{\sqrt{2}}\,(-\bfe^\ast_++\bfe^\ast_-)$ and $\bfe_{\lambda_2}=\frac{1}{\sqrt{2}}\,(\bfe^\ast_+ +\bfe^\ast_-) $ in terms of the helicity states.
For arbitrary linear polarization it follows
\begin{equation}\label{3.2.8}
\bfe_\lambda(\varphi_\lambda)\;=\;\frac{1}{\sqrt{2}}\;(e^{-i\varphi_\lambda}\bfe_+^\ast\;+\;e^{i\varphi_\lambda}\bfe_-^\ast).
\end{equation}

With (\ref{3.2.1}) and (\ref{3.1.7}) this leads to
\begin{equation}\label{3.2.9}
|f_+|^2\;=\;|f_-|^2\;=\;\frac12,\quad f_+f_-^\ast \;=\;\frac12\,e^{-2i\varphi_\lambda}\;=\;\frac12\;(\xi_3-i\xi_1), \quad f_+^\ast f_-\;=\;\frac12\;(\xi_3+i\xi_1).
\end{equation}
As an example, we perform the evaluation of (\ref{3.2.6}) explicitly for the first diagonal term,
$$\left| (a_\frac12 b_\frac12^\ast f_+\;+\;a_{-\frac12} b_{-\frac12}^\ast f_-)\;J_+\right|^2\;=\;|J_+|^2\;\left[\,|a_{\frac12}|^2\;|b_\frac12|^2\;|f_+|^2 \right.$$
\begin{equation}\label{3.2.10}
\left. +\;a_\frac12 a_{-\frac12}^\ast b_\frac12^\ast b_{-\frac12} f_+ f_-^\ast \;+\;a_{-\frac12}a_\frac12^\ast b_{-\frac12}^\ast b_\frac12 f_- f_+^\ast\;+\;
|a_{-\frac12}|^2\;|b_{-\frac12}|^2\;|f_-|^2\,\right]
\end{equation}
$$=\;\frac{1}{4}\;|J_+|^2\;\left[ 1\;+\;\zeta_{iz}\zeta_{fz}\;+\;(\zeta_{ix}\zeta_{fx} \;+\;\zeta_{iy} \zeta_{fy})\;\xi_3\;+\;(\zeta_{ix} \zeta_{fy}\;-\;\zeta_{iy}\zeta_{fx})\;\xi_1 \right].$$
Only the first term in the last line of (\ref{3.2.10}) contributes to the cross section for unpolarized particles,
due to the sums over $\zeta_i,\;\zeta_f$ and $\lambda$, which turns out to be
$$\left( \frac{d^3\sigma}{d\omega d\Omega_k d\Omega_f}\right)_0\;=\;N_0\;\frac12 \sum_{\zeta_i,\zeta_f,\lambda} \left| \sum_{m_i} a_{m_i}\sum_{m_s} b_{m_s}^\ast \sum_\sigma f_\sigma\;\tilde{M}_{fi}(\bfe_\sigma^\ast,m_i,m_s)\right|^2$$
\begin{equation}\label{3.2.11}
=\;N_0\;\left( \,|J_+|^2\;+\;|J_-|^2\;+\;|S_+|^2\;+\;|S_-|^2\,\right),
\end{equation}
where $N_0=(4\pi^2\omega p_fE_iE_f/(c^5p_i)$ is the prefactor in (\ref{3.2.2}).

For circularly polarized photons, on the other hand, (\ref{3.2.9}) has to be replaced by
$$|f_+|^2\;=\;1\;=\;\frac12\;(1+\xi_2),\quad |f_-|^2\;=\;f_+f_-^\ast \;=\; f_+^\ast f_-\;=\;0\qquad \mbox{ for } \bfe_+^\ast,$$
\begin{equation}\label{3.2.12}
|f_-|^2\;=\;1\;=\;\frac12\;(1-\xi_2), \quad |f_+|^2\;=\;f_+f_-^\ast\;=\;f_+^\ast f_-\;=\;0\qquad \mbox{ for } \bfe_-^\ast.
\end{equation}
Correspondingly, the result (3.2.10) for the first diagonal term must be replaced by
$$\frac{1}{4}\;|J_+|^2\;\left[ (1+\zeta_{iz}\zeta_{fz}\,+\,(\zeta_{iz}+\zeta_{fz})\,)\;|f_+|^2\;+\;(1+\zeta_{iz}\zeta_{fz}\,-\,(\zeta_{iz}+\zeta_{fz})\,)\;|f_-|^2\right]$$
\begin{equation}\label{3.2.13}
=\;\frac{1}{4}\;|J_+|^2\;\left[ 1+\zeta_{iz}\zeta_{fz}\;+\;(\zeta_{iz}+\zeta_{fz})\;\xi_2\right],
\end{equation}
where the distinct expressions for right- and left-circularly polarized photons have been combined into a single expression.

The comparison of (\ref{3.2.13}) with the respective expression (\ref{3.2.10}) for linearly polarized photons shows that
the $\bfxi$-independent terms are identical.
Hence, the result for arbitrarily polarized photons is obtained by adding the term in (\ref{3.2.13}), which is linear in $\xi_2$, to (\ref{3.2.10}), yielding
\begin{equation}\label{3.2.14}
\frac{1}{4}\;|J_+|^2\;\left[ 1\;+\;\zeta_{iz}\zeta_{fz}\;+\;(\zeta_{ix}\zeta_{fy}-\zeta_{iy}\zeta_{fz})\;\xi_1\;+\;(\zeta_{iz}+\zeta_{fz})\;\xi_2 \;
+ \;(\zeta_{ix}\zeta_{fx}+\zeta_{iy}\zeta_{fy})\;\xi_3\right].
\end{equation}

Proceeding with the remaining terms of (\ref{3.2.6}) in the same way, we get the following result, valid both for linearly and for circularly polarized photons,
$$\left| \sum_{m_i} a_{m_i}\sum_{m_s} b_{m_s}^\ast \sum_\sigma f_\sigma\;\tilde{M}_{fi}(\bfe_\sigma^\ast, m_i,m_s)\right|^2$$
$$=\;\frac{1}{4}\;\left( |J_+|^2\;+\;|J_-|^2\right)\;\left[ 1+\zeta_{iz}\zeta_{fz}\;+\;(\zeta_{ix}\zeta_{fx}+\zeta_{iy}\zeta_{fy})\;\xi_3\right]$$
\begin{equation}\label{3.2.15}
+\;\frac{1}{4}\;\left( |J_+|^2\;-\;|J_-|^2\right)\;\left[ (\zeta_{ix}\zeta_{fy}-\zeta_{iy}\zeta_{fx})\;\xi_1\;+\;(\zeta_{iz}+\zeta_{fz})\;\xi_2\right]
\end{equation}
$$+\;\frac{1}{4}\;\left( |S_+|^2\;+\;|S_-|^2\right)\;\left[ 1-\zeta_{iz}\zeta_{fz}\;-\;(\zeta_{ix}\zeta_{fx}-\zeta_{iy}\zeta_{fy})\;\xi_3\right]$$
$$-\;\frac{1}{4}\;\left( |S_+|^2\;-\;|S_-|^2\right)\;\left[ (\zeta_{ix}\zeta_{fy}+\zeta_{iy}\zeta_{fx})\;\xi_1\;+\;(\zeta_{iz}-\zeta_{fz})\;\xi_2\right]\;+\;R_{nd},$$
where $R_{nd}$ comprises the remaining nondiagonal terms,
$$R_{nd}\;=\;\frac12\;\left\{\mbox{ Re}(J_+J_-^\ast)\;[\zeta_{ix}\zeta_{fx}+\zeta_{iy}\zeta_{fy}\;+\;(1+\zeta_{iz}\zeta_{fz})\;\xi_3]\;+\;\mbox{Im}(J_+J_-^\ast)\;[(\zeta_{iz}+\zeta_{fz})\;\xi_1\right.$$
$$-\;(\zeta_{ix}\zeta_{fy}-\zeta_{iy}\zeta_{fx})\;\xi_2]\;+\;\mbox{Re}(S_+S_-^\ast)\;[-(\zeta_{ix}\zeta_{fx}-\zeta_{iy}\zeta_{fy})\;+\;(1-\zeta_{iz}\zeta_{fz})\;\xi_3]$$
$$+\;\mbox{Im}(S_+S_-^\ast)\;[(-\zeta_{iz}+\zeta_{fz})\;\xi_1\;+\;(\zeta_{ix}\zeta_{fy}+\zeta_{iy}\zeta_{fx})\;\xi_2]\;+\;\mbox{Re}(J_+S_+^\ast)\;[\zeta_{ix}\zeta_{fz}$$
$$-\zeta_{iz}\zeta_{fy}\,\xi_1 + \zeta_{ix}\,\xi_2-\zeta_{iz}\zeta_{fx}\,\xi_3]\;+\;\mbox{Im}(J_+S_+^\ast)\;[\zeta_{iy}-\zeta_{fx}\,\xi_1+\zeta_{iy}\zeta_{fz}\,\xi_2+\zeta_{fy}\,\xi_3]$$
$$+\;\mbox{Re}(J_-S_-^\ast)\;[\zeta_{ix}\zeta_{fz}+\zeta_{iz}\zeta_{fy}\,\xi_1-\zeta_{ix}\,\xi_2-\zeta_{iz}\zeta_{fx}\,\xi_3]\;+\;\mbox{Im} (J_-S_-^\ast)\;[\zeta_{iy}+\zeta_{fx}\,\xi_1$$
$$-\zeta_{iy}\zeta_{fz}\,\xi_2+\zeta_{fy}\,\xi_3]\;+\;\mbox{Re} (J_+S_-^\ast)\;[-\zeta_{iz}\zeta_{fx}-\zeta_{iy}\zeta_{fz}\,\xi_1-\zeta_{fx}\,\xi_2+\zeta_{ix}\zeta_{fy}\,\xi_3]$$
$$+\;\mbox{Im}(J_+S_-^\ast)\;[\zeta_{fy}+\zeta_{ix}\,\xi_1+\zeta_{iz}\zeta_{fy}\,\xi_2+\zeta_{iy}\,\xi_3]\;+\;\mbox{Re}(J_-S_+^\ast)\;[-\zeta_{iz}\zeta_{fx}+\zeta_{iy}\zeta_{fz}\,\xi_1$$
\begin{equation}\label{3.2.16}
\left. +\zeta_{fx}\,\xi_2+\zeta_{ix}\zeta_{fz}\,\xi_3]\;+\;\mbox{Im}(J_-S_+^\ast)\;[\zeta_{fy}-\zeta_{ix}\,\xi_1-\zeta_{iz}\zeta_{fy}\,\xi_2+\zeta_{iy}\,\xi_3]\right\}.
\end{equation}

This leads to the general form of the triply differential cross section,
\begin{equation}\label{3.2.17}
\frac{d^3\sigma}{d\omega d\Omega_k d\Omega_f}(\bfzeta_f,\bfzeta_i,\bfe_\lambda^\ast)\;=\;\frac{1}{4}\;\left( \frac{d^3\sigma}{d\omega d\Omega_k d\Omega_f}\right)_0\left( 1\;+\;\sum_{jkl}\tilde{C}_{jkl}\;\zeta_{ij}  \zeta_{fl} \xi_k\right).
\end{equation}
The indices $j,k,l$ run from 0 to 3, omitting $j=k=l=0$.
The missing of any of the coefficients of $\bfzeta_i,\;\bfzeta_f$ or $\bfxi$ is expressed by a zero in the corresponding subscript of $\tilde{C}_{jkl}$.
In (\ref{3.2.17}) we have introduced $\tilde{C}_{jkl}$ instead of the $C_{jkl}$ which  appear in (\ref{3.1.1}) for $l=0$, in order to write the sum in a closed form. 
The expressions for $\tilde{C}_{jkl}$ are easily read off from (\ref{3.2.16}) and are explicitly given in Appendix C.

A peculiarity of the coplanar geometry is that out of the $4^3-1=63$ possible $\tilde{C}_{jkl}$, only 31
are actually  different from zero. Moreover, the other ones are pairwise identical, and only those pertaining to unpolarized or circularly polarized photons differ from each other.

From Appendix C, the nonvanishing ones are arranged in an ascending list,
$$C_{002},\;C_{011},\;C_{013},\;C_{021},\;C_{023},\;C_{030},\;C_{032},\;C_{101},\;C_{103},\;C_{110},$$
$$C_{112},\;C_{120},\;C_{122},\;C_{131},\;C_{133},\;C_{200},\;C_{202},\;C_{211},\;C_{213},\;C_{221},$$
\begin{equation}\label{3.2.18}
C_{223},\;C_{230},\;C_{232},\;C_{301},\;C_{303},\;C_{310},\;C_{312},\;C_{320},\;C_{322},\;C_{331},\;C_{333}.
\end{equation}
Note that the next neighbours to a given polarization coefficient are always missing.
For example, if the photon polarization is not observed, an unpolarized beam electron can only acquire polarization perpendicular to the reaction plane (since  $C_{002} \neq 0,$ but $C_{001}=C_{003}=0$).

With the result that $\tilde{C}_{230}=\tilde{C}_{002}$, a physical interpretation can be given to the
polarization correlation $C_{230}$, which is not accessible experimentally for unobserved polarization of the final electron. In fact, $C_{230}$
is now identified to be the spin asymmetry when the spin of the
outgoing electron is flipped (for unpolarized initial electrons and photons),
\begin{equation}\label{3.2.19}
\tilde{C}_{230}\;=\;\tilde{C}_{002}\;=\;\frac{\sum_\lambda d^3\sigma(\bfzeta_f,\bfe_\lambda^\ast)\,-\,\sum_\lambda d^3\sigma(-\bfzeta_f,\bfe_\lambda^\ast)}
{\sum_\lambda d^3\sigma(\bfzeta_f,\bfe_\lambda^\ast)\,+\,\sum_\lambda d^3\sigma(-\bfzeta_f,\bfe_\lambda^\ast)},
\end{equation}
in correspondence to (\ref{3.1.18}) for polarized initial, but unpolarized final electrons.
Here we have used the abbreviation $d^3\sigma(\bfzeta_f,\bfe_\lambda^\ast) \equiv \frac12 \sum_{\zeta_i} \frac{d^3\sigma(\bfzetas_i,\bfzetas_f,\bfes_\lambda^\ast)}{d\omega d\Omega_k d\Omega_f}$,
now with $\bfzeta_f=(0,1,0)$ for $\varphi_s=\alpha_s=\frac{\pi}{2}$.

The difference between the $C_{jk0}$ from Tseng \cite{T02} and the $\tilde{C}_{jk0}$ introduced in (\ref{3.2.17}) is at most a sign. Comparing with (\ref{3.1.1}) one finds
\begin{equation}\label{3.2.20}
\tilde{C}_{120}=-C_{120},\quad \tilde{C}_{230}=-C_{230},\quad \tilde{C}_{200}=-C_{200},\quad \tilde{C}_{310}=-C_{310},
\end{equation}
while the remaining spin asymmetries are unchanged,
$\tilde{C}_{jk0}=C_{jk0}$ for $(jk)=(03),\;(32)\;$ and $(11)$.

\section{Outlook into noncoplanar geometry}
\setcounter{equation}{0}

When the azimuthal angle $\varphi_f$ of the momentum of the outgoing electron is neither 0 nor $\pi$, the symmetry relation (\ref{3.2.3}) is no longer true. 
Hence, in addition to $J_\pm$ and $S_\pm$, there appear the four  transition matrix elements $K_\pm$ and $T_\pm$,
\begin{equation}\label{3.3.1}
K_\pm\;=\;\tilde{M}_{fi}(\bfe_\pm^\ast,\frac12,-\frac12),\qquad T_\pm\;=\;\tilde{M}_{fi}(\bfe_\pm^\ast,-\frac12,-\frac12).
\end{equation}
As a result, all 63 polarization correlations $C_{jkl}$ do appear in the cross section, and they are all distinct. In particular, the unpolarized triply differential cross section
is now expressed as
$$\left( \frac{d^3\sigma}{d\omega d\Omega_k d\Omega_f}\right)_0\;=\;N_0\;D_0,$$
\begin{equation}\label{3.3.2}
D_0\;=\;\frac12\;\left( |J_+|^2\;+\;|J_-|^2\;+\;|S_+|^2\;+\;|S_-|^2\;+\;|K_+|^2\;+\;|K_-|^2\;+\;|T_+|^2\;+\;|T_-|^2\right).
\end{equation}
In addition to the parameters $C_{002}$ and $C_{030}$ which describe the creation of a polarized particle involving two unpolarized ones, there appear in noncoplanar geometry further parameters of this kind,
which are $C_{001},\;C_{003}$ and $C_{010},\;C_{020}$.
Thus an unpolarized electron can now create a scattered electron which is polarized in the $(x,z)$-plane,
even when the photon polarization remains unobserved,
since
$$\tilde{C}_{001}\;=\;\mbox{Re}(J_-^\ast K_-+S_+^\ast T_++S_-^\ast T_-+J_+^\ast K_+)/D_0,$$
\begin{equation}\label{3.3.3}
\tilde{C}_{003}\;=\;\frac12\;\left( |J_+|^2\;+\;|J_-|^2\;-\;|K_+|^2\;-\;|K_-|^2\;+\;|S_+|^2\;+\;|S_-|^2\;-\;|T_+|^2\;-\;|T_-|^2\right)/D_0.
\end{equation}
It is easy to check the results for coplanar geometry, using $K_\pm=-S_\mp$ and $T_\pm=J_\mp$. This gives $\tilde{C}_{001}=\tilde{C}_{003}=0.$

Another example are the parameters $\tilde{C}_{032}$ and $\tilde{C}_{200}$ which coincide in coplanar geometry. Now they are distinct,
$$\tilde{C}_{032}\;=\;\mbox{Im}(J_+^\ast K_-+S_+^\ast T_-+S_-^\ast T_++J_-^\ast K_+)/D_0,$$
\begin{equation}\label{3.3.4}
\tilde{C}_{200}\;=\;-\;\mbox{Im}(J_+^\ast S_++J_-^\ast S_- +K_+^\ast T_+ + K_-^\ast T_-)/D_0.
\end{equation}
The derivation of all noncoplanar polarization correlations in terms of $J_\pm,\;S_\pm,\;K_\pm$ and $T_\pm$ proceeds in the same way as for coplanar geometry. 

\section{Sum rules for the polarization correlations}
\setcounter{equation}{0}

Let us first consider the case of the doubly differential cross section, for which Pratt et al \cite{PMS} established the following sum rule,
\begin{equation}\label{3.4.1}
C_{32}^2+C_{12}^2+C_{20}^2 +C_{03}^2 +C_{31}^2+C_{11}^2 \,-\,C_{23}^2\;=\;1,
\end{equation}
for the polarization correlations defined in (\ref{3.1.19}), provided the outgoing electron is in a $j_f=\frac12$ state,
or more general, if the scattered electron is in a state with fixed angular
momentum $j_f$ and fixed projection $|m_f|$.
The idea of the existence of such a sum rule was based on a previous observation that for very heavy nuclei and circularly polarized photons near the short-wavelength limit, one has \cite{Jaku12}
\begin{equation}\label{3.4.2}
C_{20}^2+C_{12}^2+C_{32}^2\;\approx\;1,
\end{equation}
in analogy to a strict sum rule valid for the three polarization correlations in elastic potential scattering
(see section 3.5).

There exists another sum rule which is valid in the Born limit, that is for $Z_T \to 0.$
In that case, nearly all polarization correlations vanish, except $C_{03},\;C_{12}$ and $C_{32}$  \cite{TP73}. Therefore, for small $Z_T$, one derives from (\ref{3.4.1})  \cite{PMS},
\begin{equation}\label{3.4.3}
C_{32}^2\;+\;C_{12}^2\;+\;C_{03}^2\;\approx\;1.
\end{equation}

Let us now turn to the case of the triply differential cross section, and let us restrict ourselves to coplanar geometry. All sum rules to be 
discussed below only depend on the squares of the spin asymmetries, and we can omit the tilde since $\tilde{C}_{jkl}^2=C_{jkl}^2$.
We will show that there is a total of four sum rules,
valid for arbitrary final scattering states,
which are easily derived by using the explicit representation of $\tilde{C}_{jkl}$ in terms of $J_\pm$ and $S_\pm$ as provided in Appendix C.
Three of the sum rules relate to the unobserved polarization of just one particle.

\subsection{Unobserved polarization of the final electron}

This special case is described with 7 parameters $C_{jk0}$, and we have to prove the following sum rule, which mirrors the sum rule (\ref{3.4.1}) pertaining to the doubly differential cross section,
\begin{equation}\label{3.4.4}
C_{320}^2 +\;(C_{120}^2+C_{200}^2)\;+\;(C_{030}^2+C_{310}^2)\;+\;(C_{110}^2-C_{230}^2)\;=\;1.
\end{equation}
Multiplying by $D_0^2$, we have
$$(C_{030}^2+C_{310}^2)\;D_0^2\;+\;(C_{120}^2+C_{200}^2)\;D_0^2$$
$$=\;4\;|J_+J_-^\ast +S_-S_+^\ast|^2\;+\;4\;|J_-^\ast S_--J_+S_+^\ast|^2$$
$$=\;4\;\left[\,|J_+|^2\;|J_-|^2\;+\;|S_-|^2\;|S_+|^2\;+\;J_+J_-^\ast S_-^\ast S_+\;+\;S_-S_+^\ast J_+^\ast J_- \right.$$
\begin{equation}\label{3.4.5}
\left. +\;|J_-|^2\;|S_-|^2\;+\;|J_+|^2\;|S_+|^2\;-\;J_-^\ast S_- J_+^\ast S_+\;-\;J_+S_+^\ast J_-S_-^\ast \right].
\end{equation}

Furthermore,
$$(C_{110}^2-C_{230}^2)\;D_0^2\;=\;4\;(\mbox{Im}(J_+S_-^\ast -J_-S_+^\ast))^2\;-\;4\;(\mbox{Im}(J_+S_-^\ast + J_- S_+^\ast))^2$$
$$=\;-16\;\mbox{Im}(J_+S_-^\ast) \; \mbox{Im}(J_-S_+^\ast)\;=\;4\;(J_+S_-^\ast -J_+^\ast S_-)\,(J_-S_+^\ast - J_-^\ast S_+)$$
\begin{equation}\label{3.4.6}
=\;4\left[ J_+S_-^\ast J_-S_+^\ast-J_+S_-^\ast J_-^\ast S_+-J_+^\ast S_-J_-S_+^\ast +J_+^\ast S_-J_-^\ast S_+\right],
\end{equation}
which compensates the nondiagonal terms of (\ref{3.4.5}).
Finally,
$$(C_{320}^2-1)\;D_0^2\;=\;(\;|J_+|^2\;+\;|S_-|^2\;-\;(|J_-|^2+|S_+|^2)\;)^2\;-\;(\;|J_+|^2\;+\;|S_-|^2\;+\;|J_-|^2\;+\;|S_+|^2)^2$$
\begin{equation}\label{3.4.7}
=\;-4\;(\;|J_+|^2\;+\;|S_-|^2)(\;|J_-|^2\;+\;|S_+|^2)
\end{equation}
$$=\;-4\;(\;|J_+|^2\;|J_-|^2\;+\;|J_+|^2\;|S_+|^2\;+\;|S_-|^2\;|J_-|^2\;+\;|S_-|^2\;|S_+|^2),$$
which compensates the diagonal terms of (\ref{3.4.5}). Hence the sum rule is established.
A numerical proof, which mirrors the accuracy of the partial-wave calculations, is provided in \cite{Jaku12}.

\subsection{Unobserved polarization of the initial electron}

Here there are again 7 parameters $C_{0kl}$  involved. The second sum rule reads
\begin{equation}\label{3.4.8}
C_{023}^2\;+\;(C_{021}^2 + C_{002}^2)\;+\;(C_{030}^2+C_{013}^2)\;+\;(C_{011}^2 - C_{032}^2)\;=\;1.
\end{equation}
It simply follows from the previous sum rule (\ref{3.4.4}) if each $C_{jk0}$ is replaced by $C_{0kj}$.
Correspondingly, the proof can be copied from the
proof of (\ref{3.4.4}), if in the $\tilde{C}_{0kj}$
entering into (\ref{3.4.8}) the formal replacement $S_+ \mapsto \tilde{S}_-,\;S_- \mapsto \tilde{S}_+$ is made, while keeping $J_\pm$ unchanged.
In fact, the resulting dependence of $|\tilde{C}_{0kj}|$ on $\tilde{S}_\pm$ and  $J_\pm$ is exactly the same as the dependence of $|\tilde{C}_{jk0}|$ on $S_\pm$ and $J_\pm$.

\subsection{Unobserved photon polarization}

Also in this case there exists a sum rule,
\begin{equation}\label{3.4.9}
C_{303}^2\;+\;(C_{103}^2+C_{200}^2)\;+\;(C_{002}^2+C_{301}^2)\;+\;(C_{101}^2-C_{202}^2)\;=\;1.
\end{equation}
The proof of this third sum rule can again be carried out
with the help of the proof of (\ref{3.4.4}).
To do so, one needs the following replacement,
$S_- \mapsto i\tilde{J}_-$ and $J_-\mapsto -i \tilde{S}_-$, while $S_+$ and $J_+$ remain unchanged.
Then the resulting dependence of $\tilde{C}_{303}$ on $J_+,\;\tilde{J}_-,\;S_+$ and $\tilde{S}_-$ is identical to the dependence of $\tilde{C}_{320}$ on $J_\pm$ and $S_\pm$. This termwise correspondence between the coefficients of the two sum rules holds also for the other $C_{j0l}$ from (\ref{3.4.9}).

\subsection{Sum rule involving all linear independent parameters}

In order to derive this fourth sum rule we recall that the set $\{C_{j0l},C_{j2l}\}$, which corresponds to unpolarized or circularly polarized photons and has 15 nonzero members,
is the maximum set of linearly independent polarization correlations $C_{jkl}$.

Let us express the sum rules (\ref{3.4.4}) and (\ref{3.4.8}) in terms of this set by using the pairwise identities from Appendix C,
$$C_{320}^2+C_{120}^2+C_{200}^2+C_{202}^2+C_{122}^2+C_{322}^2-C_{002}^2\;=\;1$$ \hfill (3.4.4a)
$$C_{023}^2+C_{021}^2+C_{002}^2+C_{202}^2+C_{221}^2+C_{223}^2-C_{200}^2\;=\;1,$$ \hfill (3.4.8a)\\
and add these two equations to the third sum rule (\ref{3.4.9}). As a result, we obtain a sum rule containing each member of the above set just once,
$$C_{002}^2+C_{021}^2+C_{023}^2+C_{101}^2+C_{103}^2+C_{120}^2+C_{122}^2+C_{200}^2$$
\begin{equation}\label{3.4.10}
+C_{202}^2+C_{221}^2+C_{223}^2+C_{301}^2+C_{303}^2+C_{320}^2+C_{322}^2\;=\;3,
\end{equation}
their squares adding up to 3.
We note that the identical sum rule is obtained by adding the squares of all 15 parameters pertaining to linearly polarized photons (of the form $C_{j1l},\;C_{j3l})$,
due to the pairwise identities from Appendix C.

\section{Correspondence to the spin asymmetries in elastic scattering}
\setcounter{equation}{0}
\setcounter{figure}{0}

The basic motivation of investigating similarities between elastic electron scattering and hard brems\-strah\-lung is the particle-wave duality which holds
for photons as well as for electrons. Typical examples where the photon acts as a particle
are the photoeffect and its inverse process, the radiative recombination. Since bremsstrahlung at the short-wavelength limit (SWL) can be interpreted as radiative recombination in the limit of zero binding of the electronic final state,
it is only natural that such hard photons acquire the characteristics of a particle.

When exploring the similar behaviour of a scattered electron and an emitted photon with approximately the same momentum,
it is of advantage to look at the respective spin asymmetries which are a very sensitive tool for such investigations.
Let us therefore start by reviewing the elastic scattering process. We will assume that the collision energy does not exceed a few tens of MeV such that elastic scattering can be described
within the phase-shift analysis for potential scattering, even for nuclei with nonzero spin.

The scattering amplitude is defined by \cite{Lan4}

\begin{equation}\label{3.5.1}
f(\bfzeta_i,\bfzeta_f,p_i,\vartheta_f)\;=\;\langle  w_f|\,A\,+\,B\,\bfn \bfsigma\,|w_i\rangle,
\end{equation}
where $w_i$ and $w_f$ are the initial, respectively final, polarization spinors as defined
in sections 2.4.1 and 3.1.
In a
 coordinate system where the $z$-axis is taken along $\bfp_i$, the $y$-axis along $\bfp_i \times \bfp_f$ and the $x$-axis along $\bfe_y \times \bfp_i$,
the normal $\bfn$ to the scattering plane is equal to $\bfe_y$.
This coordinate system is chosen so as to correspond to the one used for bremsstrahlung.

We recall that the scattering amplitude is provided by the asymptotic form of the electronic scattering state $\psi_i$.
Using its partial-wave expansion (\ref{2.4.6}),
the direct (A) and the spin-flip (B) contribution to the scattering amplitude can be expressed in the following way,
$$A\;=\;\frac{1}{2ip_i}\sum_{l=0}^\infty \left[ (l+1) \,e^{2i\delta_{-l-1}}-1)\;+\;l\,(e^{2i\delta_l}-1)\right]\;P_l(\cos \vartheta_f)$$
\begin{equation}\label{3.5.2}
B\;=\;\frac{1}{2p_i}\sum_{l=1}^\infty \left( e^{2i\delta_{-l-1}}\,-\,e^{2i\delta_l}\right)\;P_l^1(\cos \vartheta_f),
\end{equation}
where $\delta_\kappa$ are the phase shifts of $\psi_i$, $P_l$ is a Legendre polynomial, $P_l^1$  an associated  Legendre function and $\vartheta_f$ is the scattering angle (between $\bfp_i$ and $\bfp_f$).

Rather than fixing the initial electronic spin and looking at its change during the scattering process, it is experimentally of advantage to vary the beam polarization while keeping the
spin polarization of the outgoing electron fixed.
In order to compare with photons in a helicity eigenstate (i.e. circularly polarized photons), this final spin polarization $\bfzeta_f$ is taken along the direction of motion $\hat{\bfp}_f$ (which will be denoted by $\zeta_\parallel=1$ and describes  right-handed electrons) or antiparallel to it ($\zeta_\parallel=-1$ for left-handed electrons). 
By choosing the initial spin polarization in turn along each of the coordinate axes, the scattering process is characterized by three polarization parameters, $S,\;R$ and $L$, which describe the spin change of the beam electron.
The cross section for electrons with initial polarization $\bfzeta_i$ attains the following form  \cite{Mo}
\begin{equation}\label{3.5.3}
\frac{d\sigma}{d\Omega_f}(\bfzeta_f,\bfzeta_i)\;=\;\frac12 \left( \frac{d\sigma}{d\Omega_f}\right)_0\;[1\;-\;R\,\zeta_\parallel(\bfzeta_i \bfe_x)\;+\;S\;(\bfzeta_i \bfe_y)\;+\;L\;\zeta_\parallel \bfzeta_i \bfe_z ].
\end{equation}
The prefactor is the differential cross section for elastic scattering of unpolarized particles, which is given by
\begin{equation}\label{3.5.4}
\left( \frac{d\sigma}{d\Omega_f}\right)_0\;=\frac12 \sum_{\zeta_i,\zeta_f} \left| f(\bfzeta_i,\bfzeta_f,p_i,\vartheta_f)\right|^2\;=\;|A|^2\;+\;|B|^2.
\end{equation}
The spin asymmetries can also be expressed in terms of $A,\;B$ with an additional dependence on  $\vartheta_f$,
$$S\;=\;\frac{2\mbox{ Re }(AB^\ast)}{|A|^2\,+\,|B|^2},\qquad
R\;=\;\frac{2 \mbox{ Im }(AB^\ast)\cos \vartheta_f\,-\,(|A|^2\,-\,|B|^2)\sin \vartheta_f}{|A|^2\,+\,|B|^2},$$
\begin{equation}\label{3.5.5}
L\;=\;\frac{(|A|^2\,-\,|B|^2)\cos \vartheta_f\,+\,2\mbox{ Im } (AB^\ast) \sin \vartheta_f}{|A|^2\,+\,|B|^2}.
\end{equation}
It is easily derived from the representation (\ref{3.5.5}) that $S,\;R$ and $L$ obey the sum rule \cite{Mo},
\begin{equation}\label{3.5.6}
S^2\;+\;R^2\;+\;L^2\;=\;1.
\end{equation}

We see that (\ref{3.5.3}) has the same structure as the doubly differential cross section (\ref{3.1.20}) for the emission of circularly polarized photons.
Thus we find a pairwise correspondence of the polarization correlations,
associating $R$ with $C_{12}$ and $S$ with $-C_{20}$ for transversely and perpendicularly polarized electrons, respectively.
  For longitudinally polarized beam electrons, one has to compare $L$ and $C_{32}$, which describe the helicity conservation in elastic scattering, respectively the helicity transfer in bremsstrahlung.
Fig.3.5.1 shows the angular dependence of both sets of parameters for a gold target.
It is obvious that the similarity increases with energy $E_e$. This is related to the fact that at high $E_e$, the electron mass can be neglected, such that $E_i \approx E_e = \omega_{\rm SWL}$, yielding the same energy of outgoing electron and photon.
The electron-photon similarity holds only for very heavy atoms.
This can be related to the fact that in the limit of high field strength ($Z_T/c \to 1$), in addition to $E_i \to \infty$, the transition matrix structure of both processes becomes alike \cite{Jaku12}.

\begin{figure}
\vspace{-1.5cm}
\centering
\begin{tabular}{cc}
\hspace{-1cm}\includegraphics[width=.7\textwidth]{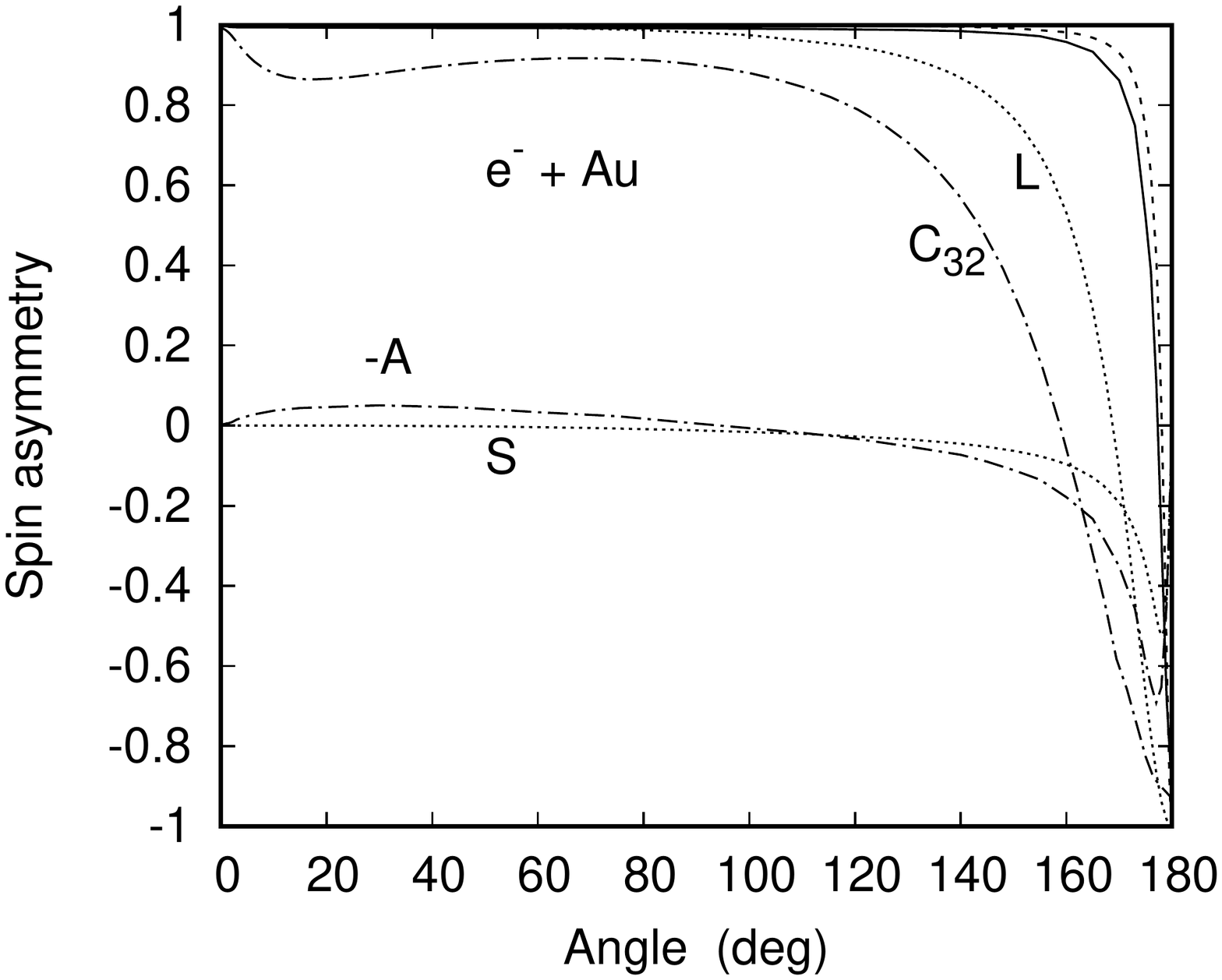}&
\hspace{-3cm} \includegraphics[width=.7\textwidth]{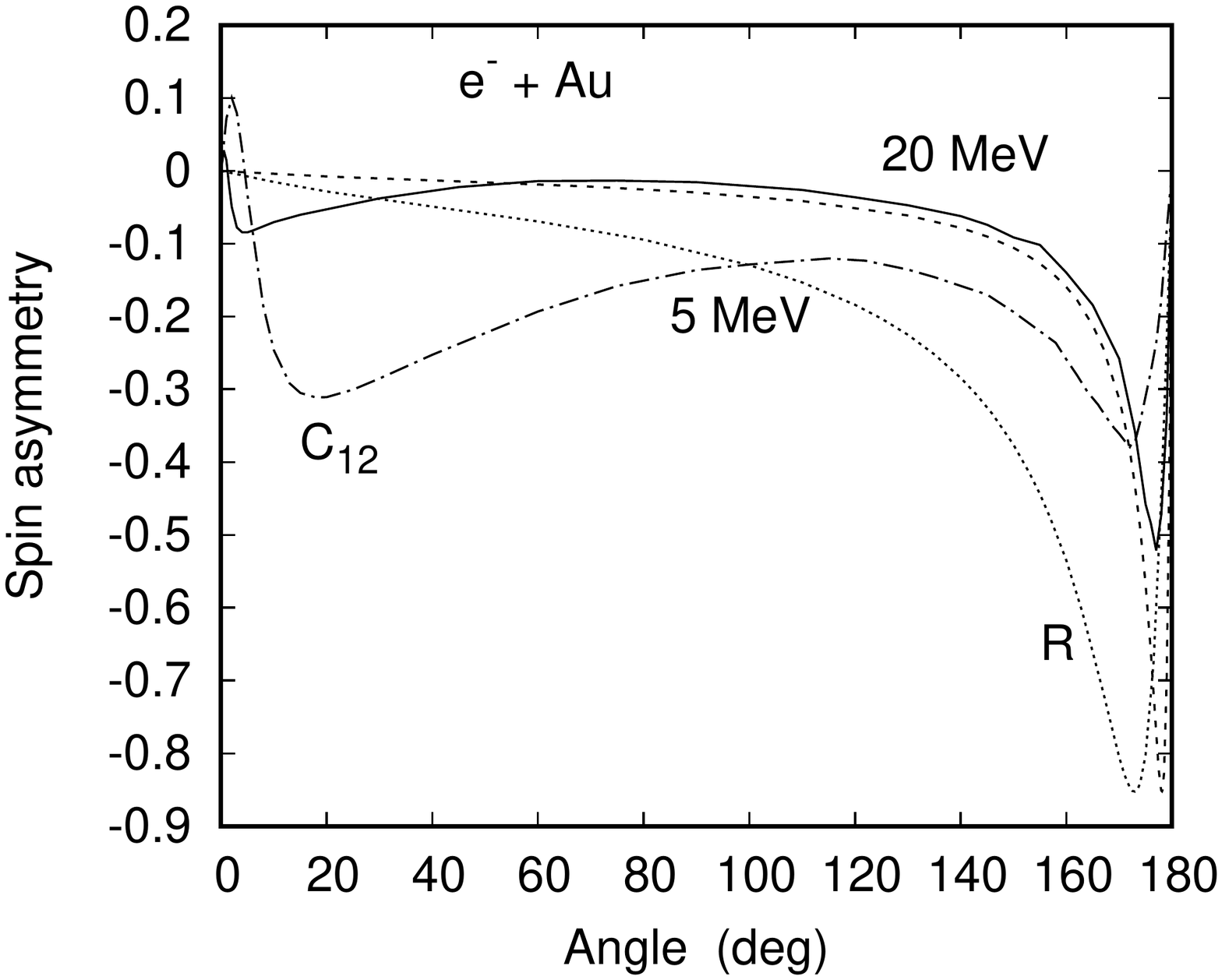}
\end{tabular}
\caption{
Angular dependence of the polarization correlations
for spin-polarized electrons colliding with a gold nucleus. Shown are the results for the bremsstahlung spin asymmetries  $C_{32},\;C_{12}$ and $A\equiv C_{20}$ at the short-wavelength limit
 in comparison with those from elastic scattering, $L,\;R$ and $S$. (The sign reversal of $A$ is due to different
definitions of the perpendicular spin asymmetries in bremsstrahlung and in elastic scattering experiments.)
Left panel: upper curves: $C_{32}$ (------------) and
$L\;(-----)$ for 20 MeV collision energy, $C_{32}\;(-\cdot - \cdot -)$ and $L\;(\cdots\cdots)$ for 3 MeV collision energy. Lower curves: $-A\;(-\cdot - \cdot -)$ and $S\,(\cdots\cdots)$ for 20 MeV. The results for $L$ and $S$ are obtained from (\ref{3.5.5}). The results for $C_{32}$ and $A$ at 20 MeV are obtained within the DSM model, and the
ones for $C_{32}$ at 3 MeV are Yerokhin's results from
the partial-wave code \cite{YS10}.
Right panel: $C_{12}\;(-\cdot - \cdot -)$ and $R\;(\cdots\cdots)$ at 5 MeV, $C_{12}$ (---------------) and $R\;(-----)$ at 20 MeV. The results for $C_{12}$ are obtained  within the DSM model, the ones for $R$ are due to (\ref{3.5.5}).
}
\end{figure}

\vspace{0.5cm}

Table 3.5. Sum rule $\Sigma(E_{e}) $ for the bremsstrahlung polarization correlations for an Au$^{79+}$ target at
collision energies  10, 15 and 20 MeV (columns 3,4,5),
and an Ag$^{47+}$ target at 10 MeV ($2^{\rm nd}$ column). The  photon emission angles $\theta_k$ range from $0^\circ$ to $ 180^\circ$.

\vspace{0.5cm}

\begin{tabular}[t]{r|l|l|l||l|}  
$\theta_k$&$\Sigma_{Ag}(10)$&$\Sigma_{Au}(10)$&$\Sigma_{Au}(15)$&$\Sigma_{Au}(20)$\\
&&&&\\ \hline
&&&&\\ 
0&0.999&0.999&1&1\\
5&0.996&0.976&0.994&0.997\\
10&0.996&0.989&0.994&0.995\\
30&0.991&0.975&0.983&0.991\\
60&0.989&0.961&0.978&0.987\\
90&0.988&0.952&0.973&0.984\\
120&0.986&0.942&0.971&0.982\\
150&0.975&0.924&0.961&0.976\\
170&0.892&0.795&0.886&0.934\\
175&0.720&0.659&0.777&0.830\\
177&0.672&0.659&0.712&0.763\\
180&0.686&0.770&0.825&0.831\\
\end{tabular}

\vspace{0.5cm}

Another test of the correspondence between the two scattering processes is provided by the sum rule. In correspondence to (\ref{3.5.6}),
one expects the three-terms sum rule (\ref{3.4.2}) for circular bremsstrahlung to be approximately valid, irrespective of the representation of the final electronic state.
Table 3.5 compares the angular dependence of the lhs of (\ref{3.4.2}), which will be termed $\Sigma(E_e)$,
for Ag and Au at collision energies of 5, 10 and 20 MeV.
It is clearly seen the $\Sigma(E_e)$ gets closer to unity the  higher $E_e$. On the other hand, it is closer to unity for the lighter elements. This is due to
the dominance of $C_{32}$ for angles up to $170^\circ$, which is lowered less for Ag than for Au at a given scattering angle.

\vspace{1.0cm}

{\Large\bf 4. Positron bremsstrahlung}
\setcounter{chapter}{4}
\setcounter{section}{0}
\setcounter{equation}{0}

\vspace{0.5cm}

Like any charged particle, also positrons will emit bremsstrahlung when they traverse an atomic field.
Investigations on high-energy positron bremsstrahlung are scarce, due to the assumption that for fast particles there should not
be much difference between electron and positron radiation. However, based on the observation of elastic lepton scattering where the high-energy diffraction structures show pronounced phase shifts between electrons and positrons
 \cite{Br91}, it is expected that similar differences occur in bremsstrahlung.
In fact, the finite-nuclear-size effects lead to bremsstrahlung diffraction structures in the same way as for elastic scattering \cite{Jaku13}.
These are particularly prominent in the spin asymmetries. With the advent of techniques for the production of polarized positron beams
 \cite{Al08}, an experimental verification should be feasible.

Theoretically, the high-energy positron bremsstrahlung spectra close to the short-wavelength limit (SWL) were already studied by Jabbur and Pratt \cite{JP63} within an analytical theory which relies on an expansion
of the final-state wavefunction (and hence of the cross section) in terms of $Z_T/c$ and where only a small number of final partial waves has to be included.
Later, the fully relativistic partial-wave theory was applied to bremsstrahlung spectra for collision energies up to 0.5 MeV  \cite{FPT,Ki86}.
A full account of the photon spectra as well as of the photon angular distribution up to 2.5 MeV was supplied by Tseng \cite{T02b}. Only recently, the polarization transfer from the positron to the photon was included for beam energies up to 1 MeV  \cite{Ye12}.
All these investigations reveal an exponential reduction of the positron bremsstrahlung as compared to electron bremsstrahlung when the SWL is approached.
This is due to the strong Coulomb repulsion between the positron and the nucleus which becomes effective at the large momentum transfers necessary to create hard photons.
On the other hand, for soft photons emitted into forward directions, the plane-wave Born approximation becomes
valid  \cite{BM54},
such that the electron-positron differences are very small \cite{Jaku18}.

\section{Positron theory}

Like for electron impact, the relativistic partial-wave formalism can be  applied to the calculation of the positron bremsstrahlung cross section.
In order to derive the scattering states of a positron one has to use charge conjugation. Given an electronic state of a bare nucleus, $\psi_{e^-}(\bfr,\bfzeta,Z_T)$, with spin polarization $\bfzeta$ and nuclear charge number $Z_T$,
the respective state for a positron is obtained by means of \cite{BD,Ros61}
\begin{equation}\label{4.1.1}
\psi_{e^+} (\bfr,\bfzeta,Z_T)\;=\;i\,\gamma^2\;\psi_{e^-}^\ast(\bfr,\bfzeta,-Z_T)
\end{equation}
with the Dirac matrix $\gamma^2\;=\; \left( \begin{array}{cc} 0 & \sigma_2\\
-\sigma_2&0
\end{array}\right) $, where $\sigma_2\,=\,\left( \begin{array}{cc} 0 &-i\\
i&0 
\end{array}\right)$.

Explicitly, the wavefunction for  an incoming electron which impinges along the $z$-direction can be written  in the following way,
\begin{equation}\label{4.1.2}
\psi^{(+)}_{i,e^-}(\bfr,\bfzeta_i,Z_T)\;=\;\sum_{m_i=\pm \frac12} a_{m_i} \sum_{\kappa_i} \sqrt{\frac{2l_i+1}{4\pi}}\;(l_i 0 \frac12 m_i\,|\,j_i m_i)\; i^{l_i}\;e^{i\delta_{\kappa_i}}
 \;\left( \begin{array}{c}
g_{\kappa_i}(r)\;Y_{j_il_im_i}(\hat{\bfr})\\
i\,f_{\kappa_i}(r)\;Y_{j_il_i'm_i}(\hat{\bfr})
\end{array}\right).
\end{equation}
Then the respective positron function will be  an outgoing state which reads
\begin{equation}\label{4.1.3}
\psi^{(-)}_{i,e^+}(\bfr,\bfzeta_i,Z_T)\;=\;i\sum_{m_i=\pm \frac12} a^\ast_{-m_i} \sum_{\kappa_i} \sqrt{\frac{2l_i+1}{4\pi}}\;(-1)^{\frac12-m_i} \;(l_i 0 \frac12 m_i\,|\,j_i m_i)
\; (-i)^{l_i}\;e^{-i\delta_{\kappa_i}}\;{f_{\kappa_i}(r)\;Y_{j_il_i'm_i}(\hat{\bfr}) \choose i g_{\kappa_i}(r)\;Y_{j_il_im_i}(\hat{\bfr})},
\end{equation}
where $g_{\kappa_i}$ and $f_{\kappa_i}$ are, respectively, the large and small components of the radial Dirac function. Note that $g_{\kappa_i}$ and $f_{\kappa_i}$ interchange their role when switching from electron to positron. 
The positron phase shifts $\delta_{\kappa_i}$ as well as $g_{\kappa_i}$ and $f_{\kappa_i}$ in (\ref{4.1.3})
result from  solutions to the Dirac equation with negative potential, $-V_T(r)$.
The other quantities appearing in (\ref{4.1.2}) and (\ref{4.1.3}) are the same as in the case of electrons (see section 2.4.1).

In the same way, the positron function which corresponds to an outgoing final electron with momentum $\bfp_f$ turns out  to be
\begin{equation}\label{4.1.4}
\psi^{(+)}_{f,e^+}(\bfr,\bfzeta_f,Z_T)=i\sum_{\kappa_f m_f}\sum_{m_l m_{s}} Y^\ast_{l_f m_l}(\hat{\bfp}_f)b^\ast_{-m_{s}}(-1)^{\frac12 -m_{s}}(l_fm_l\frac12 m_{s}\,|\,j_fm_f)
(-i)^{l_f}e^{i\delta_{\kappa_f}}{ f_{\kappa_f}(r)Y_{j_fl_f'm_f}(\hat{\bfr})\choose
i g_{\kappa_f}(r)Y_{j_fl_fm_f}(\hat{\bfr})}.
\end{equation}
 
The radiation matrix element for positron bremsstrahlung is calculated from  \cite{Ye12} (see also \cite{BD}),
\begin{equation}\label{4.1.5}
\overline{W}_{{\rm rad},e^+}(\bfzeta_f,\bfzeta_i)\;\equiv
\;-\;\frac{ie}{c}\;W_{{\rm rad},e^+}(\bfzeta_f,\bfzeta_i)\;=\;-\;\frac{ie}{c} \int d\bfr\;\psi_{i,e^+}^{(-)+}(\bfr,\bfzeta_i)\;\bfalpha\bfeps_\lambda^\ast\;e^{-i\bfks\bfrs}\;\psi_{f,e^+}^{(+)}(\bfr,\bfzeta_f),
\end{equation}
where we have retained the prefactor $ie/c$ in order to elucidate the sign change for positrons.
Correspondingly, for electrons, we can define $\overline{W}_{{\rm rad},e^-}(\bfzeta_f,\bfzeta_i)\equiv \frac{ie}{c}\,W_{\rm rad}(\bfzeta_f,\bfzeta_i)$ (with $W_{\rm rad}$ from (\ref{2.4.7})).
We shall restrict ourselves to the case where the polarization of the scattered positron is not observed
and thus has to be summed over.
According to the formalism developed for electrons \cite{T02,Jaku16}, 
 the triply differential cross section for positrons of total energy $E_i$ emitting a photon with frequency $\omega=ck$ into the solid angle $d\Omega_k$, 
while being scattered with final total energy $E_f$ into the solid angle $d\Omega_f$, is given by 
\begin{equation}\label{4.1.6}
\frac{d^3\sigma}{d\omega d\Omega_k d\Omega_f}(\bfzeta_i,\bfeps^\ast_\lambda)\;=\;\frac{4\pi^2 \omega p_fE_iE_f}{c^5p_i }\sum_{m_s=\pm\frac12} \left| F_{fi,e^+}(m_s,\bfzeta_i)\right|^2,
\end{equation}
where $F_{fi,e^+}$ is defined by means of 
\begin{equation}\label{4.1.7}
W_{{\rm rad},e^+}(\bfzeta_f,\bfzeta_i)\;=\;\frac{ic}{e}\;\overline{W}_{{\rm rad},e^+}(\bfzeta_f,\bfzeta_i)\;=\;\sum_{m_s=\pm\frac12}b_{-m_s}^\ast\;(-1)^{\frac12-m_s}\;F_{fi,e^+}(m_s,\bfzeta_i).
\end{equation}

Using the same techniques as in section 2.4.1 for the calculation of $F_{fi}$ in (\ref{2.4.19}),
the result for $F_{fi,e^+}$ is
$$F_{fi,e^+}(m_s,\bfzeta_i)\;=\;i\sum_{l_f=0}^\infty \sum_{m_l=-l_f}^{l_f} (-i)^{l_f}\;Y_{l_fm_l}^\ast(\hat{\bfp}_f)
\sum_{j_f=l_f\pm\frac12}\;(l_fm_l\frac12\,m_s\,|\,j_fm_f)$$
\begin{equation}\label{4.1.8}
\times\;\sum_{m_i=\pm\frac12} a_{-m_i}\;(-1)^{\frac12-m_i}\sum_{\kappa_i} \sqrt{2l_i+1}\;i^{l_i}\;e^{i(\delta_{\kappa_i}+\delta_{\kappa_f})}\;(l_i0\frac12\,m_i\,|\,j_im_i)\;S_{fi,e^+}.
\end{equation}
The factor $S_{fi,e^+}$ includes the sum over the photon
angular momenta $l$,
$$S_{fi,e^+}\;=\;\sum_{l=|l_i'-l_f|}^{l_i'+l_f}(-i)^l\;R_{fi}(l)\;\sum_{m_{s_f},m_{s_i}=\pm \frac12}\sqrt{\frac{(2l+1)(l-\mu)!}{(l+\mu)!}}\;P_l^\mu(\cos \theta_k)\;c_\varrho^{(\lambda)}\;W^B_{12,e^+}(l,l_f,l_i')$$
\begin{equation}\label{4.1.9}
-\;\sum_{l=|l_i-l_f'|}^{l_i+l_f'} (-i)^l\;R_{if}(l)\sum_{m_{s_f},m_{s_i}=\pm \frac12}\sqrt{\frac{(2l+1)(l-\mu)!}{(l+\mu)!}}\;P_l^\mu(\cos \theta_k)\;c_\varrho^\lambda\;\;W_{12,e^+}^B(l,l_f',l_i).
\end{equation}
Thereby $W^B_{12,e^+}$  results from the angular integration,
$$W^B_{12,e^+}(l,l_f,l_i')\;=\;\sqrt{\frac{3}{4\pi}}\;\sqrt{2l+1} \;
\sqrt{\frac{2l_f+1}{2l_i'+1}}\;(l_f 0\,l\,0|l_i'\,0)$$
\begin{equation}\label{4.1.10}
\times \;(l_i'\mu_i\frac12\,m_{s_i}|\,j_im_i)\;(l_f \mu_f \frac12 m_{s_f}\,|\,j_fm_f)\;(\frac12 m_{s_f} 1 \varrho\,|\,\frac12 m_{s_i})\;(l_f\mu_f l\mu\,|\,l_i'\mu_i).
\end{equation}
The sum over $l$ in (\ref{4.1.9})  runs again in steps of 2 due to the selection rules from the Clebsch-Gordan coefficients in $W_{12,e^+}^B$, requiring that  $l_f+l+l_i'=$ even in the first sum, and $l_f'+l+l_i=$ even
 in the second one.

The selection rules for the magnetic quantum numbers imply
\begin{equation}\label{4.1.11}
\mu\;=\;\mu_i-\mu_f,\quad \mu_f\;=\;m_f-m_{s_f},\quad \mu_i\;=\;m_i-m_{s_i},\quad \varrho\;=\; m_{s_i}-m_{s_f},\quad m_l\;=\;m_f-m_s.
\end{equation}

The radial integrals $R_{fi}$ and $R_{if}$ are given by
\begin{equation}\label{4.1.12}
{R_{fi}(l) \choose R_{if}(l)}\;=\;\int_0^\infty r^2\,dr\;j_l(kr)\;{g_{\kappa_f}(r)\; f_{\kappa_i}(r) \choose f_{\kappa_f}(r)\;g_{\kappa_i}(r)}
\end{equation}
and agree with the ones for electron scattering (except for the negative charge number in the defining equation of the radial functions).

The interrelation between electron and positron bremsstrahlung cross sections turns out to be given by the formal identity, 
\begin{equation}\label{4.1.13}
\sum_{m_s}\left| F_{fi,e^+}(m_s,\bfzeta_i)\right|^2\;=\;\sum_{m_s}\left| F_{fi,e^-}(m_s,\bfzeta_i)\right|^2,
\end{equation}
taken into consideration that the respective radial integrals as well as the phase shifts differ in the sign of $Z_T$.

Eq.(\ref{4.1.13}) can be proved by changing in (\ref{4.1.6}) with (\ref{4.1.9})
simultaneously the sign of all magnetic quantum numbers except for $\varrho$ and $\mu$, and by making use of the symmetry properties of the Clebsch-Gordan coefficients and of the spherical harmonic functions.

The spin asymmetry for positrons is defined in the same way as for electrons. Thus the respective formulae from chapter 3, including the sum rules, are also valid for positron impact. 

\section{Results for positron versus electron impact}
\setcounter{equation}{0}

For brevity we restrict ourselves to the triply differential cross section in coplanar geometry, which is most likely to be covered by future experiments.
Fig.4.2.1 shows the angular dependence of bremsstrahlung emitted by 5 MeV and 15 MeV leptons colliding with a $^{208}$Pb target,
scattering into the backward direction, and losing $3/4$ of its kinetic energy.
The shape of the angular distribution, displayed in the figure, is characteristic for high-energy brems\-strah\-lung.
For both lepton species, the maximum photon intensity is radiated into a small cone around the beam axis, and the focusing near $\theta_k=0$ increases with collision energy.
This feature is explained in section 2.4.3.

\setcounter{figure}{0}
\begin{figure}
\vspace{-1.5cm}
\includegraphics[width=13cm]{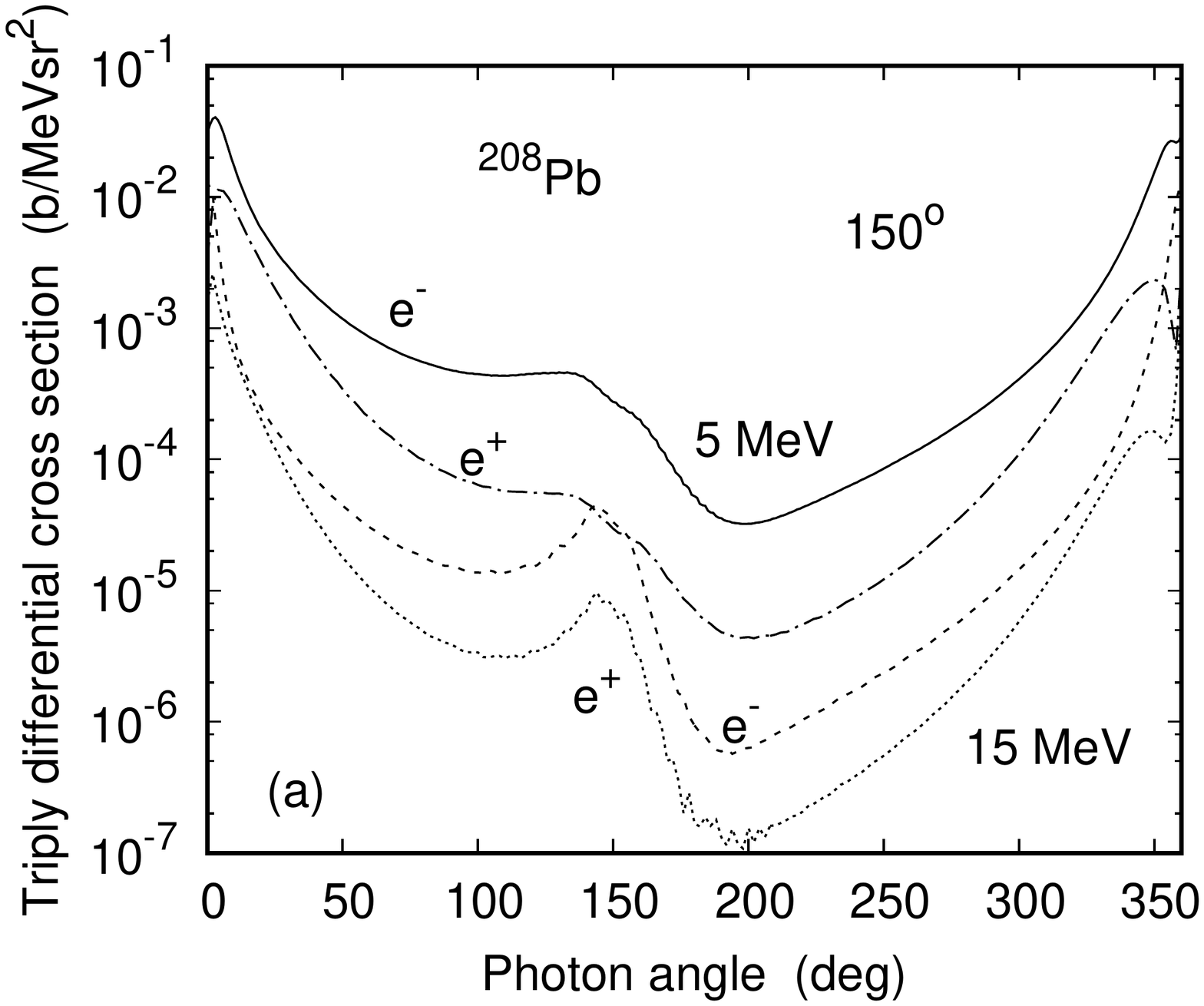}
\caption{
Triply differential bremsstrahlung cross section $d^3\sigma/d\omega d\Omega_k d\Omega_f$ from electrons and positrons colliding at $E_e=5$ MeV and 15 MeV with $^{208}$Pb as a function of photon angle. The lepton angles are $\vartheta_f=150^\circ$ and $\varphi_f=0$.
The photon frequency is $\omega=0.75\; E_e$.
Electrons: --------, 5 MeV; $-----$, 15 MeV. Positrons: $-\cdot - \cdot -$, 5 MeV; $\cdots\cdots$, 15 MeV. The wiggles near $180^\circ$ are due to numerics.}
\end{figure}

\vspace{0.2cm}

There appears a second maximum in the angular distribution when the photon is
ejected into the same direction as the lepton.
This peak gets more pronounced when either the beam energy is increased, or when the photon frequency gets lower.
For very high collision energies it  splits into a sharp double peak at forward scattering angles and low $\omega$ (see, e.g.,  \cite[Fig.4.20]{HN04}).

Positron cross sections  usually fall below those for electrons because the high momentum transfer to the nucleus, 
necessary to emit hard photons into the backward hemisphere, requires close lepton-nucleus collisions
which are suppressed by the repulsive positron-nucleus interaction.
Included in the calculations are nuclear size effects, resulting from a potential which considers the nuclear charge distribution. These effects come into play at backward photon angles, and they are clearly visible at 15 MeV (\cite{Jaku16}, see also Fig.2.7.1).

The frequency dependence of the photon  intensity is displayed in Fig.4.2.2 for 3.5 MeV leptons colliding with gold. Considered are small emission angles of photon and lepton,
but the two particles are emitted
into opposite sides of the beam axis.
Justified by the applicability of the plane-wave Born approximation at small angles when the photon frequency is low, that theory is used to interpret the structures
near
2.5 MeV as resulting from interference \cite{Jaku18} between the two parts of the transition amplitude (\ref{2.1.5}) which correspond to photon emission before, respectively after, the scattering from the nuclear field.
The strong decrease of the positron intensity near the high-energy end of the spectrum is also clearly seen.

\begin{figure}
\vspace{-1.5cm}
\includegraphics[width=13cm]{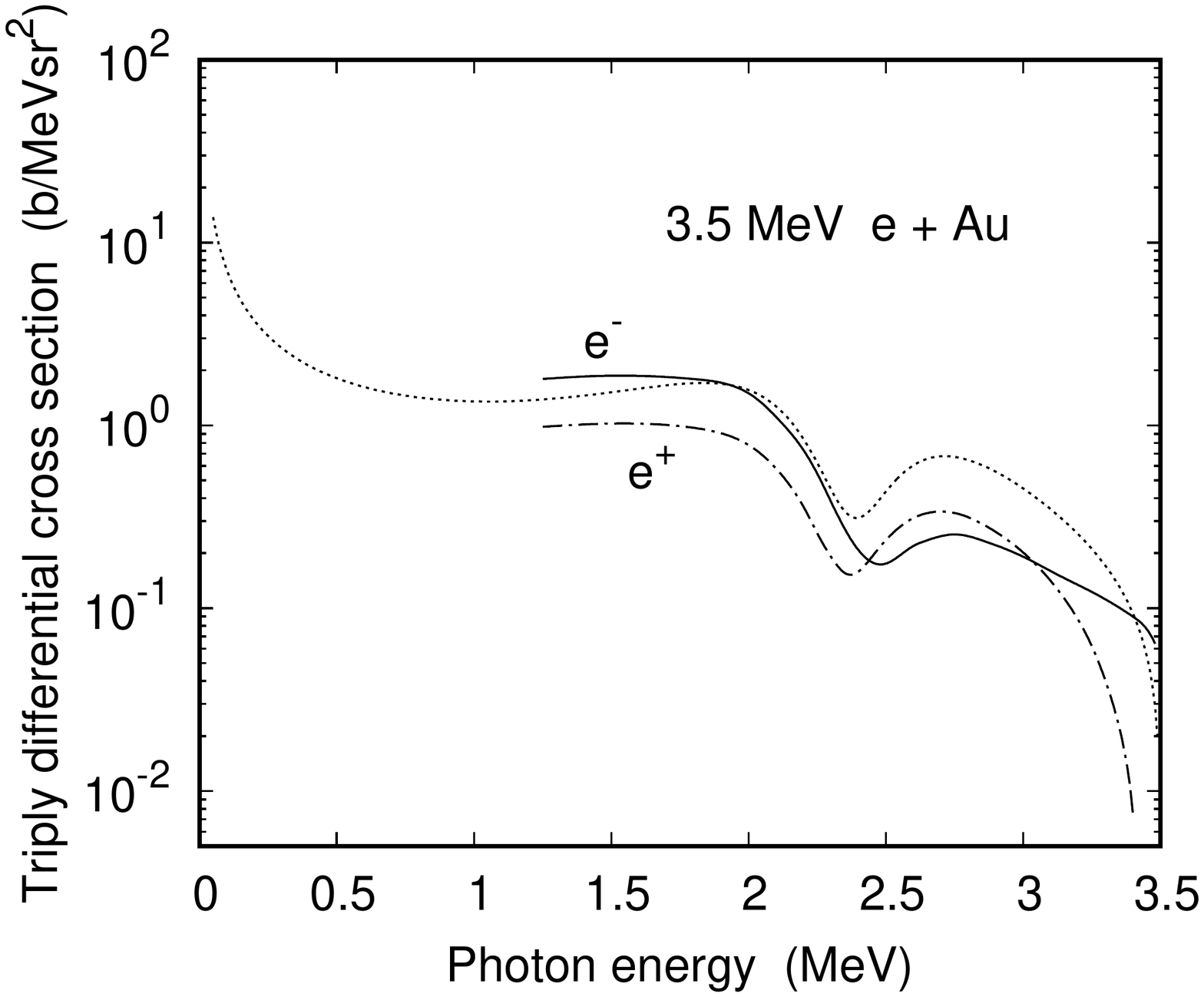}
\caption{
Triply differential cross section from 3.5 MeV electrons and positrons colliding with gold as a function of photon frequency.
The photon angle is $\theta_k=20^\circ$, the lepton angles are $\vartheta_f=30^\circ$ and $\varphi_f=180^\circ$.
$\cdots\cdots$, PWBA; ----------, electron scattering: $-\cdot - \cdot -$, positron scattering.}
\end{figure}

As an example for the angular dependence of the polarization correlations, we have displayed in Fig.4.2.3 the circular polarization correlations $C_{320}$, $C_{120}$ and $C_{200}$ from 3.5 MeV leptons
colliding with $^{208}$Pb. Since the  spin asymmetry increases with photon frequency and scattering angle, we have fixed $\omega$ to 3/4 of the collision energy $E_e$, and  have taken a backward scattering angle.
$C_{320}$ as well as $C_{120}$  have  angular dependencies which are similar for the two leptons (see Fig.4.2.3(a)).
However, for the perpendicular spin asymmetry $C_{200}$,
shown in Fig.4.2.3(b),
there is a sign change when switching form electrons to positrons.
In this context we recall that $C_{200}=0$ in PWBA, with a linear increase in $Z_T$ for small $Z_T$  \cite{TP73}.
In contrast, the other two spin asymmetries are finite in PWBA and remain so for small $Z_T$ irrespective of the lepton species.

\begin{figure}
\vspace{-1.5cm}
\centering
\begin{tabular}{cc}
\hspace{-1cm}\includegraphics[width=.7\textwidth]{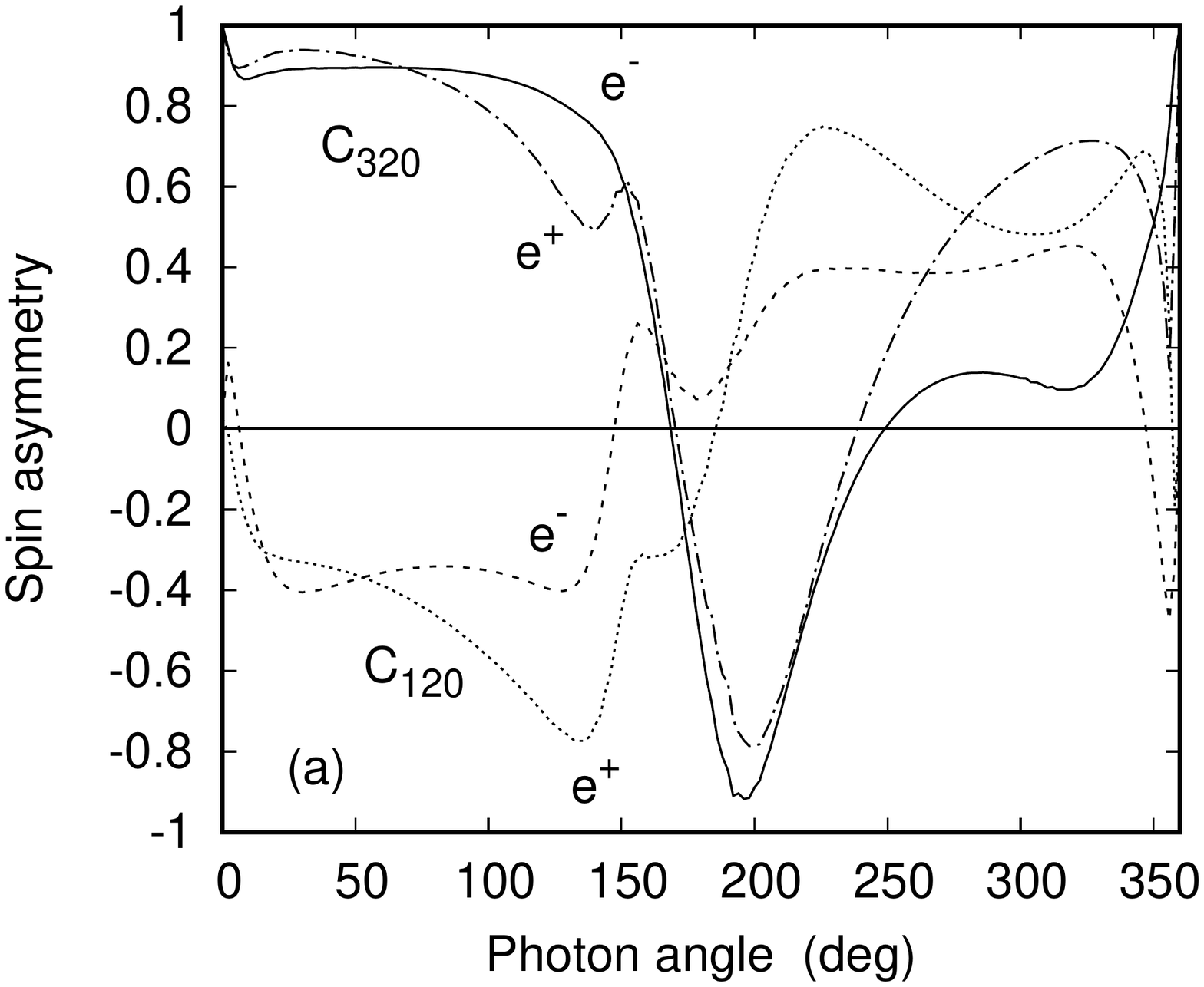}&
\hspace{-3cm} \includegraphics[width=.7\textwidth]{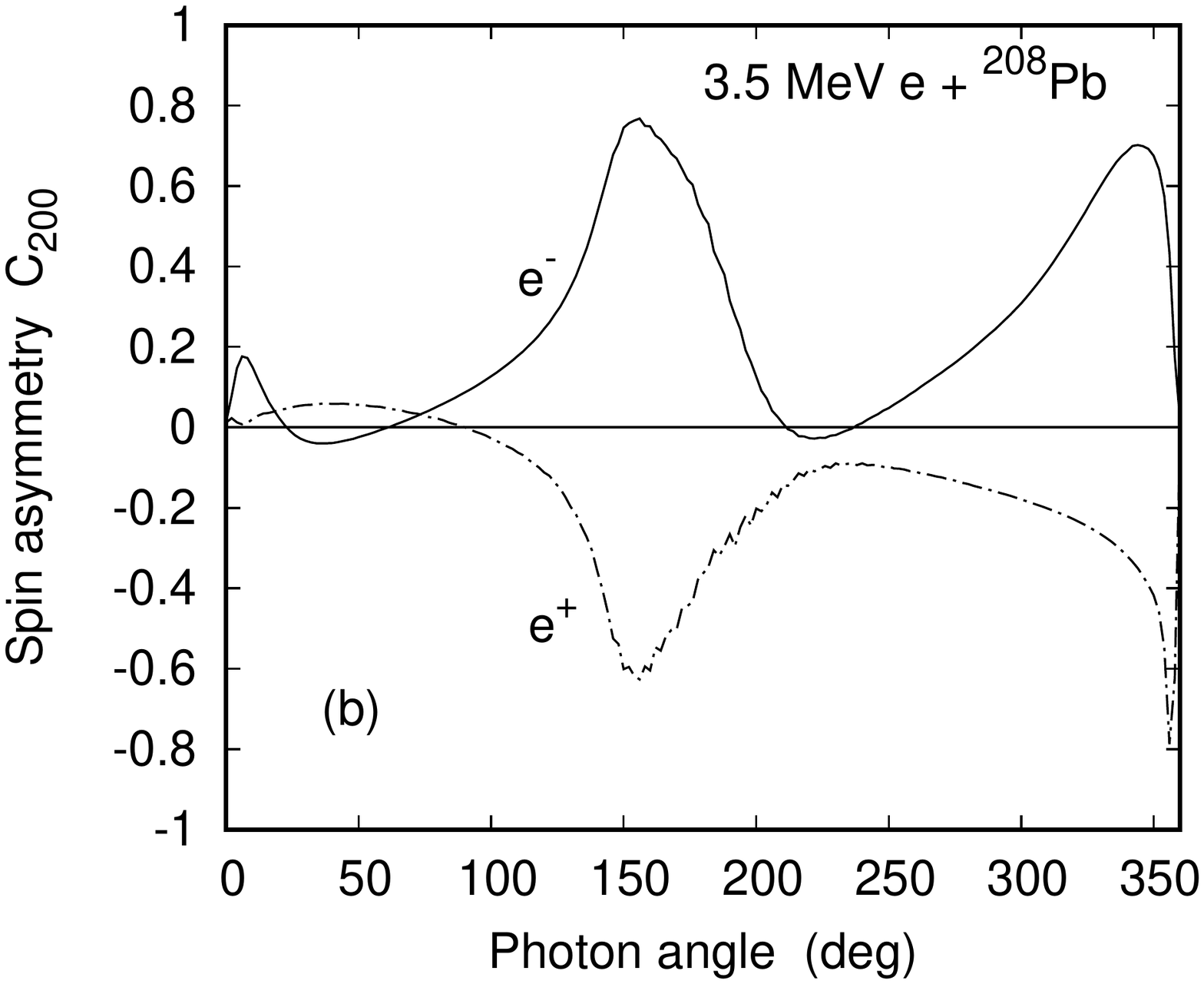}
\end{tabular}
\caption{
Polarization correlations (a) $C_{320}$ and $C_{120}$, and (b) $C_{200}$ from 3.5 MeV leptons colliding with $^{208}$Pb as a function of photon angle. The lepton angles are $\vartheta_f=150^\circ$ and $\varphi_f=0$, and the photon frequency is $\omega=3/4\;E_e = 2.625$ MeV.
(a): $C_{320}$ for electrons (-------------) and positrons ($-\cdot -\cdot -$), $C_{120}$ for electrons ($-----)$ and positrons $(\cdots\cdots)$.
(b): ---------------, electrons; $-\cdot - \cdot -$, positrons.
Wiggles near $150^\circ - 180^\circ$ are due to numerics.
}
\end{figure}

\vspace{1.5cm}

\newpage 

{\Large\bf 5. Experiment in comparison with theory}
\setcounter{chapter}{5}
\setcounter{section}{0}
\setcounter{equation}{0}
\setcounter{figure}{0}


\section{Cross sections}

Early experiments on electron bremsstrahlung cross sections, preferably at low collision energies, started in the beginning
of the fourties of last century.
An overview of these early experiments is provided in the article by Koch and Motz \cite{KM59}.
Somewhat later, there were more systematic measurements at energies in the MeV region  \cite{RD67,Re68,Da68,AZ66,Ai66}.
Recently, however, the subject was taken up again, providing bremsstrahlung spectra for collision energies near and below 0.1 MeV  \cite{Ga17,Ga18}.

\subsection{Unobserved final electrons}

In all  experiments mentioned above, only the angular or spectral distribution of the emitted photons were recorded, while the scattered electrons, as well as the polarization degrees of freedom, were disregarded.

Starting with the early low-energy experiments and their theoretical interpretation, we show in
Fig.5.1.1  the angular distribution of photons from 0.38 MeV electrons colliding with a gold target.
Except for the forward angles there is good agreement between experiment  and the partial-wave calculations  \cite{TP71}.
The Sommerfeld-Maue theory fails at all angles for this heavy target and the comparatively low collision energy.
Screening by the atomic electrons is important for photon angles $\theta_k \lesssim 40^\circ.$

\vspace{0.2cm}

\begin{figure}
\vspace{-1.5cm}
\includegraphics[width=13cm]{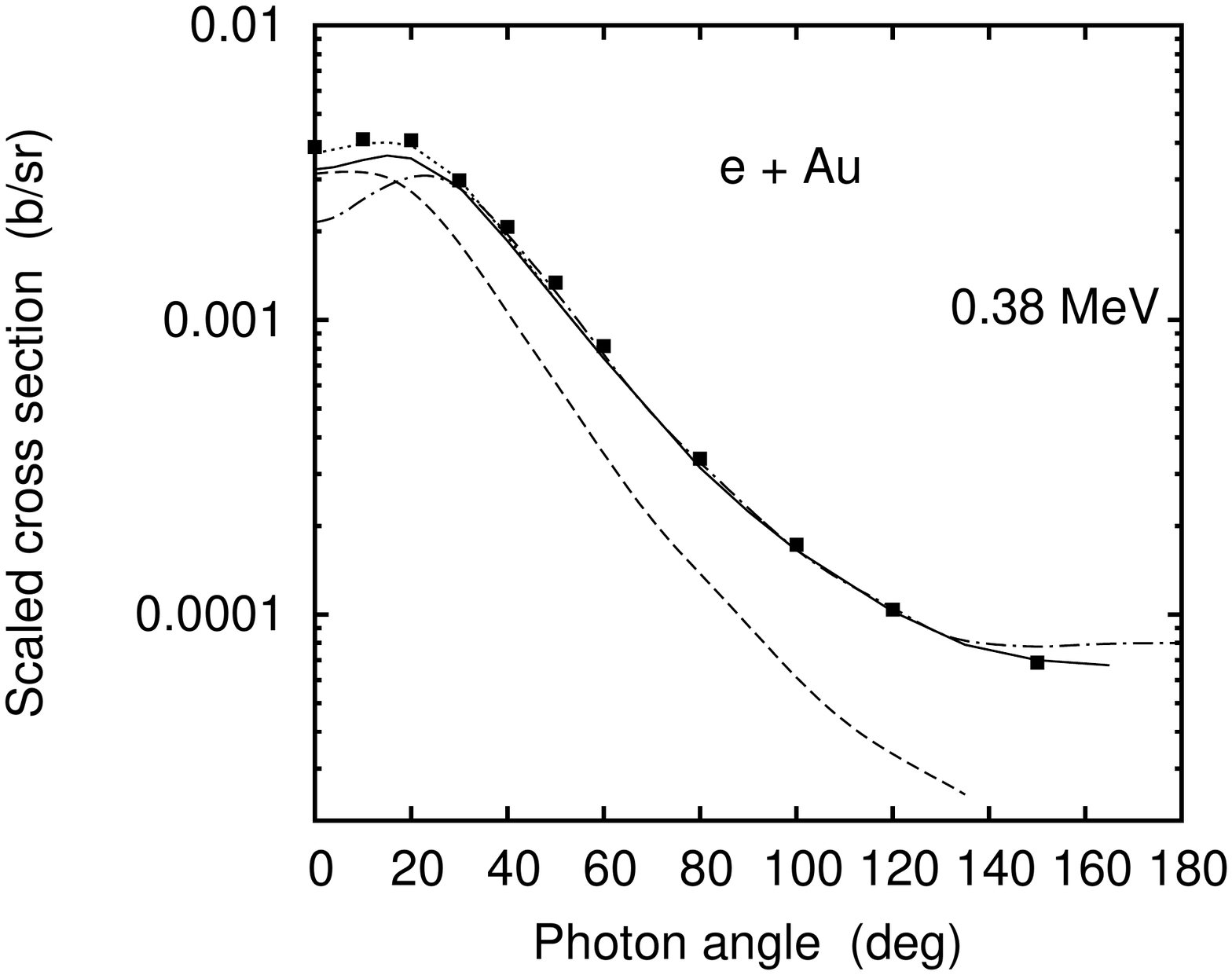}
\caption{
Scaled doubly differential cross section $\frac{\omega}{Z_T^2}\,\frac{d^2\sigma}{d\omega d\Omega_k}$ for bremsstrahlung
emission from 0.38 MeV electrons colliding with Au $(Z_T=79)$ as a function of photon angle $
\theta_k$.
The photon frequency is $\omega=0.228$ MeV.
Partial-wave theory including screening (-------------)
and for a point-like nucleus $(\cdots \cdots $) as well as the Sommerfeld-Maue theory ($-----$, for a point-like nucleus) from Tseng and Pratt \cite{TP71}. Earlier partial-wave results from Brysk et al ($-\cdot-\cdot-$ \cite{BZP69}) are also shown.
The experimental data $(\blacksquare)$ are from Aiginger \cite{Ai66}. 
}
\end{figure}

Fig.5.1.2 shows the photon angular distribution from 1 MeV electrons colliding with aluminum.
At this higher collision energy, the maximum has shifted to zero degrees.
For this collision system, there exist two sets of experiments, and the partial-wave as well as the SM theory perform well for this light target.

\begin{figure}
\vspace{-1.5cm}
\includegraphics[width=13cm]{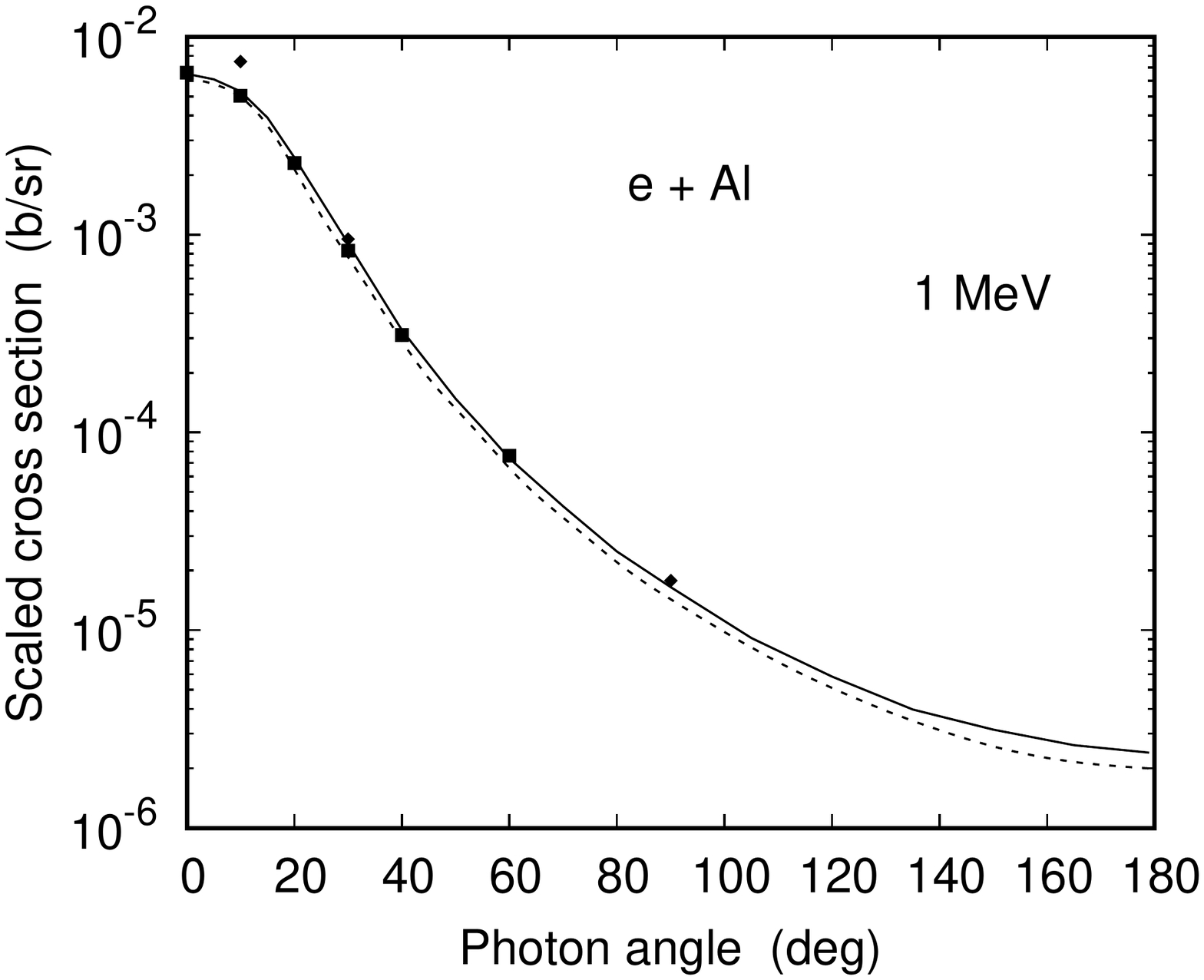}
\caption{
Scaled doubly  differential bremsstrahlung cross section $\frac{\omega}{Z_T^2}\,\frac{d^2\sigma}{d\omega d\Omega_k}$  from 1 MeV electrons colliding with Al ($Z_T=13)$ as a function of photon angle $\theta_k$.
The photon frequency is $\omega=0.7$ MeV.
Partial-wave results including screening (------------),
Sommerfeld-Maue results ($-----$, for a pointlike nucleus), taken from Tseng and Pratt \cite{TP71}. Experiment: $(\blacklozenge)$, Motz \cite{Mo55}; ($\blacksquare)$, Rester and Dance \cite{RD67}.
}
\end{figure}

The new  experiments  by Garc\'{i}a-Alvarez et al \cite{Ga17,Ga18}
provide  the dependence of the doubly differential cross section on photon frequency and collision energy for targets in the range $6\leq Z_T\leq 79$.
As an example, Fig.5.1.3 displays the frequency dependence for 0.1 MeV electrons colliding with Te and Ta at a photon angle of $90^\circ$.
The partial-wave theory explains the data except at the lowest frequencies where the polarizability and the shell structure of the target atom come into play
(polarization bremsstrahlung, see \cite{Ko01}).
There exists a comparison of further systems investigated by Garc\'{i}a-Alvarez \cite{Ga18} with tabulated partial-wave results \cite{SB85}, based on the Tseng and Pratt calculations \cite{Ga18}, 
as well as with partial-wave results from a new low-energy code put forth by Po\v{s}kus \cite{Po19}.

\begin{figure}
\vspace{-1.5cm}
\includegraphics[width=13cm]{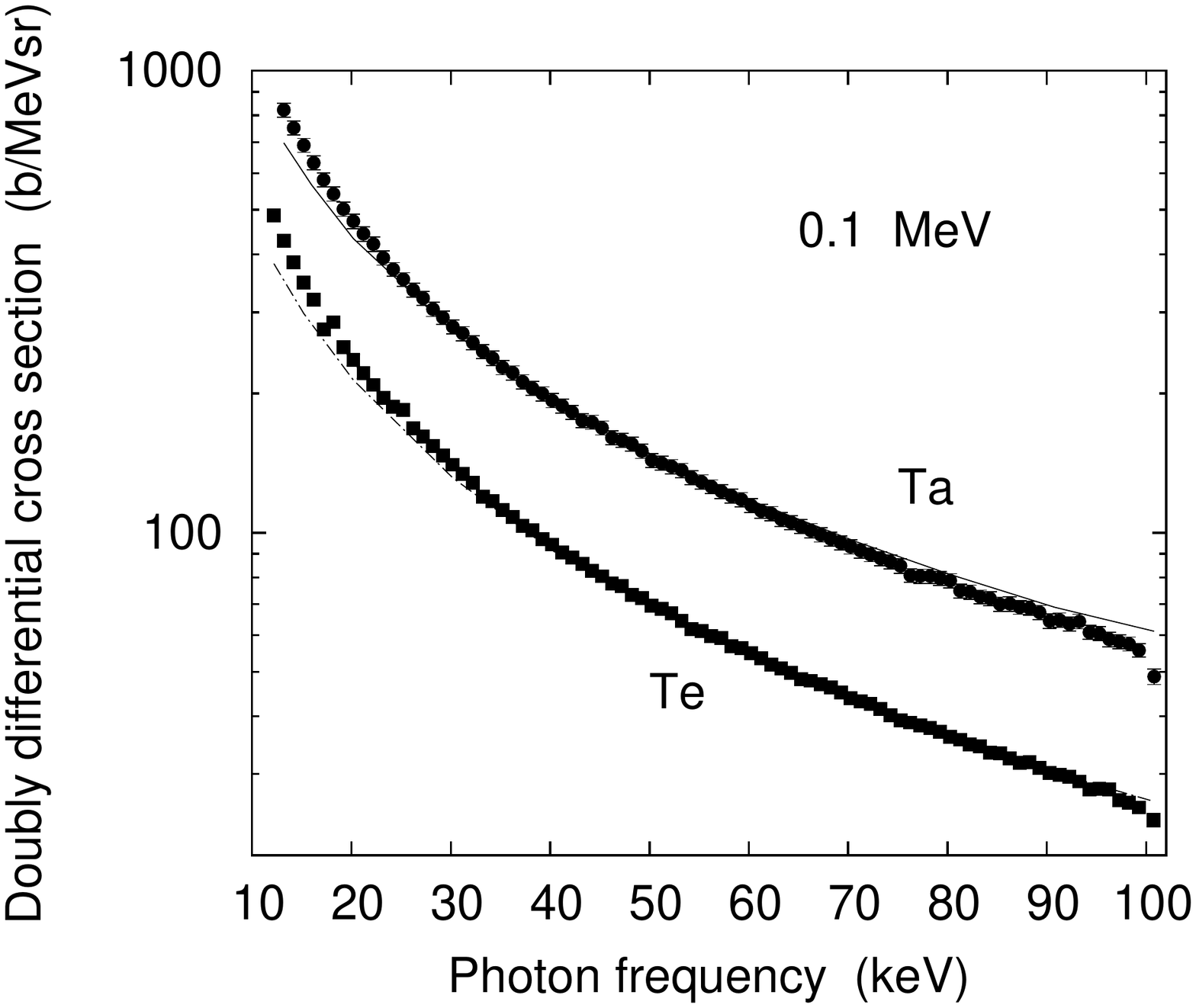}
\caption{
Frequency dependence of the bremsstrahlung doubly differential cross section in collision of 0.1 MeV electrons  with Te ($Z_T=52)$ and Ta ($Z_T=73$) at a photon angle of $90^\circ$.
Shown are the Dirac partial-wave results for Ta (-------------) and Te $-\cdot-\cdot -)$, including screening, in comparison with the experimental data of Garc\'{i}a-Alvarez et al ($\bullet$, Ta; $\blacksquare$, Te \cite{Ga18}).
The error bars are included. For Te, they are within the size of the symbols.
}
\end{figure}

Proceeding to higher beam energies around a few MeV, the frequency dependence of bremsstrahlung from 2.5 MeV electrons colliding with a gold target is provided in Fig.5.1.4.
The photon emission angle is $\theta_k=30^\circ$.
For this collision system, screening effects are visible at $\omega <0.5$ MeV.
Included are, besides the DW results, also the results from the NLO-SM theory.
The latter theory provides a qualitative description of the photon spectra, even for a gold target, 
and also for angles in the backward hemisphere as shown in  recent work \cite{JM19}.
For $\theta_k \lesssim 20^\circ$ or medium-heavy targets, the agreement with experiment, respectively  with the DW theory
is even quantitative.

\begin{figure}
\vspace{-1.5cm}
\includegraphics[width=8cm]{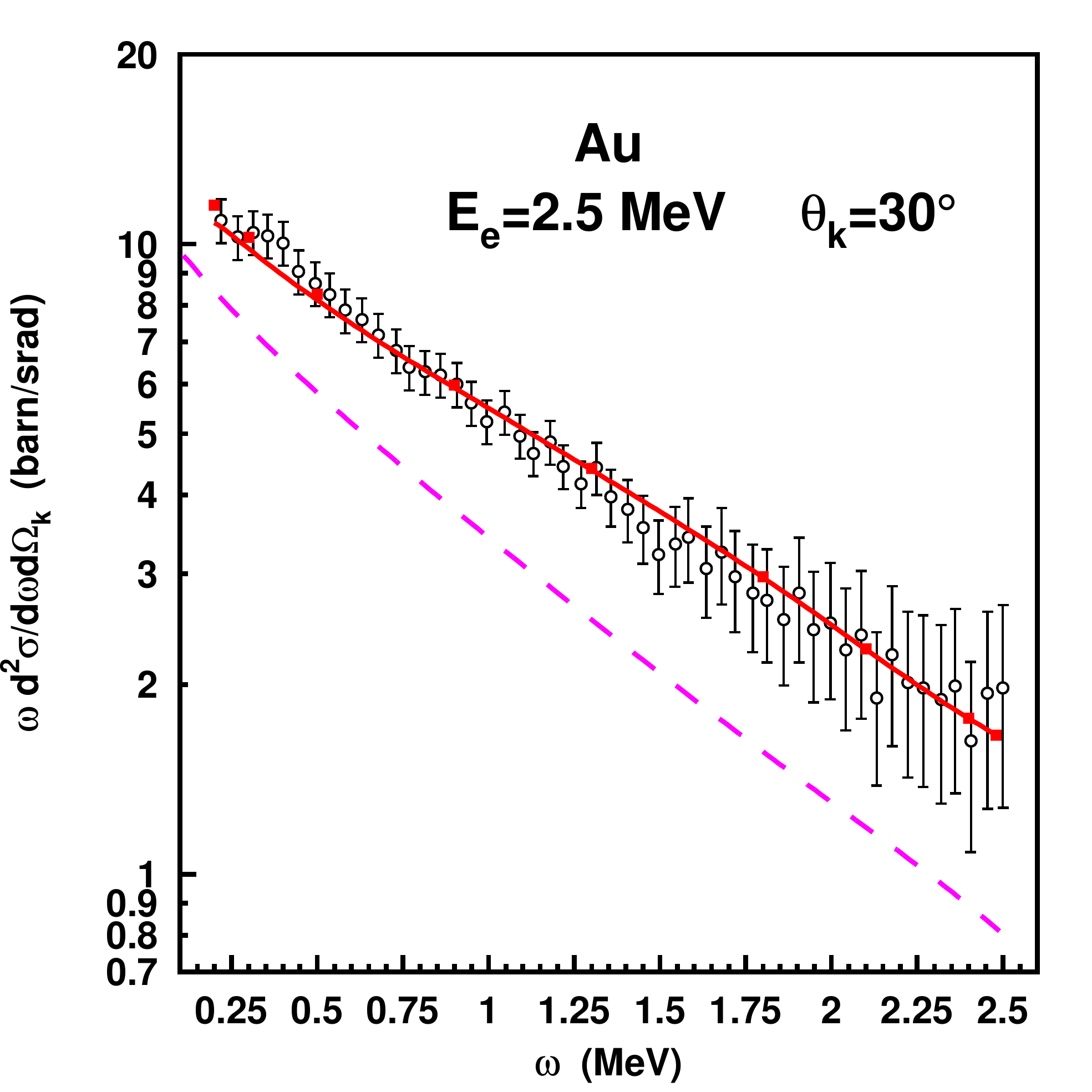}
\caption{
Bremsstrahlung intensity $\omega \frac{d^2\sigma}{d\omega d\Omega_k}$ from 2.5 MeV electrons colliding with Au $(Z_T=79)$ at $\theta_k=30^\circ$ as a function of frequency $\omega$.
Shown are results from the unscreened DW theory ($\blacksquare)$, the screened DW theory (by applying the OMW additivity rule, -------------) as well as
from the screened NLO-SM theory $(-----$ \cite{MJ17}).
The experimental data ($\circ$) are from Rester and Dance \cite{RD67}.
}
\end{figure}

Fig.5.1.5 shows the photon spectrum at a collision energy of 9.66 MeV, which is the highest energy for which experimental data, differential in angle {\it and} frequency, are available.
For the forward angle of $\theta_k=5.9^\circ$ the partial-wave calculations suffer from severe convergence problems for $\omega <5$ MeV.
Clearly, the underprediction of experiment near and below 5 MeV is caused by an insufficient number of partial waves (at $\omega=5.4$ MeV, $|\kappa_i|\leq 220$ and $|\kappa_f|\leq 55$ are required for convergence,
which adds to a total of $5.3 \times 10^6$ terms in the transition amplitude).
The SM as well as the NLO-SM theories perform quite well at this small angle.
Actually the data follow rather the unscreened theory for $\omega <5$ MeV; this overprediction of the screened results
by experiment may also here be due to the influence of polarization bremsstrahlung in the data.

\setcounter{figure}{4}
\begin{figure}
\vspace{-1.5cm}
\includegraphics[width=8cm]{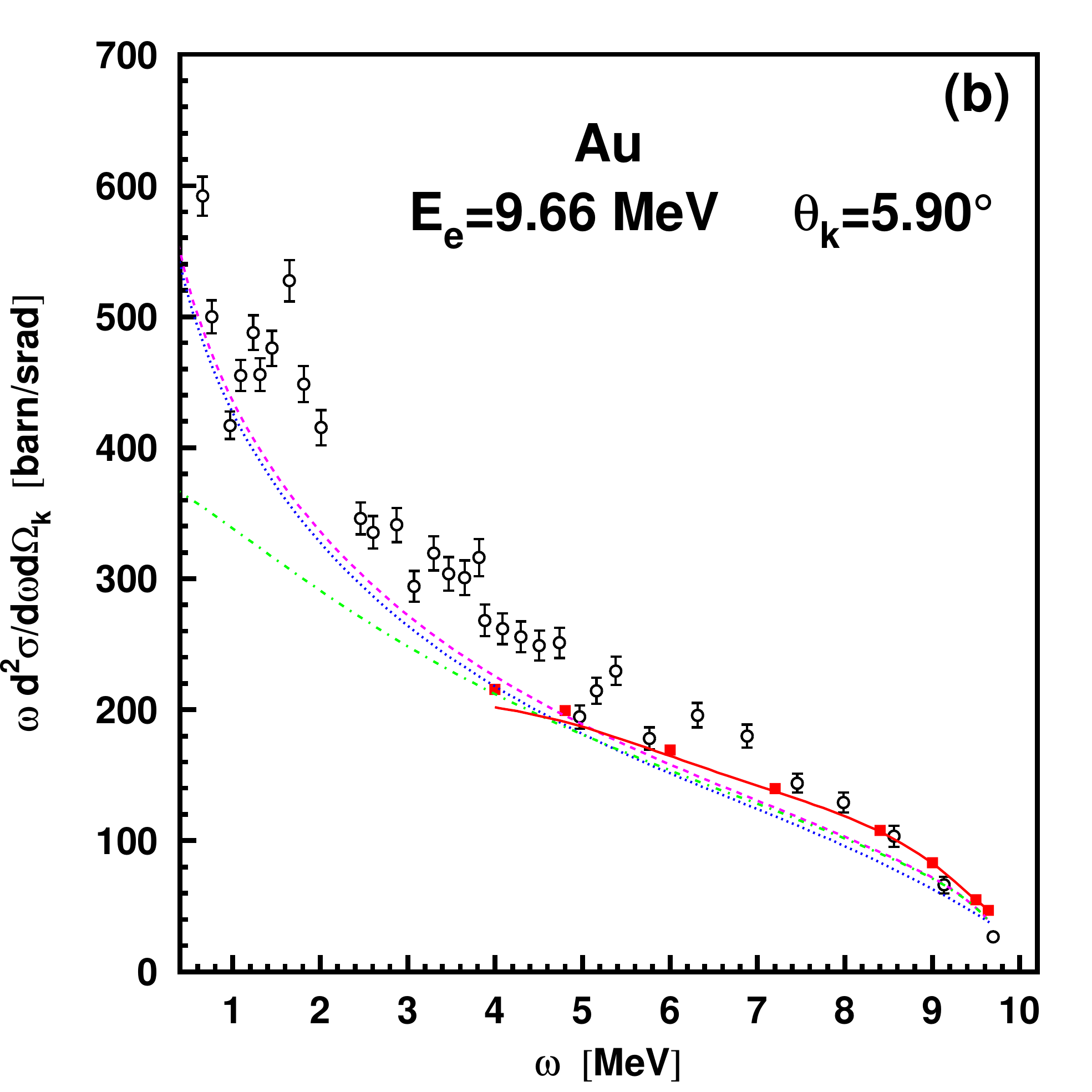}
\caption{
Bremsstrahlung intensity $\omega\,d^2\sigma/d\omega d\Omega_k$ from 9.66 MeV electrons colliding with Au at a photon angle of $\theta_k=5.9^\circ$
as a function of frequency $\omega$.
Shown are results from the DW theory without $(\blacksquare$) and with screening (---------------, using the OMW additivity rule)
as well as from the unscreened SM ($\cdots\cdots$) and NLO-SM $(-----)$ theory.
Screened results from the NLO-SM theory $(-\cdot - \cdot -$) are included \cite{MJ17}. The experimental data ($\circ$) are from Starfelt and Koch \cite{SK56}.
}
\end{figure}

In Fig.5.1.6 the photon angular distribution at a frequency of 1.3 MeV is shown for 2.5 MeV electrons colliding with tin.
As compared to the angular distributions from Figs.5.1.1 and 5.1.2 the increase of the forward focusing is clearly seen
when proceeding from 0.38 MeV to 2.5 MeV.
On the other hand, for a gold target at $\theta_k=30^\circ$ and $\omega/E_e=0.6$,
the cross section has decreased by a factor of 0.03 when increasing the collision energy  from 0.38 MeV to 2.5 MeV \cite{MJ17}.

\begin{figure}
\vspace{-1.5cm}
\includegraphics[width=8cm]{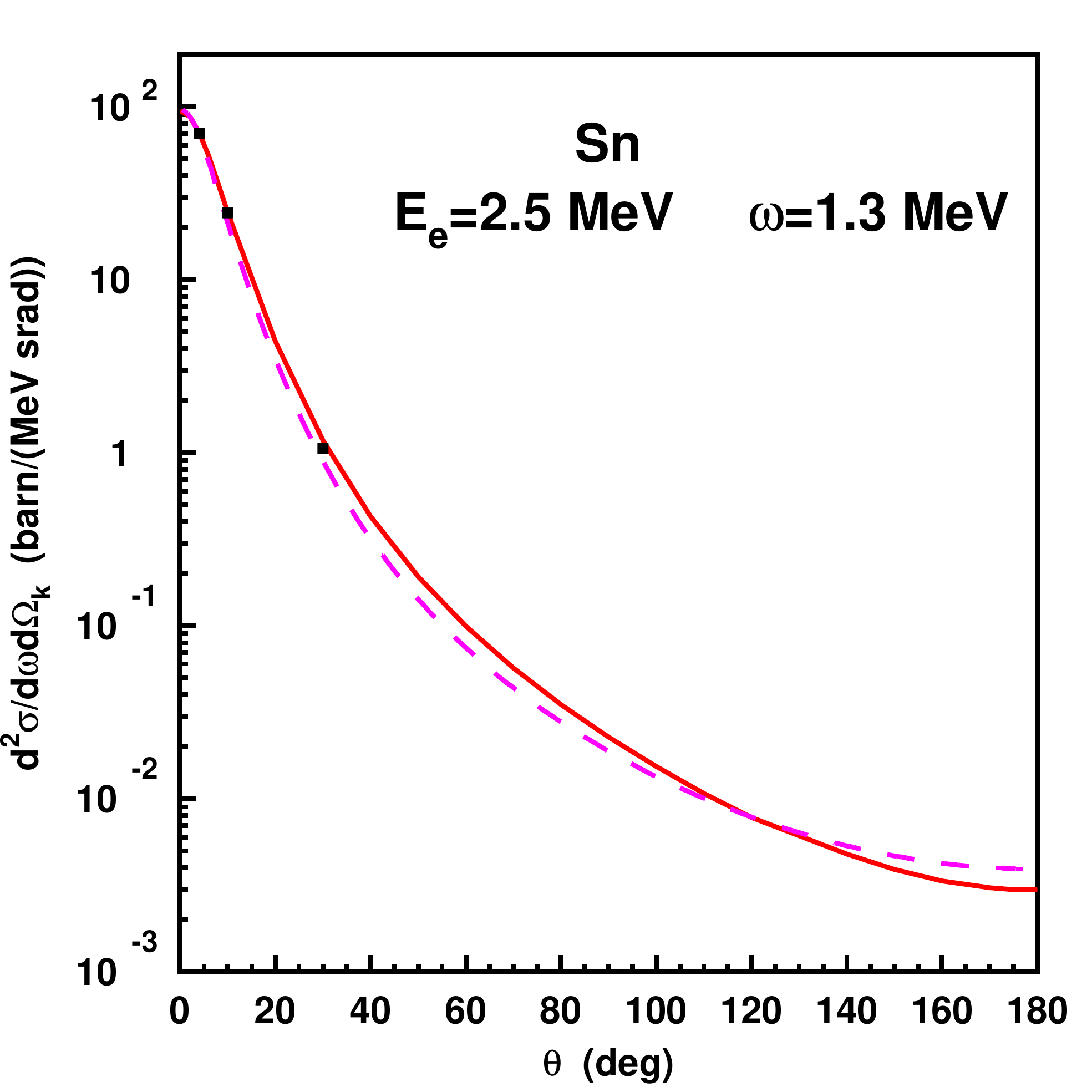}
\caption{
Doubly differential bremsstrahlung cross section from 2.5 MeV electrons colliding with Sn ($Z_T=50$) as a function of photon angle $\theta_k$.
The frequency is $\omega=1.3$ MeV. Shown are results from the screened DW (---------------) and from the screened  NLO-SM ($-----$) theory \cite{M20}, together with experimental data from Rester and Dance ($\blacksquare$ \cite{RD67}, error bars being within the size of the symbols).
}
\end{figure}

At beam energies above 50 MeV, the bremsstrahlung is so strongly peaked in the foremost direction (see, e.g. Fig.2.5.2), that an angular resolution is no longer possible.
Therefore, only the integrated spectral distribution,
\begin{equation}\label{5.1.1}
\frac{d\sigma}{d\omega}\;=\;2\pi \int_0^\pi \sin \theta_k\;d\theta_k\;\frac{d^2\sigma}{d\omega d\Omega_k},
\end{equation}
is accessible to experiment.
Such measurements were performed at 24 MeV and 34 MeV by Barber et al \cite{Ba55}, at 500 MeV by Brown \cite{Br56}, and recently by Mangiarotti at al \cite{Ma19}.
While the early experiments were done for a single photon frequency, the new measurements cover the lower half of the photon spectrum.
However, these measurements were not performed on an absolute scale, and the intensity for a copper target ($Z_T=29$) is used as a reference value for the heavier targets.

Fig.5.1.7 shows the corresponding ratio of the singly differential cross sections as a function of target nuclear charge number $Z_T$.
The Sommerfeld-Maue theory performs well in describing the experimental decrease with $Z_T$.
The NLO-SM theory gives nearly the same results \cite{Ma19}, demonstrating fast convergence of the perturbative series at photon frequencies well below the tip.
The failure of the inconsistent next-to-next-to-leading-order results is again obvious.

\begin{figure}
\vspace{-1.5cm}
\includegraphics[width=8cm]{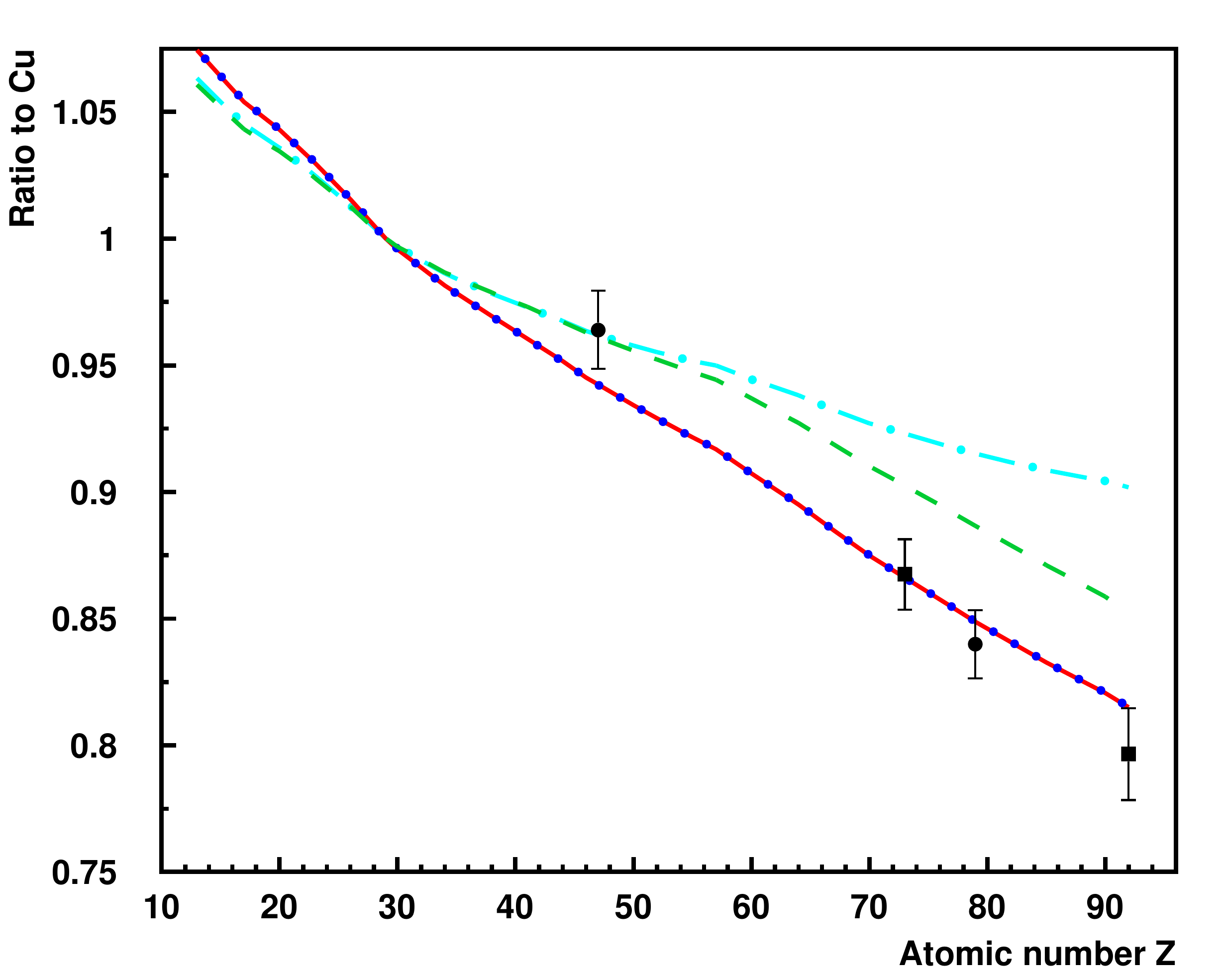}
\caption{
Ratio $\frac{29^2}{Z_T^2}\;\frac{d\sigma/d\omega(Z_T)}{d\sigma/d\omega(29)}$  as a function of nuclear charge $Z_T$.
Shown are the experimental data of Brown ($\blacksquare$, \cite{Br56}) at a photon frequency of $\omega=234$ MeV and the data of Mangiarotti at al ($\bullet$, \cite{Ma19}) which are independent of $\omega$ in the investigated region (5 MeV $\leq \omega \leq 250$ MeV).
Also shown are results  from the SM theory (---------------), from the NLO-SM theory ($\cdots\cdots$), from the NNLO-SM model ($-----$) and from the PWBA ($-\cdot -\cdot -$).The calculations are performed at $\omega=234$ MeV \cite{Ma19} and screening is included. 
}
\end{figure}

\subsection{Coincidence measurements}

Experiments where the photon intensity at a fixed angle is measured in coincidence with the scattered electron,
and subsequently put on an absolute scale, are quite rare.
Early coincidence experiments were performed by Aehlig and coworkers \cite{AS72,AMS77} who did record the photon intensity, and later by Nakel and his group \cite{HN04},
who in many cases normalized their data to the SM theory.
All these measurements were done in coplanar geometry, where the electron detector is placed in the reaction plane which is defined by the beam axis and the photon spectrometer.

\setcounter{figure}{7}
\begin{figure}
\vspace{-1.5cm}
\includegraphics[width=13cm]{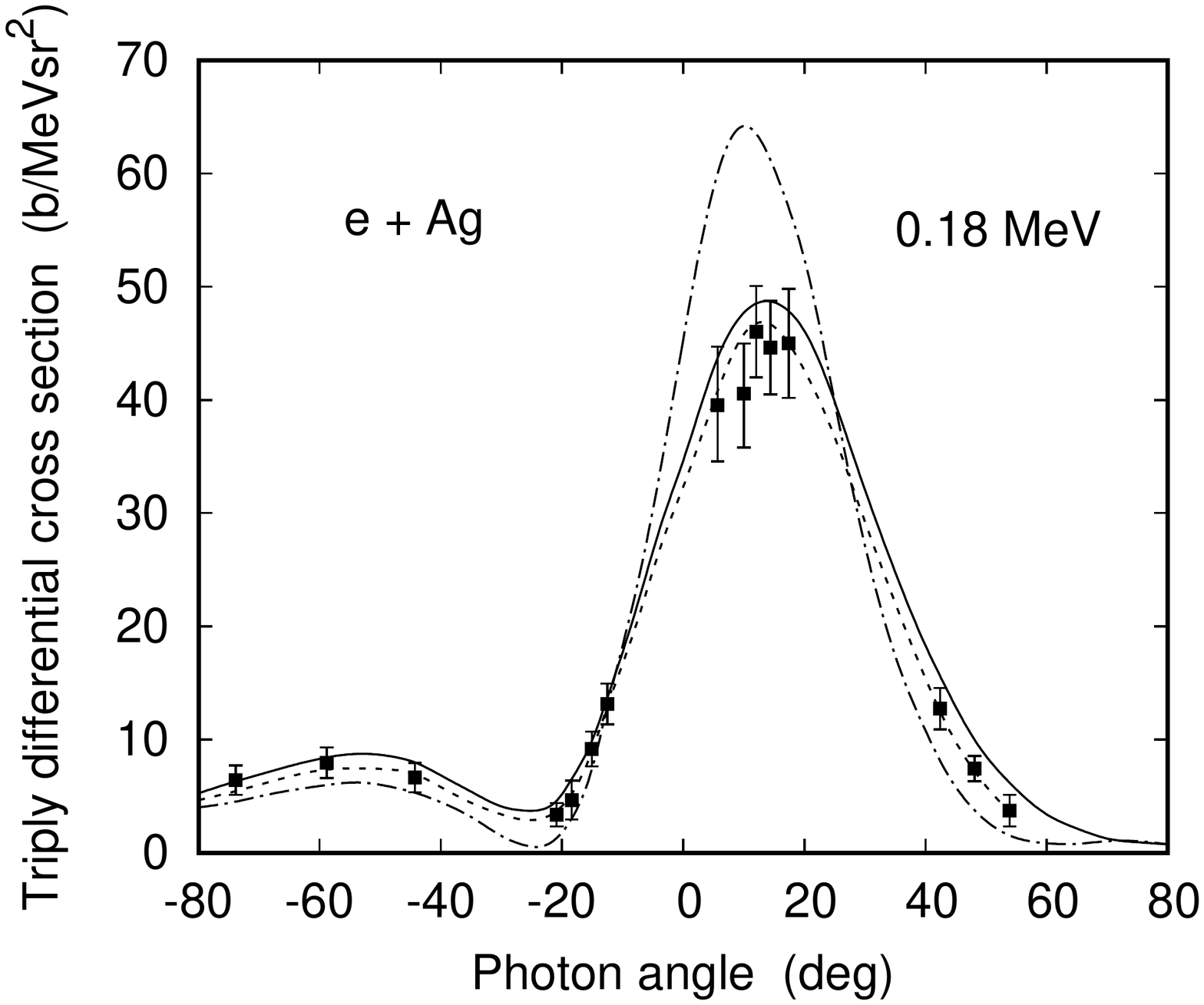}
\hspace{-2cm}
\caption{
Triply differential cross section for bremsstrahlung emitted in collisions of 180 keV electrons with Ag $ (Z_T=47)$ as a function of photon angle $\theta_k$.
The photon frequency is $\omega = 80$ keV,  the scattering angle is $\vartheta_f=30^\circ$ and the azimuthal angle $\varphi_f$ is $0^\circ$ (for $\theta_k>0$),
respectively $180^\circ$ (plotted with reversed sign of $\theta_k$).
Shown are results, including screening,  from the DW theories  by Shaffer et al ($-\cdot - \cdot -)$ \cite{Sh96}) and by Keller and Dreizler (---------------- \cite{KD97}),
as well as results from the Sommerfeld-Maue theory ($----$ \cite{AS72}).
The experimental data $(\blacksquare)$ are from Aehlig and Scheer \cite{AS72}. 
}
\end{figure}

Fig.5.1.8 shows the angular distribution of the photon intensity from 0.18 MeV electrons colliding with silver 
at a photon frequency of $\omega=0.08$ MeV. It is seen that the photon angle $\theta_k\approx 15^\circ$ defining the maximum of the cross section is close to the scattering angle $\vartheta_f=30^\circ$, 
while a second maximum appears near $-60^\circ$ ($=300^\circ$ in the nomenclature of Fig.2.7.1).
At such a moderate collision velocity the photon intensity is not yet peaked in beam direction, which remains true for the doubly differential cross section (see Fig.5.1.1).
In cases where the scattering angle is in the backward hemisphere, not yet investigated experimentally, the coincident photon angular distribution
shows two separate peaks, one near $\vartheta_f$ and the other at foremost angles (see Fig.2.7.1).
For the medium-heavy silver target and forward angles, not only  the partial-wave theory by Keller and Dreizler \cite{KD97} or by Tseng \cite{T02}, but also the SM theory,
represent the measured cross section quite well, while an earlier partial-wave calculation \cite{Sh96} overestimates the maximum.

\begin{figure}
\vspace{-1.5cm}
\includegraphics[width=13cm]{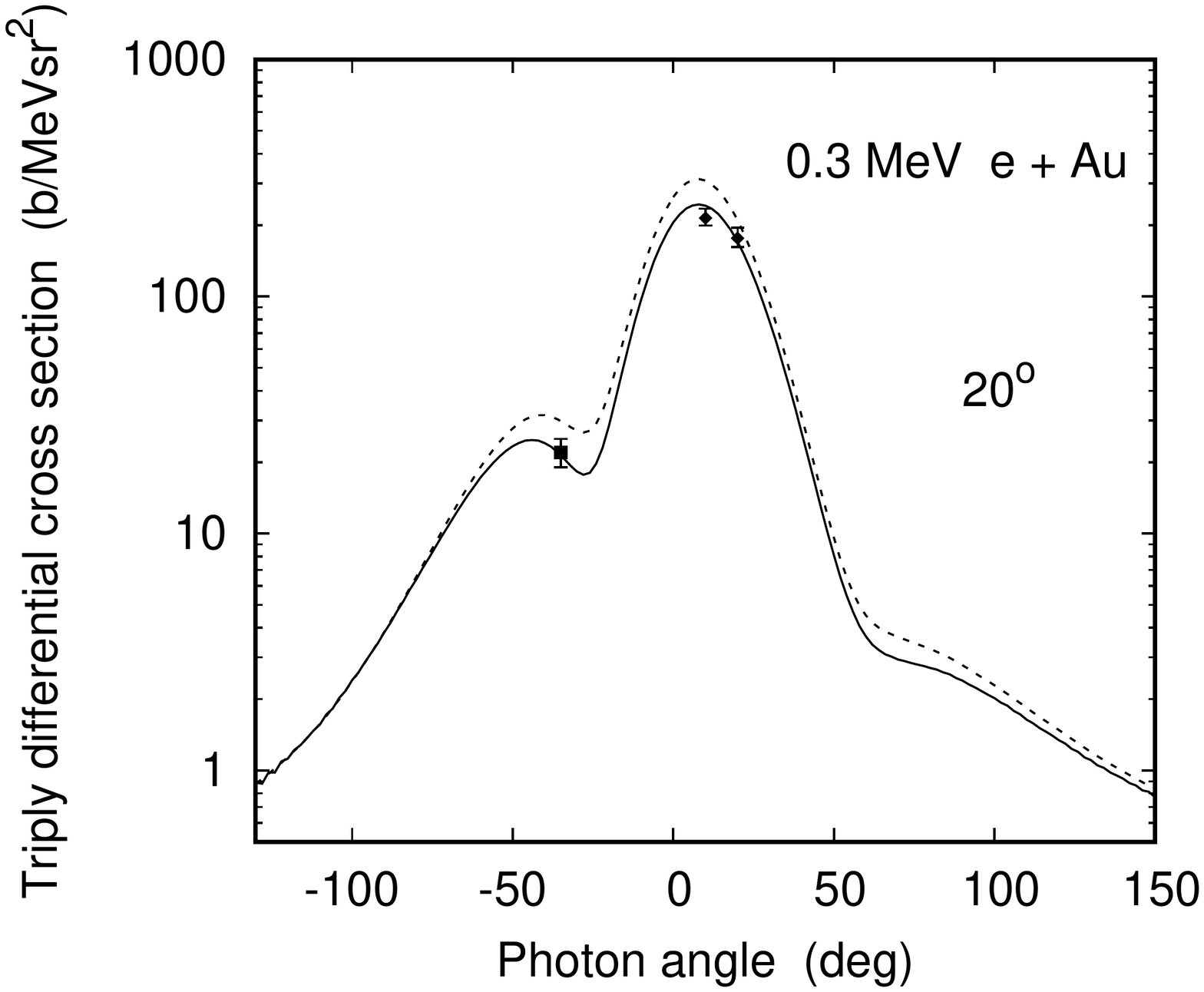}
\caption{
Triply differential cross section for bremsstrahlung from 0.3 MeV electrons colliding with Au as a function of photon angle.
The frequency is $\omega=0.1 $ MeV and the scattering angle is $\vartheta_f=20^\circ$, the azimuthat angle $\varphi_f=0^\circ$, respectively $180^\circ$. 
Results from the Dirac partial-wave theory (------------------, including screening; $-----$, point nucleus) are compared with the measurements of Aehlig et al ($\blacklozenge$ \cite{AMS77}) and Geisenhofer and Nakel ($\blacksquare$ \cite{GN96}).
}
\end{figure}

We proceed to a heavier target (Au) and a slighly higher collision energy (0.3 MeV) and show in Fig.5.1.9
the corresponding angular distribution.
For this system, the spin asymmetry $C_{002}$ was also measured (see Fig.5.2.10).
While the absolute intensity was not  recorded simultaneously with the measurement  of $C_{200}, $ there exist a few data by Aehlig et al \cite{AMS77} and by Geisenhofer and Nakel \cite{GN96}, taken from their spectral distribution.
At the forward angles screening is important, lowering the triply differential cross section by up to 50 percent, as compared to the Dirac partial-wave results for a Coulomb potential with $Z_T=79$.
Clearly, the experimental data support the consideration of screening.

The coincident photon spectrum for this collision system is displayed in Fig.5.1.10 for a forward photon angle lying in the same $(\theta_k=20^\circ$), respectively opposite $(\theta_k=-35^\circ$) half-plane as the scattered electron ($\vartheta_f=20^\circ$). Comparison is made with the measurements from Aehlig et al \cite{AMS77} and those from Nakel and collaborators \cite{KN82}
who later repeated this experiment \cite{GN96} (only these later data are shown in the figure).
For $\theta_k=20^\circ$, there is a smooth decrease of the photon intensity with frequency, represented fairly well both by the partial-wave and by the Sommerfeld-Maue theory.
However, for back-to-back ejection ($\theta_k=-35^\circ$), there appears an interference structure in the spectrum with a maximum near 80 keV, which is confirmed experimentally. The SM theory predicts in addition a minimum near 20 keV. However,  due to an unstable oscillatory behaviour, we did not find it possible
 to obtain 
reliable DW results at such low frequencies.
The interference structure at small angles is more prominent at higher
collision energies, where an interpretation in the framework of the Born approximation can be made (see section 4.2 and Fig.4.2.2).

\begin{figure}
\vspace{-1.5cm}
\centering
\begin{tabular}{cc}
\hspace{-1cm}\includegraphics[width=.7\textwidth]{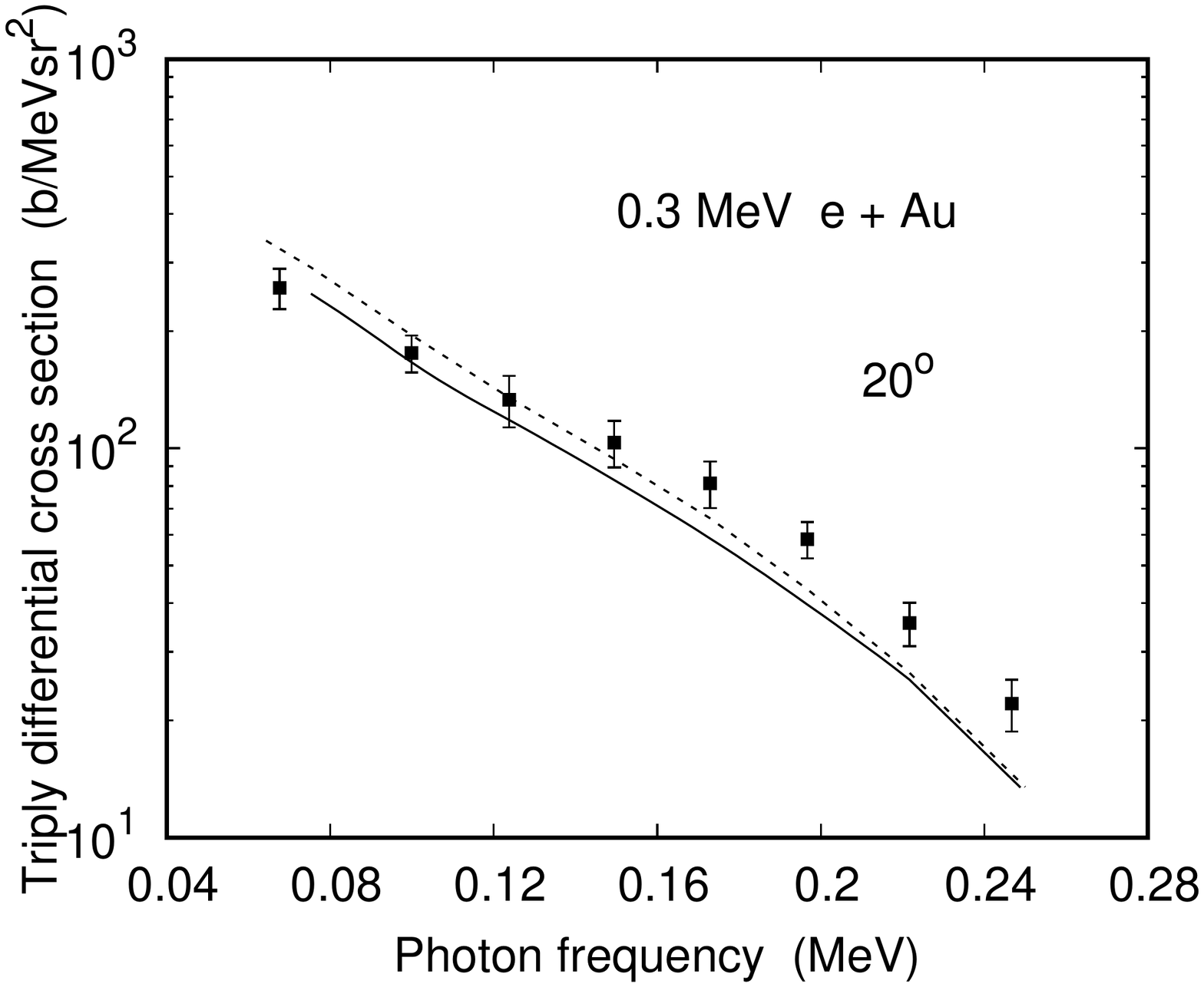}&
\hspace{-3cm} \includegraphics[width=.7\textwidth]{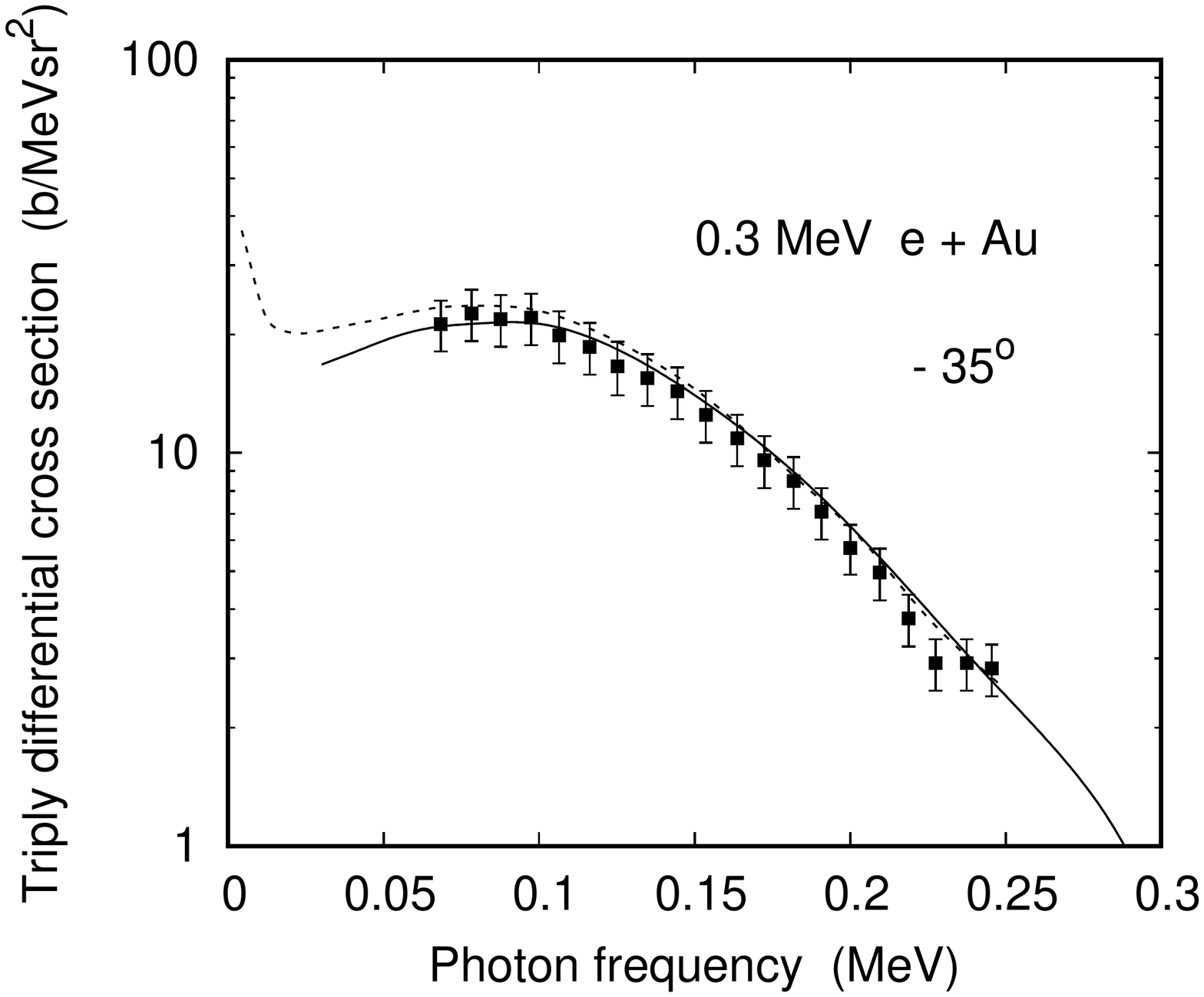}
\end{tabular}
\caption{Triply differential cross section for bremsstrahlung from 0.3 MeV electrons colliding with Au $(Z_T=79)$
as a function of photon frequency.
The scattering angle is $\vartheta_f=20^\circ$ and the photon angle $\theta_k$ is (left) $20^\circ$ and (right) $-35^\circ$ (the latter being equivalent to $325^\circ$).
Left: SM ($----$) and partial-wave results  (---------------) including screening from Keller and Dreizler \cite{KD97}; experimental data ($\blacksquare$) from Aehlig et al \cite{AMS77}.
Right: Dirac partial-wave results including screening (------------). The experimental data ($\blacksquare$) and the results from the screened SM theory according to Haug \cite{Ha96} $(-----$) are taken from Geisenhofer and Nakel \cite{GN96}.
}
\end{figure}

\section{Spin asymmetries}

\setcounter{equation}{0}

Measurements of the spin asymmetry only require relative intensities and thus there is no need for an absolute normalization
of the photon spectra or the angular distributions.
Moreover, due to the sensitivity to relative phases in the various contributions to the cross section, the information extracted from the spin asymmetries is superior to the one obtained from mere cross section measurements.
Experiments on the polarization degrees of freedom started also in the middle of the last century.
However, at that time it was not feasible to produce an intense polarized electron beam.
So the first polarization correlations investigated were the linear polarization $P_1$ and the circular polarization transfer $P_3$.
$P_1$ can be measured for unpolarized electrons (see, e.g. \cite{MP60,KE73}),
and it has been thoroughly investigated over the years.
In particular, it has the advantage that it is the only nonvanishing spin asymmetry in the nonrelativistic limit,
and even more, it remains finite at ultrahigh energies.
For $P_3$, longitudinally polarized electrons are needed. Such electrons are available in nature. They are, for example, emitted in the $\beta$-decay of heavy elements
and can be used in bremsstrahlung studies  \cite{SG58}.
Like $P_1$, also $P_3$ 
remains finite at ultrahigh energies. 
Moreover, $P_1$ and $P_3$ can be large even for light target materials  \cite{TP73}.

With the advent of efficient accelerators and strong polarized sources,
 further polarization correlations could be measured.
However, all  these experiments have in common that
they are basically proof-of-principle tests,
rather than systematic investigations.
This is particularly true for the recent  experiments published  after 2010. Section 5.2.1 gives a complete record of these measurements.

\subsection{Unobserved final electrons}

In this subsection we consider the polarization correlations from measurements where only the emitted
photon is recorded, but not the scattered electron.
The fact that $P_1 \equiv C_{03}$ is accessible for unpolarized electron sources, follows immediately from the formula (\ref{3.1.19}).
Averaging over the spin polarization $\bfzeta_i$ of the impinging electron, one is left with
\begin{equation}\label{5.2.1}
\frac{d^2\sigma}{d\omega d\Omega_k}(\bfe_\lambda^\ast)\;=\; \frac12\;\left( \frac{d^2\sigma}{d\omega d\Omega_k}\right)_0\;(1\;+\;C_{03}\xi_3),
\end{equation}
with $\xi_3=\cos 2\varphi_\lambda\;$ (see (\ref{3.1.7})).
The spin asymmetry $P_1$ can be measured by recording the intensity $I_\parallel$ of the emitted photons polarized in the reaction $(x,z)$-plane
along $\bfe_{\lambda_2}=(-\cos \theta_k,0,\sin \theta_k)$ (i.e. $\varphi_\lambda=0,$ such that $\xi_3=1$),
as well as the intensity $I_\perp$ of the photons polarized perpendicular to the reaction plane along $\bfe_{\lambda_1}=(0,1,0)$ (i.e. $\varphi_\lambda=\frac{\pi}{2},$ yielding $\xi_3=-1)$.
$I_\parallel$ and $I_\perp$ were measured by positioning a Ge(Li) detector in the reaction plane, respectively perpendicular to the reaction plane,
by rotating the polarimeter about the collimator axis \cite{KE73}.
Thus $P_1$ follows from
\begin{equation}\label{5.2.2}
P_1\;=\;\frac{I_\parallel \,-\,I_\perp}{I_\parallel \,+\,I_\perp}.
\end{equation}
Fig.5.2.1 shows the angular dependence of $P_1$ for 0.1 MeV electrons colliding with gold at a frequency near the high-energy end of the spectrum, in comparison with partial-wave results using a screened  target potential.

\setcounter{figure}{0}
\begin{figure}
\vspace{-1.5cm}
\includegraphics[width=13cm]{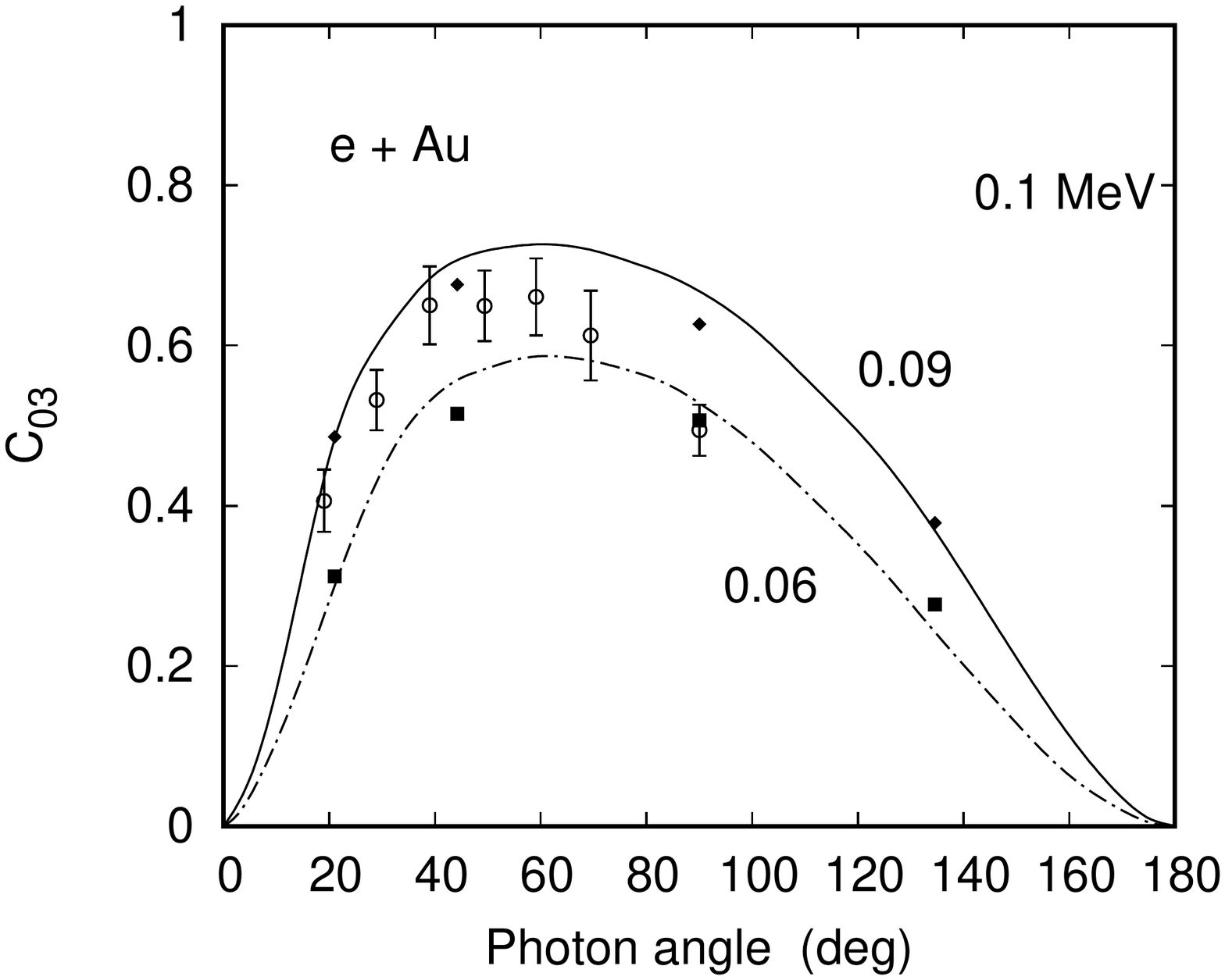}
\caption{
Linear polarization correlation $C_{03}$ for bremsstrahlung emitted in 0.1 MeV e+Au collisions with frequencies 0.06 MeV and 0.09 MeV, as a function of photon angle $\theta_k$.
Shown are experimental results from Motz and Placious \cite{MP60} at 0.09 MeV $(\circ)$ and from Kuckuck and Ebert \cite {KE73} at 0.06 MeV ($\blacksquare$) and 0.09 MeV ($\blacklozenge)$,
as well as partial-wave results (including screening) from Tseng and Pratt \cite{TP73} ($-\cdot - \cdot -$, $\omega =0.06$ MeV; -----------, $\omega=0.09$ MeV).
}
\end{figure}

At $0^\circ$ and $180^\circ$ the spin asymmetry vanishes because the emitted photon is aligned with the beam axis, such that the reaction plane degenerates into a straight line.
The resulting cylindrical symmetry around the beam axis leads to equal intensities, i.e. $I_\parallel = I_\perp$.
It is also confirmed experimentally that for (weak-) relativistic collision velocities the spin asymmetry attains its largest value not at $90^\circ$ (as
would be the case for impact energies in the keV region \cite{TP73}, where  the radiation intensity shows a dipole pattern \cite{TP71}), but at somewhat smaller angles.
Furthermore, $P_1$  increases with frequency. 

The first measurement of the circular polarization correlation $C_{32}$  profited from the longitudinal polarization of  $\beta$-rays emitted during the decay of
 a $^{90}$Sr + $^{90}$Y source. Bremsstrahlung photons were produced by means of the interaction of these electrons with a 
composite target (consisting of Ni, Cu and Fe \cite{Go57}).
The measurement of the photon polarization was performed by means of Compton scattering from oriented electrons which are available in magnetized iron.
By switching the direction of the magnetic field, the intensity
$I(\bfe_-^\ast)$ of left-handed, respectly the one, $I(\bfe_+^\ast)$, of right-handed photons could be recorded. $C_{32}$ then follows from
\begin{equation}\label{5.2.3}
C_{32}\;=\;\frac{I(\bfe_+^\ast)\,-\,I(\bfe_-^\ast)}{I(\bfe_+^\ast)\,+\,I(\bfe_-^\ast)}.
\end{equation}

Fig.5.2.2 shows the resulting spin asymmetry arising from electrons 
emitted
either by $^{90}$Sr (with $E_e=0.535 $ MeV) or by $^{90}$Y (with $E_e=2.24$ MeV) as a function of photon frequency.

\begin{figure}
\vspace{-1.5cm}
\includegraphics[width=13cm]{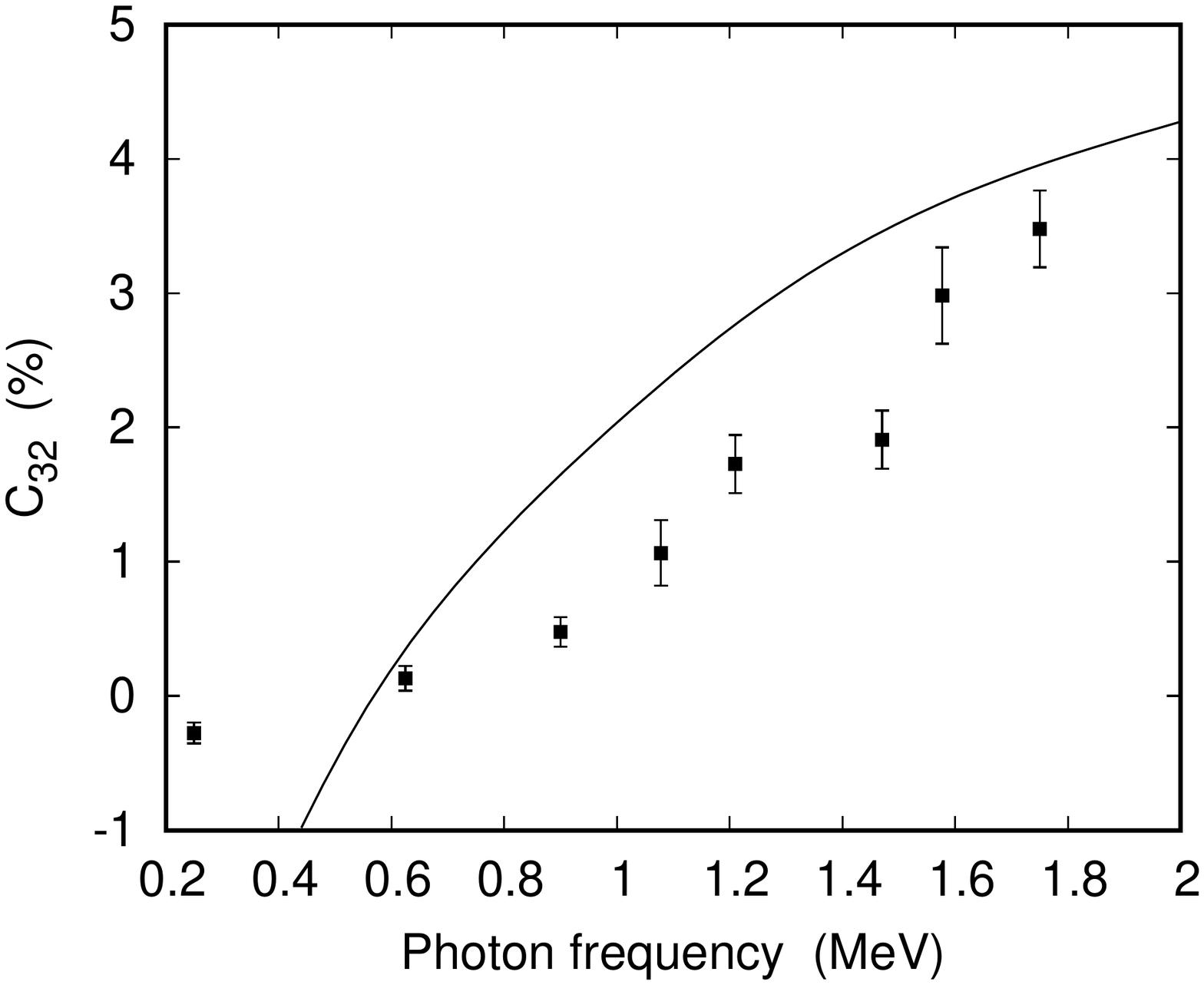}
\caption{
Circular polarization correlation $C_{32}$ (in percent) for bremsstrahlung in forward direction induced by $\beta$-rays from $^{90}$Sr + $^{90}$Y, impinging on a Monel target, as a function of photon frequency.
The full line represents the computed magnetic response, the symbols $(\blacksquare)$ are results from the measurement 
(Goldhaber et al \cite{Go57}).
}
\end{figure}

\vspace{0.2cm}

The measurement of the  polarization correlation $C_{20}$ requires an electron beam which is polarized perpendicular to the reaction plane.
Such a beam can be produced by irradiating a GaAsP crystal
with circularly polarized light from a HeNe laser.
The hereby emitted electrons are longitudinally polarized, 
and it was necessary to turn their polarization vector. In the early experiments this was achieved 
 by means
of a $90^\circ$ deflection of the electrons in an electrostatic sector field. 
After that, the electrons were accelerated in a van de Graaf accelerator to a maximum energy of up to 500 keV
(see, e.g. \cite{MN90}).

It is important to note that an electron beam produced in such a way is never completely polarized. Instead, the so-called degree of polarization $P_e$ is introduced, which is a measure of the fraction of polarized electrons in the beam.
The  initial electron spin polarization, 
as well as the respective polarization correlation, enter linearly into the cross section (more precisely, in terms of the product $C_{ij}\cdot \zeta_k$, see (\ref{3.1.19})).
Therefore a degree of polarization which is smaller than unity can be accounted for by either reducing the $k$-th coordinate $\zeta_k$ of $\bfzeta$ by the factor $P_e$, or equivalently,
by reducing the measured spin asymmetry  by
the factor  $P_e$ as compared to the theoretically
calculated value.
This implies that an experiment for $C_{20}$ yields
\begin{equation}\label{5.2.4}
P_e\;C_{20}\;=\;\frac{\sum_\lambda d^2\sigma(\bfzeta_i,\bfe_\lambda^\ast)\,-\,\sum_\lambda d^2\sigma(-\bfzeta_i,\bfe_\lambda^\ast)}{\sum_\lambda d^2\sigma(\bfzeta_i,\bfe_\lambda^\ast)\,+\,\sum_\lambda d^2\sigma(-\bfzeta_i,\bfe_\lambda^\ast)}
\end{equation}
in terms of the photon intensity for spin-up, respectively for spin-down polarized electrons.
In the measurements, this spin-flip is achieved by changing the helicity of the laser light.

\begin{figure}
\vspace{-1.5cm}
\centering
\begin{tabular}{cc}
\hspace{-1cm}\includegraphics[width=.7\textwidth]{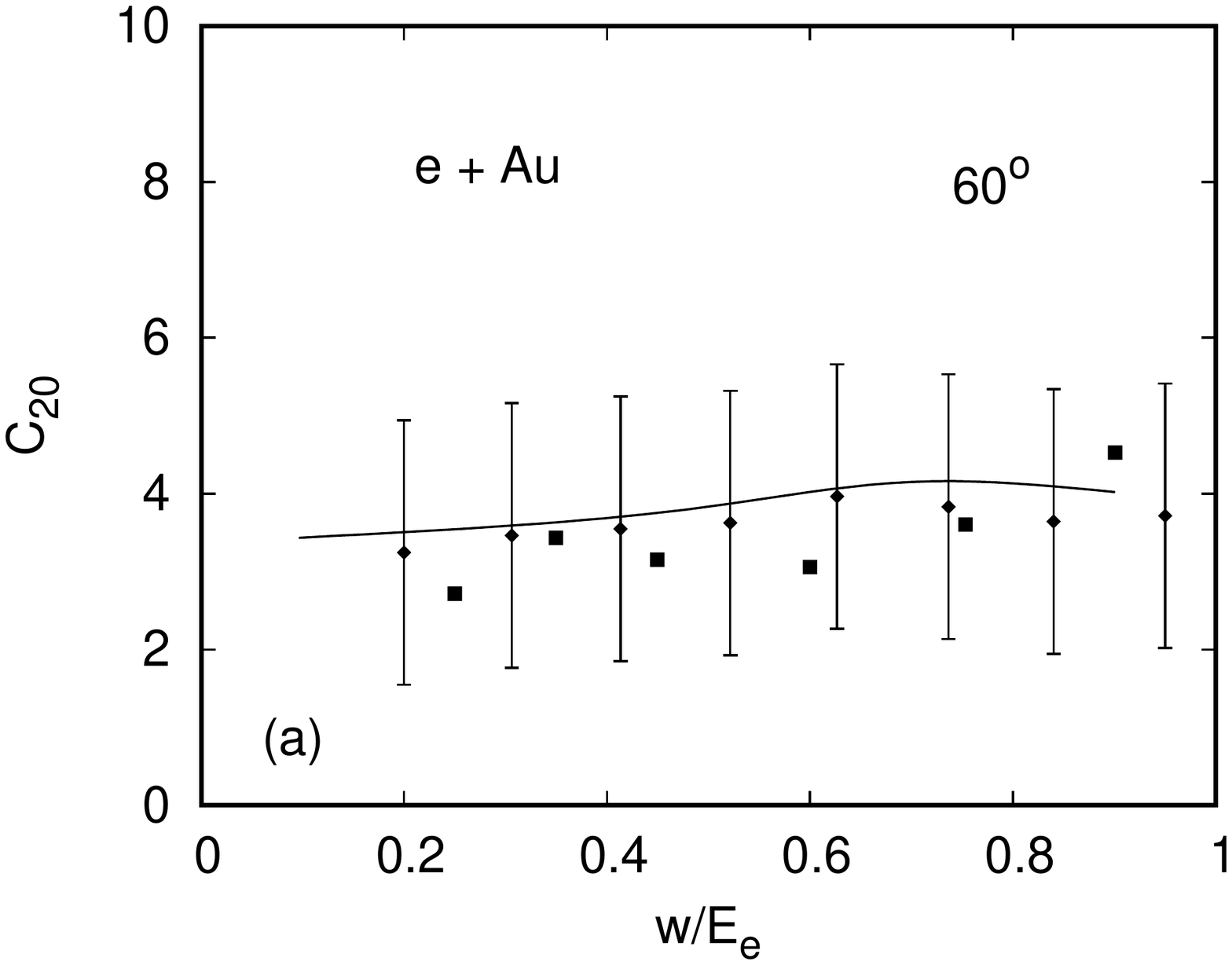}&
\hspace{-3cm} \includegraphics[width=.7\textwidth]{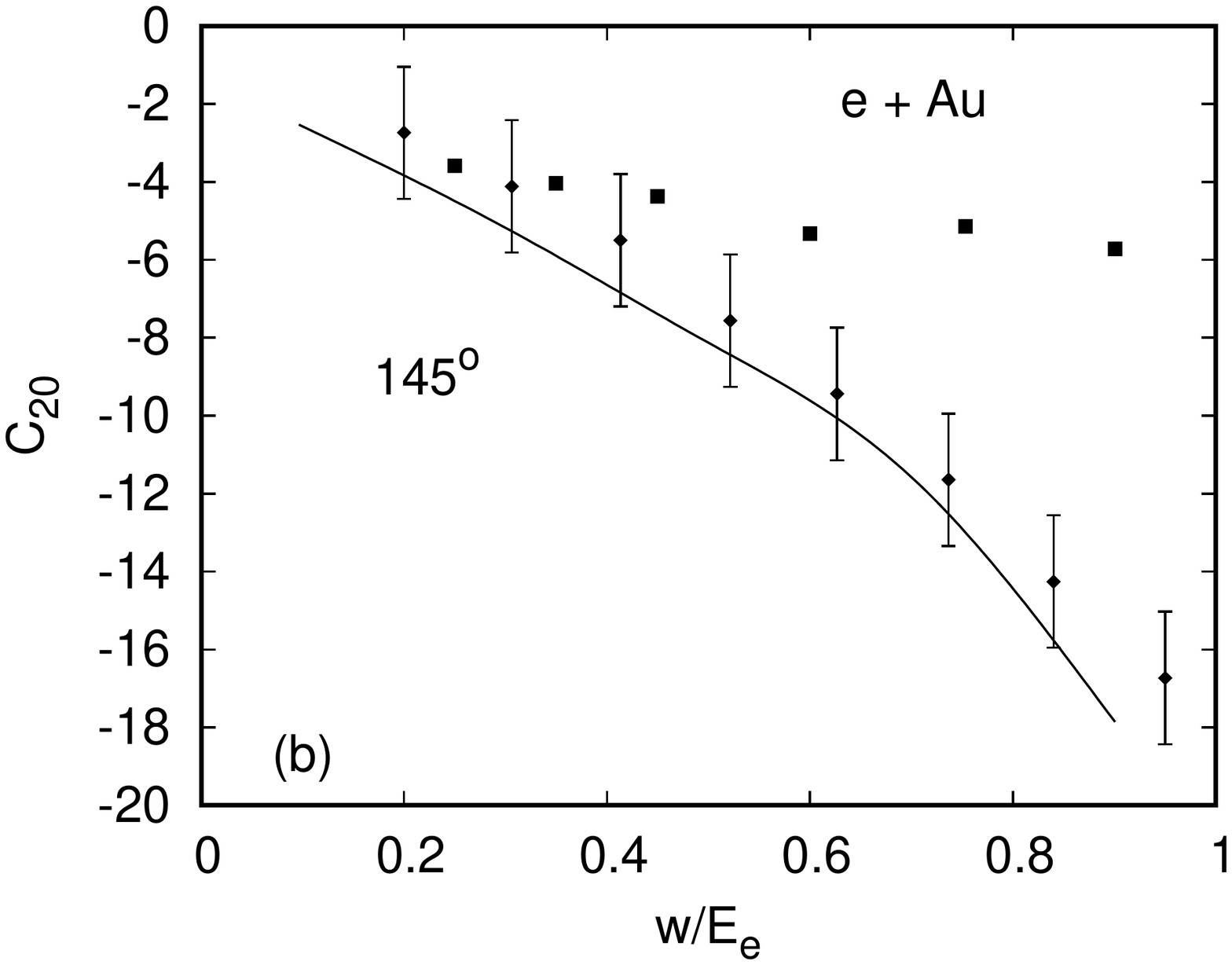}
\end{tabular}
\caption{
Spin asymmetry $C_{20}$ from 0.128 MeV electrons colliding with Au (a) at $\theta_k=60^\circ$  and (b) at $145^\circ$ as a function of the ratio $\omega/E_e$.
Shown are the experimental data from Mergl and Nakel \cite{MN90} $(\blacklozenge)$ and from Schaefer et al \cite{Sch82} $(\blacksquare)$, as well as the partial-wave results from Tseng and Pratt (--------------, as described in \cite{TP73}).
}
\end{figure}

Fig.5.2.3 displays $C_{20}$ resulting from bremsstrahlung by 0.128 MeV electrons colliding with gold at two selected photon angles for frequencies between 0.026 MeV and 0.122 MeV.
The more recent data from Mergl and Nakel \cite{MN90} agree well with the partial-wave results and show the
strong increase of the spin asymmetry  with photon angle and for the backward angle also the increase  with frequency.
Note that the sign of $C_{20}$ depends on the choice of coordinate system and bears no physics.

The further polarization correlations, $C_{12},\;C_{11}$ and $C_{31}$, were only investigated much later, when high-intensity polarized electron beams and more advanced polarimeters became available.
The change from a longitudinally polarized beam to a transversely polarized one was now accomplished with the help of a Wien-filter based spin rotator  \cite{SE77}. Thus any beam deflection can be avoided.
Moreover, such a filter allows for a continuous variation of the tilt angle $\alpha_s$  between the beam axis and the polarization vector $\bfzeta_i$.
For a fixed $\alpha_s$, $C_{12}$ and $C_{32}$ are mixed according to (in analogy to  (\ref{3.1.17}))
\begin{equation}\label{5.2.5}
P_{3}(\alpha_s)\;=\;P_e\;C_{32}\;\cos \alpha_s\;-\;P_e\;C_{12}\;\sin\alpha_s.
\end{equation}

\begin{figure}
\vspace{-1.5cm}
\includegraphics[width=13cm]{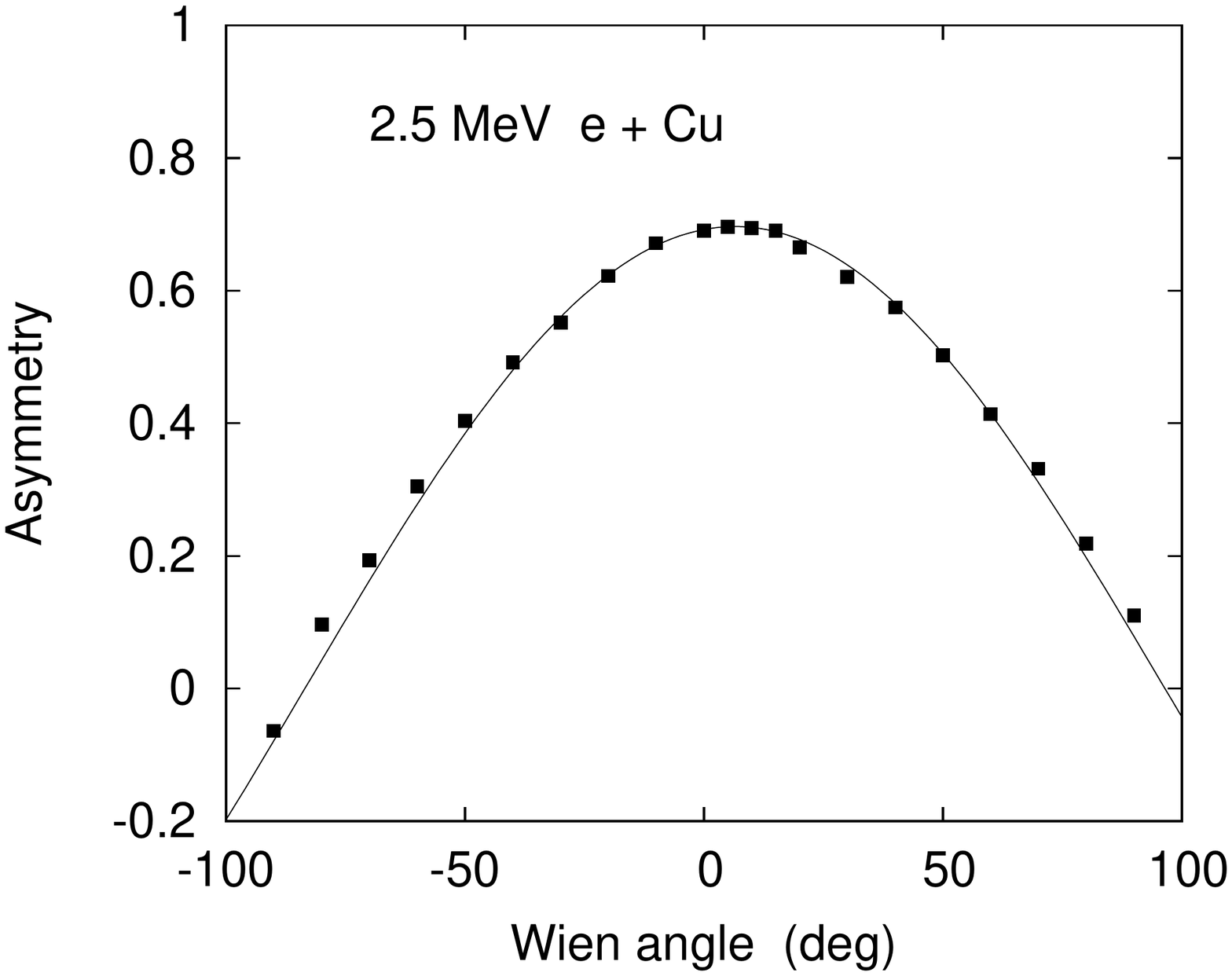}
\caption{
Asymmetry $P_3(\alpha_s)$ for bremsstrahlung from 2.5 MeV electrons colliding
with Cu at $\theta_k=12^\circ$ as a function of the Wien angle $\alpha_s$.
Shown are theoretical results for an average frequency of $\omega=1.75 $ MeV, using (\ref{5.2.5}) with $P_e=1$.
The measured points ($\blacksquare)$  are communicated by Barday and are normalized to theory in the peak maximum.
}
\end{figure}

Fig.5.2.4 displays the dependence of the measured asymmetry on $\alpha_s$ for 2.5 MeV spin-polarized electrons colliding with copper at a photon angle of $12^\circ \pm 6^\circ$.
Designed for testing a transmission Compton polarimeter for accurate in-beam polarization measurements, the photon frequency was not resolved (more precisely, all photons in the range 1 MeV $\leq \omega \leq 2.5$ MeV
 were recorded). Moreover, no absolute measurement was made, but only the  angle $\bar{\alpha}_s$ relating to
the maximum of the $P_3(\alpha_s)$-distribution  could
be determined. The experimental setup is described in Barday et al \cite{Ba11}.

Theoretically, the  angle $\bar{\alpha}_s$ can be obtained from the derivative of $P_3(\alpha_s)$ with respect to $\alpha_s$,
\begin{equation}\label{5.2.6}
P_3'(\alpha_s)\;=\;P_e\,\left[ -C_{32} \sin \alpha_s - C_{12} \cos \alpha_s \right]\;=\;0,
\end{equation}
such that
\begin{equation}\label{5.2.7}
\bar{\alpha}_s\;=\;\arctan \left( \frac{-C_{12}}{C_{32}}\right),
\end{equation}
irrespective of $P_e$.
Hence it provides a measure of the ratio between $C_{12}$ and $C_{32}$. Also shown in the figure are theoretical results within the Dirac partial-wave theory, including screening, for an average frequency of $\omega=1.75$ MeV and $\theta_k=12^\circ$.
The theoretical value is $\bar{\alpha}_s = \arctan \left( \frac{-0.0787}{0.692}\right)\,= 6.5^\circ$, which is to be compared with the experimental value of  $5.7^\circ$.

\begin{figure}
\vspace{-1.5cm}
\includegraphics[width=10cm]{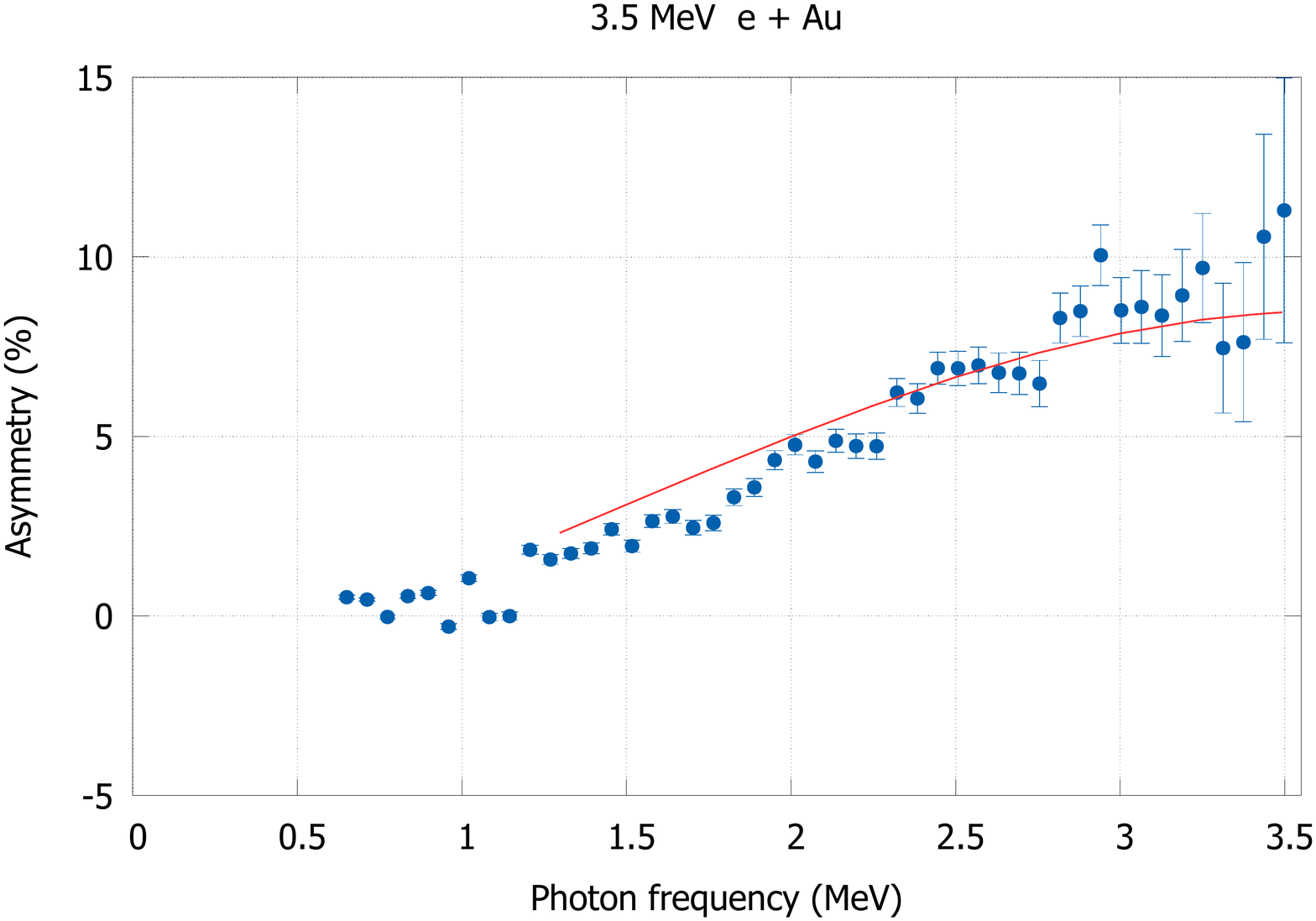}
\caption{
Asymmetry  (in percent) of bremsstrahlung emitted from 3.5 MeV longitudinally polarized electrons colliding with  Au at $\theta_k=0^\circ$ as a function of photon frequency. The experimental data  are from Nillius \cite{Ni20}.
Comparison is made with calculations  from the  DW theory including screening (----------).
}
\end{figure}

In a subsequent experimental investigation, the spectral resolution of the spin asymmetry was achieved for 3.5 MeV electrons colliding with a lead target  \cite{NA11}, however only
for photons emitted in the beam direction where $C_{12}=0$.
Fig.5.2.5 provides 
results from a more recent and more accurate experiment  \cite{Ni20}.  
Shown is the frequency dependence of the asymmetry resulting from spin-flipping  3.5 MeV longitudinally
polarized electrons when colliding with an Au target.
Comparison is made with theoretical results, 
  based on $C_{32}$ as  calculated from the screened Dirac partial-wave theory. This theory gives a  good description of the experimental data.
It should be noted that for $C_{32}$ also the Sommerfeld-Maue theory performs well (see \cite{Jaku11b}). It underestimates the partial-wave theory by less than 3 percent in the whole frequency range.
Earlier results from a high-energy approximation by
Olsen and Maximon \cite{OM59}, compared to the $0^\circ$-experiments
on a Pb target, overestimate the asymmetry considerably
 \cite{NA11}.

In further experiments, the extraction of both the frequency dependence and the absolute values for $C_{32}$ and $C_{12}$ became possible.
Fig.5.2.6 shows  results from longitudinally and transversely polarized electrons colliding with gold at an energy of 3.5 MeV and a photon angle
of $\theta_k=21^\circ$. The experimental setup
is described in \cite{NA16}. In order to achieve better statistics, in addition to flipping the initial electron spin, two detectors, placed symmetrically at $21^\circ$ on each side of the beam line, were used to determine the asymmetry.
The difference in intensity which is recorded in each of the two detectors (for a fixed initial spin) is related to $C_{ij}$
in the same way as the difference obtained with only one detector upon flipping the spin. This equivalence is easily seen from a virtual rotation of the experimental setup
by $180^\circ$ around the beam axis.
Comparison is made with partial-wave calculations,
and there is a qualitative agreement with the measurements both for $C_{32}$ and $C_{12}$.

\begin{figure}
\vspace{-1.5cm}
\centering
\begin{tabular}{cc}
\hspace{-2.5cm}\includegraphics[width=.6\textwidth]{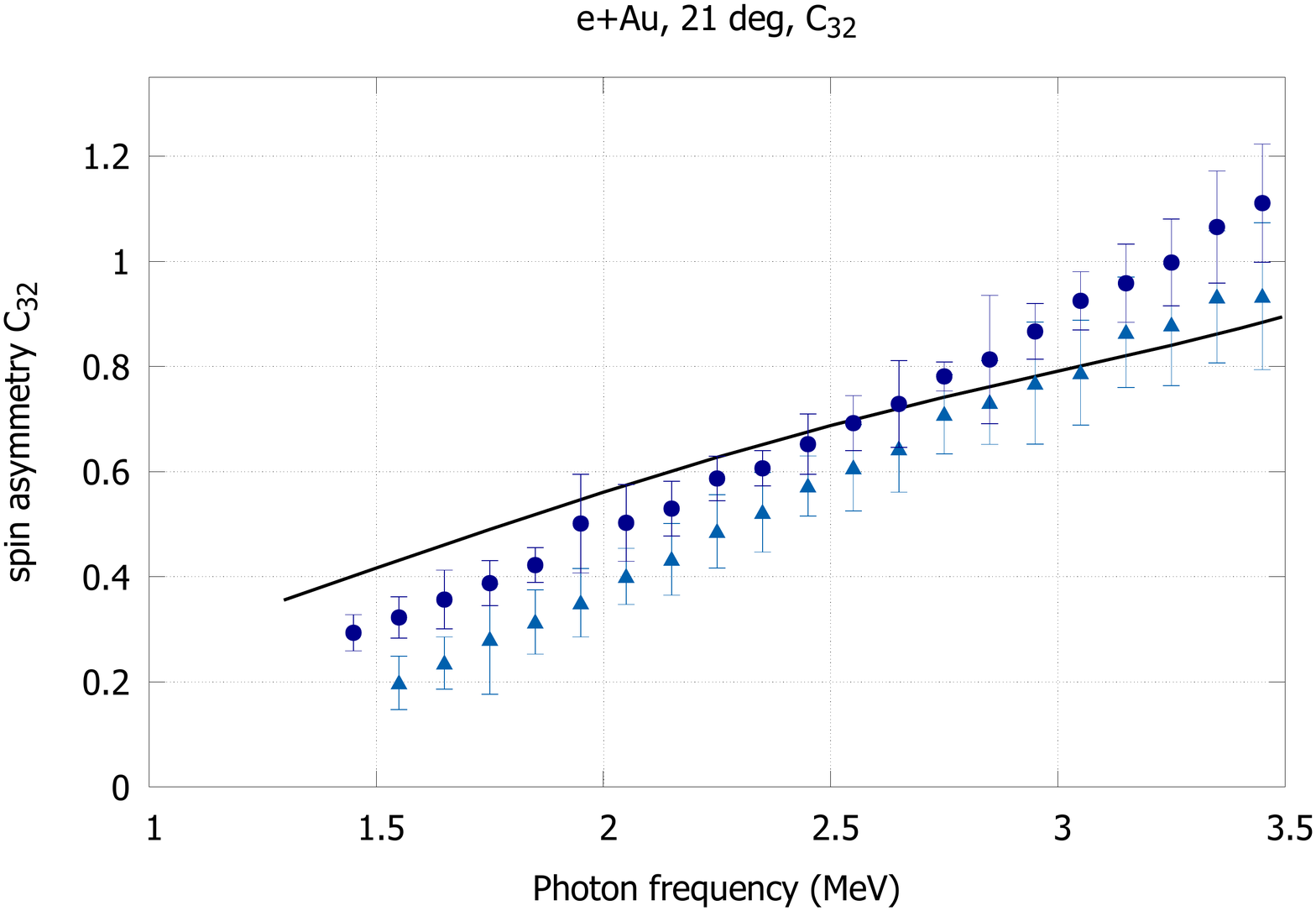}&
\hspace{-0.6cm} \includegraphics[width=.6\textwidth]{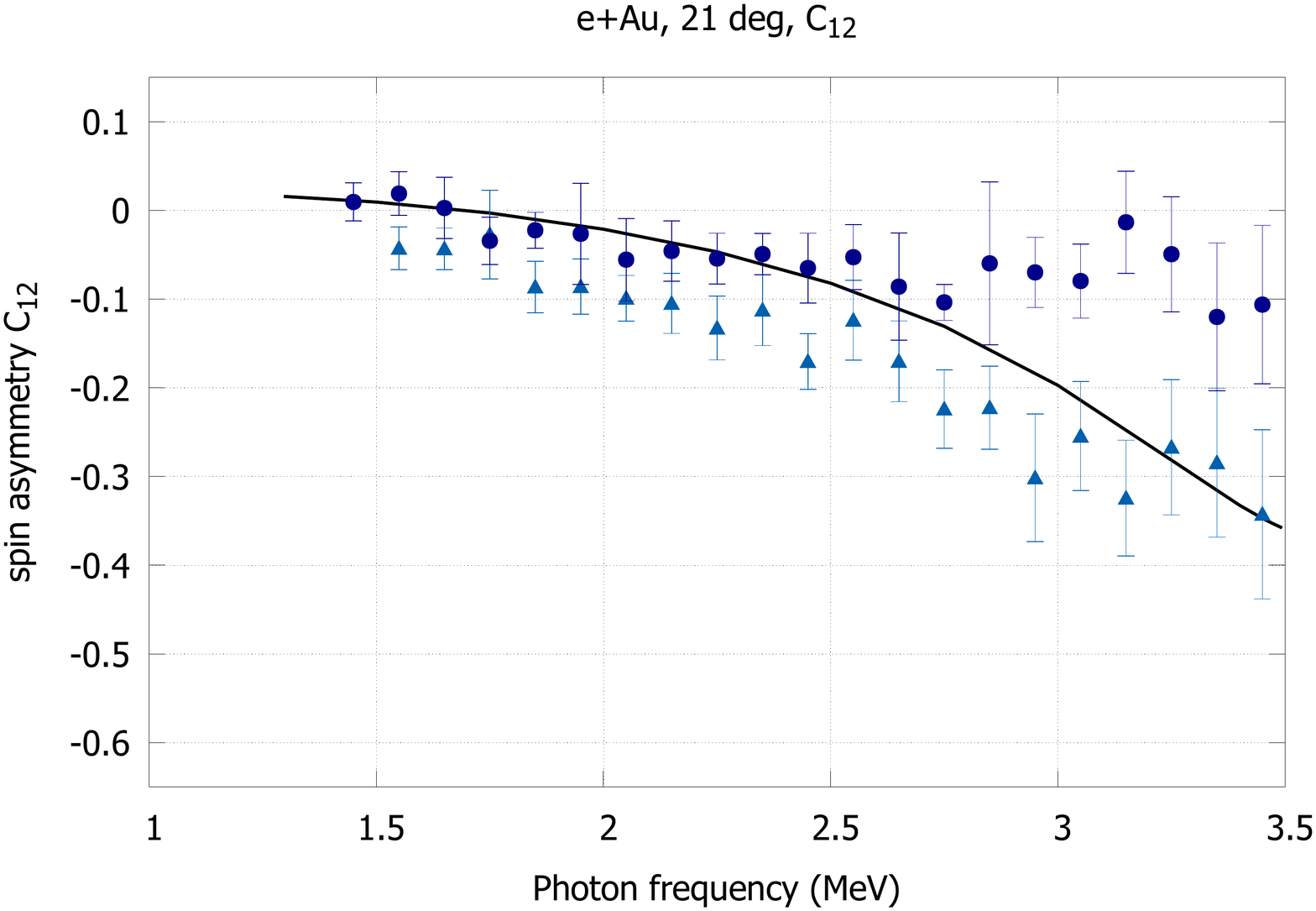}
\end{tabular}
\caption{
Polarization correlations $C_{32}$ (left) and $C_{12}$ (right) for bremsstrahlung from 3.5 MeV  polarized electrons colliding with Au at $\theta_k=21^\circ$ as a function of photon frequency.
The two sets of experimental points are measured with the right-hand (upper symbols), respectively, left-hand (lower symbols) detector (Nillius \cite{Ni20}).
Comparison is made with results from the Dirac partial-wave theory for a gold nucleus (-----------).
}
\end{figure}

At approximately the same time when $C_{12}$ was investigated, the first measurements of the linear polarization $P_2$ were carried out, both for longitudinally and transversely polarized electron beams.
However, instead of varying the Wien angle $\alpha_s$,
this angle was fixed to $0^\circ$ or $90^\circ$, while a novel
position-sensitive detector was employed \cite{Ta11,Ma12}.
In front of the detector a second target was placed, which induces Compton scattering of the emitted photons.
This Compton scattering intensity is sensitive to the 
photon polarization  \cite{KN29}. Such a detector device allows for a simultaneous determination of the position 
 (i.e. of the azimuthal angle $\varphi$ of the photon impact point relative to the reaction plane) and of the energy of the Compton-scattered photon.
From the $\varphi$-distribution the degree of linear polarization $P_L$ and the orientation of the polarization axis, described by the tilt angle $\chi$ with respect to the reaction plane, can be inferred.
The degree of linear polarization is related to $P_1$ and $P_2$ by means of
\begin{equation}\label{5.2.8}
P_L\;=\;\sqrt{P_1^2\,+\,P_2^2},
\end{equation}
while the tilt ange is a measure of the ratio between
$P_2$ and $P_1$,
\begin{equation}\label{5.2.9}
\tan 2\chi\;=\;\frac{P_2}{P_1},
\end{equation}
where we recall that $P_1=C_{03}$.

In order to derive (\ref{5.2.9}) we note that $\chi$ is related to the angle $\varphi_\lambda$ introduced in section 3.1.
With the polarization vector $\bfzeta_i$ in the $(x,z)$-plane, we have $\bfzeta_i \bfe_x = \sin \alpha_s$ and $\bfzeta_i \bfe_z = \cos \alpha_s$.
From (\ref{3.1.21}) we obtain the doubly differential cross section for linearly polarized photons,
accounting for $P_e\neq 1$,
\begin{equation}\label{5.2.10}
\frac{d^2\sigma}{d\omega d\Omega_k}(\alpha_s,\bfe_\lambda^\ast(\varphi_\lambda))\;=\;\frac12 \left( \frac{d^2\sigma}{d\omega d\Omega_k}\right)_0 \left[ 1+C_{03}\cos(2\varphi_\lambda)+P_e\,(C_{11} \,\sin \alpha_s -C_{31}\,\cos \alpha_s)\,\sin(2\varphi_\lambda)\right].
\end{equation}
The corresponding polarization $P(\alpha_s,\varphi_\lambda)$ is calculated from (\ref{3.1.11}) by using that
$\cos(2\varphi_\lambda +\pi)=-\cos(2\varphi_\lambda)$ and
$ \sin (2\varphi_\lambda + \pi)=-\sin(2\varphi_\lambda),$
$$ P(\alpha_s,\varphi_\lambda)\;=\;\frac{d^2\sigma(\alpha_s,\bfe_\lambda^\ast(\varphi_\lambda))-d^2\sigma(\alpha_s,\bfe_\lambda^\ast(\varphi_\lambda+\frac{\pi}{2}))}{d^2\sigma(
\alpha_s,\bfe_\lambda^\ast(\varphi_\lambda))+d^2\sigma(\alpha_s,\bfe_\lambda^\ast(\varphi_\lambda+\frac{\pi}{2}))}$$
\begin{equation}\label{5.2.11}
=\;C_{03}\cos(2\varphi_\lambda)\;+\;P_e\;(C_{11}\,\sin \alpha_s - C_{31}\,\cos \alpha_s)\;\sin(2\varphi_\lambda).
\end{equation}

According to (\ref{3.1.13}) we define $P(\alpha_s,\frac{\pi}{4})\equiv P_2(\alpha_s)=P_e(C_{11} \sin \alpha_s - C_{31} \cos \alpha_s)$.
We determine the angle $\bar{\varphi}_\lambda$ where $P(\alpha_s,\varphi_\lambda)$ has its maximum by calculating the derivative of $P(\alpha_s,\varphi_\lambda)$ with respect to $\varphi_\lambda$,
$$P'(\alpha_s,\varphi_\lambda)\;=\;-2C_{03}\sin(2\varphi_\lambda)\;+\;2\,P_e\;(C_{11}\,\sin \alpha_s - C_{31}\,\cos \alpha_s)\;\cos(2\varphi_\lambda)$$
\begin{equation}\label{5.2.12}
=\;-\;2P_1\,\sin (2\varphi_\lambda)\;+2P_2(\alpha_s)\;\cos(2\varphi_\lambda)\;=\;0,
\end{equation}
such that 
\begin{equation}\label{5.2.13}
\tan (2\bar{\varphi}_{\lambda}(\alpha_s))\;=\;\frac{P_2(\alpha_s)}{P_1}.
\end{equation}
This result implies that we have the identification $\chi(\alpha_s)=\bar{\varphi}_\lambda(\alpha_s).$
From $\tan (2\chi)=\frac{P_2}{P_1}$ we obtain 
$\cos(2\chi)=(1+\tan^2 (2\chi))^{-1/2}=P_1/\sqrt{P_1^2+P_2^2}$
and $\sin(2\chi)=\tan(2\chi)(1+\tan^2(2\chi))^{-1/2}=P_2/\sqrt{P_1^2+P_2^2}.$
The polarization which corresponds to the angle $\chi(\alpha_s)$ follows from (\ref{5.2.11}),
$$P(\alpha_s,\chi(\alpha_s))=P_1\cos(2\chi(\alpha_s))+P_2\sin(2\chi(\alpha_s))$$
\begin{equation}\label{5.2.15}
=\;\frac{P_1^2}{\sqrt{P_1^2+P_2^2(\alpha_s)}}\;+\;\frac{P_2^2(\alpha_s)}{\sqrt{P_1^2+P_2^2(\alpha_s)}}\;=\sqrt{P_1^2+P_2^2(\alpha_s)}\; \equiv\; P_L(\alpha_s).
\end{equation}
Hence $P_L$ in (\ref{5.2.8}) is the polarization in the maximum. The polarization vector leading to this maximum polarization is,
 according to the definition below (\ref{3.1.4}),
given by 
$\bfe_\lambda(\chi(\alpha_s))=\sin \chi(\alpha_s) \,\bfe_{\lambda_1}+\cos \chi(\alpha_s)\,\bfe_{\lambda_2}$.

The measurement of $\chi$ and $P_L$ at $\alpha_s=0^\circ$ and $90^\circ$ provides the simultaneous determination of $C_{03},\;C_{11}$ and $C_{31}$ by inverting (\ref{5.2.8}) and (\ref{5.2.9}),
\begin{equation}\label{5.2.16}
P_1=\frac{P_L(\alpha_s)}{\sqrt{1+\tan^2 2\chi(\alpha_s)}},\qquad 
P_2(\alpha_s)= \; \frac{P_L(\alpha_s)\,\tan 2\chi(\alpha_s)}{\sqrt{1+\tan^2 2\chi(\alpha_s)}}
\end{equation}
upon using $P_eC_{31}=-P_2(0),\;P_eC_{11}=P_2(90^\circ).$

\begin{figure}
\vspace{-1.5cm}
\centering
\begin{tabular}{cc}
\hspace{-1cm}\includegraphics[width=.7\textwidth]{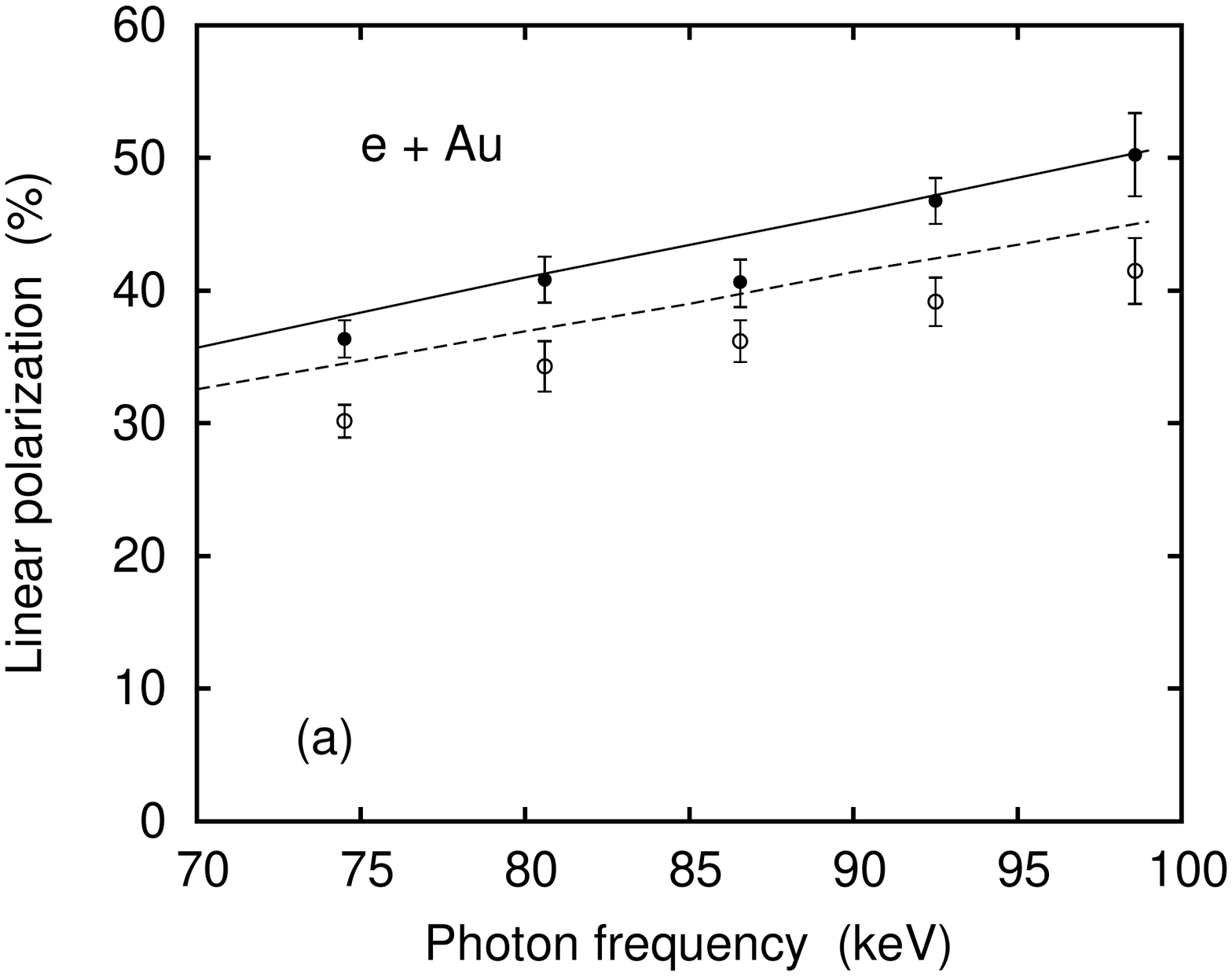}&
\hspace{-3cm} \includegraphics[width=.7\textwidth]{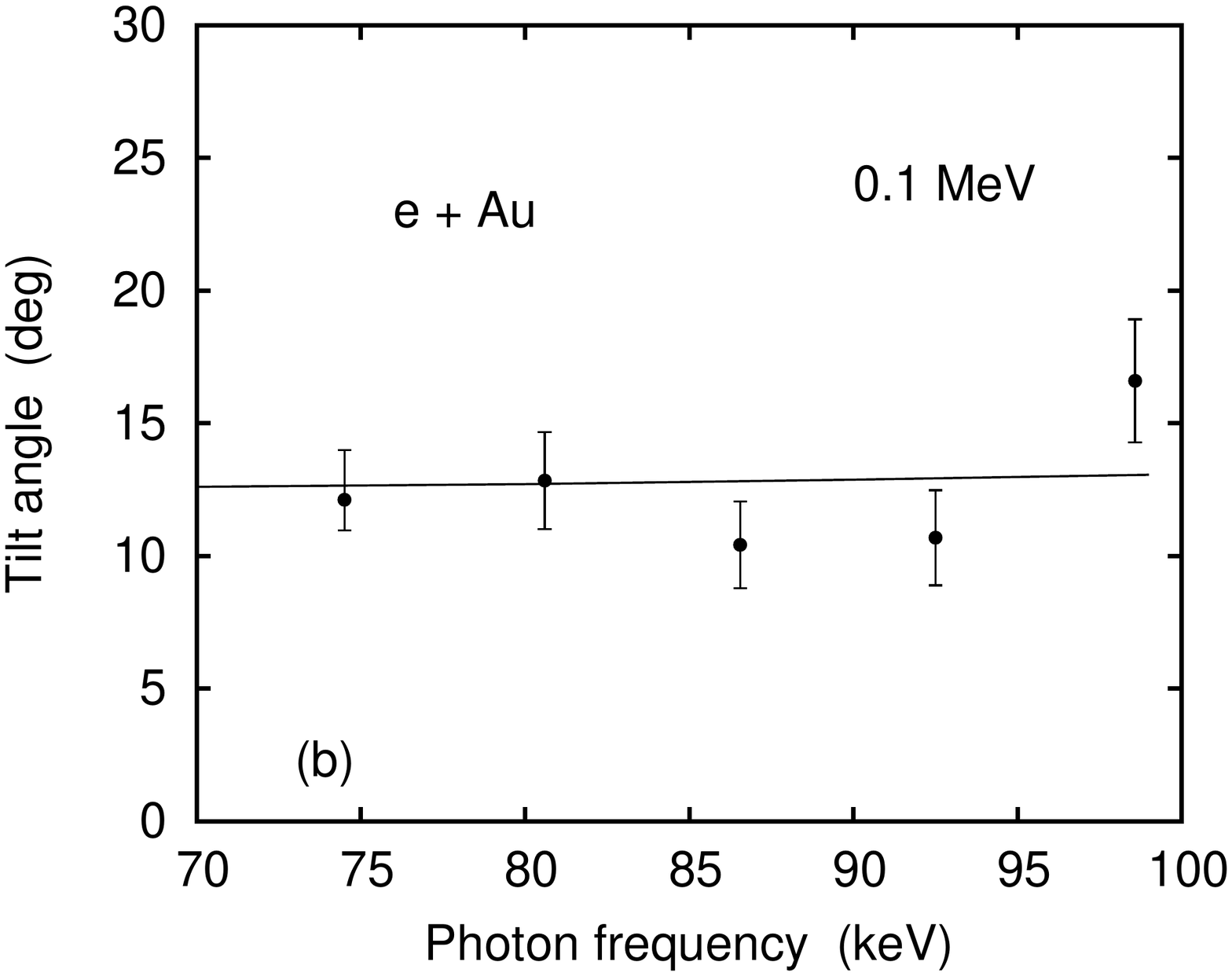}
\end{tabular}
\caption{
(a) Degree $P_L$ of linear polarization and (b) tilt angle $\chi$ for bremsstrahlung from 100 keV transversely polarized electrons colliding with Au
as a function of photon frequency $\omega$.
The photon emission angle is $130^\circ$. 
Shown are experimental results from M\"{a}rtin et al ($\bullet$, \cite{Ma12}) together with calculations by Yerokhin and Surzhykov
 using the  partial-wave theory (---------).
Included in (a) are the results for an unpolarized beam ($\circ$, experiment; $----$, theory). 
}
\end{figure}

\begin{figure}
\vspace{-1.5cm}
\includegraphics[width=13cm]{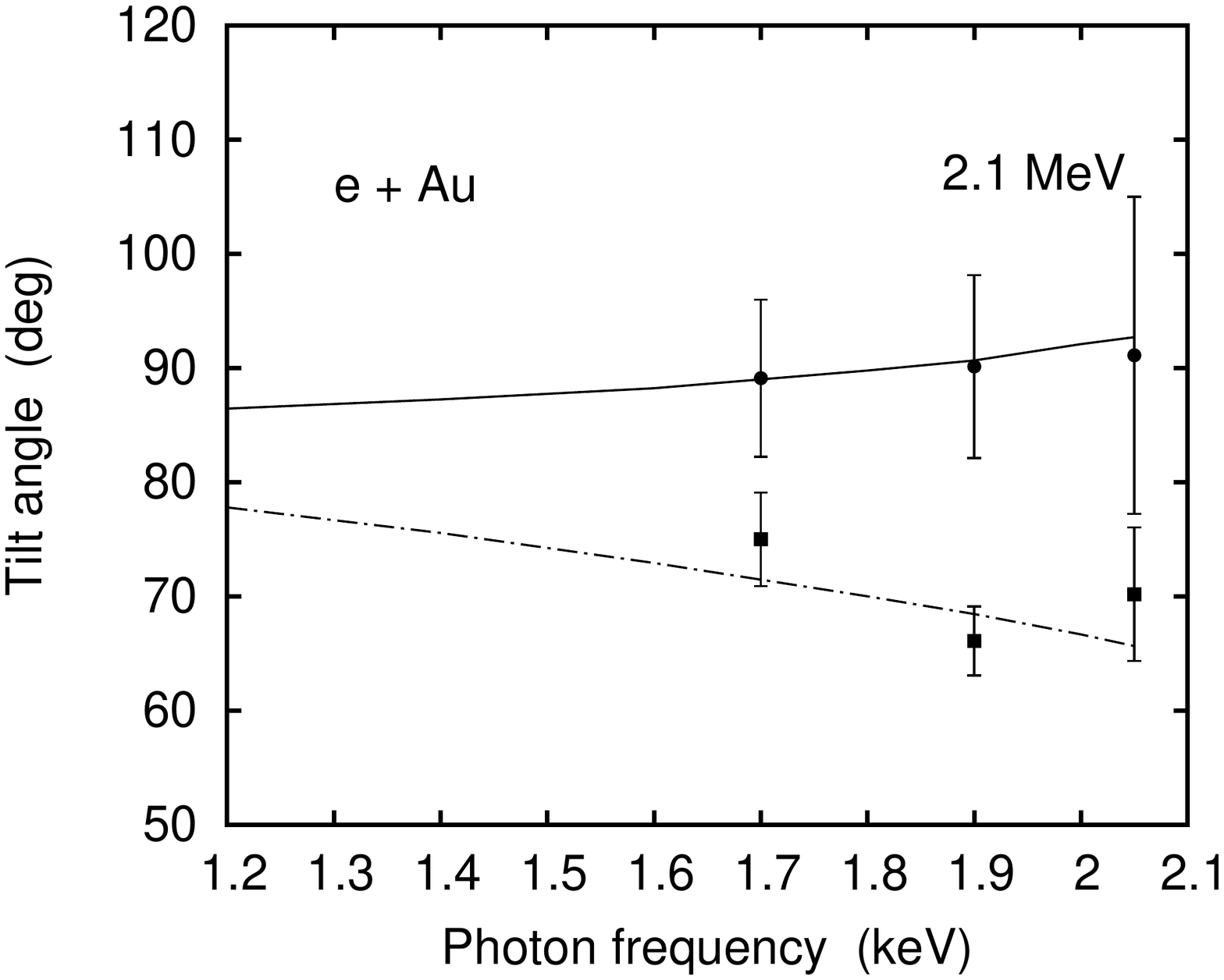}
\caption{
Tilt angle $\chi$ for bremsstrahlung from 2.1 MeV longitudinally and transversely polarized electrons colliding with Au
as a function of photon frequency $\omega$. The photon angle is $\theta_k=90^\circ$ and the degree of polarization is $P_e=0.8$.
Shown are the experimental results from Kovtun et al \cite{Ko15} for longitudinal ($\blacksquare$) and transverse ($\bullet$) polarization as well as  calculations within the partial-wave theory by
Yerokhin and Surzhykov ($-\cdot - \cdot -$, longitudinal polarization; ---------, transverse polarization).
}
\end{figure}

Results for $P_L$ (with degree of beam polarization $P_e=0.76$) and $\chi$ from 0.1 MeV transversely polarized electrons colliding with gold are shown in Fig.5.2.7, together with the results for an unpolarized beam (where $P_2=0$).
It is seen that $P_L$ increases with $\omega$, while $\chi$ is nearly independent of $\omega$ for such a low impact energy, as predicted by the partial-wave theory.

Experiments on the same collision system, but with a different experimental setup and a higher precision were carried out for longitudinally polarized electrons. 
The chosen photon angle of $90^\circ$ leads to a considerably smaller tilt angle ($\chi=2.1^\circ$  for the two considered frequency intervals $\omega \in (0.09,0.096)$ MeV and $\omega \in (0.096,0.1)$ MeV  \cite{Ta11}).

Table 5.2.1 shows the experimental results for $\chi$, including the ones for a transversely polarized beam, at $\theta_k=90^\circ$ and an average frequency of 0.0955 MeV
in comparison with Dirac partial-wave results.

\vspace{0.5cm}
Table 5.2.1\\
Tilt angle $\chi$ for bremsstrahlung from 0.1 MeV longitudinally ($\parallel$) and transversely ($\perp$) polarized
electrons colliding with a gold target.
The experimental data (Tashenov et al \cite{Ta11,Ta13}) are for $\omega=0.0955 \pm 0.0024$ MeV and $\theta_k=90^\circ \pm 3^\circ$. They are corrected for the degree of beam polarization ($P_e=0.75 \pm 0.4)$.
The calculations are performed within the (unscreened) Dirac partial wave theory for $\theta_k=90^\circ$ and $\omega = 0.0955$ MeV according to (\ref{5.2.9}). 

\begin{tabular}[t]{c||r|r|}  
Tilt angle & Experiment & Theory\\
&&\\ \hline
&&\\ 
$\chi_\perp(P_e=1)$ & $-6.48^\circ \pm 0.8^\circ$& $-6.76^\circ$\\
$\chi_\parallel(P_e=1)$ & $2.09^\circ \pm 0.4^\circ$& $2.02^\circ$\\
&&\\ \hline
\end{tabular}

\vspace{0.2cm}

Later, $\chi$ was also measured as a function of the Wien angle $\alpha_s$, and the perpendicular polarization correlation $C_{20}$ was recorded in the same experiment  \cite{Ta13}.

The transverse polarization correlations, such as $P_2$, increase in modulus with collision energy up to a few MeV. One has to keep in mind that these spin asymmetries are due to the purely relativistic spin-spin interaction between electron and target. This interaction is 
particularly large in strong fields (i.e. for heavy targets) and at high impact energies when the electron gets close to the 
target nucleus.
For these reasons, the experiment described above was later repeated at a beam energy of 2.1 MeV  \cite{Ko15}.

Fig.5.2.8 shows the resulting frequency dependence of $\chi$ for both longitudinally and transversely polarized electrons colliding with gold.
It is seen that at 2.1 MeV, the tilt angle reaches very large values, up to $90^\circ$.
Calculations within the partial-wave theory indicate
that $\chi$ decreases with $\omega$ for longitudinal polarization, while it increases with $\omega$ for transverse polarization. The experimental data are compatible with such a behaviour.

\subsection{Coincidence observations}

There are a series of polarization experiments where the bremsstrahlung photon is recorded in coincidence with the scattered electron.
Nearly all such experiments were carried out by Nakel and his coworkers, and a comprehensive overview is provided in  \cite{HN04}.

The first investigations concerned the linear polarization $P_1$ which now is identified with $C_{030}$, since the  scattered electron, but not its polarization, is recorded.
The measurements were performed with a beam of unpolarized electrons, typically at an energy of $0.1-0.5$ MeV.
A coplanar geometry was chosen with the direction 
of the scattered electron kept fixed, while the photon angle was varied.
Such coincidence experiments have the advantage that the photon frequency can accurately be determined from the energy loss
of the radiating electron.

\begin{figure}
\vspace{-1.5cm}
\centering
\begin{tabular}{cc}
\hspace{-1cm}\includegraphics[width=.7\textwidth]{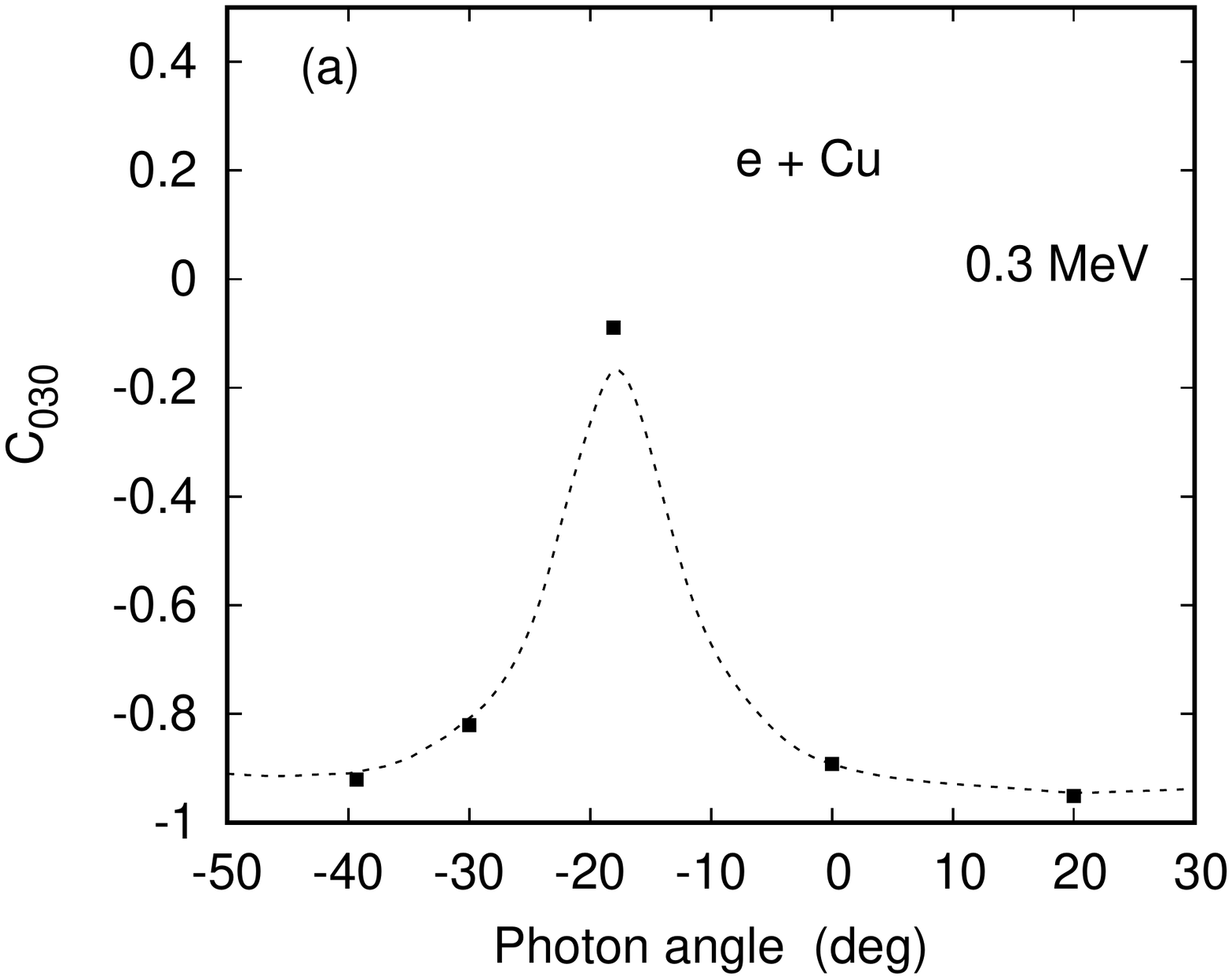}&
\hspace{-3cm} \includegraphics[width=.7\textwidth]{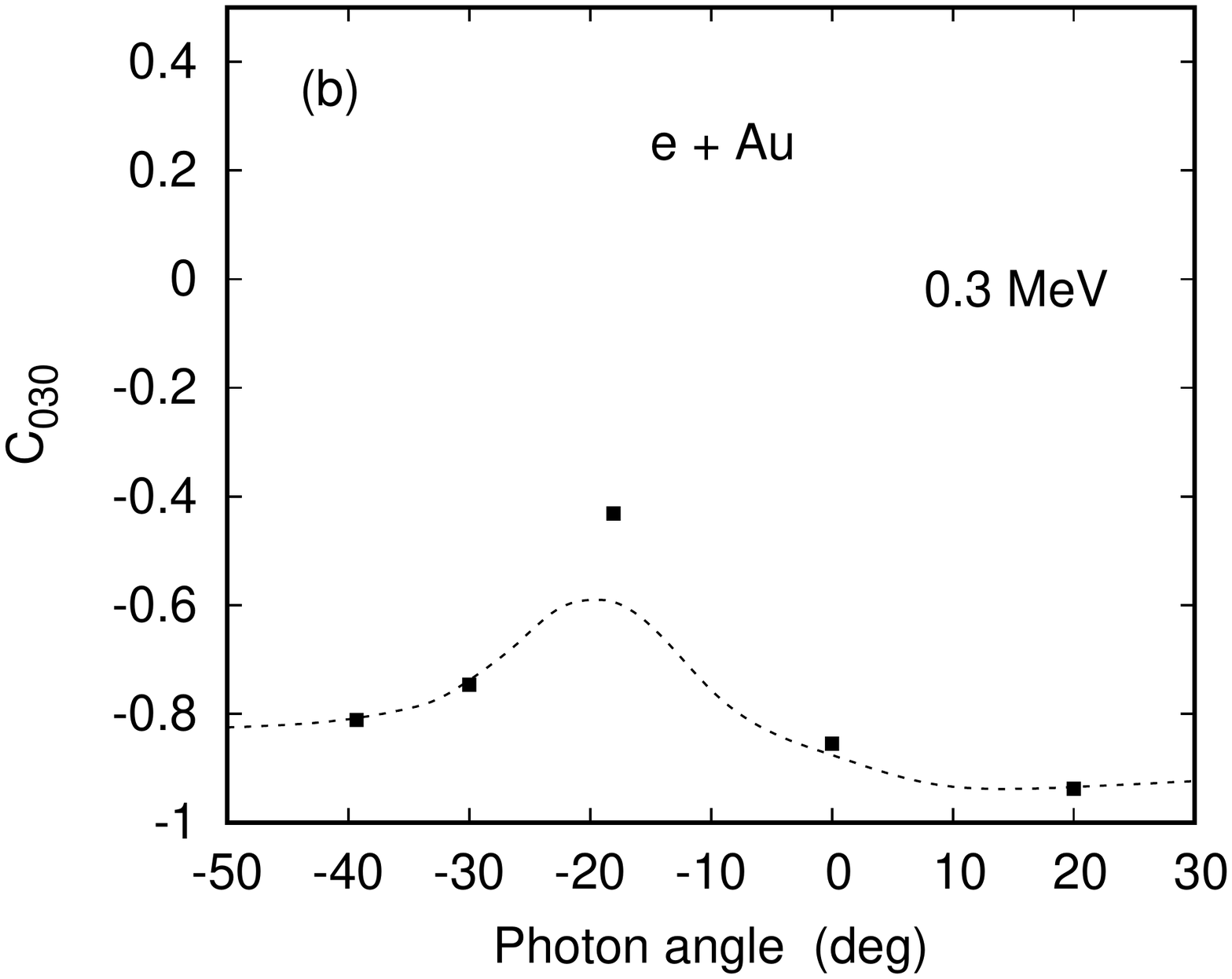}
\end{tabular}
\caption{
Linear polarization $C_{030}$ of bremsstrahlung emitted from 0.3 MeV unpolarized electrons
in collision (a) with Cu  and (b) with Au  as a function of photon angle $\theta_k$.
The electron final energy and angle are, respectively, 0.14 MeV and $\vartheta_f=20^\circ$.
Shown are measurements from Bleier and Nakel ($\blacksquare)$ \cite{BN84}, as well as calculations by Haug within the Sommerfeld-Maue theory ($----$) as described in \cite{EH69}.
}
\end{figure}

Fig.5.2.9 shows the angular distribution of $P_1$ for 0.3 MeV incoming and 0.14 MeV outgoing electrons scattered into the forward hemisphere.
Experimental results are shown for a copper and a gold target.
According to the sign-inverted formula (\ref{5.2.2}) as defined in this experiment, $P_1=-1$ corresponds to the case where all photons are polarized in the reaction plane.
A small fraction of perpendicularly polarized photons is only visible if the intensity $I_\parallel$ is strongly reduced, as is the case near a minimum of the triply differential cross section.
Consequently, a reduction of unit polarization is only found near an angle of $\theta_k=-\vartheta_f=20^\circ$, where the triply differential cross section has its minimum \cite{BN84}. This reduction is less for the heavier target.

Partial-wave results for this geometry give only a small reduction of unit polarization (by 5\%) near $\theta_k=-20^\circ$.
However, the calculation by Haug, shown in Fig.5.2.9,  takes into account all experimental implications,
including multiple scattering where contributions from noncoplanar geometry
come into play. Particularly this effect leads to a strong depolarization \cite{BN84}.

A second set of experiments was made with an electron beam polarized perpendicular to the reaction plane.
This allowed for the determination of $C_{200}$, using again the coplanar geometry.
Fig.5.2.10 shows the angular dependence of this spin asymmetry from 0.3 MeV electrons colliding with gold.
The large excursion of $C_{200}$ near $\theta_k=-30^\circ$ corresponds also here to a strong minimum
in the cross section.
Theory predicts  a second excursion of $C_{200}$ near $60^\circ$, where the photon intensity has another shallow minimum (see Fig.5.1.9).

\begin{figure}
\vspace{-1.5cm}
\includegraphics[width=13cm]{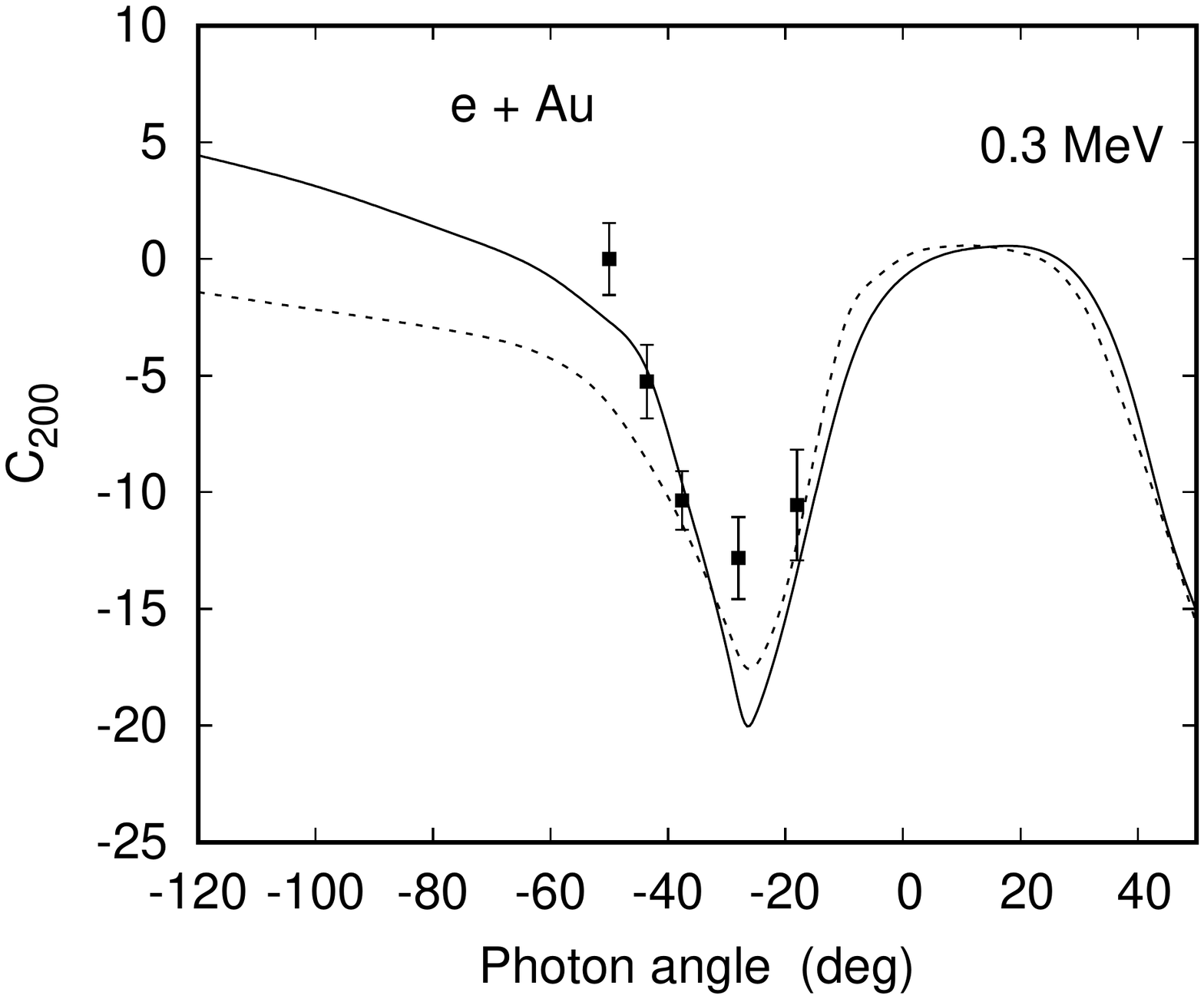}
\caption{
Spin asymmetry $C_{200}$ for bremsstrahlung emitted from 0.3 MeV perpendicularly polarized electrons colliding with Au 
as a function of photon angle $\theta_k$.
The electron is scattered at an angle of $\vartheta_f =20^\circ$ with final energy 0.2 MeV.
The experimental data ($\blacksquare$) are from Mergl et al \cite{Me92}, the full line is the result from a partial-wave calculation (Keller and Dreizler \cite{KD97}),
and the dashed line results from the Sommerfeld-Maue theory (Haug \cite{Ha96}).
}
\end{figure}

Apart from the partial-wave results by Keller and Dreizler \cite{KD97} shown in the figure, there exist further partial-wave calculations by
 Tseng \cite{T02}  using a different atomic potential. 
Although screening is here of little importance, there are substantial discrepancies between the two results as shown in \cite{T02}.

We note that of the seven nonvanishing polarization correlations in coplanar geometry, the aforementioned ones
($C_{030}$ and $C_{200}$) are the only ones investigated experimentally so far.
Neither exist any measurements in noncoplanar geometries. 
\vspace{0.5cm} 

{\Large\bf 6. Summary}

\vspace{0.2cm}


In this review we have concentrated on the presentation  of the current bremsstrahlung theories such as the general Dirac partial-wave theory and its asymptotic approximations for ultrarelativistic collisions and photons near the short-wavelength limit.
We have also considered analytical approaches which go beyond the Sommerfeld-Maue theory, originating either from a quantum mechanical or from a quasiclassical consideration.

A further focus was put on the polarization correlations between impinging electron and emitted photon or scattered electron, introducing sum rules.
The validity of the PWBA for heavy targets in the case of soft photons was elucidated with the help of simultaneously investigating positron and electron bremsstrahlung,
which also served as a test for the $Z_T$-dependence of the spin asymmtry.

The Dirac partial-wave theory was used to explain recent experiments on the polarization correlations, covering a collision energy region from $0.1-3.5$ MeV.
Also new measurements of the differential photon spectra
were included.
It was further demonstrated that at an impact energy of 500 MeV, where the Dirac partial-wave theory is no longer applicable, the analytical higher-order theories are able to explain the recently measured inclusive photon spectra.

Bremsstrahlung has to be seen in context with related processes which are not touched upon in this review.
To these belong the radiative capture of a loosely bound target electron into the continuum of a heavy, highly stripped projectile.
Such inverse kinematics is in fact the only tool for measuring a photon at the high-energy end of the spectrum in coincidence with a scattered electron of near-zero kinetic energy.
Related to hard bremsstrahlung is further the time-reversed process, the photoeffect.
Also pair creation by an energetic photon in the field of a high-$Z_T$ target can be described with a theory much alike the one for electron bremsstrahlung.
The investigation of such processes helps to shed further light onto the phenomena of radiation physics.

\vspace{0.5cm}

{\large\bf Acknowledgment}

It is a pleasure to thank Alessio Mangiarotti for an inspiring and fruitful collaboration. My thanks go also to Andrey Surzhykov and Vladimir Yerokhin for many discussions and
for their help with setting up the partial-wave code,
to Fabian Nillius for communicating unpublished experimental data, and to Fazlul Haque for providing static screened potentials.


\renewcommand{\theequation}{\Alph{section}.\arabic{equation}}



\section*{Appendix A: $\;\;$ Proof of convergence of the SM function to the exact solution $\psi$ as $E \to \infty$}
\setcounter{equation}{0}
\setcounter{section}{1}
\renewcommand{\thesection}{\Alph{section}}
\setcounter{subsection}{0}

In section 2.2 it is shown that inserting the Sommerfeld-Maue function into the Dirac equation leads to the remainder $R^{SM}$ from (\ref{2.2.7}), which can be estimated for $p
\geq c$ (i.e. $\eta < \frac{2Z_T}{c}$) by
\begin{equation}\label{E.1}
|R^{SM}|\;\leq\; C_0\;\frac{Z_T^2}{r}\;\left| _1F_1(1+i\eta,2,z)\right|\;\left|\sin \frac{\theta}{2}\right|,
\end{equation}
where $C_0$ is some constant, and $z=2ipr \sin^2\frac{\theta}{2}.$

For large $|z|$, corresponding to large $p$ or $r$ at fixed angle $\theta_0>0$,
the confluent hypergeometric function can be expanded in the following way \cite[p.508]{AS64}, retaining only the incoming wave,
\begin{equation}\label{E.2}
_1F_1(1+i \eta,2,z)\;=\;\Gamma(2)\;e^{i\pi-\pi \eta}\;\frac{e^{-(1+i\eta)\ln z}}{\Gamma(1-i\eta)}\;\left[ 1\;+\;O\left( \frac{1}{z}\right)\right],
\end{equation}
with the estimate
\begin{equation}\label{E.3}
\left| _1F_1(1+i\eta,2,z)\right|\;\leq\;\left| e^{-\pi \eta}\;\frac{e^{-(1+i\eta)\ln z}}{\Gamma(1-i\eta)}\right|\;+\;C_2\left| e^{-\pi \eta}\;\frac{e^{-(1+i\eta)\ln z}}{\Gamma(1-i\eta)\;z}\right|,
\end{equation}
$C_2$ being some constant.

Fix $r=R_0$. Then for $r\geq R_0$ and $\theta \geq \theta_0$, using
\begin{equation}\label{E.4}
\left| e^{-\pi \eta}\;\frac{z^{-(1+i\eta)}}{\Gamma(1-i\eta)}\right|
\;=
\;e^{-\pi \eta/2}\;\frac{1}{2pr \sin^2\frac{\theta}{2}}\;\frac{1}{|\Gamma(1-i\eta)|},
\end{equation}
we estimate the remainder, introducing new constants $C_1$ and $C_3$,
$$|R^{SM}|\;\leq\;C_0\;\frac{Z_T^2}{R_0}\;\left| \sin \frac{\theta}{2}\right|\;\frac{e^{-\pi \eta/2}}{|\Gamma(1-i\eta)|}\;\left[ \frac{1}{2pR_0 \sin^2\frac{\theta}{2}}\;+\;\frac{C_2}{(2pR_0\sin^2 \frac{\theta}{2})^2}\right]$$
\begin{equation}\label{E.5}
\leq\;C_1\;\frac{Z_T^2}{R_0}\;\frac{1}{pR_0 \sin \frac{\theta_0}{2}}\;+\;C_3\;\frac{Z_T^2}{R_0}\;\frac{1}{p^2R_0^2 \sin^3 \frac{\theta_0}{2}}.
\end{equation}

For sufficiently high momentum $p$, each of the two terms can be made smaller than $\frac{\epsilon}{2}$. In particular,
$$C_1\;\frac{Z_T^2}{pR_0^2\sin \frac{\theta_0}{2}}\;<\;\frac{\epsilon}{2}\;\Longleftrightarrow\;p\;>\;p_0\;=\;\frac{2C_1 Z_T^2}{\epsilon R_0^2 \sin \frac{\theta_0}{2}}$$
\begin{equation}\label{E>6}
C_3\;\frac{Z_T^2}{p^2R_0^3\sin^3\frac{\theta_o}{2}}\;<\;\frac{\epsilon}{2}\;\Longleftrightarrow\;p\;>\;p_1\;=\sqrt{\frac{2C_3 Z_T^2}{\epsilon R_0^3 \sin^3 \frac{\theta_0}{2}}},
\end{equation}
such that
\begin{equation}\label{E.7}
|R^{SM}|\;<\;\epsilon\qquad\quad \mbox{ for }\; p\;>\;p_{\rm max} = \max\{p_0,p_1,c\},
\end{equation}
or equivalently, for $E>E_{\rm max}=\sqrt{p_{\rm max}^2 c^2+c^4}.\;$ This proves the convergence of
\begin{equation}\label{E.8}
\psi^{SM} \;\longrightarrow\;\psi\qquad \mbox{for } E \;\to \;\infty.
\end{equation}
However, since $E_{\rm max}$ depends both on $R_0$ and $\theta_0$, the convergence is not uniform.

\section*{Appendix B: $\;\;$ Numerical details for the calculation of radial integrals in the partial-wave theory}
\setcounter{equation}{0}
\setcounter{section}{2}
\setcounter{subsection}{0}

\vspace{0.2cm}

The evaluation of the radial integrals in the partial-wave theory requires for each triple $(\kappa_f,\kappa_i,l)$
of final-state angular momentum quantum number $\kappa_f$, initial-state quantum number $\kappa_i$ and photon angular momentum $l$
the calculation of two integrals,
\begin{equation}\label{A.1}
R_{fi}(l)\;=\;\int_0^\infty r^2dr\;g_{\kappa_f}(r)\;f_{\kappa_i}(r)\;j_l(kr),
\end{equation}
\begin{equation}\label{A.2}
R_{if}(l)\;=\;\int_0^\infty r^2dr\;f_{\kappa_f}(r)\;g_{\kappa_i}(r)\;j_l(kr),
\end{equation}
where $g_\kappa$ and $f_\kappa$ are, respectively, the large and small components of the radial Dirac scattering function, and $j_l$ is a spherical Bessel function.

Since $g_\kappa$ and $f_\kappa$ behave asymptotically like modulated plane waves, the integrand of (\ref{A.1}) or (\ref{A.2}) is a strongly oscillating function at large distances $r$. 
Moreover, all functions, $g_\kappa,\;f_\kappa$ and $j_l$, decrease weakly  with $r$ (like $1/r$) for $r \to \infty$, such that the total integrand decreases only like $1/r$.
This makes the evaluation of $R_{fi}$ and $R_{if}$ rather challenging.

Let us consider the case of a potential with a Coulombic tail (either a nuclear potential
or an ionic potential), and let the asymptotic potential behave according to $-Z_s/r$ with $Z_s \leq Z_T$ where $Z_T$ is the nuclear charge number.

The calculation is simplified by the fact that a numerical solution of the Dirac equation is not required
if either r is large, or if $\kappa$ is high. Both cases  correspond to large distances from the nuclear center where the short-range part of  the potential
has become negligibly small. Under such conditions we can profit from the analytically known solutions to a point-nucleus Coulomb field (the Coulomb-Dirac functions). 
For large distances (at small $\kappa$ where the short-range phase shift $\delta_\kappa$  is nonzero), the solutions $g_\kappa$ and $f_\kappa$ can be
represented in terms of a superposition of the regular and irregular Coulomb-Dirac functions to the charge number $Z_s$  \cite{Sal95},
\begin{equation}\label{A.3}
{g_\kappa \choose f_\kappa} \;=\; \cos \delta_\kappa\;{g_\kappa \choose f_\kappa}_{\rm reg} \;+\; \sin \delta_\kappa\;{g_\kappa \choose f_\kappa}_{\rm irr} \qquad \mbox{ for $r$ large},
\end{equation}
where $\delta_\kappa$ is determined by matching the numerical inner solution of the Dirac equation to the large-$r$ representation (\ref{A.3}) at some distance outside the range of the short-range part of the potential.
For large $\kappa$, one has $\delta_\kappa =0$ such that the scattering state can for arbitrary $r$ be taken as the regular Coulomb-Dirac function, ${g_\kappa \choose f_\kappa} = {g_\kappa \choose f_\kappa}_{\rm reg}$.

An efficient method to deal with strongly oscillating integrands is a deformation of the real integration path into the complex plane.
Making use of the representation (\ref{A.3}),
each of these functions can be split into two components, one behaving asymptotically like $e^{ipr}$ and the other one like $e^{-ipr}$, where $p=p_i$ for the initial and $p=p_f$ for the final state, such that
\begin{equation}\label{A.4}
{g_\kappa \choose f_\kappa} \;=\;{g_\kappa^+ \choose f_\kappa^+ } \;+\;{g_\kappa^- \choose f_\kappa^-}.
\end{equation}
For the first summand in (\ref{A.4}) which behaves like $e^{ipr}$, the integration path from some distance $R_m$ (large enough such that (\ref{A.3}) is valid) to $\infty$ is deformed along the line $l_+ = R_m+iy,\;y>0$ and closed along the infinitely far semicircle in the upper half plane.
For the second summand the deformation is along the line $l_-=R_m-iy,\;y>0$ and is closed in the lower half plane. With this choice, the integrands along
$l_+$ and $l_-$ are exponentially decreasing, while there is no contribution from the infinitely far semicircles.
This is easily shown if we make use
of the fact that also the Bessel
function can be decomposed into a pair of Hankel functions, $j_l(kr)=\frac12(h_l^{(1)}(kr) + h_l^{(2)}(kr))$, each behaving asymptotically like $e^{ikr}$ and $e^{-ikr}$, respectively \cite{AS64}.
Moreover, from energy  conservation,
$E_i=E_f+kc$, one gets
\begin{equation}\label{A.5}
\Delta k \equiv p_i\;-\;p_f\;-\;k\;>\;0,
\end{equation}
which follows from the auxiliary expression $\Delta k \cdot \frac{c}{k} (p_i+p_f)=E_i+E_f-c(p_i+p_f)$ which is positive provided the electron rest energy is retained. The inclusion of recoil reduces $p_f$ and does not affect the validity of (\ref{A.5}).
Therefore, $g_{\kappa_i}$ and $f_{\kappa_i}$ provide the leading exponents, and it is only necessary to decompose
$f_{\kappa_i}$ and $g_{\kappa_i}$ for not too high energies.

With this in mind, the complex-plane rotation method transforms (\ref{A.1}) into
$$ R_{fi}(l)\;=\;\int_0^{R_m} r^2dr\;g_{\kappa_f}(r)\;f_{\kappa_i}(r)\;j_l(kr)\;+\left.\;i\int_0^\infty dy\;r^2 g_{\kappa_f}(r)\;f_{\kappa_i}^{(+)}(r)\;j_l(kr)\right|_{r=R_m+iy}$$
\begin{equation}\label{A.6}
\left. -\;i\int_0^\infty dy\;r^2g_{\kappa_f}(r)\;f_{\kappa_i}^{(-)}(r)\;j_l(kr)\right|_{r=R_m-iy}
\end{equation}
$$=\;\int_0^{R_m} r^2dr\;g_{\kappa_f}(r)\;f_{\kappa_i}(r)\;j_l(kr)\;+\left.\;2 \mbox{ Re }\left\{i\int_0^\infty dy\;r^2 g_{\kappa_f}(r)\;f_{\kappa_i}^{(+)}(r)\;j_l(kr)\right\} \right|_{r=R_m+iy},$$
where the second equality results from $f_{\kappa_i}^{(-)}=f_{\kappa_i}^{(+)\ast}$ because $f_\kappa$ is real-valued.

For high $p_i$ (i.e. $E_i$ in the MeV region)
there may occur exponential overflow in the separate contributions to $g_{\kappa_f}$ or $f_{\kappa_i}^{(+)}$ when $y$ is large. In that case, the infinite integral is split at some $y_m$ and is approximated by
\begin{equation}\label{A.7}
 2 \mbox{ Re }\left\{ i\int_0^{y_m} dy\;r^2g_{\kappa_f}(r)\;f_{\kappa_i}^{(+)}(r)\;j_l(kr)\right\}
\;+\;\mbox{ Re } \left\{ i\int_{y_m}^{y_{\rm max}} dy\;r^2\;\tilde{g}_{\kappa_f}^{(-)}(r)\;\tilde{f}_{\kappa_i}^{(+)}(r)\;\tilde{h}_l^{(2)}(kr)\;e^{i \Delta k\cdot r}\right\},
\end{equation}
where $y_{\rm max} \sim 8/\Delta k$ and  the tilde denotes the omission of the exponential factors, e.g. $g_{\kappa_f}^{(-)}(r)=\tilde{g}_{\kappa_f}^{(-)}(r)\,e^{-ip_fr}$.
The splitting value $y_m$ should be taken large enough such that the retained leading term with the weakest asymptotic decrease gives a good approximation.

There are several possibilities to represent the Coulomb-Dirac waves such that a decomposition according to (\ref{A.4}) can be made.
One of them 
involves the nonrelativistic Coulomb waves  \cite{Sal95}. For these functions there exists an efficient series expansion for large $r$, if $\kappa$ is not too large ($|\kappa| \lesssim 100$).  
Denoting by $F_\gamma(\eta,pr)$ the regular and by $G_\gamma(\eta,pr)$ the irregular Coulomb waves with $\eta=Z_sE/(pc^2)$, the regular and irregular Coulomb-Dirac functions are given by
$$ {g_{\kappa,{\rm reg}}(r) \choose g_{\kappa,{\rm irr}}(r)}\;=\;\frac{N_\kappa}{pr}\left[ (\kappa +\gamma)\sqrt{\gamma^2+\eta^2}\;pc\;{ F_\gamma(\eta,pr) \choose G_\gamma(\eta,pr)}\;-\;
\frac{Z_s}{c}\;(\gamma c^2-\kappa E)\;{F_{\gamma -1}(\eta,pr) \choose G_{\gamma -1}(\eta,pr)}\right],$$
\begin{equation}\label{A.12}
{f_{\kappa,{\rm reg}}(r) \choose f_{\kappa,{\rm irr}}(r)}\;=\;-\;\frac{N_\kappa}{pr}\left[ -\,\frac{Z_s}{c}\;\sqrt{\gamma^2+\eta^2}\;pc\;{F_\gamma(\eta,pr) \choose G_\gamma(\eta,pr)}\;+\;
(\kappa + \gamma)\;(\gamma c^2-\kappa E)\;{F_{\gamma -1}(\eta,pr) \choose G_{\gamma -1}(\eta,pr)}\right],
\end{equation}
where $\gamma=\sqrt{\kappa^2-(Z_s/c)^2}$ and the normalization constant is given by
\begin{equation}\label{A.13}
N_\kappa \;=\;\sqrt{\frac{E+c^2}{\pi E}}\;\frac{\mbox{ sign }\kappa}{\gamma\;\sqrt{(Z_s/c)^2(E+c^2)^2+(\kappa +\gamma)^2(pc)^2}}.
\end{equation}
We note that the reduction factor $\sqrt{(E-c^2)/(E+c^2)}$ of the small component is inherent in this representation.
In order to allow for an (\ref{A.4})-type splitting, the functions $F_\gamma$ and $G_\gamma$ are for large distances written in the following way \cite{AS64},
$$F_\gamma \;=\;\tilde{g}\;\cos \theta_\gamma\;+\;\tilde{f}\;\sin \theta_\gamma,\qquad G_\gamma \;=\; \tilde{f}\;\cos \theta_\gamma \;-\;\tilde{g}\;\sin \theta_\gamma,$$
\begin{equation}\label{A.14}
\theta_\gamma(r)\;=\;pr +\eta \ln(2pr)-\gamma \pi/2\,+ \mbox{ arg } \Gamma(\gamma +1 -i\eta).
\end{equation}

Then the radial Dirac function (\ref{A.3}) is given by
$${g_\kappa(r) \choose f_\kappa(r)}\;=\;\frac{M_\kappa}{pr}\;\left\{ i\,e^{i\delta_\kappa}\frac12 \left[ {-c_1 \choose d_1}\;C_\gamma(r)\;+\;{c_2 \choose d_2}\;C_{\gamma-1}(r)\right] \right.$$
\begin{equation}\label{A.15}
+\left. \;i\,e^{-i\delta_\kappa}\frac12\left[ {c_1 \choose -d_1}\;C_\gamma^\ast(r) \;+\;{-c_2 \choose -d_2}\;C^\ast_{\gamma-1}(r)\right] \right\},
\end{equation}
where $C_\gamma$ and the coefficients $c_1,\;c_2$ and $d_1,\;d_2$ are defined by
$$C_\gamma(r)\;=\;e^{i \theta_\gamma(r)}\;(\tilde{f}(pr)\;+\;i\,\tilde{g}(pr)),$$
\begin{equation}\label{A.16}
c_1\;=\;(\kappa +\gamma)\;\sqrt{\gamma^2+\eta^2}\;pc,\qquad c_2\;=\;\frac{Z_s}{c}\;(\gamma c^2-\kappa E),
\end{equation}
$$d_1\;=\;-\;\frac{Z_s}{c}\;\sqrt{\gamma^2+\eta^2}\;pc,\qquad d_2\;=\;(\kappa +\gamma)\;(\gamma c^2-\kappa E).$$
Since the functions $\tilde{f}$ and $\tilde{g}$  are only needed for large arguments $pr$, one can use the series expansions
$$\tilde{f}(pr) \;+\;i\,\tilde{g}(pr)\;=\;1\;+\;\frac{(-i\eta -\gamma)\,(-i\eta +\gamma+1)}{2ipr}$$
\begin{equation}\label{A.17}
+\;\frac{(-i\eta -\gamma)\,(-i\eta +\gamma +1)\,(-i\eta -\gamma +1)\,(-i\eta +\gamma +2)}{2!\;(2ipr)^2}\;+\;\cdots,
\end{equation}
which converges rapidly for large r and not too large $\gamma$, respectively, $\kappa$.
The representation (\ref{A.15}) is readily continued into the complex plane.
Since $e^{i\theta_\gamma(r)} \sim e^{ipr}$ and $e^{-i \theta_\gamma(r)} \sim e^{-ipr},$
the functions defined in the first and second line of (\ref{A.15}) are, respectively, identified with
${g_\kappa^+ \choose f_\kappa^+}$ and ${g_\kappa^- \choose f_\kappa^-}$ from (\ref{A.4}).

Alternatively, the Coulomb-Dirac waves can be represented in terms of Whittaker functions of the second kind, $W_{\alpha,\gamma}(r)$  \cite{YSha99,YS10}. 
These functions can easily be continued into the complex plane, and they behave asymptotically like $W_{\alpha,\gamma}(z) \sim e^{-z/2}$ for $z \in {\Bbb C}$. They can either be expressed in terms of series expansions  \cite{YSha99} or by means of an integral representation \cite{AS64}. These functions are of advantage for large quantum numbers, $|\kappa| \gtrsim 100$, where the irregular solutions are not needed.

For potentials $V(r)$ which decay faster than $1/r$ for $r \to \infty$, representing e.g. a neutral atom,
there exists some distance $R_m$ such that $rV(r)\approx 0$ for $r>R_m$. Thus
 the  functions $g_\kappa$ and $f_\kappa$ can at large distances be represented in terms of a superposition of the solutions to the free radial Dirac equation, the spherical Bessel ($j_l$) and Neumann ($n_l$) functions. Details are provided in  \cite{YS10}.

\newpage

\section*{Appendix C: $\;\;$ Polarization correlations in coplanar geometry}
\setcounter{equation}{0}
\setcounter{section}{3}
\setcounter{subsection}{0}

\vspace{0.2cm}

We provide all nonvanishing polarization correlations $\tilde{C}_{jkl}$ (defined in such a way that they all add positively in the sum (\ref{3.2.17})
for the triply differential cross section).

With the unpolarized cross section of (\ref{3.2.11})
abbreviated by
$$\left( \frac{d^3\sigma}{d\omega d\Omega_k d\Omega_f}\right)_0\;=\;\frac{4\pi^2\omega \,p_f E_iE_f}{c^5\,p_i}\;D_0,$$
\begin{equation}\label{B.1}
D_0\;=\;|J_+|^2\;+\;|J_-|^2\;+\;|S_+|^2\;+\;|S_-|^2,
\end{equation}
the $\tilde{C}_{jkl}$ are expressed in terms of the matrix elements $J_\pm,\;S_\pm$ as defined in (\ref{3.2.5}).
We also define $\tilde{C}_{000}=1.$
From (\ref{3.2.2}) with (\ref{3.2.15}) and (\ref{3.2.16}) in comparison with (\ref{3.2.17}) we derive,
starting with $\tilde{C}_{0kl}$ for unobserved $\zeta_i$,
$$ \tilde{C}_{000}\;=\;(\;|J_+|^2\;+\;|J_-|^2\;+\;|S_+|^2\;+\;|S_-|^2)/D_0\;=\;1\;=\;\tilde{C}_{232},$$
$$\tilde{C}_{023}\;=\;(\;|J_+|^2\;-\;|J_-|^2\;+\;|S_+|^2\;-\;|S_-|^2)/D_0\;=\;-\tilde{C}_{211},$$
$$\tilde{C}_{021}\;=\;2\mbox{ Re}(-J_+S_-^\ast \;+\;J_-S_+^\ast)/D_0\;=\;\tilde{C}_{213},$$
$$\tilde{C}_{002}\;=\;2\mbox{ Im}(J_+S_-^\ast\;+\;J_-S_+^\ast)/D_0\;=\;\tilde{C}_{230},$$
\begin{equation}\label{B.2}
\tilde{C}_{030}\;=\;2\mbox{ Re}(J_+J_-^\ast \;+\;S_+S_-^\ast)/D_0\;=\;\tilde{C}_{202},
\end{equation}
$$\tilde{C}_{013}\;=\;2\mbox{ Im}(J_+J_-^\ast\;+\;S_+S_-^\ast)/D_0\;=\;\tilde{C}_{221},$$
$$\tilde{C}_{011}\;=\;2\mbox{ Im}(J_+^\ast S_+\;+\;J_-S_-^\ast)/D_0\;=\;-\tilde{C}_{223},$$
$$\tilde{C}_{032}\;=\;2\mbox{ Im}(-J_+^\ast S_+\;+\;J_-S_-^\ast)/D_0\;=\;\tilde{C}_{200},$$
and further continuing with $\tilde{C}_{j0l}$ for unobserved $\bfxi$,
$$\tilde{C}_{301}\;=\;-2\mbox{ Re}(J_+S_-^\ast\;+\;J_-S_+^\ast)/D_0\;=\;-\tilde{C}_{133},$$
$$\tilde{C}_{101}\;=\;2\mbox{ Re}(J_+J_-^\ast\;-\;S_+S_-^\ast)/D_0\;=\;\tilde{C}_{333},$$
\begin{equation}\label{B.3}
\tilde{C}_{103}\;=\;2\mbox{ Re}(J_+S_+^\ast\;+\;J_-S_-^\ast)/D_0\;=\;-\tilde{C}_{331},
\end{equation}
$$\tilde{C}_{303}\;=\;(\;|J_+|^2\;+\;|J_-|^2\;-\;|S_+|^2\;-\;|S_-|^2)/D_0\;=\;\tilde{C}_{131}.$$
Finally, we list the remaining $\tilde{C}_{jk0}$ for unobserved $\bfzeta_f$,
$$\tilde{C}_{120}\;=\;2\mbox{ Re}(J_+S_+^\ast\;-\;J_-S_-^\ast)/D_0\;=\;-\tilde{C}_{312},$$
$$\tilde{C}_{310}\;=\;-2\mbox{ Im}(-J_+J_-^\ast \;+\;S_+S_-^\ast)/D_0\;=\;-\tilde{C}_{122},$$
\begin{equation}
\label{B.4}
\tilde{C}_{110}\;=\;2\mbox{ Im}(J_+S_-^\ast\;-\;J_-S_+^\ast)/D_0\;=\;\tilde{C}_{322},
\end{equation}
$$\tilde{C}_{320}\;=\;(\;|J_+|^2\;-\;|J_-|^2\;-\;|S_+|^2\;+\;|S_-|^2)/D_0\;=\;\tilde{C}_{112}.$$
As a total, there are $\frac12 \cdot 64 -1=31$ nonvanishing polarization correlations.

\vspace{1cm}


\begin{thebibliography}{99}

\bibitem{AS64} M.Abramowitz and I.A.Stegun, {\it Handbook of Mathematical Functions} (Dover Publications, New York, 1964).
\bibitem{AB62} A.I.Achieser and W.B.Berestezki, {\it Quanten-Elektrodynamik} (Harri Deutsch, Frankfurt/M, 1962), \S 29.
\bibitem{AMS77} A.Aehlig, L.Metzger and M.Scheer, Z. Phys. A {\bf 281}, 205 (1977).
\bibitem{AS72} A.Aehlig and M.Scheer, Z. Phys. {\bf 250}, 235 (1972). 

\bibitem{Ai66} H.Aiginger, Z. Phys. {\bf 197}, 8 (1966).

\bibitem{AZ66} H.Aiginger and H.Zinke, Acta Phys. Austriaca {\bf 23}, 76 (1966).

\bibitem{Al08} G.Alexander et al, Phys. Rev. Lett. {\bf 100}, 210801 (2008).

\bibitem{BGK} V.V.Balashov, A.N.Grum-Grzhimailo and N.M.Kabachnik, {\it Polarization and Correlation Phenomena in Atomic Collisions} (Kluwer Academic, New York, 2000).

\bibitem{Ba55} W.C.Barber, A.I.Berman, K.L.Brown and W.D.George, Phys. Rev. {\bf 99}, 59 (1955).

\bibitem{Ba11} R.Barday, K.Aulenbacher, P.Bangert, J.Enders, A.G\"{o}\"{o}k, D.H.\ja, F.Nillius, A.Surzhykov and V.A.Yerokhin, J. Phys. Conf. Series {\bf 298}, 133 (2011).

\bibitem{BL58} R.A.Berg and C.N.Lindner, Phys. Rev. {\bf 112}, 2072 (1958).

\bibitem{BH34} H.A.Bethe and W.Heitler, Proc. Roy. Soc. A {\bf 146}, 83 (1934).

\bibitem{BM54} H.A.Bethe and L.C.Maximon, Phys. Rev. {\bf 93}, 768 (1954).

\bibitem{Lan4} V.B.Berestetskii, E.M.Lifshitz and L.P.Pitaevskii,  {\it Quantum Electrodynamics}  Course of Theoretical Physics Vol.4, $2^{nd}$ edition (Elsevier, Oxford, 1982).

\bibitem{BD} J.D.Bjorken and S.D.Drell, {\it Relativistic Quantum Mechanics} (McGraw-Hill, New York, 1964).

\bibitem{BN84} W.Bleier and W.Nakel, Phys. Rev. A {\bf 30}, 607 (1984).

\bibitem{Bo72} E.Borie, Il Nuovo Cim. A {\bf 11}, 969 (1972).

\bibitem{Br91} V.Breton et al, Phys. Rev. Lett. {\bf 66}, 572 (1991).

\bibitem{Br56} K.L.Brown, Phys. Rev. {\bf 103}, 243 (1956).

\bibitem{BZP69} H.Brysk, C.D.Zerby and S.K.Penny, Phys. Rev. {\bf 180}, 104 (1969).

\bibitem{Da68} W.E.Dance, D.H.Rester, B.J.Farmer, J.H.Johnson and L.L.Baggerly, J. Appl. Phys. {\bf 39}, 2881 (1968).

\bibitem{Da28} C.G.Darwin, Proc. Roy. Soc. London A {\bf 118}, 654 (1928).


\bibitem{PM10} A.Di Piazza and A.I.Milstein, Phys. Rev. A {\bf 82}, 042106 (2010).

\bibitem{PM12} A.Di Piazza and A.I.Milstein, Phys. Rev. A {\bf 85}, 042107 (2012).
 
\bibitem{DS84} T.W.Donnelly and I.Sick, Rev. Mod. Phys. {\bf 56}, 461 (1984).

\bibitem{Dr52} S.D.Drell, Phys. Rev. {\bf 87}, 753 (1952).

\bibitem{Ed60} A.R.Edmonds, {\it Angular Momentum in Quantum Mechanics} $2^{nd}$  edition (Princeton University Press, Princeton, 1960).


\bibitem{EH69} G.Elwert and E.Haug, Phys. Rev. {\bf 183}, 90 (1969).

\bibitem{FPT} I.J.Feng, R.H.Pratt and H.K.Tseng, Phys. Rev. A {\bf 24}, 1358 (1981).

\bibitem{Fu34} W.H.Furry, Phys. Rev. {\bf 46}, 391 (1934).

\bibitem{Ga17} J.A.Garc\'{i}a-Alvarez et al, J. Phys. B {\bf 50}, 155003 (2017).

\bibitem{Ga18} J.A.Garc\'{i}a-Alvarez, J.M.Fern\'{a}ndez-Varea, V.R.Vanin and N.L.Maidana, J. Phys. B {\bf 51}, 225003 (2018).

\bibitem{GN96} E.Geisenhofer and W.Nakel, Z. Phys. D {\bf 37}, 123 (1996).

\bibitem{GP64} E.S.Ginsberg and R.H.Pratt, Phys. Rev. {\bf 134}, B773 (1964).

\bibitem{Go57} M.Goldhaber, L.Grodzins and A.W.Sunyar, Phys. Rev. {\bf 106}, 826 (1957).

\bibitem{Go28} W.Gordon, Z. Phys. {\bf 48}, 180 (1928).

\bibitem{Grad} I.S.Gradshteyn and I.M.Ryzhik, {\it Table of Integrals, Series and Products} (Academic Press, New York, 1965).

\bibitem{Ha63} R.Hagedorn, {\it Relativistic Kinematics} (Benjamin, New York, 1963), \S5.

\bibitem{H18b} A.K.F.Haque, Private Communication (2018).

\bibitem{H18} A.K.F.Haque, M.A.Uddin, D.H.\ja $\;$ and B.C.Saha, J. Phys. B {\bf 51}, 175202 (2018).

\bibitem{Ha69} E.Haug, Phys. Rev. {\bf 188}, 63 (1969).

\bibitem{Ha96} E.H.Haug, Z. Phys. D {\bf 7}, 9 (1996).

\bibitem{H10} E.Haug, Eur. Phys. J. D {\bf 58}, 297 (2010).

\bibitem{HN04} E.Haug and W.Nakel, {\it The Elementary Process of Bremsstrahlung} (World Scientific, Singapore, 2004).

\bibitem{H54} W.Heitler, {\it The Quantum Theory of Radiation}, $3^{\rm rd}$ edition (Oxford University Press, Oxford, 1954).

\bibitem{Ho58} R.Hofstadter, F.Bumiller and M.R.Yearin, Rev. Mod. Phys. {\bf 30}, 482 (1958).

\bibitem{H16} M.I.Hossain, A.K.F.Haque, M.A.R.Patoary, M.A.Uddin and A.K.Basak, Eur. Phys. J. D {\bf 70} (2), 41 (2016).

\bibitem{HR66} D.F.Hubbard and M.E.Rose, Nucl. Phys. {\bf 84}, 337 (1966).


\bibitem{JP63} R.J.Jabbur and R.H.Pratt, Phys. Rev. {\bf 129}, 184 (1963).

\bibitem{JP64} R.J.Jabbur and R.H.Pratt, Phys. Rev. {\bf 133}, B1090 (1964).


\bibitem{Jaku10} D.H.\ja, Phys. Rev. A {\bf 82}, 042714 (2010).

\bibitem{Jaku11b} D.H.\ja, Phys. Lett. A {\bf 375}, 1671 (2011).


\bibitem{Jaku12} D.H.\ja, Phys. Rev. A {\bf 85}, 042714 (2012).

\bibitem{Jaku13} D.H.\ja, Phys. Lett. A {\bf 377}, 1885 (2013).

\bibitem{Jaku16} D.H.\ja, Phys. Rev. A {\bf93}, 052716 (2016).

\bibitem{Jaku17} D.H.\ja, arXiv:1610.09131 [physics.atom-ph] (2016); Eur. Phys. J. D {\bf 71}: 209 (2017).

\bibitem{Jaku18} D.H.\ja, J. Phys. B {\bf 51}, 055001 (2018).

\bibitem{Jaku20} D.H.\ja, J. Phys. G {\bf 47}, 075102 (2020)

\bibitem{JM19} D.H.\ja $\,$ and A.Mangiarotti, Phys. Rev. A 100, 032703 (2019).

\bibitem{JS11} D.H.\ja $\,$ and  A.Surzhykov, Eur. Phys. J. D {\bf 62}, 177 (2011).

\bibitem{JY13} D.H.\ja $\,$ and V.A.Yerokhin, Eur. Phys. J. D {\bf 67}:4 (2013).

\bibitem{Jo83} C.J.Joachain, {\it Quantum Collision Theory}, $3^{\rm rd}$ edition (North Holland, Amsterdam, 1983).

\bibitem{KD97} S.Keller and R.M.Dreizler, J. Phys. B {\bf 30}, 3257 (1997).

\bibitem{Ki86} L.Kim, R.H.Pratt, S.M.Seltzer and M.J.Berger, Phys. Rev. A {\bf 33}, 3002 (1986).

\bibitem{Ki83} L.Kissel, C.A.Quarles and R.H.Pratt, At. Data Nucl. Data Tables {\bf 28}, 381 (1983).

\bibitem{Kl77} S.Klarsfeld, Phys. Lett. {\bf 66}B, 86 (1977).

\bibitem{KN29} O.Klein and Y.Nishina, Z. Phys. {\bf 52}, 853 (1929).

\bibitem{KM59} H.W.Koch and J.W.Motz, Rev. Mod. Phys. {\bf 31}, 920 (1959).

\bibitem{KN82} M.Komma and W.Nakel, J. Phys. B {\bf 15}, 1433 (1982).

\bibitem{Ko01} A.V.Korol, O.I.Obolensky, A.V.Solov'yov and I.A.Solovjev, J.Phys. B {\bf 34}, 1589 (2001).

\bibitem{Ko15} O.Kovtun, V.Tioukine, A.Surzhykov, V.A.Yerokhin, B.Cederwall and S.Tashenov, Phys. Rev. A {\bf 92}, 062707 (2015).

\bibitem{KM15} P.A.Krachkov and A.I.Milstein, Phys. Rev. A {\bf 91}, 032106 (2015).

\bibitem{KLM16} P.A.Krachkov, R.N.Lee and A.I.Milstein, Physics Uspekhi {\bf 59}, 619 (2016).

\bibitem{KE73} R.W.Kuckuck and P.J.Ebert, Phys. Rev. A {\bf 7}, 456 (1973). 

\bibitem{Lee00} R.N.Lee, A.I.Milstein and V.M.Strakhovenko, JETP {\bf 90}, 66 (2000).

\bibitem{Lee04} R.N.Lee, A.I.Milstein, V.M.Strakhovenko and O.Ya.Schwarz, JETP {\bf 100}, 1 (2005).

\bibitem{Lu97} M.W.Lucas, D.H.\ja, M.Kuzel and K.O.Groeneveld, Int. J. Mod. Phys. A {\bf 12}, 305 (1997).

\bibitem{M20} A.Mangiarotti, Private Communication

\bibitem{MJ17} A.Mangiarotti and D.H.\ja, Phys. Rev. A {\bf 96}, 042701 (2017).

\bibitem{MM17} A.Mangiarotti and M.N.Martins, Radiat. Phys. Chem. {\bf 141}, 312 (2017).

\bibitem{Ma19} A.Mangiarotti, W.Lauth, D.H.\ja, P.Klag, A.A.Malafronte, M.N.Martins, C.F.Nielsen and U.I.Uggerh\o j, Phys. Lett. B {\bf 815}, 136113 (2021).

\bibitem{Ma12} R.M\"{a}rtin et al, Phys. Rev. Lett. {\bf 108}, 264801 (2012).

\bibitem{MN90} E.Mergl and W.Nakel, Z. Phys. D {\bf 17}, 271 (1990).

\bibitem{Me92} E.Mergl, H.Th.Prinz, C.D.Schr\"{o}ter and W.Nakel, Phys. Rev. Lett. {\bf 69}, 901 (1992).

\bibitem{Mo55} J.W.Motz, Phys. Rev. {\bf 100}, 1560 (1955).

\bibitem{Mo} J.W.Motz, H.Olsen and H.W.Koch, Rev. Mod. Phys. {\bf 36}, 881 (1964).

\bibitem{MP60} J.W.Motz and R.C.Placious, Nuovo Cimento {\bf 15}, 571 (1960).


\bibitem{MYS} R.A.M\"{u}ller, V.A.Yerokhin and A.Surzhykov, Phys. Rev. A {\bf 90}, 032707 (2014).

\bibitem{Na66} W.Nakel, Phys. Lett. {\bf 22}, 614 (1966).

\bibitem{NA11} F.Nillius and K.Aulenbacher, J. Phys. Conf. Series {\bf 298}, 012024 (2011).

\bibitem{NA16} F.Nillius and K.Aulenbacher, Proceedings of Science: PSTP2015, 031 (2016).

\bibitem{Ni20} F.Nillius, PhD Thesis, Mainz (2021).

\bibitem{No54} A.Nordsieck, Phys. Rev. {\bf 93}, 785 (1954).

\bibitem{OM59} H.Olsen and L.C.Maximon, Phys. Rev. {\bf 114}, 887 (1959).

\bibitem{OMW57} H.Olsen, L.C.Maximon and H.Wergeland, Phys. Rev. {\bf 106}, 27 (1957).

\bibitem{Po19} A.Po\v{s}kus, At. Data Nucl. Data Tables {\bf 129-130}, 101277 (2019).

\bibitem{P60} R.H.Pratt, Phys. Rev. {\bf 120}, 1717 (1960).

\bibitem{PMS} R.H.Pratt, R.A.M\"{u}ller and A.Surzhykov, Phys. Rev. A {\bf 93}, 053421 (2016).

\bibitem{PT75} R.H.Pratt and H.K.Tseng, Phys. Rev. A {\bf 11}, 1797 (1975).

\bibitem{P77} R.H.Pratt, H.K.Tseng, C.M.Lee, L.Kissel,
C.MacCallum and M.Riley, At. Data Nucl. Data Tables {\bf 20}, 175 (1977).

\bibitem{Re68} D.H.Rester, Nucl. Phys. A {\bf 118}, 129 (1968).

\bibitem{RD67} D.H.Rester and W.E.Dance, Phys. Rev. {\bf 161}, 85 (1967).
 \bibitem{RDP72} G.Roche, C.Ducos and J.Proriol, Phys. Rev. A {\bf 5}, 2403 (1972).

\bibitem{Ros61} M.E.Rose, {\it Relativistic Electron Theory} (Wiley, New York, 1961) \S26.

\bibitem{RJ64} J.D.Rozics and W.R.Johnson,  Phys. Rev {\bf 135}, B56 (1964).

\bibitem{SE77} M.Salomaa and H.A.Enge, Nucl. Instr. Meth. {\bf 145}, 279 (1977).

\bibitem{Sal95} F.Salvat, J.M.Fern\'{a}ndez-Varea and W.Williamson Jr., Comput. Phys. Commun. {\bf 90}, 151 (1995).

\bibitem{Sal87} F.Salvat, J.D.Mart\'{i}nez, R.Mayol and J.Parellada, Phys. Rev. A {\bf 36}, 467 (1987).

\bibitem{Sch82} H.R.Schaefer, W.von Drachenfels and W.Paul, Z. Phys. A {\bf 305}, 213 (1982).

\bibitem{SG58} H.Schopper and S.Galster, Nucl. Phys. {\bf 6}, 125 (1958).

\bibitem{SL68} H.D.Schulz and G.Lutz, Phys. Rev. {\bf 167}, 1280 (1968)

\bibitem{SB85} S.M.Seltzer and M.J.Berger, Nucl. Instr. Meth. B {\bf 12}, 95 (1985).

\bibitem{Sh96} C.D.Shaffer, X.M.Tong and R.H.Pratt, Phys. Rev. A {\bf 53}, 4158 (1996).

\bibitem{Si69} R.H.Siemann, W.W.Ash, K.Berkelman, D.L.Hartill, C.A.Lichtenstein and R.M.Littauer, Phys. Rev. Lett. {\bf 22}, 421 (1969).

\bibitem{SM35} A.Sommerfeld and A.W.Maue, Ann. Phys. {\bf 22}, 629 (1935).

\bibitem{SK56} N.Starfelt and H.W.Koch, Phys. Rev. {\bf 102}, 1598 (1956).

\bibitem{Ta11} S.Tashenov, T.B\"{a}ck, R.Barday, B.Cederwall, J.Enders, A.Khaplanov, Yu. Poltoratska, K.-U.Sch\"{a}ssburger and A.Surzhykov, Phys. Rev. Lett. {\bf 107}, 173201 (2011).

\bibitem{Ta13} S.Tashenov et al, Phys. Rev. A {\bf 87}, 022707 (2013).

\bibitem{T97} H.K.Tseng, Phys. Rev. A {\bf 56}, 2868 (1997).

\bibitem{T02} H.K.Tseng, J. Phys. B {\bf 35}, 1129 (2002).
 
\bibitem{T02b} H.K.Tseng, Chin. J. Phys. {\bf 40}, 168 (2002).

\bibitem{TP71} H.K.Tseng  and R.H.Pratt,  Phys. Rev. A {\bf 3}, 100 (1971).

\bibitem{TP73} H.K.Tseng and R.H.Pratt, Phys. Rev. A {\bf 7}, 1502 (1973).


\bibitem{Ue35} E.A.Uehling, Phys. Rev. {\bf 48}, 55 (1935).

\bibitem{YSha99} V.A.Yerokhin and V.M.Shabaev, Phys. Rev. A {\bf 60}, 800 (1999).

\bibitem{YS10} V.A.Yerokhin and A.Surzhykov, Phys. Rev. A {\bf 82}, 062702 (2010).

\bibitem{Ye12} V.A.Yerokhin, A.Surzhykov, R.M\"{a}rtin, S.Tashenov and G.Weber, Phys. Rev. A {\bf 86}, 032708 (2012).


\end{thebibliography}
\end{document}